\documentclass[12pt,twoside,openany]{book}

\usepackage{lipsum}

\usepackage[top = 3cm, bottom = 3cm, left = 2.5cm, right = 2.5cm]{geometry}

\usepackage{amsmath,amssymb}
\usepackage[pdftex]{color,graphicx}
\usepackage{calligra} 
\usepackage{color}
\definecolor{indigo}{RGB}{0,0,120}
\usepackage[colorlinks=true, linkcolor=indigo, citecolor=blue, urlcolor=indigo]{hyperref}
\usepackage{rotating}

\usepackage{fancyhdr}

\usepackage{accents}
\newcommand\thickbar[1]{\accentset{\rule{.4em}{1.1pt}}{#1}}

\newcommand\abstractname{Abstract}

\makeatletter
\if@titlepage
  \newenvironment{abstract}{%
      \titlepage
      \null\vfil
      \@beginparpenalty\@lowpenalty
      \begin{center}%
        \bfseries \abstractname
        \@endparpenalty\@M
      \end{center}}%
     {\par\vfil\null\endtitlepage}
\else
  
\fi
\makeatother

\def\fl{\noindent}
\newcommand{\bra}{\langle}
\newcommand{\ket}{\rangle}

\newcommand{\tl}[1]{\tilde{#1}}
\newcommand{\pdr}{\partial}
\newcommand{\grad}{{\bf \nabla}}
\newcommand{\beq}{\begin{equation}}
\newcommand{\eeq}{\end{equation}}
\newcommand{\beqs}{\begin{eqnarray}}
\newcommand{\eeqs}{\end{eqnarray}}
\newcommand{\half}{\frac{1}{2}}
\newcommand{\ov}[1]{\frac{1}{#1}}

\newcommand{\DD}[2]{\frac {d #1}{d #2}}
\newcommand{\dd}[2]{\frac {\partial #1}{\partial #2}}
\newcommand{\imply}{\Rightarrow}
\newcommand{\pt}{\circ}
\newcommand{\T}{{\cal T}}
\newcommand{\calF}{{\cal F}}

\newcommand{\sech}{\,\text{sech}}
\newcommand{\cn}{\,\text{cn}}

\def\al{\alpha}		\def\g{\gamma} 		 		\def\del{\delta}	\def\D{\Delta}		\def\eps{\epsilon} 	\def\la{\lambda}		\def\sig{\sigma}	\def\tht{\theta}	\def\om{\omega}		
\def\ka{\kappa}
\def\tr{\;{\rm tr}\;}

\newcount\colveccount  
\newcommand*\colvec[1]{\global\colveccount#1  \begin{pmatrix} \colvecnext} \def\colvecnext#1{#1 \global\advance\colveccount-1
        \ifnum\colveccount>0 \\ \expandafter\colvecnext
        \else \end{pmatrix} \fi}
        
\DeclareMathAlphabet{\mathcalligra}{T1}{calligra}{m}{n}
\DeclareFontShape{T1}{calligra}{m}{n}{<->s*[2.2]callig15}{}

\newcommand{\bfv}{{\bf v}}

\newcommand{\bfx}{{\bf x}}
\newcommand{\bfr}{{\bf r}}

\newcommand{\bfc}{{\bf c}}

\newcommand{\bfF}{{\bf F}}

\newcommand{\bfw}{{\bf w}}

\newcommand{\bfPhi}{{\bf \Phi}}
\newcommand{\bfU}{{\bf U}}
\newcommand{\bfn}{{\bf n}}
\newcommand{\bfhatn}{{\bf \hat n}}
\newcommand{\bfP}{{\bf P}}
\newcommand{\bfG}{{\bf G}}
\newcommand{\bfa}{{\bf a}}

\newcommand{\bfy}{{\bf y}}

\newcommand{\Lagr}{{\mathcal L}}
\newcommand{\calG}{{\mathcal G}}

\newcommand{\lambdabar}{{\mkern0.75mu\mathchar '26\mkern -9.75mu\lambda}}

\usepackage{subcaption}
\newcommand\thesisname{Field theoretic viewpoints on certain fluid mechanical phenomena}
\newcommand\myname{Sachin~Shyam~Phatak}
\newcommand\guidename{Dr.~Govind~S.~Krishnaswami }
\newcommand\datename{Submitted: May 2020\\
Defended: October 6, 2020
}

\begin{document}

\thispagestyle{empty}

\frontmatter

\begin{center}

\vspace*{5mm}

{\Large \bf Field theoretic viewpoints on certain fluid mechanical phenomena}
\vspace{1cm}
\\{\Large by}\\ 
\vspace{1cm}
{\Large \textbf{Sachin Shyam Phatak}} 

\vspace{2cm}
\large 

{\it A thesis submitted in partial fulfilment of the requirements for \\ the degree of Doctor of Philosophy in Physics} \\
\vspace*{0.8cm}
to  
\vspace*{0.8cm} 

Chennai Mathematical Institute

\begin{figure}[h]
\begin{center}
 \includegraphics[width = 5cm]{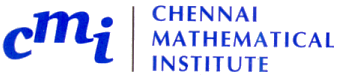}
 \end{center}
\end{figure}

\vspace{0.1cm}

Plot H1, SIPCOT IT Park, Siruseri,\\
Kelambakkam, Tamil Nadu 603103, \\
India

\vspace{2cm}

\datename

\end{center}
\normalsize


\newpage
\thispagestyle{empty}
\large
\begin{flushleft}
Advisor:

Dr. Govind S Krishnaswami, \emph{Chennai Mathematical Institute.}\\ 
\vspace*{1cm}
Doctoral Committee Members: 
\begin{enumerate}
\item Dr. Alok Laddha, \emph{Chennai Mathematical Institute.}\\
\item Dr. R Shankar, \emph{Institute of Mathematical Sciences, Chennai.}
\end{enumerate}
\end{flushleft}

\newpage

\chapter*{\center Declaration}
\normalsize

This thesis titled \emph{\thesisname} is a presentation of my original research work, carried out with my collaborators and under the guidance of \guidename  at Chennai Mathematical Institute. This work has not formed the basis for the award of any degree, diploma, associateship, fellowship or other titles in Chennai Mathematical Institute or any other university or institution of higher education.

\vspace*{5mm}
\begin{flushright}
\myname\\
May 2020
\end{flushright}

\vspace*{30mm}

In my capacity as the supervisor of the candidate's thesis, I certify that the above statements are true to the best of my knowledge.
\vspace*{5mm}
\begin{flushright}
\guidename\\
May 2020
\end{flushright}



\chapter*{Acknowledgments}
\addcontentsline{toc}{chapter}{Acknowledgments}

To me, this thesis is the culmination of a long journey through various ups and downs. I had a major life change during the course of my PhD, in which I decided to move out of Chennai and take a break in the interest of my mental health among other reasons. My PhD supervisor Prof. Govind S Krishnaswami was extremely supportive throughout. During this break, I set up a software consultancy company and was on the verge of dropping out of my PhD, but Prof. Govind convinced me otherwise, helped me get back on track and stay on track despite multiple constraints that my new job imposed on my time. He even traveled many times to my office and home in Bangalore to work with me. These meetings would last through the day, for at least a couple of days, and were always very productive. They were also a lot of fun - we had good food, played a board game or two, joked around with my friends, and got to know each other better. Interestingly, I was in the first undergraduate course he taught at CMI as well as his first graduate course, and was the first PhD student he decided to advise. I sincerely thank Prof. Govind for his friendship, support, guidance and kindness.

I'd like to thank Sonakshi Sachdev and Prof. A Thyagaraja for their collaboration. I thank Professors Alok Laddha and R Shankar who are on my doctoral committee for their encouragement and helpful suggestions. Thanks are also due to Prof. J K Bhattacharjee and Prof. V P Nair for their detailed comments and insightful questions on the thesis. I'd also like to thank the late Prof. Sriram Nambiar for his friendship and encouragement, and for following up with me when I moved out of Chennai. Thanks are also due to Professors R Jagannathan, G Rajasekaran, H S Mani, K G Arun, T R Govindarajan, K Narayan, V V Sreedhar, R Parthasarathy, the rest of the physics department and the academic staff, as well as the mess, security, housekeeping and administrative staff at CMI. I'd like to specially thank Rajeshwari ma'am and Sripathi sir for being very helpful throughout.

I would like to acknowledge financial support from CMI, the Science and Engineering Research Board, Govt. of India, the Infosys Foundation and the JN Tata Trust.

CMI gave me some amazing friends, and I'd like to thank them for their support and advice, interesting discussions, fun, and for allowing me to annoy them at times - Aditya Kela, Anil Nair, Avadhut Purohit, Bishal Deb, Gopakumar Mohandas, Himalaya Senapati, Navaneeth Mohan, Prabhat Kumar Jha, Prakash Saivasan, Puhup Manas, Sarjick Bakshi, Shibasis Roy, Shyamlal Karra, Soumendra Ganguly, Subramani Muthukrishnan, Sushrut Karmalkar and Varun Ramanathan.

Special thanks are due to Sucharitha Vasu and Suhas H V, who've been my friends through thick and thin, and helped me by being there, listening and caring.

To keep my spirits up during my PhD, I volunteered at a classroom in St. Francis Xavier Matriculation School, Velachery, that was managed by Teach for India. I conducted hands-on science workshops for kids once a week and had an absolute ball teaching. I'd like to thank Teach for India for the amazing volunteering opportunity and the wonderful kids for accepting me as their teacher.

I thank my parents Shyam and Deepa Phatak, and my brother Sundeep Phatak for their support throughout.

I'd like to thank my team at Beneathatree - Srinidhi Prahlad, Shashank K S, Rohit Shetty, Girish Pallagatti, Anil Nair and Avadhut Purohit, without whom I'd never have been able to sail through the ups and downs of this long journey. They say friends are the family we choose. I've been lucky to have Srinidhi and Shashank as my family. Words can't do justice to the thanks I owe them both, but I'll try. If it weren't for Srinidhi's wisdom, his brilliant managerial skills and his capacity to remain logical in the toughest of times, I would never have been able to keep my motivation up to work on this PhD. If it weren't for Shashank's meticulous and loving care, resourcefulness, and industriousness, I'd never have been able to continue my PhD after moving to Bangalore. 

I'd like to specially thank Srinidhi's mother Shylaja B S, and grandmother Tarabai for making me part of their family, for their love, and their unwavering and unconditional support. 

Finally, I'd like to thank the non-humans who've been my companions at various points of time - Brass the dog, Pickett the praying mantis, the free pigeons Violet, Blue, Koti and Choti, and currently Breeze the cat - who helped me regain the joy of life simply by their existence.



\newpage
\normalsize
\chapter*{\center Abstract}
\addcontentsline{toc}{chapter}{Abstract}


In this thesis we study field theoretic viewpoints on certain fluid mechanical phenomena.

In the Higgs mechanism, mediators of the weak force acquire masses by interacting with the scalar condensate, leading to a vector boson mass matrix. On the other hand, a rigid body accelerated through an inviscid, incompressible and irrotational fluid feels an opposing force linearly related to its acceleration, via an added-mass tensor. We uncover a striking physical analogy between the two effects and propose a dictionary relating them. The correspondence turns the gauge Lie algebra into the space of directions in which the body can move, encodes the pattern of gauge symmetry breaking in the shape of an associated body and relates symmetries of the body to those of the scalar vacuum manifold. The new viewpoint raises interesting questions, notably on the fluid analogs of the broken symmetry and Higgs particle, and the field-theoretic analogue of the added mass of a composite body.

Ideal gas dynamics can develop shock-like singularities which are typically regularized through viscosity. In one dimension (1d), shocks can also be conservatively smoothed via Korteweg-de Vries (KdV) dispersion. In the second part, we develop a minimal conservative regularization of 3d adiabatic flow of a gas with exponent $\gamma$, by adding a capillarity energy $\beta_* (\nabla \rho)^2/\rho$ to the Hamiltonian. This leads to a nonlinear body force with 3 derivatives of density, while preserving the conservation laws of mass and entropy. The regularized model admits dispersive sound, solitary and periodic traveling waves, but no steady continuous shock-like solutions. Nevertheless, in 1d, for $\gamma = 2$, numerical solutions in periodic domains show recurrence and avoidance of gradient catastrophes via pairs of solitary waves displaying phase-shift scattering. This is explained via an equivalence between our model (for homentropic potential flow in any dimension) and a defocussing nonlinear Schr\"odinger (NLS) equation (cubic for $\gamma = 2$), with $\beta_*$ mimicking $\hbar^2$. Thus, our model is a generalization of both the single field KdV and NLS equations to adiabatic gas dynamics in any dimension.

\chapter*{List of publications}
\addcontentsline{toc}{chapter}{List of publications}

This thesis is based on the following publications.

\begin{itemize}
	\item Krishnaswami G S and Phatak S S, {\it Higgs Mechanism and the Added Mass Effect}, \href{https://doi.org/10.1098/rspa.2014.0803}{Proc. R. Soc. A {\bf 471}, 20140803 (2015)}, \href{https://arxiv.org/abs/1407.2689}{[arXiv:1407.2689]}.
	\item Krishnaswami G S and Phatak S S, {\it The Added Mass Effect and the Higgs Mechanism: How Accelerated Bodies and Elementary Particles Can Gain Inertia}, \href{https://doi.org/10.1007/s12045-020-0936-8}{Resonance {\bf 25}(2), 191 (2020)}, \href{https://arxiv.org/abs/2005.04620}{[arXiv:2005.04620]}.
	\item Krishnaswami G S, Phatak S S, Sachdev S and Thyagaraja A, {\it Nonlinear dispersive regularization of inviscid gas dynamics}, \href{https://doi.org/10.1063/1.5133720}{AIP Advances {\bf 10}, 025303 (2020)}, \href{https://arxiv.org/abs/1910.07836}{[arXiv:1910.07836]}.
\end{itemize}


\small

\tableofcontents

\normalsize



\mainmatter

\chapter{Introduction}

The first part of this thesis develops a novel physical analogy between the added mass effect for rigid bodies accelerated through ideal nonrelativistic fluid flows and the Higgs mechanism for mass generation in the gauge field theories of particle physics. The second part of this thesis concerns a conservative regularization of shock-like singularities in inviscid gas dynamics and a remarkable connection to the nonlinear Schr\"odinger field. Thus, both parts of this thesis explore connections between fluid dynamics and other field theories. While the first is a physical correspondence with relativistic field theories, the second is a precise equivalence with a nonrelativistic field theory.

\section{Added mass effect and the Higgs mechanism}

Just as photons mediate the electromagnetic force, the $W^{\pm}$ and $Z^0$ gauge bosons mediate the weak nuclear force, responsible, among other things, for beta decay. The mass of each of these force carriers is inversely proportional to the range of the corresponding force. For instance, the Coulomb potential ($\propto 1/r$) between electric charges has an effectively infinite range corresponding to the masslessness of photons. Similarly the range of the strong nuclear force between protons and neutrons is given by the range of the Yukawa potential ($\propto e^{-r/\lambdabar}/r$). The latter is of order the reduced Compton wavelength $\left(\lambdabar = \hbar/m c \approx 10^{-15}~{\rm m}\right)$ of $\pi$-mesons $(m \approx 0.139$~GeV$/c^2)$ which mediate the strong force. 

On the other hand, the weak nuclear force is very short ranged ($\approx 10^{-17} ~{\rm m}$), which requires the $W^{\pm}$ and $Z^0$ gauge bosons to be very massive ($80$-$91$GeV$/c^2$; by contrast, the proton mass is only $0.938$GeV$/c^2$). However, including mass terms for the $W^{\pm}$ and $Z^0$ particles in the gauge theory of weak interactions \cite{weinberg, salam, Ryder}, explicitly breaks the `gauge' symmetry of the action and unfortunately destroys the predictive power (`renormalizability') of the theory. The Higgs mechanism solves this problem by a process of `spontaneous' rather than `explicit' breaking of the gauge symmetry.\footnote{While the Higgs mechanism {\it does} explain the masses of the $W$ and $Z$ bosons as well as the quarks (e.g. up and down) and leptons such as the electron and muon, it is not relevant to explaining why the proton or neutron are so much more massive than the quarks that make them up. The latter has to do with the binding energy of gluons.} It was proposed in the work of several physicists including Higgs \cite{HiggsPaper}, Brout, Englert \cite{EnglertBroutPaper}, Guralnik, Hagen and Kibble \cite{GuralnikHagenKibble} in 1964, building on earlier work of Anderson \cite{Anderson} in superconductivity\footnote{Photons in vacuum are massless and travel at the speed $c$ of light. Each component of the electric and magnetic field in an EM wave satisfies the d'Alembert wave equation $\ov{c^2}\frac{\partial^2 \chi}{\partial t^2} - \grad^2 \chi = 0$. These waves are transversely polarized, since in the absence of charges ${\grad \cdot {\bf E}} = \grad \cdot {\bf B} = 0$, so that these fields (${\bf E}(\bfr, t) = {\bf E}_{\bf k} e^{i ({\bf k}\cdot \bfr - \omega t)}$ where $\omega = c |{\bf k}|$) are orthogonal to the direction ${\bf \hat k}$ of propagation (${\bf \hat k} \cdot {\bf E}_{\bf k} = {\bf \hat k} \cdot {\bf B}_{\bf k} =0$). In superconductors and plasmas however, photons become massive - a `mass' term enters the wave equation: $\ov{c^2}\frac{\partial^2 \chi}{\partial t^2} - \grad^2 \chi + m^2 \chi= 0$. Consequently, photons travel at a speed less than $c$ and display both transverse and longitudinal polarizations. This photon mass can be explained by an `abelian' version of the Higgs mechanism. The Meissner effect is a physical manifestation of this photon mass: magnetic fields are expelled from a superconductor except over a thin surface layer whose thickness is given by the London penetration depth. Similarly, in plasmas the electric field of a test charge is screened beyond the Debye screening length \cite{raichoudhury}. Both the penetration depth and screening length are inversely proportional to the photon mass so that they diverge in vacuum.}. The $W^{\pm}$ and $Z^0$ bosons are nominally massless, but appear to be massive due to interactions with a (`Higgs') scalar field whose condensate permeates all of space like a fluid (the strength of this condensate is measured by the vacuum expectation value (vev) of the scalar field $\approx 246$GeV$/c^2$). As a consequence of this spontaneous symmetry breaking, gauge invariance is not lost and the predictive power of the theory is restored. The non-zero vacuum value of the scalar field leads, effectively, to mass terms in the Lagrangian density bilinear in the gauge fields $M_{ab} A_{\mu}^a A_\mu^b$. Here $A_\mu^a$ are the components of the gauge boson vector potential, with $a$ labelling a basis for the Lie algebra of the gauge group (such as $SU(2)$) and $\mu = 0,1,2,3$ labelling the spacetime coordinates. This leads to a gauge boson mass-squared matrix $M_{ab}$. The eigenvalues of $M_{ab}$ are the squares of the masses of the vector bosons (such as the $W^\pm, Z^0$ and the photon). As is well known, the Higgs mechanism has received experimental support through the discovery of the Higgs boson \cite{ATLAS,CMS} which is the lightest excitation of the scalar field.

It is tempting to look for analogies between the Higgs mechanism and forces felt by bodies moving through fluids, to complement standard examples of (abelian) mass generation for photons in a superconductor or plasma. Fluid analogies are often unsatisfactory, since they suggest resistive effects which are not present in the Higgs mechanism. However, McClements and Thyagaraja \cite{mc-clements-thyagaraja} recently pointed out that a dissipationless fluid analog for the Higgs mechanism could be provided by the added-mass effect \cite{Lamb}. The latter has to do with rigid bodies accelerated in an ideal fluid experiencing an opposing force proportional to their acceleration. More precisely, to impart an acceleration $\bfa = \dot \bfU(t)$ to a body of mass $m$ (executing translational motion with velocity $\bfU(t)$) immersed in an inviscid, incompressible and irrotational fluid, one must apply a force exceeding $m {\bf a}$, since energy must also be pumped into the induced fluid flow. The added force $F^{\rm add}_i =  \mu_{ij} a_j(t)$ is proportional to the acceleration, but could point in a different direction, as determined by the added-mass tensor $\mu_{ij}$. $\mu_{ij}$ depends on the fluid and shape of the body, but not on its mass distribution, unlike its inertia tensor. For example, the added-mass tensor of a sphere is $\del_{ij}$ times half the mass of displaced fluid. So an air bubble accelerated in water `weighs' about $\frac{\rho_{\rm water}}{2 \rho_{\rm air}} \approx 400$ times its actual mass. The added-mass effect is different from buoyancy: when the bubble is accelerated horizontally, it feels a  horizontal opposing {\em acceleration reaction force} $\bfG = - \bfF^{\rm add}$ aside from an {\it upward} buoyant force which is independent of its acceleration $\bfa$ and equal to the weight of fluid displaced. Thus, unlike the buoyant force which is always directed opposite to gravity, $\bfG$ depends on the acceleration of the body as well as its shape. Moreover, buoyancy is present in hydrostatics, while the added mass effect is a purely hydrodynamic phenomenon. Thus, air bubbles would rise 400 times faster in water if buoyancy were the only force acting on them. For similar reasons, submarines and airships must carry more fuel than one might expect, even after accounting for viscous effects and form drag\footnote{Form drag has to do with loss of energy to infinity: waves can propagate and carry energy to infinity even in a flow without viscosity.}. In fact, one would have to apply a larger force while playing volleyball under water than in air or in outer space. As a consequence of the added mass effect, bodies gain inertia when they are accelerated through a fluid. On the other hand, as emphasised by d'Alembert (see \cite{Batchelor}, sections 5.11 and 6.4), no force is required to move a body at {\it constant} velocity through an incompressible, inviscid and irrotational fluid.

The added mass effect is also different from lift \cite{Batchelor, raichoudhury}. For instance, suppose a long circular cylinder is uniformly translated horizontally (perpendicular to its axis) at speed $U$ through fluid of density $\rho$ asymptotically at rest. Suppose it is also spinning about its axis. Then the cylinder experiences an upward lift force per unit length, of magnitude $\rho U \Gamma$ where $\Gamma$ is the circulation around the cylinder. This lift force is not proportional to the body's acceleration and can be produced even when the angular and translational velocities of the cylinder are constant. Thus, lift is not an acceleration-reaction force. In addition, the added mass force is present even in inviscid, irrotational flow without circulation with impenetrable boundary conditions, while the generation of lift requires circulation and vorticity in a viscous boundary layer with no-slip boundary conditions.

The added mass effect goes at least as far back as the work of Green \cite{Green}, Bessel, Stokes, Poisson and others. In his 1850 paper \emph{On the effect of the internal friction of fluids on the motion of pendulums} \cite{Stokes}, Stokes is mainly concerned with, ``the correction usually termed the reduction to a vacuum'' of a pendulum swinging in air. He credits Bessel with the discovery of an additional effect, over and above buoyancy, which appears to alter the inertia of the pendulum swinging in air. According to Bessel, this added mass was proportional to the mass of the fluid displaced by the body. Stokes says, ``Bessel represented the increase of inertia by that of a mass equal to $k$ times the mass of the fluid displaced, which must be supposed to be added to the inertia of the body itself.'' In this same paper, Stokes also refers to Colonel Sabine, who directly measured the effect of air by comparing the motion of a pendulum in air with that in a large vacuum chamber. The added inertia for the pendulum in air was deduced to be $0.655$ times the mass of air displaced by the pendulum. Stokes attributes the first mathematical derivation of the added mass of a sphere to Poisson, who discovered that the mass of a swinging pendulum is augmented by half the mass of displaced fluid.

The added mass effect is quite different from viscous drag. The former is a frictionless effect that gives rise to an opposing force proportional to the {\it acceleration} of the body, thus adding to the mass or inertia $m$ of the body. This additional inertia $\mu$ is called its added or virtual mass. The added mass depends on the shape of the body, the direction of acceleration relative to the body as well as on the density $\rho$ of the surrounding fluid. For example, the added mass of a sphere is one-half the mass of fluid displaced by it. Unlike the moment of inertia of a rigid body, its added mass does not depend on the distribution of mass within the body; in fact it is independent of the mass of the body, but grows with the density of the fluid. The more familiar viscous drag on a body is an opposing frictional force that depends on its \emph{speed}: for slow motion it is proportional to the speed, but at high velocities it can be proportional to the square of the speed. The added mass effect is also different from buoyancy. The latter always opposes gravity and unlike the added mass effect, is present even when the body is stationary.

In this thesis, after introducing the added mass and Higgs mechanisms, we develop a novel and precise physical analogy between the two. It is not a mathematical duality like the high temperature-low temperature Kramers-Wannier duality in the Ising model \cite{kramers-wannier} or the AdS/CFT gauge-string duality \cite{maldacena}, but provides a fascinatingly new viewpoint on fluid-mechanical and gauge-theoretic phenomena. We discover a way of associating a rigid body to a pattern of spontaneous symmetry breaking (SSB). We call this the {\it Higgs Added-Mass (HAM) correspondence}, it applies both to abelian and non-abelian gauge models. Consider a $3+1$ dimensional Yang-Mills theory with $d$-dimensional gauge group $G$, which spontaneously breaks to a subgroup $H$ when coupled to scalars $\phi$ in a specified representation of $G$, subject to a given $G$-invariant potential $V$. The correspondence relates this to a rigid body accelerated (for simplicity) through a non-relativistic, inviscid, incompressible (constant density) irrotational fluid which is asymptotically at rest in $\mathbb{R}^d$. The Lie algebra $\underline{G}$ plays the role of the space through which fluid flows (with the body centered at the origin). Thus, the dimension $d$ of the space in which the fluid flows is equal to the dimension of the Lie algebra of the gauge group $G$. In particular, the ($3+1$) space-time dimension of the gauge theory is unrelated to $d$. The fluid is the analog of the scalar field, while the rigid body plays the role of the vector bosons. The scalar field medium is characterized by a constant non-vanishing vacuum expectation value (condensate) of the scalar field which is the analog of the constant fluid density. Just as accelerating the body in different directions could result in different added masses, various generators of the Lie algebra correspond to vector bosons with possibly different masses. In particular, a circular disk moving in three dimensions (3d), when accelerated along its plane has no added mass, though it {\it does} have an added mass for motion perpendicular to its surface. This corresponds to having two massless photons and one massive vector boson in a spontaneously broken gauge theory with a three dimensional gauge group ($U(1) \times U(1) \times U(1)$ with a single complex scalar subject to the standard quartic Mexican hat potential -- see \S~\ref{s:ssb-patters-rigid-bodies}). Similarly, we find the rigid body that corresponds to the $SU(2) \times U(1)$ electroweak gauge theory. The body is a 3d manifold (a cylindrical shell with cross-section given by an ellipsoid of revolution) moving in a fluid filling $\mathbb{R}^4$. 

Thus, our correspondence allows us to visualize the pattern of gauge symmetry breaking in a simple way. Moreover, there is a broken symmetry even in the added mass effect. For example, when a sphere moves at constant velocity through a fluid, the pressure is symmetric about its equatorial plane (i.e., the pressure distribution on the front and rear hemispheres are the same). When the sphere is accelerated, this symmetry is broken. There is a greater pressure on the front hemisphere, leading to an opposing acceleration reaction force. Moreover, we propose a fluid analog for the Higgs particle. The correspondence proceeds through the respective mass matrices, and relates symmetries on either side, as exemplified by numerous examples that we present.

More precisely, the HAM correspondence, in its simplest form, is between the following two physical systems.

\begin{enumerate}
 \item A simply connected rigid body which gains an added mass when it is accelerated through a fluid filling a $d$-dimensional volume. We assume for simplicity that the flow is inviscid, irrotational and incompressible. The fluid is assumed to extend to infinity in all directions and to be asymptotically at rest.

 \item A classical $3+1$ dimensional\footnote{A similar correspondence could be developed for $2+1$ dimensional gauge theories as well} Yang-Mills gauge theory, with a gauge symmetry group $G$, coupled to scalar fields $\phi$. The scalars are assumed to transform under a specified representation of $G$ and are subject to a $G$-invariant scalar potential $V(\phi)$. Some of the gauge bosons can become massive through spontaneous symmetry breakdown of $G$ to a residual symmetry group $H$.
\end{enumerate}

We say that a gauge theory {\em corresponds} to a certain rigid body if the eigenvalues of the gauge boson mass-squared matrix $M_{ab}$ are the squares of the eigenvalues of the rigid body added mass tensor $\mu_{ij}$. Note that the added mass eigenvalues do not in general determine the rigid body. A cube and a sphere (of appropriate side and radius) can have the same added mass eigenvalues, just as a cube and a sphere are both spherical tops and may have the same inertia tensors. Similarly, the gauge boson mass-squared matrix $M_{ab}$ may not uniquely determine the gauge theory. So at this level of precision, the correspondence is between a class of gauge theories and a class of rigid bodies.

On the other hand, what do the added masses and gauge boson masses depend on? The added mass tensor of a rigid body accelerated through an ideal fluid depends on the rigid body only through the shape of its surface, not on its mass distribution. But the added mass {\em does} depend on the density of fluid, which is assumed constant. As for the masses of gauge bosons, they depend on the representation of the group $G$ under which the scalars transform, as well as on the scalar potential, and therefore on the residual symmetry group $H$ as well. For example, a gauge group $G = U(1) \times U(1) \times U(1)$ with the scalar field $\phi$ in 1d, 2d and 3d representations, leads to different spectra of vector boson masses. Each representation corresponds to a different rigid body: an elliptical disk, a hollow elliptical cylinder and an ellipsoid (see \S~\ref{s:ssb-patters-rigid-bodies}).

We emphasize that the HAM correspondence is {\em not} proposed as a `duality' like the AdS/CFT correspondence. In other words, we do not suggest that one can solve the gauge theory by doing a fluid mechanical calculation or vice versa. However, the HAM correspondence does provide a rough dictionary between various concepts, parameters and equations in the added mass and Higgs mechanisms, and provides a new viewpoint on each subject, both of which have self-contained formulations. It also raises some interesting questions both in fluid mechanics and in gauge theory which we discuss in Chapter \ref{s:discussion-ham}.

\section[Regularization of inviscid gas dynamics]{Nonlinear dispersive regularization of inviscid gas dynamics}
\label{s:gas-dyn-intro}

In the second part of this thesis, we study a conservative regularization of inviscid gas dynamics. Gas dynamics has been an active area of research with applications to high-speed flows, aerodynamics and astrophysics. The equations of ideal compressible flow are known to encounter shock-like singularities with discontinuities in density, pressure or velocity \cite{whitham}. These singularities are often resolved by the inclusion of viscosity. However, as the KdV equation $(u_t + u u_x = \eps u_{xxx})$ illustrates, such singularities in the 1d Hopf (or kinematic wave) equation $u_t + u u_x = 0$ can also be regularized conservatively via dispersion \cite{ablowitz}, as in dispersive shock wave theory (see \cite{whitham,biondini-etal,El-Hoefer,ali-kalisch-1} and references therein) with applications to undular bores in shallow water and blast waves in Bose-Einstein condensates.

In this thesis, we develop a minimal conservative regularization of ideal gas dynamics, which we refer to as R-gas dynamics. The regularization involves the introduction of a new body-force term in the gas-dynamic equations, which prevents the development of large gradients in density, pressure or velocity (ultraviolet catastrophes), while preserving the rotation, translation and Galilean symmetries of ideal gas dynamics.

Somewhat analogous conservative `rheological' regularizations of vortical singularities in ideal Eulerian hydrodynamics, magnetohydrodynamics and two-fluid plasmas have been developed in \cite{thyagaraja,govind-sonakshi-thyagaraja-pop,govind-sonakshi-thyagaraja-2-fluid, sonakshi-thesis}. The current work may be regarded as a way of extending the single-field KdV equation to include the dynamics of density, velocity and pressure and also to dimensions higher than one. There is of course a well-known generalization of KdV to 2d, the Kadomtsev-Petviashvili (KP) equation \cite{kp}. However, unlike KP, our regularized equations are rotation-invariant and valid in any dimension. 

It is well known \cite{gardner} that the dispersive regularization term in the KdV equation $u_t - 6uu_x + u_{3x} = 0$ arises from the gradient energy term in the Hamiltonian $H = \int(u^3 + (1/2)u_x^2)\, dx$, upon use of the Poisson brackets $\{u(x), u(y)\} = \pdr_x \del(x-y)$. In fact, KdV does not conserve mechanical and capillarity energies separately \cite{ali-kalisch-3, karczewska-rozmej-infeld}. By analogy with this, we obtain our regularized model by augmenting the Hamiltonian of ideal adiabatic flow of a gas with polytropic exponent $\gamma$, by a density gradient energy $\beta(\rho) (\grad\rho)^2$. Such a term arose in the work of van der Waals and Korteweg \cite{vdW,korteweg,dunn-serrin,gorban-karlin} in the context of capillarity, but can be important even away from interfaces in any region of rapid density variation, especially when dissipative effects are small, such as in weak shocks, cold atomic gases, superfluids and collisionless plasmas. It has also been used to model liquid-vapor phase transitions and in the thermomechanics of interstitial working \cite{dunn-serrin}. We argue that the simplest choice of capillarity coefficient that leads (using the standard Poisson brackets (\ref{e:PB-3d})) to local conservation laws for mass, momentum, energy and entropy (with the standard mass, momentum and entropy densities) is $\beta(\rho) = \beta_*/\rho$ where $\beta_*$ is a constant. By contrast, the apparently simpler option of taking $\beta(\rho)$ constant leads, in 1d, to a KdV-like linear dispersive term $\rho_{xxx}$ in the velocity equation, but results in a momentum equation that, unlike KdV \cite{ali-kalisch-3}, is {\it not} in conservation form for the standard momentum density $\rho u$. A consequence of the constitutive law $\beta = \beta_*/\rho$ is that the ideal momentum flux $\rho u^2 + p$ is augmented by a stress $-\beta_* (\rho_{xx} - \rho_x^2/\rho)$ corresponding to a Kortweg-type \emph{grade} 3 elastic material \cite{dunn-serrin,gorban-karlin}. This leads to new nonlinear terms in the momentum equation with third derivatives of $\rho$, somewhat reminiscent of KdV. One of the effects of these nonlinear dispersive terms is to allow for `upstream influence' \cite{benjamin} which is forbidden by the hyperbolic equations of inviscid gas dynamics under supersonic conditions. Interestingly, our regularization term is also related to the quantum mechanical Bohm potential \cite{Bohm-1} and Gross quantum pressure (p.476 of \cite{gross}) encountered in superfluids. Moreover, unlike KdV, our equations extend in a natural way to any dimension. 

The regularized equations admit cubically dispersive sound waves, solitary waves and periodic traveling waves, but no steady continuous shock-like solutions satisfying the Rankine-Hugoniot conditions. Nevertheless, in 1d, for $\gamma = 2$, numerical solutions show recurrent behaviour in periodic domains, while the spectral distribution of energy shows a rapid decay with mode number. What is more, the gradient catastrophe (for initial conditions that would otherwise lead to discontinuities) is averted through the formation of pairs of solitary waves which can display approximate phase-shift scattering. This is explained via an equivalence between our regularized equations [in the special case of constant specific entropy ($p / \rho^\g = {\rm constant}$) potential flow $\bfv = \grad \phi$] and a defocussing nonlinear Schr\"odinger equation (NLSE) with $\beta_*$ playing the role of $\hbar^2$. This equivalence, which proceeds via the Madelung transformation \cite{madelung} $\psi = \sqrt{\rho} \exp\left( i\phi/2\sqrt{\beta_*} \right)$, may be regarded as a conservative analog of the Cole-Hopf transformation for Burgers, applies in any dimension, and results in a defocusing NLSE with $|\psi|^{2(\g - 1)} \psi$ nonlinearity so that one obtains the celebrated cubic NLSE for $\gamma = 2$. In 1D, the latter is known to admit an infinite number of conservation laws and display recurrence\footnote{It is noteworthy that the {\it quantum} version of the 1d cubic NLSE (Lieb-Liniger model) has recently been given a hydrodynamical description (generalized hydrodynamics \cite{bertini, castro-doyon-yoshimura}) with infinitely many local conservation laws and has been used to model 1d gases of ultracold Rubidium atoms which retain memory of their initial state \cite{schemmer}.}. Thus, our regularized equations may be viewed as a generalization of both the single field KdV and nonlinear Schr\"odinger equations to the adiabatic dynamics of density, velocity, pressure and entropy in a gas in any dimension.

\section{Organization of this thesis}
\label{s:thesis-organization}

We begin in \S\ref{s:added-mass-effect} by introducing the added mass effect. The one- and two-dimensional cases are introduced through examples in Sections \ref{s:1d-added-mass-effect} and \ref{s:added-mass-cylinder-2d}. Sections \ref{s:added-mass-3d} and \ref{s:pressure-force-added-mass-effect} contain a general treatment of the added mass effect in three dimensions, where we derive the added mass tensor and find the added force on a body. Then, in \S\ref{s:examples-added-mass-tensors} we give several examples of added mass tensors of one, two and three dimensional rigid bodies in 2d and 3d flows. \S\ref{s:d-geq-3-added-mass-effect} uses a multipole expansion to generalize the formula for the acceleration reaction force and the added mass tensor to flows in $d \geq 3$ dimensions. In \S\ref{s:abelian-higgs} we describe the abelian version of the Higgs mechanism, followed by an example of the non-abelian case in \S\ref{s:non-abelian-higgs} where we also obtain the corresponding vector boson mass-squared matrix. In \S\ref{s:ideal-SSB-pattern} we clarify how more than one SSB pattern can lead to the same mass-squared matrix and introduce the notion of an `ideal' SSB pattern. \S\ref{s:ham-correspondence} contains the Higgs added mass (HAM) correspondence along with numerous examples. We conclude with a discussion in \S\ref{s:discussion-ham} followed by several appendices. Appendix \ref{a:symmetry-of-mu} describes the boundary conditions that ensure the symmetry of the added mass tensor for potential flow. In Appendix \ref{a:derivation-energy-added-mass} we provide an alternate approach to derive the added mass effect (in $d$ dimensions) using the energy of induced flow. In Appendix \ref{a:derivation-mom-added-mass} we derive the added mass tensor by considering a suitably regularized flow momentum and discuss some of the subtleties in the choice of the regularizing outer boundary. In Appendix \ref{a:ellipsoid-added-mass} we discuss the added mass tensor of an ellipsoid moving in 3d and the limiting cases of an elliptical and a circular disk also moving in 3d. In Appendix \ref{a:2d-added-mass-effect} we derive the added mass tensor in 2d using a multipole expansion and use techniques of conformal mapping to derive the added mass tensor for an elliptical disk moving in 2d. In Appendix \ref{a:extension-added-mass-effect-compressible} we briefly touch upon the extension of the added mass effect to compressible potential flow. The unsteady Bernoulli equation is used to obtain an expression for the acceleration-reaction force and added mass tensor which could vary with time and location of the body due to density variations. In Appendix \ref{a:ham-extension-fermions} we attempt to extend the Higgs-added mass correspondence to fermions coupled to scalars, though without gauge fields. In Appendix \ref{a:casimir-added-mass} we mention an intriguing parallel between the added mass effect and the Casimir effect for two parallel moving plates.

We begin the second part of the thesis in \S \ref{s:3d-hamil-form-R-gas-dyn} by giving the Lagrangian (in terms of Clebsch variables) and Hamiltonian formulations and equations of motion (EOM) of adiabatic R-gas dynamics in 3d. The mass, momentum, energy and entropy equations are all expressed in conservation form. In \S \ref{s:formulation-r-gas-dynm}, we specialize to 1d and discuss the special case of constant entropy (isentropic/barotropic) flow  in which case the velocity equation also acquires a conservation form. Sound waves are discussed in \S \ref{s:dispersive-sound-waves} and shown to be governed at long wavelengths by a cubic dispersion relation similar to that of the linearized KdV equation. In \S \ref{s:steady-trav-quadrature}, the local conservation laws are used to reduce the determination of steady and traveling wave solutions in 1d to a single quadrature of a generalization of the Ermakov-Pinney equation. A mechanical analogy and phase plane analysis is used to show that the only such non-constant bounded solutions are cavitons (in density) and periodic waves.  While these results hold for any value of $\g$, for $\g = 2$, closed-form $\text{sech}^2$ and cnoidal wave solutions are obtained, physically interpreted and compared with the corresponding KdV solutions. Aside from overall scales, steady solutions are parametrized by a pair of dimensionless shape parameters: a Mach number and a curvature. A parabolic embedding and a virial theorem for steady flows are given in Appendix \ref{a:parabolic-embed-LJ-id}. In \S \ref{s:weak-form}, the weak form of the R-gas dynamic equations is given, and in \S\ref{s:patched-shock-weak-sol} an attempt is made to find a steady shock-like profile by patching half a caviton with a constant solution. However, it is shown that there are no such continuous profiles that satisfy all the Rankine-Hugoniot conditions, though it may be possible to satisfy the mass flux condition alone. To study more general time-dependent solutions of R-gas dynamics and the evolution of initial conditions that could lead to shock-like discontinuities, we set up in \S \ref{s:IVP-numerical}, a semi-implicit spectral numerical scheme for the isentropic R-gas dynamic equations with periodic boundary conditions (BCs) in 1d. For $\g = 2$, our numerical solutions indicate that our regularization evades the gradient catastrophe through the formation of a pair of solitary waves at the top and bottom of a velocity profile with steep negative gradient. Though we do not observe a KdV-like solitary wave train, these solitary waves can suffer collisions and approximately re-emerge with a phase shift. We also observe a rapid decay of energy with mode number and recurrent behavior with Rayleigh quotient fluctuating between bounded limits, indicating an effectively finite number of active Fourier modes. In \S \ref{s:r-gas-to-nlse} we use a canonical transformation to reformulate 3d adiabatic R-gas dynamics in terms of a complex scalar field coupled to an entropy field and three Clebsch potentials. For isentropic potential flows, this formulation shows that R-gas dynamics for any $\g$ reduces to a defocusing 3d NLSE. In \S \ref{s:caviton-to-nlse}, the regularized Bernoulli equation is used to show that steady R-gas dynamic solutions map to solutions of NLSE with harmonic time dependence, with the $\g = 2$ caviton in 1d corresponding to the dark soliton of the cubic NLSE. In \S \ref{s:rayleigh-quotient-NLSE-conserved-quantities} we relate the conserved quantities and bounded Rayleigh quotient of NLSE to their R-gas dynamic analogues. This connection lends credence to our numerical observations, since the cubic NLSE with periodic BCs in 1d is known to possess an infinity of conserved quantities in involution \cite{faddeev-takhtajan}. We also note in \S \ref{s:neg-pressure-vortex-filament} that the {\it negative} pressure $\g = 2$ isentropic R-gas dynamic equations in 1d are equivalent to the vortex filament and Heisenberg magnetic chain equations. We conclude with a discussion in \S \ref{s:discussion-r-gas-dyn}, followed by four appendices. Appendix \ref{a:vect-fld-phase-portrait} classifies steady solutions of R-gas dynamics for various values of the conserved fluxes by a phase portrait analysis of an appropriate vector field. In Appendix \ref{a:hamil-lagr-for-steady-sol} we develop a canonical formalism for steady R-gas dynamics. In Appendix \ref{a:parabolic-embed-LJ-id} we provide a parabolic embedding which could be used to find steady solutions numerically and Lagrange-Jacobi identities which provide valuable checks on numerics used to obtain steady solutions. Finally in Appendix \ref{a:numerical-scheme} we detail the semi-implicit spectral scheme used to numerically solve the initial value problem for 1d R-gas dynamics with periodic boundary conditions.

\chapter{Added mass effect and the Higgs mechanism}
\label{c:ham}

\section{The added mass effect}
\label{s:added-mass-effect}

The added mass effect may be understood through the simplest of ideal flows. Consider purely translational motion\footnote{It could be a challenging task for an external agent to ensure that an irregularly shaped body executes purely translational motion, i.e. to ensure that there are no unbalanced torques about its centre of mass that cause the body to rotate.} of a rigid body of mass $m$ in an inviscid, incompressible and irrotational fluid at rest in 3-dimensional space. To impart an acceleration $\bf a = \dot {\bf U}$ to it, an external agent must apply a force $\bfF^{\rm ext}$ exceeding $m {\bf a}$. Newton's second law relates the components of this force to those of its acceleration:
	\beq
	 F_i^{\rm ext} = m \: a_i + F^{\rm add}_i \quad \text{where} \quad i =1,2,3.
	\eeq
Here $F^{\rm add}_i = \sum_{j=1}^3 \mu_{ij} \: a_j$. Part of the externally applied force goes into producing a fluid flow. The added force $\bfF^{\rm add}$ is proportional to acceleration, but could point in a different direction, depending on the shape of the body. The constant $3 \times 3$ symmetric matrix $\mu_{ij}$ that relates acceleration to the added force is the `added-mass tensor'. Unlike the inertia tensor of a rigid body, $\mu_{ij}$ is independent of the distribution of mass in the body, though it depends on the fluid and the shape of the body. The added mass tensor $\mu_{ij}$ for a sphere is a multiple of the identity matrix: $\mu_{ij} = \mu \del_{ij}$. In other words, the added mass $\mu$ of a sphere is the same in all directions. When a sphere is accelerated horizontally in water, it feels a horizontal opposing acceleration-reaction force ${\bf G} = - {\bf F}^{\rm add} = - \mu {\bf a}$ aside from an upward buoyant force, an opposing viscous force etc. It turns out that the added mass grows roughly with the cross sectional area presented by the accelerated body. For instance, a flat plate has no added mass when accelerated along its plane.

\subsection{One dimensional flow along a circle}
\label{s:1d-added-mass-effect}

\begin{figure}
 \centering
 \includegraphics[scale=0.33]{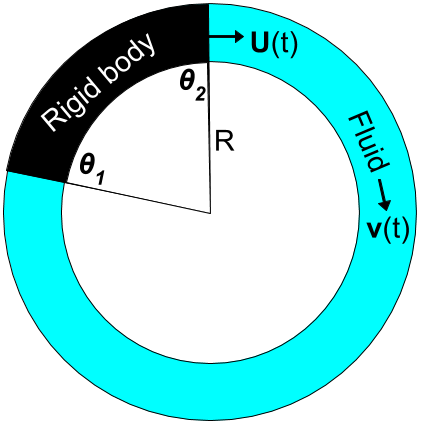} 
 \caption{Arc-shaped rigid body accelerated through fluid along a circle} \label{fig-arc-in-circular-fluid}
 \end{figure}

We begin by introducing the added mass effect through a simple example of incompressible flow in \emph{one-dimension}. Incompressible here means the fluid always has the same density everywhere. Consequently, the velocity $v$ of the fluid must be the same everywhere, though it could depend on time. Suppose an arc-shaped rigid body of length $L$ moves along the rim of a circular channel of radius $R$ (see Fig.~\ref{fig-arc-in-circular-fluid}). We will suppose that the ends of the body are at the angular positions $\theta_1(t)$ and $\theta_2(t)$ so that $R (\theta_2 - \theta_1) = L$. The fluid occupying the rest of the circumference of the channel has velocity $v(t)$ tangent to the circle at all angular positions $\theta$.

Now imagine an external agent moving the body at speed $U(t)$ so that its ends have the common speed $R \dot \tht_1 = R \dot \tht_2 = U(t)$. Since the fluid cannot enter the body, it must have the same speed as the end-points of the body: $v(\tht_1,t) = v(\tht_2,t) = U(t)$. Thus, the fluid instantaneously acquires the velocity of the body everywhere: $v(t) = U(t)$. Signals can be communicated instantaneously since the speed of sound is infinite in an incompressible flow.

In the absence of the fluid, an external agent would have to supply a force $M \dot U$ to accelerate the body. To find the additional force required in the presence of the fluid, we consider the kinetic energy of the fluid:
	\beq
	E = \half \int_{\rm fluid} \rho \: v^2(t) \:  R \, d \tht = \half \rho v^2(t) (2\pi R - L).
	\eeq
The rate of change of flow kinetic energy $\dot E$ is the rate at which energy must be pumped into the flow:
\beq
	\dot E = \rho \: v(t) \: \dot v(t) \: (2\pi R - L) = \rho \: U(t) \: \dot U(t) \: (2\pi R - L).
\eeq
$\dot E$ must equal the extra power supplied by the external agent, i.e., $\dot E = F^{\rm add} U(t)$. Thus, the additional force
\beq
	F^{\rm add} = \rho \: \dot U(t) \: (2\pi R - L)
\eeq
is proportional to the body's acceleration. $G = -F^{\rm add}$ is called the `acceleration reaction force'. The constant of proportionality
\beq
	\mu = \rho(2 \pi R - L)
\eeq
is called the added mass. Notice that $\mu$ is equal to the mass of fluid. This is peculiar to flows in one dimension. Had we taken the fluid to occupy an infinitely long channel (instead of a circle), the added mass would have been infinite. Furthermore, the added mass is proportional to the fluid density $\rho$ and depends on the shape of the body, but is independent of the body's mass. Crucially, the added force is proportional to the body's acceleration as opposed to its velocity. In particular, a body moving uniformly would not acquire an added mass.\footnote{The result that uniformly moving finite bodies in an unbounded steady potential flow do not feel any opposing force is a peculiarity of inviscid hydrodynamics called the d'Alembert paradox \cite{Batchelor}. Strictly speaking, this result is valid only in the absence of `vortex sheets' and `free streamlines'. In commonly encountered fluids, viscous forces introduce dissipation and surface waves carry away energy to `infinity' so that an external force is required even to move a body at constant velocity.}

We next turn to bodies accelerated through two- and three-dimensional flows, which are much richer than the one-dimensional example above. An intrepid reader who cannot wait to explore the analogy with the Higgs mechanism may proceed directly to \S \ref{s:ham-correspondence}.

\subsection{Two-dimensional flow around a cylinder}
\label{s:added-mass-cylinder-2d}

\begin{figure}
 \centering
 \includegraphics[scale=0.5]{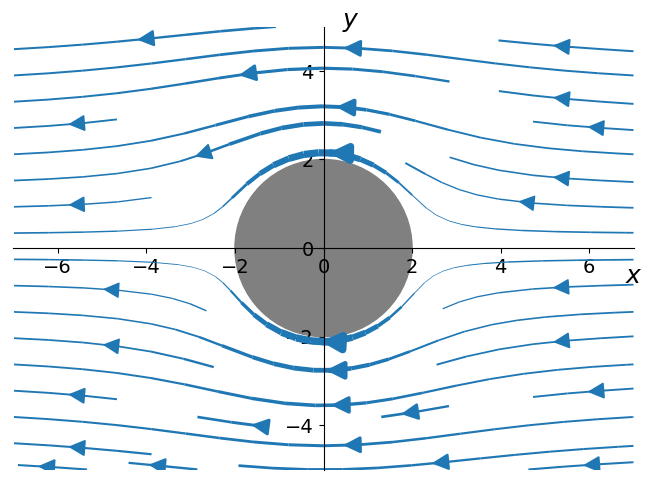} 
 \caption{Snapshot of velocity field in the $x$-$y$ plane for potential flow past a stationary right circular cylinder of radius $a = 2$ with axis along $z$. Asymptotically $\bfv \to -\hat x$ ($U = 1$). The thickness of flow lines grows with fluid speed.} \label{fig-flow-around-cylinder}
 \end{figure}

We next consider inviscid flow perpendicular to an infinitely long right circular cylinder of radius $a$ with axis along $\hat z$, as shown in Fig.~\ref{fig-flow-around-cylinder}. We will find the added mass per unit length of the cylinder by determining the velocity field of the fluid flowing around it.

Although the fluid moves in 3d space, the flow is assumed to be quasi two-dimensional with translation invariance in the $z$-direction. Thus, we take the $z$-component of the flow velocity to vanish so that $\bfv$ points in the $x$-$y$ plane. For simplicity, we further take the flow to be irrotational ($\grad \times \bfv = 0$) which allows us to write $\bfv = \grad \phi$ in terms of a velocity potential $\phi$. We also take the flow to be incompressible ($\grad \cdot \bfv = 0$, this is reasonable as long as the flow speed is much less that that of sound) which requires the velocity potential to satisfy Laplace's equation $\grad^2 \phi = 0$. To find $\phi$, we must supplement Laplace's equation with boundary conditions. The fluid cannot enter the body, so $\bfhatn \cdot \bfv = \bfhatn \cdot \grad \phi = 0$ on the surface of the body. Here $\bfhatn$ is the outward-pointing unit normal on the body's surface. Additionally, we suppose that far away from the body the fluid moves uniformly: $\bfv \rightarrow -U \hat x$. In other words, the cylinder is at rest while the fluid moves leftward past it.

Of course, we are interested in a cylinder accelerating through an otherwise stationary fluid. However, it is easier to solve Laplace's equation in a region with fixed boundaries. So, we begin by considering flow around a stationary cylinder. After finding the velocity field of this flow, we will apply a Galilean boost and move to a frame where the cylinder moves at velocity $U \hat x$. By making $U$ time-dependent, we will find the acceleration-reaction force and added mass of the cylinder.

Our first task is to solve Laplace's equation in the $x$-$y$ plane subject to the impenetrability boundary condition ${\bfhatn \cdot \grad \phi} = \frac{\pdr \phi}{\pdr r} = 0$ at $r = a$ and the asymptotic condition $\phi \to -U r \cos \theta$ as $r \to \infty$ (so that $\bfv \to -U\hat x$ as $r \to \infty$). Here, the cylinder is assumed centered at the origin about which the plane-polar coordinates $r = \sqrt{x^2 + y^2}$ and $\theta = \tan^{-1}(y/x)$ are defined. Laplace's equation in these coordinates takes the form
\beq
	\grad^2 \phi = \frac{1}{r}\frac{\pdr}{\pdr r} \left( r \frac{\pdr \phi}{\pdr r} \right) + \frac{1}{r^2} \frac{\pdr^2 \phi}{\pdr\theta^2} = 0.
	\label{eqn-laplace-cylinder}
\eeq
We will solve this linear equation by separation of variables and the superposition principle. Let us suppose that $\phi$ is a product $\phi(r, \theta) = R(r) \Theta(\theta)$ where $\Theta(\theta + 2\pi) = \Theta(\theta)$ is periodic around the cylinder. Upon division by $R \,\Theta$ the partial differential equation (\ref{eqn-laplace-cylinder}) becomes a pair of ordinary differential equations. Indeed, we must have
\beq
	\ov{R} r \frac{d}{dr} \left(r \frac{dR}{dr}  \right) = -\ov{\Theta} \DD{^2\Theta}{\theta^2} = n^2 = {\rm constant}.
\eeq
The separation constant $n^2$ must be positive for $\Theta$ to be periodic. In fact, the equation $\Theta'' = -n^2 \Theta$ has solutions
\beq
	\Theta_n(\theta) = A_n \cos n\theta + B_n \sin n\theta.
	\label{eqn-theta-periodicity-cylinder}
\eeq
Periodicity and linear independence then require $n$ to be a non-negative integer. For each such $n$, the radial equation has the solution
\beq
	R_n(r) =
	\begin{cases}
		C_n r^n + \frac{D_n}{r^n} 	& {\rm for} \;\; n > 0 \;\; {\rm and}\\
		C_0 + D_0 \ln r 			& {\rm for} \;\; n = 0.		
	\end{cases}
\eeq
Here $A_n, B_n, C_n$ and $D_n$ are constants of integration. By the superposition principle, the general solution of (\ref{eqn-laplace-cylinder}) is
\beq
	\phi(r, \theta) = (C_0 + D_0 \ln r) + \sum_{n=1}^\infty \left( C_n r^n + \frac{D_n}{r^n} \right) \left( A_n \cos n\theta + B_n \sin n\theta \right).
\eeq
The asymptotic boundary condition implies that $C_0 = D_0 = 0$, $C_1 = -U$ and $A_1 = 1$ while $C_n = A_n = B_n = 0$ for all other $n$. The impenetrability condition gives $D_1 = -U a^2$. Thus, the velocity potential for flow around the cylinder is
\beq
	\phi(r, \theta) = -U \cos\theta \left(r + \frac{a^2}{r} \right).
\eeq
The corresponding velocity field is $\bfv = \grad \phi = -U \hat x + U \frac{a^2}{r^2} \left( \cos \theta\, \hat r + \sin \theta \,\hat \theta \right)$.

Now we make a Galilean boost to a frame where the fluid is stationary at infinity, but the cylinder moves rightward with velocity $U \hat x$. The resulting velocity field around the moving cylinder is
\beq
	\bfv' = \bfv + U \hat x = U \frac{a^2}{r'^2} \left( \cos \theta'\, \hat r' + \sin \theta' \, \hat \theta' \right),
\eeq
where $r'$ and $\theta'$ are defined relative to the instantaneous centre of the cylinder.

To investigate the added mass effect, we suppose an external agent wishes to accelerate the cylinder (of mass $M$ per unit length) at the rate $\dot {\bf U} = \dot U \hat x$. Part of the energy supplied goes into the kinetic energy of the flow and is manifested as the added mass of the cylinder. Just as the kinetic energy of the cylinder per unit length ($\half M U^2$) is quadratic in $U$, so is that of the flow:
\beq
	K_{\rm flow} = \half \sig \int_0^{2\pi}\int_a^\infty \left(\bfv'\right)^2 r' \, dr'\, d\theta' = \half \sig \iint \frac{U^2 a^4}{r'^3}dr'\,d\theta' = \half \sig \pi a^2 U^2 \equiv \half \mu U^2.
\eeq
 Here $\sig$ is the density of the fluid per unit area in the $r,\theta$ plane. Thus the total kinetic energy per unit length supplied by the agent is
\beq
	K_{\rm total} = \half \left( M + \mu \right) U^2.
\eeq
The associated power supplied is $\dot K_{\rm total} = (M + \mu) \dot {\bf U} \cdot {\bf U} \equiv {\bf F^{\rm ext}} \cdot {\bf U}$. Thus, a force ${\bf F^{\rm ext}} = \left( M + \mu \right) \dot {\bf U}$ is required to accelerate the body at ${\bf\dot U}$. The mass of the cylinder therefore appears to be augmented by an added mass per unit length $\mu = \sig \pi a^2$. We notice that the added mass of the cylinder is equal to the mass of fluid displaced, though this is not always the case as we will learn in \S~\ref{s:added-mass-3d}.

More generally, it can be shown that if the cylinder had an elliptical rather than circular cross section, the added mass is different for acceleration along the two semi-axes (see Appendix \ref{a:mu-elliptical-disk}). If $a_1$ and $a_2$ are the lengths of the two semi-axes, the added masses are $\mu_{1,2} = \sig \pi a_{2,1}^2$ for motion along the corresponding semi-axis. Thus, the added mass is smaller when the cylinder presents a smaller cross section.

\subsection{Added mass tensor in 3d}
\label{s:added-mass-3d}

We have seen that to accelerate a cylinder of mass $m$ perpendicular to its axis in an ideal fluid, an external agent must provide a force $\bfF^{\rm add} = \mu \dot \bfU$ in addition to the inertial force $m \dot \bfU$ where $\bfU$ is the velocity of the cylinder. More generally, the body's acceleration need not be directed along ${\bf F}^{\rm add}$, though we will see that the two are linearly related: $F^{\rm add}_i = \sum_{j=1}^3 \mu_{ij} \dot U_j$. For flows in 3 dimensions, the added mass tensor $\mu_{ij}$ is a real, symmetric, $3 \times 3$ matrix with positive eigenvalues. It turns out that $\mu_{ij}$ is proportional to the (constant) density $\rho$ of the fluid and depends on the shape of the rigid body. However, unlike the inertia tensor\footnote{The inertia tensor of a rigid body is the $3\times 3$ matrix $I_{ij} = \int \rho(\bfr) \, (r^2 \del_{ij} - r_i r_j) \, d^3r$ where $\rho(\bfr)$ is its mass density and the integral extends over points $\bfr$ in the rigid body.} $I_{ij}$ of a rigid body, $\mu_{ij}$ is independent of its mass distribution. For example, for a sphere of radius $a$, the added mass tensor is a multiple of the identity, $\mu_{ij} = \frac{2}{3}\pi a^3 \rho\, \del_{ij}$. In other words, the added mass of a sphere is half the mass of fluid displaced irrespective of the direction of acceleration which is always along $\bfF^{\rm add}$. In particular, for an air bubble in water, the added mass is about $\half \rho_{\rm water}/\rho_{\rm air} \approx 400$ times its actual mass. More generally, for bodies less symmetrical than a sphere, the added mass tensor need not be a multiple of the identity and the added masses along different `principal' directions can be different.

\begin{figure}[h]
\centering
\begin{subfigure}{.5\textwidth}
  \centering
  \includegraphics[width=.8\linewidth]{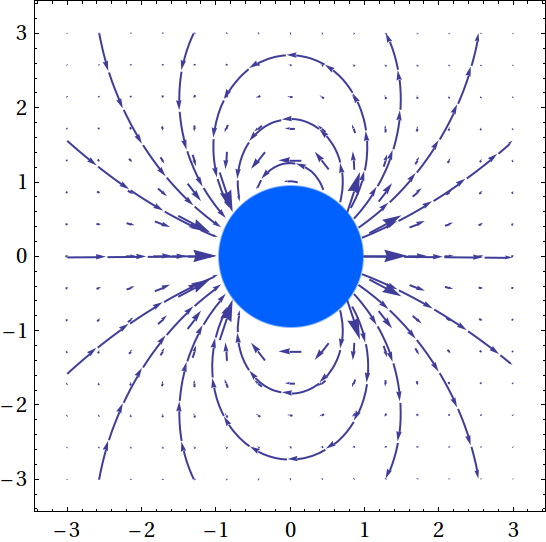}
  \caption{}
  \label{fig:sub1}
\end{subfigure}%
\begin{subfigure}{.5\textwidth}
  \centering
  \includegraphics[width=.8\linewidth]{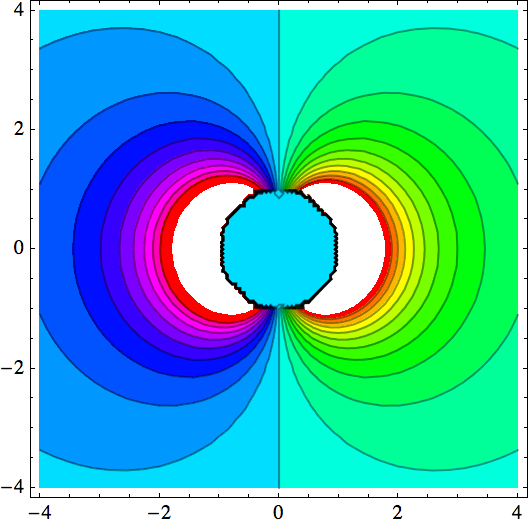}
  \caption{}
  \label{fig:sub2}
\end{subfigure}
\caption{(a) Integral curves of velocity vector field for incompressible, irrotational and inviscid fluid flow around a spherical rigid body moving rightwards and (b) equipotential surfaces of the flow potential for the same fluid flow}
\label{f:potential-flow-sphere}
\end{figure}

Let us illustrate a simple consequence of the added mass being a tensor rather than a scalar. Suppose our rigid body is irregularly shaped and has an added mass tensor $\mu_{ij}$ with non-zero off-diagonal entries. In order to accelerate it along the $\hat x$ direction, we must apply a force $\bfF = m\dot U \hat x + \bfF^{\rm add}$, where $\bfF^{\rm add} = \dot U \left( \mu_{11}\hat x + \mu_{21} \hat y + \mu_{31} \hat z\right)$. Thus, to accelerate the body along $\hat x$, we would have to supply an added force in a different direction. On the other hand, even an irregularly shaped rigid body always has (at least) three `principal axes'. They have the property that the added mass tensor is diagonal $({\rm diag}(\mu_1, \mu_2, \mu_3))$ when expressed in the principal axis basis. Thus, for instance, a force along the second principal axis produces an acceleration in the {\em same} direction with added mass $\mu_2$. The principal axes are the eigenvectors of $\mu_{ij}$ and $\mu_{1,2,3}$ are the corresponding eigenvalues.

In this Section we derive the added mass effect in 3-dimensional flows and express $\mu_{ij}$ an an integral over the surface of the body \cite{Batchelor}. As before, consider inviscid, incompressible and irrotational flow around a rigid body (assumed to be simply connected) in a large container. For simplicity, we assume that the external agent accelerates the body along a straight line without rotating it. The fluid is assumed to be at rest far from the body ($\bfv \to 0$ as $|\bfr| \to \infty$) and its velocity ${\bf v}$ is expressed in terms of a potential $\bfv = \grad \phi$. Due to incompressibility, $\phi$ must satisfy Laplace's equation $\grad \cdot \bfv = \grad^2 \phi = 0$. Impenetrability requires the boundary condition (BC) $\grad \phi \cdot \bfhatn = {\bf U}(t) \cdot \bfhatn$ on the body's surface where $\bfhatn$ is the unit outward normal.

The information in $\phi$ may be conveniently packaged in a `potential vector field' $\bfPhi(\bfr, t)$. To see this, notice that Laplace's equation and the BC is a system of inhomogeneous linear equations of the form $L \phi = b$ where $b$ is linear in $\bfU$, with solution $\phi = L^{-1}b$. Thus, $\phi$ must be linear in $\bfU$ and may be expressed as $\phi = \bfU \cdot \bfPhi$. Here $\bfPhi(\bfr, t)$ is independent of $\bfU$ and can depend only on the position of the observation point $\bfr$ relative to the body's surface. Being rigid and in rectilinear motion, the surface of the body is determined by the location of a marked point in the body, which may be chosen say, as the center of volume ${\bf r}_0$. Thus, 
\beq
\phi(\bfr, t) = {\bf U(t)} \cdot {\bf \Phi}({\bf r} - {\bf r}_0(t)).
\label{e:potential-vector-field}
\eeq
For example (see Fig.~\ref{f:potential-flow-sphere}), for a sphere of radius $a$ centered at the origin at time $t_0$,
\beq
\bfPhi({\bf r}, t_0) = -\frac{1}{2} a^3 \frac{\bf \hat{r}}{r^2} \;\; {\rm and} \;\; \phi(\bfr, t_0) = -\frac{1}{2} a^3 \bfU \cdot \frac{\bf \hat{r}}{r^2}.
\eeq

\subsection{Finding the added mass tensor from the fluid pressure on the body}
\label{s:pressure-force-added-mass-effect}

To obtain the pressure force on the body, we use a generalization of Bernoulli's equation $(p + \half \rho v^2 = {\rm constant})$ to time-dependent potential flows with constant density (see Box 3 of \cite{ham-resonance}):
\beq
p + \half \rho v^2 + \rho \dd{\phi}{t} \: = \: B(t)
\label{e-bernoulli-eqn}
\eeq
where $B(t)$ is a function of time alone. This may be used to write the total pressure force on the body as an integral over its surface $A$:
\beq
{\bf F} \; = \; - \int_{A} p \, {\bfhatn} \, dA = \rho \int_{A} \left( \frac{\partial \phi }{\partial t} + \frac{1}{2} v^2 \right) {\bfhatn} \, dA.
\eeq
The Bernoulli constant $B(t)$ does not contribute as the integral $\int_A \bfhatn \, dA$ vanishes over the closed surface $A$ of the body. Using the factorization $\phi = {\bf U} \cdot {\bf \Phi}$, we write $\bfF$ as a sum of an acceleration reaction force $\bf G$ and an acceleration-independent ${\bf G}'$
	\beq
	{\bf F} = \rho \int_{A} {\bf\dot{U}} \cdot {\bf \Phi} \: {\bfhatn} \: dA
+ \int_{A} \left[ \half \rho v^2 - \rho {\bf U} \cdot {\bf v} \right] \: {\bfhatn} \: dA \equiv \bfG + \bfG'.
	\label{e:G-G-prime}
	\eeq
Here we used ${\bf \Phi} = {\bf \Phi}(\bfr - \bfr_0(t))$ where $\bfr_0(t)$ is a fixed point in the moving rigid body to express 
\begin{equation}
\bfU \cdot \frac{\partial}{\partial t} {\bf \Phi}\left(\bfr - \bfr_0(t)\right) =  \grad(\bfU \cdot \bfPhi) \frac{\partial}{\partial t} (\bfr - \bfr_0(t)) = \grad \phi \cdot \left( -\frac{\partial \bfr_0(t)}{\partial t} \right) = \grad \phi \cdot (-\bfU) = -\bfv \cdot \bfU.
\end{equation}
The non-acceleration reaction force ${\bf G}'$ vanishes in fluids asymptotically at rest in $\mathbb{R}^3$ \cite{Batchelor}. Using a multipole expansion for $\phi$, one estimates that $\bfG'$ can be at most of order $1/R$ in a large container of size $R$ (see Appendix \ref{a:non-acceleration-reaction-force}). It is as if fluid can hit the container and return to push the body. In what follows, we ignore this boundary effect. When acceleration due to gravity $\bf g$ is included, ${\bf G}'$ features a buoyant term $- \rho {\rm Vol}_{\rm body} \, {\bf g}$ equal to the weight of fluid displaced, which we also suppress. Thus, the acceleration reaction force is
\begin{equation}
G_i = -F_i^{\rm add} = - \mu_{ij} \dot U_j \quad \textmd{where}\quad 
\mu_{ij} = - \rho \int_{A} \Phi_j \: n_i \: dA.
\label{e:added-mass-tensor}
\end{equation}
The added-mass tensor $\mu_{ij}$ is a direction-weighted average of the potential vector field $\bf \Phi$ over the body surface. It is proportional to the fluid density and depends on the shape of the body surface. $\mu_{ij}$ may be shown to be time-independent\footnote{ To see that $\mu_{ij}$ is a constant tensor even as the body moves in the fluid, we note that $\Phi_i(\bfr,t) = \Phi_i({\bf r} - \bfr_0(t))$. Thus, $\Phi_i$ at a marked point on the body surface does not change with time. Consequently, the value of the integral in (\ref{e:added-mass-tensor}) is independent of time.} and symmetric (see Appendix \ref{a:symmetry-of-mu}).

The rate at which energy is pumped into the fluid is 
\beq
{\bf F^{\rm add}} \cdot {\bf U}(t) = \sum_{i,j=1}^3\mu_{ij} \dot U_j U_i = \DD{}{t}\left(\sum_{i,j=1}^3\half \mu_{ij} U_i U_j\right).
\eeq
Thus the flow kinetic energy may be expressed entirely in terms of the body's velocity and added-mass tensor:
\beq
K = \half \int_V \rho v^2 \: dV \; = \; \sum_{i,j=1}^3\frac{1}{2}\mu_{ij} U_i U_j.
\eeq
It follows that the added-mass tensor $\mu_{ij}$ is a positive matrix (its eigenvalues are non-negative since $K \geq 0$).

\subsection{Examples of added mass tensors}
\label{s:examples-added-mass-tensors}

\begin{figure}
	\centering
	\includegraphics[scale=0.4]{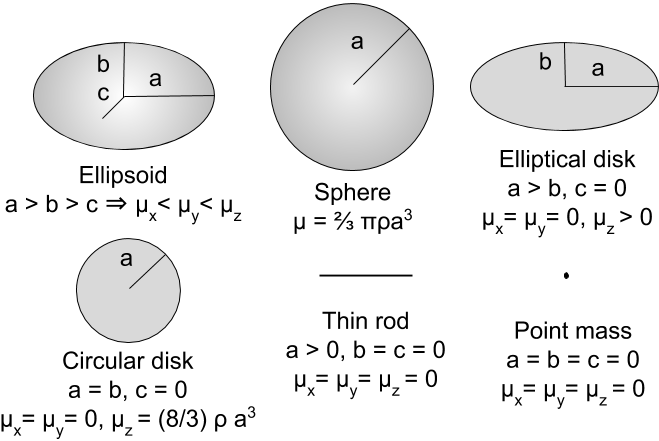}
	\caption{Examples of rigid bodies (an ellipsoid and its limiting cases) and their principal added masses.} \label{f:bodies-and-added-masses}
\end{figure}

To a particle physicist, mass generation in a medium sounds like the Higgs mechanism, and an added-mass tensor is reminiscent of a mass matrix. To uncover a precise correspondence between these phenomena, it helps to have explicit examples. By solving potential flow around rigid bodies, one obtains their added-mass tensors (see Fig.~\ref{f:bodies-and-added-masses}). We will relate these rigid bodies and their added-mass tensors to specific patterns of spontaneous gauge symmetry breaking in \S~\ref{s:ssb-patters-rigid-bodies}. 

For a $2$-sphere of radius $a$, $\mu_{ij} = \frac{2}{3} \pi a^3 \rho \del_{ij}$ is isotropic. The added-mass of a sphere is half the mass of fluid displaced, irrespective of the direction of acceleration. For an ellipsoid $\frac{x^2}{a^2} + \frac{y^2}{b^2} + \frac{z^2}{c^2} =  1$, $\mu_{ij}$ is diagonal in the principal axis basis. If $a > b > c$, then the eigenvalues satisfy $\mu_x < \mu_y < \mu_z$. Roughly, added-mass grows with cross-sectional area presented by the accelerating body. In its principal axis basis (see Appendix \ref{a:ellipsoid-added-mass})
	\beq
	(\mu_x, \mu_y, \mu_z) = \frac{4}{3}\pi\, abc\, \rho \, \left(\frac{\alpha}{2-\alpha}, \frac{\beta}{2-\beta}, \frac{\gamma}{2-\gamma} \right),
	\eeq
where 
    \beq
    \alpha = abc\, \int_{0}^{\infty} (a^2 + \lambda)^{-1} \D^{-1} d\lambda \quad
    \text{with} \quad
    \Delta = \sqrt{(a^2 + \lambda)(b^2 + \lambda)(c^2 + \lambda)}
    \eeq
and cyclic permutations thereof. In particular, for an ellipsoid of revolution with $a = b$, the corresponding pair of added-mass eigenvalues coincide $\mu_x = \mu_y$. On the other hand, by taking $c \to 0$ we get an elliptic disk, for which two added-mass eigenvalues $\mu_x$ and $\mu_y$ vanish. These correspond to acceleration along its plane. With impenetrable boundary conditions, an elliptic disk does not displace fluid or feel an added-mass when accelerated along its plane. The third eigenvalue $\mu_z$, for acceleration perpendicular to its plane, is $\frac{4}{3}\pi \rho ab^2 E( 1 - b^2/a^2 )^{-1}$, where $E(m)$ is the complete elliptic integral of the second kind. Taking $a=b$, the principal added-masses of a circular disk are $(0,0,(8/3)\rho a^3)$. Shrinking the elliptical disk further, a thin rod of length $2a$ has no added-mass. Irrespective of which way it is moved, it does not displace fluid with impenetrable boundary conditions. The same is true of a point mass or any body whose dimension is less than that of the flow domain by at least two (codimension $\geq 2$). For an infinite right circular cylinder, the added-mass per unit length for acceleration perpendicular to its axis is equal to the mass of fluid displaced. If the axis of the cylinder is along $z$, then the added-mass tensor per unit length is $\frac{\mu_{ij}}{L} = \pi a^2 \rho\; {\rm diag}(1,1,0)$, where $a$ is its radius. Though these examples pertain to three dimensional flows, the added mass effect generalizes to rigid bodies accelerated through plane flows (see \S\ref{s:added-mass-cylinder-2d}, Appendix \ref{a:2d-added-mass-effect} and \cite{Batchelor}) as well as flows in $4$ and higher dimensions. For example, an elliptical disk with semi-axes $a, b$ accelerated through planar potential flow has an added mass tensor $\mu_{ij} = \pi \sigma a b \del_{ij}$ where $\sigma$ is the (constant) mass of fluid per unit area. In \S~\ref{s:d-geq-3-added-mass-effect} we develop the formalism for the added mass effect in $d \geq 3$ dimensions. This will be used in the following sections where we relate the added mass effect in $d$-dimensional flows to spontaneous breaking of a $d$-dimensional gauge group $G$.

\subsection{Added mass effect in $d \geq 3$ dimensions}
\label{s:d-geq-3-added-mass-effect}


The Higgs-added mass (HAM) correspondence relates spontaneous breaking of a $d$-dimensional gauge group $G$ to the added mass effect in $d$-dimensional fluids. Since there is no restriction on the dimension of $G$, our correspondence requires an extension of the standard added mass effect \cite{Batchelor,Kambe} to flows in $d \geq 4$, which we give here. Consider incompressible potential flow in $\mathbb{R}^d$ around a simply connected rigid body moving with velocity $\bfU(t)$. We assume that the body executes purely translational motion and that $\bfv \to 0$ asymptotically. The velocity potential satisfies the Laplace equation $\grad^2 \phi = 0$ subject to impenetrable boundary conditions on the body surface: $\bfn \cdot \grad \phi = \bfn \cdot \bfU$. With the origin located inside the body, $\phi$ admits a multipole expansion
	\beq
	\phi(\bfr) = c/r^{d-2} + c_i \pdr_i (1/r^{d-2}) + c_{ij} \partial_i \partial_j ( 1/r^{d-2}) + \ldots \label{e:multipole-expansion}
	\eeq
in terms of Green's function of the laplacian (and its derivatives)
	\beq
	g(r) = -\frac{\Gamma(d/2)}{2\pi^{d/2}(d-2)} \frac{1}{r^{d-2}}
	\eeq
satisfying $\grad^2 g(r) = \del^d(\bfr)$. As in the Cauchy contour integral formula, the multipole tensor coefficients (which are linear in $\bfU$) may be expressed as integrals of $\phi$ and its derivatives over the body surface $A$,
	\beqs
 	c &=& \frac{\Gamma(\frac{d}{2})}{2\pi^{d/2}(d-2)} \oint_{A} {\bf n} \cdot \grad \phi(\bfr) \, dA, \quad 
 	c_i = \frac{\Gamma(\frac{d}{2})}{2\pi^{d/2}(d-2)} \, \oint_{A} \left[ (\bfn \cdot \grad \phi) r_i - \phi n_i \right]\,dA, \cr
	c_{ij} &=& \frac{\Gamma(\frac{d}{2})}{2\pi^{d/2}(d-2)} \oint_{A} \left[ (\bfn \cdot \grad \phi) r_i r_j - \phi (n_i r_j + n_j r_i) \right]\, dA, \quad \ldots 
	\label{e-multipole-moments}\label{e-multipole-coefficients}
	\eeqs
For incompressible flow without sources, the monopole coefficient $c \equiv 0$. As in the $3$d case, the impenetrable boundary condition constrains $\phi$ to be linear in $\bfU$, which allows us to write it as $\phi = {\bf \Phi}\cdot\bfU$. The potential vector field ${\bf \Phi}(\bfr, t) = {\bf \Phi}(\bfr - \bfr_0(t))$ is independent of $\bf U$. Here $\bfr_0(t)$ is a convenient reference point fixed in the body. As in \S \ref{s:pressure-force-added-mass-effect}, we use Bernoulli's equation (\ref{e-bernoulli-eqn}) to write the pressure-force on the body surface $A$ in terms of $\phi$, and use the factorization $\phi = {\bfPhi}\cdot\bfU$ to write the force as the sum of an acceleration reaction $\bf G$ and a non-acceleration force ${\bf G}'$, as in (\ref{e:G-G-prime}). From the multipole expansion $\phi \sim 1/r^{d-1}$ and it follows that $\bfG'$ vanishes when the flow domain is all of $\mathbb{R}^d$. Thus we get the same formula (as in 3d) for the added mass tensor $\mu_{ij}$ from the acceleration-reaction force:
	\beq
	G_i \;=\; \rho\, \dot{U_j} \oint_{A} \Phi_j\, n_i \, dA \;\equiv\; -\mu_{ij}\,\dot{U_j} \quad \Rightarrow \quad \mu_{ij} = -\rho \oint_{A} \Phi_j\, n_i \, dA.
	\eeq
Despite appearances, $\mu_{ij}$ only depends on the dipole term in $\phi$. The linearity of the boundary condition in $\bf U$ implies that $c_i = d_{ij} U_j$ is linear in $\bf U$. The constant {\it source doublet} or {\it dipole tensor} $d_{ij}$ depends only on the shape of the body. Using (\ref{e-multipole-coefficients}) for $c_i$ and the boundary condition on the surface, we obtain
	\beqs
	c_i &=& d_{ij}U_j = \frac{\Gamma(\frac{d}{2})}{2(d-2)\pi^{d/2}} \oint_{A} \left[ (\bfn\cdot\bfU)r_i - \phi n_i \right]\, dA \cr 
	&=& \frac{\Gamma(\frac{d}{2})}{2(d-2)\pi^{d/2}} U_j \left[ \int_{{\rm body}} \partial_j r_i\, dV - \oint_{A} \Phi_j n_i\, dA \right] \cr
	&=& \frac{\Gamma(\frac{d}{2})}{2(d-2)\pi^{d/2}} \left[ V_{\rm body} \del_{ij} + \frac{\mu_{ij}}{\rho} \right]U_j.
	\label{e:source-doublet-vector}
	\eeqs
Since this is valid for any velocity $\bf U$ we arrive at a relation between $\mu_{ij}$ and the dipole tensor
	\beq
	\mu_{ij} = \rho \left[ \frac{2(d-2) \, \pi^{d/2}}{\Gamma(d/2)} \: d_{ij} - V_{\rm body} \del_{ij} \right].
	\eeq
This expression for $\mu_{ij}$ shows that it only depends on the dipole part of $\phi$. It does not involve integrals and gives a simple way of computing $\mu_{ij}$ once the dipole term in $\phi$ is known. Let us illustrate this with the example of a $(d-1)$-dimensional sphere $S^{d-1}_a$ of radius $a$, moving through fluid in $\mathbb{R}^d$. A moving sphere instantaneously centered at the origin induces a dipole flow field with potential $ \phi = c_i \pdr_i r^{2-d} = -(d-2) r^{-d} \:  {\bf c} \cdot {\bf r}$. The multipole tensors $c_{ij}, c_{ijk}, \ldots$ are {\rm constant} tensors of rank $> 1$, linear in $\bfU$. Spherical symmetry of the body denies us any other vector/tensor from which to construct them, so they must vanish. The dipole coefficient $\bf c$ may be self-consistently determined by inserting this formula for $\phi$ in (\ref{e-multipole-moments}). One obtains
	\beq
	c_i = \frac{a^d}{(d-1)(d-2)} U_i \qquad
	\text{or} \qquad
	d_{ij} = \frac{a^d}{(d-1)(d-2)} \, \del_{ij}.
	\label{e-sphere-dipole-moment} 
	\eeq
Hence, the added mass tensor for a $(d-1)$-sphere of radius $a$ moving in $\mathbb{R}^d$ is
	\beq
	\mu_{ij}^{\rm sphere} \;=\; \rho \frac{2 \pi^{d/2} a^d}{d(d-1) \Gamma\left( \frac{d}{2}\right)} \del_{ij}
	= \frac{\text{(Mass of fluid displaced)}}{(d-1)} \del_{ij}.
	\eeq
This reduces to the well-known results (\S\ref{s:examples-added-mass-tensors}) for planar or 3d flow around a disk or $2$-sphere. In \S\ref{s:discussion-ham}, we speculate on the possible relevance of a suitable $d \to \infty$ limit.

\section{The Higgs mechanism}
\label{s:higgs-mechanism}

The weak force, responsible for $\beta$-decay is short ranged. Hence, the mediators of this force, the $W$ and $Z$ bosons, must be massive. But naively inserting a mass term for these bosons in the Lagrangian for the theory of weak interactions leads to loss of gauge invariance and renormalizability. The Higgs mechanism solves this problem. While it is the non-abelian version of the Higgs mechanism which is relevant to the weak gauge bosons, we begin here with the simpler abelian version.

\subsection{Abelian Higgs Model}
\label{s:abelian-higgs}

\begin{figure}
\begin{center}
\includegraphics[scale=0.75]{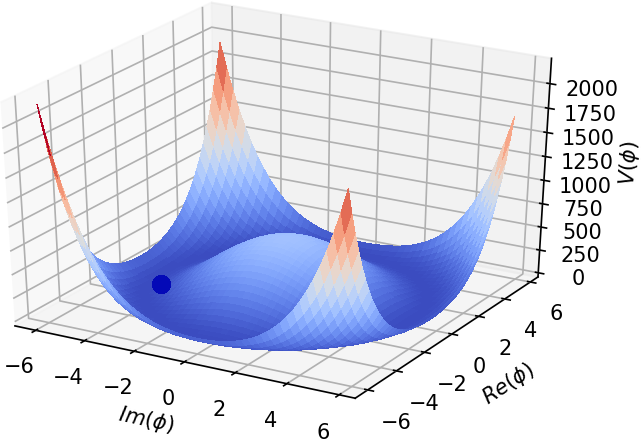}
\caption{Spontaneous symmetry breaking in the Higgs mechanism is achieved through a potential $V(\phi) = \lambda (|\phi|^2 - a^2)$ for the complex scalar field $\phi$ shown above. Here the vacuum expectation value (vev) of the scalar field is $a = 5$ and the scalar `self-coupling' is $\lambda = 1$. Though $V$ is circularly symmetric, a `particle' (the blue dot) in its minimum energy state must lie at a particular point along the circular valley, thereby `spontaneously' breaking the rotation symmetry. The lowest-lying excitation is a massless `Goldstone mode' corresponding to a particle rolling arbitrarily slowly along the circular valley. The lowest-lying (massive) radial excitation corresponds to the Higgs particle.}
\label{f:mexican-hat-potential}
\end{center}
\end{figure}

In the abelian Higgs model, one minimally couples the massless gauge field $A_{\mu}$ to a complex scalar field $\phi$ with a Mexican hat potential (see Fig.~\ref{f:mexican-hat-potential}). The choice of the ground state spontaneously breaks the symmetry of the potential and $\phi$ acquires a non-zero vacuum expectation value. The interaction term $\phi^{*}\phi A^{\mu}A_{\mu}$ between the scalar and the gauge field becomes an effective mass term for the gauge field $A_{\mu}$.

Following the treatment of this topic by Coleman \cite{Coleman} and Ryder \cite{Ryder}, we demonstrate the above in the case of the abelian gauge symmetry group $U(1)$. We start with the Lagrangian
	\beq
	\Lagr = (\partial_{\mu} +  ieA_{\mu})\,\phi^*\,(\partial^{\mu} -  ieA^{\mu})\,\phi\,-\,m^2\phi^*\phi\,-\,\lambda(\phi^*\phi)^2\,-\,\frac{1}{4}F_{\mu\nu}F^{\mu\nu}.
	\eeq
To give mass to $A_{\mu}$, we tune the parameter $m^2$ so that $m^2 < 0$. The minimum of the potential then occurs at $|\phi| = {-m^2}/{2\lambda} \equiv a$ so that the vacuum expectation value of $\phi$ is $a$. We now rewrite the Lagrangian in terms of scalar fields $\phi_1$ and $\phi_2$, having zero vacuum expectation value, defined as follows: 
	\beq
	\phi \equiv a + \frac{\phi_1 + i\phi_2}{\sqrt{2}}.
	\eeq
In terms of these new fields, the Lagrangian is
	{\small
	\beq
	\Lagr = -\frac{1}{4}F_{\mu\nu}F^{\mu\nu}\, + \, e^2a^2 A^{\mu}A_{\mu} \, + \, \frac{1}{2}(\partial_{\mu}\phi_1)^2 \, + \, \half(\partial_{\mu}\phi_2)^2 \, - \, 2\lambda a^2\phi_1^2 \, + \, \sqrt{2}eaA^{\mu}\partial_{\mu}\phi_2 \, + \, \mbox{\small cubic \& quartic.}
	\eeq}
We initially had 2 field degrees of freedom coming from the massless vector boson and 2 more coming from the complex scalar field, to give a total of 4 degrees of freedom. From the above Lagrangian, we find that currently, there are a total of 5 degrees of freedom - 3 coming from the massive vector boson and 2 more coming from the scalar fields $\phi_1$ and $\phi_2$. Hence, there is one spurious degree of freedom. This spurious degree of freedom can be gauged out, however. Under an infinitesimal rotation $\phi \to e^{i\Lambda}\phi \approx (1 + i\Lambda) \phi$, the scalar fields transform as follows:
	\beq
	\phi_1 \rightarrow \phi_1^{\prime} = \phi_1 - \Lambda\phi_2\quad \text{and}\quad
 	\phi_2 \rightarrow \phi_2^{\prime} = \phi_2 + \Lambda\phi_1 + \sqrt{2}\Lambda a
	\eeq
We now choose $\Lambda$ such that $\phi_2^{\prime} \equiv 0$. Rewriting the Lagrangian in terms of the transformed fields, we obtain
	\beq
	\Lagr = -\frac{1}{4}F_{\mu\nu}F^{\mu\nu}\, + \, e^2a^2 A^{\mu}A_{\mu} \, + \, \frac{1}{2}(\partial_{\mu}\phi_1)^2 \, - \, 2\lambda a^2\phi_1^2 \, + \, (\mbox{\small cubic \& quartic interaction terms})
	\eeq
where we have dropped the primes for convenience. We now see that the the scalar field $\phi_2$ does not appear in the Lagrangian at all, and hence, the spurious extra degree of freedom has been removed. Thus, in effect, the vector boson $A_{\mu}$ behaves as if it has a mass $m_{A} = \sqrt{2} ea$. A byproduct of this mechanism is another massive scalar particle (the `radial' excitation $\phi_1$), called the Higgs boson, whose mass is $m_{H} = 2\sqrt{\lambda}a$.

\subsection{Higgs mechanism in a non-abelian gauge theory}
\label{s:non-abelian-higgs}

Let $G = SO(3)$ be the full gauge symmetry group, and let $\phi$ transform in the fundamental representation of $G$ as a three component real vector $\phi \rightarrow e^{t_a\theta_a}\phi$, where $\phi = (\phi_1, \phi_2, \phi_3)^t$ and the generators of the $SO(3)$ Lie algebra are chosen as
	\beq
 t_1 = \colvec{3}{0 & 0 & 0}{0 & 0 & -1}{0 & 1 & 0},\quad t_2 = \colvec{3}{0 & 0 & 1}{0 & 0 & 0}{-1 & 0 & 0}\quad \text{and}\quad t_3 = \colvec{3}{0 & -1 & 0}{1 & 0 & 0}{0 & 0 & 0}.
	\eeq
Gauge transformations rotate the $\phi$ vector in the internal space at each point $x$. The gauge field is $W_\mu = W_\mu^a t_a$ and the covariant derivative is
\begin{equation}
 D_{\mu}\phi = (\partial_\mu - g W_\mu^a t_a)\phi
\end{equation}
where $g$ is the coupling constant. The square of the covariant derivative is
\begin{equation}
 (D_\mu \phi)^t D^\mu \phi = ||\partial_\mu \phi||^2 + g^2 (\phi^t t_a^t t_b \phi) W_\mu^a W^{\mu b} + \cdots
\end{equation}
where we have omitted some cubic interactions. The Lagrangian density is
	\beq
	{\cal L} = - \ov{4} \tr F_{\mu \nu} F^{\mu \nu} +  (D_\mu \phi)^t D^\mu \phi - V(\phi)
	\eeq
where the field strength is
	\beq
	F_{\mu \nu} = \pdr_\mu W_\nu - \pdr_\nu W_\mu - g [W_\mu, W_\nu].
	\eeq
This Lagrangian describes (without spontaneous symmetry breaking (SSB), $m^2 < 0$), three massless gauge fields $W_\mu^1, W_\mu^2, W_\mu^3$ (each possessing two transversely polarized field degrees of freedom) interacting with a $3$ component real scalar field. So the Lagrangian describes $9$ field degrees of freedom.

To study SSB, we take the scalar potential to be the usual Higgs potential:
\begin{equation}
 V(\phi) = V(\phi^t \phi) = -m^2 \phi^t \phi + \lambda (\phi^t \phi)^2
\end{equation}
so that when $m^2 > 0$, the vacuum manifold (obtained by solving $V'(\phi^t \phi) = 0$) is a sphere $S^2_\eta$ of radius $\eta = ||\phi|| = \sqrt{\frac{m^2}{2\lambda}} \neq 0$. We may choose any point on this sphere to be our vacuum. For convenience, we choose
	\beq
	\langle \phi \rangle = \colvec{3}{0}{0}{\eta}.
	\eeq
This vacuum breaks the $SO(3)$ symmetry. Had it been a global symmetry, we would have two massless Goldstone bosons corresponding to translations along the 2d vacuum manifold and one massive scalar corresponding to fluctuations in the radial direction. In the spontaneously broken gauge theory, we expect that these two Goldstone modes are `eaten' by two of the photons making them massive, leaving behind one massless photon and one massive Higgs scalar.

The departure of $\phi$ from its vacuum value can in general be non-zero in all its components. But using the $SO(3)$ symmetry, we can gauge away the first two components (use the gauge symmetry to orient $\phi$ in the third direction everywhere) so that
\begin{equation}
 \phi(x) = \colvec{3}{0}{0}{\eta} + \colvec{3}{0}{0}{\psi(x)}
\end{equation}
where $\psi$ is a real scalar field with zero vacuum expectation value. We write the square of the covariant derivative in terms of $\psi$
\begin{equation}
 (D_\mu \phi)^t D^\mu \phi = (\partial_\mu \psi)^2 + g^2 \psi^2 (t_a^t t_b)_{33} W_\mu^a W^{\mu b} + \cdots
\end{equation}
where we have omitted cubic and quartic interactions. The terms bilinear in the gauge fields show that the gauge fields have become massive. We find that $(t_1^t t_1)_{33} = (t_2^t t_2)_{33} = 1$ and $(t_a^t t_b)_{33} = 0$ for all other choices of $a,b$. Hence, the mass-squared matrix for the gauge fields in the $W_\mu^a$ basis is
\begin{equation}
 M_{ab} = g^2\eta^2\colvec{3}{1&0&0}{0&1&0}{0&0&0}
\end{equation}
which shows that the gauge bosons $W_\mu^1$ and $W_\mu^2$ are massive with mass $g\eta$, whereas $W^3_\mu$ is massless. It is worth noting that choosing our vacuum, and hence the scalar field departure from vacuum, to lie along the first or second direction instead of the third, would have resulted in having $W_\mu^1$ or $W_\mu^2$ being massless instead of $W_\mu^3$.

The mass of the Higgs scalar $\psi$ can be determined by expanding the scalar potential to second order in $\psi$:
\begin{equation}
 V(\phi^t\phi) = -m^2 (\eta + \psi)^2 + \lambda (\eta + \psi)^4 = 2m^2\psi^2 + \textmd{\small constants, higher order terms}.
\end{equation}
We therefore read off the mass of Higgs scalar $\psi$ to be $\sqrt{2}m$.

\subsection{More than one SSB pattern having the same mass-squared matrix}
\label{s:ideal-SSB-pattern}

As we shall see in \S\ref{s:ssb-patters-rigid-bodies}, two rigid bodies with different shapes could have the same added mass tensor. We will see here that a similar phenomenon occurs in the Higgs mechanism, where more than one SSB pattern can have the same mass-squared matrix. We illustrate this with an example. Let $G = U(1)^2$, with coupling constants $g_1, g_2$ and a complex doublet of scalars $\phi = (\phi_1, \phi_2)^t$, the $i^{\rm th}$ scalar bearing charge $q_{ij}$ with respect to the $j^{\rm th}$ $U(1)$ factor in $G$. Let the scalar potential be $V = V_1(|\phi_1|) + V_2(|\phi_2|)$ where $V_i(|\phi|) = -m_i^2|\phi|^2 + \lambda_i |\phi|^4$. The vevs of the scalars are $\langle \phi_1 \rangle = \sqrt{m^2_1/2\lambda_1} = \eta_1$ and $\langle \phi_2 \rangle = \sqrt{m^2_2/2\lambda_2} = \eta_2$. The general mass-squared matrix for vector bosons in this SSB pattern is
\begin{equation}
 M_{ab} = 2 \sum_{j=1}^2 \eta_j^2 q_{ja}g_a q_{jb} g_b.
\end{equation}
Consider the SSB pattern where the coupling constants $g_1 = g_2 = g$, the vevs $\eta_1 = \eta_2 = \eta$ and the charge matrix is a multiple of the identity: $q_{ij} = q \del_{ij}$. The mass-sqared matrix for this SSB pattern is
\begin{equation}
 M_{ab} = 2\eta^2 g^2 q^2 \del_{ab}.
\end{equation}
Now consider another SSB pattern with unequal coupling constants $g_1, g_2$ and unequal vevs $\eta_1, \eta_2$, but satisfying the equation $\eta_1g_1 = \eta_2  g_2 = \eta g$ where $\eta$ and $g$ are the vev and coupling constant of the previous SSB pattern. Let the charge matrix be the same as in the previous SSB pattern, $q_{ij} = q \del_{ij}$. The mass-squared matrix in this case is
\begin{equation}
 M_{ab} = 2 \eta_a^2 g_a^2 q^2 \del_{ab}
\end{equation}
which is the same as the previous mass-squared matrix, due to the equation satisfied by $\eta_1, \eta_2, g_1, g_2$. Hence, we have shown that two SSB patterns could correspond to the same mass-squared matrix. Out of the above two SSB patterns, the former, with the lesser number of parameters, can be considered to be more ``ideal'' than the latter.

\section{The Higgs-added mass correspondence}
\label{s:ham-correspondence}

Recalling our discussion of the added mass effect, gauge bosons acquiring masses via the Higgs mechanism sounds like the virtual masses of a rigid body accelerated through a fluid along its principal directions. Moreover, the gauge boson mass matrix is reminiscent of the added mass tensor. In fact, inspired in part by \cite{thyagaraja}, we have uncovered \cite{ham-prsa} a delightful analogy between these two physical phenomena which we call the `Higgs added mass correspondence', that we now proceed to describe.

\subsection{Spontaneous symmetry breaking patterns and their rigid bodies}
\label{s:ssb-patters-rigid-bodies}

As discussed in \S\ref{s:abelian-higgs}, in the simplest version of the Higgs mechanism, a $G = $U$(1)$ gauge field $A$ in $3+1$ space-time dimensions is coupled to a complex scalar $\phi$ with potential $V(\phi) = - m^2 |\phi|^2 + \la |\phi|^4$, ($m^2 , \la > 0$) and Lagrangian
	\beq
	{\cal L} = \half ({\bf E}^2 - {\bf B}^2) + |(\pdr_\mu - i g A_\mu) \phi|^2 -  V(\phi).
	\eeq
The space of scalar vacua $\cal M$ (global minima of $V$) is a circle of radius $\eta = \sqrt{m^2/2\lambda}$. If U$(1)$ were a global symmetry we would have one angular Goldstone mode. A non-zero vacuum expectation value (vev) $\bra \phi \ket = \eta$ leads to complete spontaneous breaking of the symmetry group $G$. If $\phi = (\eta + \rho) e^{i \chi/\eta}$, we may gauge away $\chi$ and get a mass term $g^2 \eta^2 A^2$ for the photon (which has `eaten' the Goldstone mode), and a radial scalar mass term $m^2 \rho^2$ corresponding to the Higgs particle. In general \cite{non-Abelian-Higgs-Kibble}, $G$ breaks to a residual symmetry group $H$ whose generators annihilate the vacuum and $g^2 \eta^2 A^2$ is replaced by gauge boson mass terms $\half M_{ab} A_\mu^a A_\mu^b$. {\it We say that a spontaneously broken gauge theory corresponds to a rigid body, if vector boson masses and added-mass eigenvalues coincide.} In particular, the dimension of $G$ must equal that of the flow domain. We begin with some examples of SSB patterns and associated rigid bodies. In these examples, the space of scalar vacua ${\cal M}$ is the quotient $G/H$. They reveal a relation between symmetries of $G/H$ and of a corresponding {\it ideal} rigid body. By an ideal rigid body, we mean one with maximal symmetry group among those with identical added-mass eigenvalues: for example, a round sphere of appropriate radius, instead of a cube.

\begin{enumerate}

\item Consider an SO$(3)$ gauge theory minimally coupled to a triplet of real scalars interacting via the above potential $V$. ${\cal M}$ is a $2$-sphere of radius $\eta$ resulting in two Goldstone modes. They are eaten by $2$ of the $3$ gauge bosons leaving one massless photon. The mass-squared matrix $M$ is $2 g^2 \eta^2 \: {\rm diag}(1,1,0)$. $G=$SO$(3)$ breaks to $H = $SO$(2)$. The corresponding rigid body moves in fluid filling  three dimensional Euclidean space, since $\dim G = 3$. The rigid body must have one zero and two equal added-mass eigenvalues to correspond to the mass matrix $M$. An ideal rigid body that does the job is a hollow cylindrical shell, say $S^1 \times [-1,1]$. Such a shell has no added-mass when accelerated along its axis. Due to its circular cross section, the added-masses are equal and non-zero for acceleration in all directions normal to the axis.

\item Similarly, an SO$(n)$ gauge theory coupled to $n$-component real scalars spontaneously breaks to $H=$SO$(n-1)$. The vacuum manifold ${\cal M}$ is a sphere S$^{n-1}$ of radius $\eta$. We get $n-1$ vector bosons of mass $\sqrt{2} g \eta$ and $n_\gamma = {(n-1)(n-2)/2}$ massless photons. A corresponding ideal rigid body moving through fluid filling $\mathbb{R}^{\half (n^2 - n)}$ is the product $S^{n-2} \times B^{n_\g}$, generalizing the cylindrical shell $S^1 \times B^1$ when $n=3$. Here $B^{n_\g}$ is a unit ball $|{\bf x}| \leq 1$  for ${\bf x} \in \mathbb{R}^{n_\g}$. This ideal rigid body has equal non-zero added-masses when accelerated along the first $n-1$ directions and no added-mass in the remaining $n_\gamma$ {\em flat} directions. We call $S^{n-2}$ its curved factor and the unit ball $B^{n_\g}$ its flat factor. $B^1$ is the unit interval while $B^2$ is the unit disk, etc. It is easily seen that for $n = 3$ and $4$, acceleration along the direction of the interval or in the plane of the disk displaces no fluid, the same holds for $n \geq 5$.

\item For SU$(2)$ gauge fields coupled to a complex scalar doublet with the same potential $V$, ${\cal M}$ is a $3$-sphere of radius $\eta$. All $3$ gauge bosons are equally massive. The mass-squared matrix is $M_{ab} = (g^2 \eta^2/2) \: \delta_{ab}$ and SU$(2)$ breaks completely. A corresponding ideal rigid body is a $2$-sphere of radius $a= (3 g \eta/2 \pi \sqrt{2} \rho)^{1/3}$ moving through a fluid in three dimensions. The same group with scalars in other representations could lead to different SSB patterns and rigid bodies. With adjoint scalars, SU$(2) \to$ U$(1)$ with two equally massive vectors, corresponding to a hollow cylindrical shell moving in 3d.

\item In unbroken gauge theories, all gauge bosons remain massless. Such a theory with $d$-dimensional gauge group, corresponds to a point particle (or one of codimension more than one) moving through $\mathbb{R}^d$, which has no added mass. For instance, SU$(2)$ coupled to a complex scalar triplet in the potential $V = m^2 |\phi|^2 + \lambda |\phi|^4$ with $m^2 > 0$, remains unbroken and corresponds to a point particle moving through $\mathbb{R}^3$.

\item SU$(3)$ with fundamental scalars breaks to SU$(2)$ and ${\cal M} = S^5$. There are $3$ massless photons, $4$ vector bosons of mass $g \eta/\sqrt{2}$ and a heavier singlet of mass $\sqrt{2} g \eta/\sqrt{3}$. The corresponding ideal rigid body moves in $\mathbb{R}^8$. Its curved factor is a 4d ellipsoid $\sum_{i=1}^4 \frac{x_i^2}{a^2} + \frac{x_5^2}{b^2} = 1$ with $b < a$. The unit ball $B^3$ is its flat factor, which gives rise to three vanishing added-mass eigenvalues $\mu_6=\mu_7=\mu_8 =0$. Acceleration along the first five coordinates $x_1, \ldots, x_4, x_5$ leads to added-mass eigenvalues $\mu_1 = \ldots = \mu_4 < \mu_5$ since the semi-axes satisfy $a > b$ (higher added-mass when larger cross-section presented).

\item A U(1) gauge theory coupled to a complex scalar with charge $g n$ ($\phi \to e^{i n \tht(x)} \phi$) breaks completely in the above potential $V$, leaving one vector boson with mass $\sqrt{2} g n \eta$. The corresponding rigid body can be regarded as an arc of a circle moving through fluid flowing around the circumference, as in \S~\ref{s:1d-added-mass-effect}.

\item Another illustrative class of theories have $G = $U$(1)^d$ with couplings $g_1, \ldots, g_d$ and $p$ complex scalars in a reducible representation ($p \leq d$ ensures all Goldstone modes are eaten). We assume the scalar $\phi_j$ has charge $q_{jk}$ under the $k^{\rm th}$ U$(1)$ factor and transforms as $\phi_j \to e^{i q_{jk} \tht_k(x)} \phi_j$. They are subject to the potential $\sum_{i=1}^{p} \left(- m_i^2 |\phi_i|^2 + \la_i |\phi_i|^4\right)$. If $\eta_i = (m_i^2/2\lambda_i)^{1/2}$, the vacuum manifold is a $p$-torus, the product of circles of radii $\eta_i$: ${\cal M} = S^1_{\eta_1}\times\ldots\times S^1_{\eta_p}$. There are $p$ Goldstone modes and the mass-squared matrix $M_{ab} = 2 \sum_{j=1}^p \eta_j^2 q_{ja} g_a q_{jb} g_b$ is a sum of $p$ rank-one matrices and generically has $d - p$ zero eigenvalues; $G=U(1)^d$ breaks to $U(1)^{d-p}$. A corresponding ideal rigid body moving in $\mathbb{R}^d$ generalizes the cylinder with elliptical cross-section. It is a product of a (curved) ellipsoid with a (flat) unit ball: $\{ \sum_{i=1}^{p}x_i^2/a^2_i = 1\} \times B^{d-p}$. For pairwise unequal $a_i$, it has distinct non-zero added-mass eigenvalues when accelerated along $x_1, \ldots, x_p$ and none along its $d-p$ flat directions. E.g., a U$(1)^3$ theory with a complex doublet in the above reducible representation breaks to U$(1)$. The corresponding rigid body is a cylinder with elliptical cross-section moving in $\mathbb{R}^3$. On the other hand, with $3$-component complex scalars, U$(1)^3$ completely breaks leaving $3$ massive vector bosons with generically distinct masses. A corresponding ideal rigid body is an ellipsoid moving through fluid filling $\mathbb{R}^3$.

\item It is interesting to identify the rigid body corresponding to electroweak symmetry breaking. Here $G =$ SU$(2)_L \times$ U$(1)_Y$ and $H = $U$(1)_{\rm Q}$ with a massless photon and $m_{W^+} = m_{W^-} < m_Z$. The corresponding rigid body must move through fluid filling $\mathbb{R}^4$, and have principal added-masses $\mu_1 = \mu_2 < \mu_3$, $\mu_4 = 0$. An ideal rigid body generalizes a hollow cylinder. It is the 3d hypersurface $\{\sum_{i=1}^3 x^2_i/a^2_i = 1\} \times [-1,1]$ with $a_1 = a_2 > a_3 > 0$, embedded in $\mathbb{R}^4$. It has an ellipsoid of revolution as cross-section. When accelerated along $x_4$, it displaces no fluid, but has equal added-masses when accelerated along $x_1$ and $x_2$.

\end{enumerate}

More generally, we may associate an ideal rigid body to any pattern $G \to H$ of SSB, through its vector boson mass-squared matrix $M_{ab}$. $M_{ab}$ can always be block diagonalized into a $p \times p$ non-degenerate block (whose eigenvalues $m_1^2, \ldots, m_p^2$ are the squares of the masses of the massive vector bosons) and a $(d-p)\times(d-p)$ zero matrix corresponding to massless photons, where $\dim G = d$ and $\dim H = d-p$. A corresponding ideal rigid body is a product of curved and flat factors. To the non-degenerate part of $M_{ab}$ we associate a `curved' $p-1$ dimensional ellipsoid $\frac{x_1^2}{a_1^2} + \ldots + \frac{x_p^2}{a_p^2} = 1$. The semi-axis lengths $a_i$ are fixed by the vector boson masses. The `flat' factor of the body can be taken as a $(d-p)$ dimensional unit ball $B^{d-p}: \left\{ x_{p+1}, \ldots , x_d \: | \: x_{p+1}^2 + \ldots + x_{d}^2 \leq 1 \right\}$. For $p=d-1$ it is an interval and for $p = d-2$ it is a unit disk etc. Motion along the flat directions $x_{p+1} \ldots x_d$ does not displace fluid, leading to $d-p$ zero added mass eigenvalues while acceleration in the first $p$ directions leads to $p$ non-zero added mass eigenvalues. If the vector boson masses are ordered as $0 < m_1 = m_2 = \ldots = m_{p_1} < m_{p_1 +1} = \ldots = m_{p_1 + p_2} < \ldots < m_{p - p_r + 1} = \ldots = m_p$, then the corresponding semi-axes of the ellipsoid satisfy $0 > a_1 = a_2 = \ldots = a_{p_1} > a_{p_1 +1} = \ldots = a_{p_1 + p_2} > \ldots > a_{p - p_r + 1} = \ldots = a_p$ since the added mass grows with cross-sectional area presented. Here we have allowed for degeneracies among the masses, so that there are $r$ distinct non-zero masses with degeneracies $p_1, \ldots, p_r$ and $p = p_1 + \ldots + p_r$. To find an explicit formula for the semi-axes $a_i$ in terms of the vector boson masses and fluid density $\rho$, we would need to solve the potential flow equations around this rigid body.

\subsubsection{Symmetries of $G/H$ and of the rigid Body}
\label{s-symm-GmodH-and-rigid-body}

In all these examples, the ideal rigid body corresponding to a given pattern of symmetry breaking is a product of curved and flat factors, with added-mass for acceleration along the former. The flat factor could be taken as an interval/disk/ball of dimension $\dim H$. The vacuum manifold ${\cal M} = G/H$ could be endowed with a non-degenerate metric determined by the vector boson mass-squared matrix $M_{ab}$, since in all these examples, the number of Goldstone modes $p = \dim {\cal M}$ is equal to the number of massive vector bosons. $M_{ab}$ is in general degenerate, but may be block diagonalized into a non-degenerate $p \times p$ block and a zero matrix (corresponding to residual symmetries in $\underline{H}$). The non-degenerate part defines a metric $g$ on the quotient $G/H$. $G/H$ is a homogeneous space, so consider any point $m$ and define its `group of symmetries' $\cal G$ as the subgroup of $O(p)$ that fixes the metric at $m$, i.e, $R^t R = I, R^t g R = g$. So $\cal G$ are orthogonal symmetries of the metric in the tangent space $T_m(G/H)$. By homogeneity, $\cal G$ is independent of the chosen point $m$. Then $\cal G$ coincides with the group of rotation and reflection symmetries of the curved factor of the corresponding ideal rigid body. So the group $\cal G$ consists of symmetries of both the vector boson `mass metric' and the Euclidean metric in the flow domain inhabited by the rigid body. Let us illustrate this equality of symmetry groups in the above examples, the results are summarized in Table 1. To identify the group of symmetries in each case, we go to a basis in which the mass metric $g$ at a given point $m$ on $G/H$ is diagonal $g = {\rm diag}(\la_1, \ldots, \la_p)$. The eigenvalues are ordered as 
	\beq
	0 < \la_1 = \ldots = \la_{p_1} < \la_{p_1 + 1} = \ldots = \la_{p_1+p_2} < \ldots < \la_{p-p_{r}+1} = \ldots = \la_{p}
	\eeq
with $p = p_1 + \ldots + p_r$. Then one checks that the subgroup of O$(p)$ that commutes with $g$ is O$(p_1) \times$ O$(p_2) \times \cdots \times$ O$(p_r)$, with O$(1) = \mathbb{Z}_2$.

\begin{enumerate}

\item[(A)] If $G = $SU$(2)$, then ${\cal M} = S^3$ with round metric (all three eigenvalues equal), and the group of symmetries ${\cal G} = $O$(3)$ is maximal. The corresponding ideal rigid body $S^2$ has the same isometry group O$(3)$. 

\item[(B)] If $G= $SO$(p+1)$, ${\cal M} = $S$^{p}$ is round and ${\cal G} =$ O$(p)$, coinciding with the isometry group of the curved factor ${\rm S}^{p-1}$ of the corresponding rigid body.

\item[(C)] Suppose $G = $ U$(1)^d$ with $p$ scalars as above. Then ${\cal M}$ is a $p$-torus, generically with circles of distinct radii $\eta_i$. The symmetry group at $m \in {\cal M}$ is generated by reflections about $m$ along the $p$ circumferences, so ${\cal G} = (\mathbb{Z}_2)^p$. $\cal G$ coincides with the symmetry group (generated by $x_i \to - x_i$) of the ellipsoid factor $\{ \sum_{i=1}^p x_i^2/a_i^2 = 1 \}$ in the corresponding ideal rigid body. If two radii $\eta_1, \eta_2$ coincide, then ${\cal G} = $O$(2) \times (\mathbb{Z}_2)^{p-2}$ which agrees with the symmetries of an ellipsoid of revolution $(x_1^2/a_1^2 + x_2^2/a_1^2 + x_3^2/a_3^2 + \cdots + x_p^2/a_p^2 = 1)$, which is the curved factor of the corresponding ideal rigid body. 

\item[(D)] If $G =$ SU$(2) \times $U$(1)$ of the electroweak standard model, then ${\cal M} = $S$^3$. The metric is not round, as $m_W \ne m_Z$. ${\cal G} = $O$(2) \times \mathbb{Z}_2$, coinciding with the symmetry group of the curved factor of the corresponding rigid body.

\item[(E)] If $G = $SU$(3)$ then ${\cal M} = $S$^5$ with non-round metric and the symmetry group on either side is O$(4) \times \mathbb{Z}_2$ corresponding to the five non-zero added-masses $\mu_1 = \ldots = \mu_4 < \mu_5$.

\end{enumerate}

\begin{sidewaystable}
\tiny
\begin{center}

\title{Table of patterns of SSB and rigid bodies}

\begin{tabular}{| c | c | c | c | c | c | c | c | c | c | c |} \hline
\textbf{Gauge group $G$} & \textbf{Repn.} & \textbf{Potential} & \textbf{Conditions} & \textbf{Vac. Mfld.} & \textbf{residual H} &  \textbf{Vector bosons} & \textbf{Scalars} & \textbf{Fluid} & \textbf{Rigid Body} & $\calG$
\\ \hline

U(1) & 1d cx & $V(\phi)$ & $g n \eta \ne 0$ & $S_\eta^1$ & $\{1\}$ & $m_Z = g n \eta$ & $m_H = \sqrt{2}m$ & $S^1$ & $[\tht_1,\tht_2]$ & $Z_2$
\\ \hline

U(1) & 2d cx & $V_1(\phi_1) + V_2(\phi_2)$ & $g n_i \eta \ne 0$ & $S_{\eta_1}^1 \times S_{\eta_2}^1$ & $\{1\}$ & \begin{tabular}[c]{@{}c@{}}\scriptsize $ m_Z^2 = $\\ \scriptsize $g^2 (n_1^2 \eta_1^2 + n_2^2 \eta_2^2) $ \end{tabular}& \begin{tabular}[c]{@{}c@{}} $m_g = 0, $\\$ m_{H_{1,2}} = \sqrt{2}m_{1,2}$ \end{tabular} & $S^1$ & $ [\tht_1,\tht_2]$ & $Z_2$
\\ \hline

U$(1) \times $U$(1)$ & 1d cx & $V(\phi)$ & $\eta g n_i \ne 0$ & $S_{\eta}^1$ & $U(1)$ & \scriptsize $m_\gamma = 0, m_Z \ne 0$ & $m_H = \sqrt{2}m$ & R$^2$ & Rod & $Z_2$
\\ \hline

U$(1) \times $U$(1)$ & 2d cx & $V_1(\phi_1) + V_2(\phi_2)$ & $n_1 n_4 \ne n_2 n_3$ & $S_{\eta_1}^1 \times S_{\eta_2}^1$ & $\{1\}$ & $Z, Z'$ & $m_{H_{1,2}} = \sqrt{2}m_{1,2}$ & R$^2$ & Elliptical disk & $Z_2 \times Z_2$
\\ \hline

U$(1) \times $U$(1)$ & 2d cx & $V_1(\phi_1) + V_2(\phi_2)$ & $n_1 n_4 = n_2 n_3$ & $S_{\eta_1}^1 \times S_{\eta_2}^1$ & U$(1)$ & $Z, \gamma$ & \begin{tabular}[c]{@{}c@{}} $m_g = 0, $\\$ m_{H_{1,2}} = \sqrt{2}m_{1,2}$ \end{tabular} & R$^2$ & Rod & $Z_2$
\\ \hline

U(1)$^3$ & 1d cx & $V(\phi)$ & $\eta \: {\bf ng} \ne {\bf 0}$ & $S_{\eta}^1$ & U(1) $\times$ U(1) & $\gamma, \gamma',Z$ & $m_{H} = \sqrt{2}m$ & R$^3$ & Elliptical disk & flip over $Z_2$
\\ \hline

U(1)$^3$ & 2d cx & $V_1(\phi_1) + V_2(\phi_2)$ & \begin{tabular}[c]{@{}c@{}}$n_1 n_5 \ne n_2 n_4$ \\or $n_2 n_6 \ne n_3 n_5$ \end{tabular}& $S_{\eta_1}^1 \times S_{\eta_2}^1$ & U(1) & $\gamma$, $Z$, $Z'$ & $m_{H_{1,2}} = \sqrt{2}m_{1,2}$ & R$^3$ & \begin{tabular}[c]{@{}c@{}} Hollow\\ elliptical cylinder\end{tabular} & $Z_2 \times Z_2$ \\ \hline

U(1)$^3$ & 3d cx & \begin{tabular}[c]{@{}c@{}}$V_1(\phi_1) +$\\$V_2(\phi_2) + V_3(\phi_3)$\end{tabular} & \begin{tabular}[c]{@{}c@{}}Mass matrix\\rank $3$\end{tabular}& \begin{tabular}[c]{@{}c@{}}$S_{\eta_1}^1 \times S_{\eta_2}^1$ \\ $\times S_{\eta_3}^1$\end{tabular} & $\{1\}$ & $Z$, $Z'$, $Z''$ & $m_{H_{1,2,3}} = \sqrt{2}m_{1,2,3}$ & R$^3$ & Ellipsoid & $(Z_2)^3$
\\ \hline

SU(2) & 2d cx & $ V(\phi)$ & $g \eta \ne 0$ & $S_{\eta}^3$ & $\{1\}$ & \scriptsize $m_Z = m_{Z'} = m_{Z''}$ & $m_{H} = \sqrt{2}m$ & R$^3$ & Sphere & $O(3)$
\\ \hline

SU$(2)$ & 2d cx & $ m^2 |\phi|^2 + \la |\phi|^4$ & $\eta = 0$ or $g=0$ & $\{ \phi = 0 \}$ & SU$(2)$ & \scriptsize $\gamma, \gamma', \gamma''$ & $4$ Higgs, $m_{H} = m$ & R$^3$ & Point particle & O(3)
\\ \hline

Any G of dim $d$ & $d$d cx & $m^2 |\phi|^2 + \la |\phi|^4$ & $\eta = 0$ or $g=0$ & $\{ \phi = 0 \}$ & G & $\g_1,\ldots,\g_d$ & $2d$ Higgs, $m_H = m$ & R$^d$ & Point particle & O$(d)$
\\ \hline

SO(3) & 3d rl & $ V(\phi)$ & $g \eta \ne 0$ & $S_{\eta}^2$ & SO(2) & \scriptsize $\gamma, m_Z = m_{Z'}$ & $m_{H} = \sqrt{2}m$ & R$^3$ & \begin{tabular}[c]{@{}c@{}} Hollow\\ circular cylinder\end{tabular} & O$(2)$
\\ \hline

$SU(2)_L \times U(1)_Y$ & 2d cx & $ V(\phi)$ & $\eta g \ne 0$ & $S_{\eta}^3$ & U$(1)_{\rm EM}$ & $W^\pm, Z^0, \gamma$ & $m_{H} = \sqrt{2}m$ & R$^4$ & Ellipsoid $\times [a,b]$ & O$(2) \times Z_2$
\\ \hline

$SU(3)$ & 3d cx & $V(\phi)$ & $g\eta \neq 0$ & $S^5_\eta$ & $SU(2)$ & \begin{tabular}[c]{@{}c@{}} $\g_1,\g_2,\g_3$,\\ $m_{Z_1} = \cdots = m_{Z_4}$\\ $< m_{Z_5}$\end{tabular} & $m_H = \sqrt{2}m$ & R$^8$ & \begin{tabular}[c]{@{}c@{}} $\left\{\sum_{i=1}^4 \frac{x_i^2}{a^2} + \frac{x_5^2}{b^2}\right\}$\\ $\times$ 3-ball $B^3$\end{tabular} & $O(4)\times Z_2$
\\ \hline 

\end{tabular}
\end{center}

\caption{Patterns of spontaneous symmetry breaking and corresponding rigid bodies for various gauge groups $G$ and scalar field representations (real - rl or complex - cx). The vacuum manifold $\cal M$, residual symmetry group $H$, fluid flow domain, ideal rigid body and group of symmetries $\cal G$ of the curved factor of the body are listed. The Higgs potential in the case of a point particle is $V = m^2 |\phi|^2 + \lambda |\phi|^4$, while in all other cases $V = -m^2 |\phi|^2 + \lambda |\phi|^4$; $V_i(\phi) = -m_i^2|\phi|^2 + \lambda_i|\phi|^4$, $g_i \ne 0$ are the gauge couplings for the various simple factors in gauge group $G$. These results hold for generic values of charges $q_{ij}$, vev $\eta_i$ and gauge couplings. $S^n_\eta$ denotes an $n$-sphere of radius $\eta$.}\label{SidewaysTable}

\end{sidewaystable}
\normalsize

\subsection{More analogies between the Higgs mechanism and the added mass effect}

\begin{table}[h]
\small
\begin{center}
\begin{tabular}{| p{8cm} | p{8cm} |} \hline
{\textbf{Added-Mass Effect}} & {\textbf{Higgs Mechanism}} \\ \hline
 
Rigid body & Gauge bosons \\ \hline
Fluid & Scalar field \\ \hline
Space occupied by fluid & Gauge Lie algebra \\ \hline

Dimension of container & $\dim G = $ number of gauge bosons \\ \hline

Constant fluid density $\rho$ & vev $\bra \phi \ket$ of Higgs scalar condensate \\ \hline

Direction of acceleration of rigid body & Direction in space spanned by gauge bosons \\ \hline

Added-mass tensor $\mu_{ij}$ & Gauge boson mass matrix $M_{ab}$ \\ \hline

Squares of eigenvalues of added mass tensor $\mu_{ij}$ & Eigenvalues of vector boson mass$^2$ matrix \\ \hline

Acceleration along flat face of rigid body & Massless photon \\ \hline

Zero modes of $\mu_{ij}$. E.g. No added mass when a thin plate is accelerated along flat surface. $m = \#$(independent directions along which accelerated motion gives no added mass) & Zero modes of vector boson mass$^2$ matrix. E.g. massless photons in directions of residual gauge symmetry $H$; $m = \dim(H)$. \\ \hline

Spherical rigid body moving in 3d & SU$(2) \to \{ 1 \}$, doublet; equal-mass gauge bosons \\ \hline

Hollow cylindrical shell in 3d & SO$(3) \to$ SO$(2)$, triplet; 2 equal-mass bosons, photon \\ \hline

Ellipsoid of revolution: semi-axes $r_1 = r_2 > r_3$ & System of $W^{\pm}, Z^0$ bosons with $m_{W^{\pm}} < m_Z$ \\ \hline

Acceleration vector of body moving in $d$ dimensions: no added force component $F_e$ if acceleration component $a_e = 0$ where $\hat e$ is an eigen-direction of $\mu_{ij}$ & Vector of gauge couplings $g_i$ for $G = U(1)^d$: Gauge boson $W_1$ has no added mass if coupling $g_1  = 0$.  \\ \hline

Linear dimensions of rigid body & Charges of scalars $n_i$ under $U(1)$ factors of gauge group $U(1)^d$ \\ \hline

3 ways to have zero added mass: acceleration $a_i \to 0$, $\rho \to 0$ and shrink the body to a point. & 3 ways for gauge group to be unbroken: couplings $g_i \to 0$, vev $\bra \phi \ket \to 0$, make scalars uncharged. \\ \hline

Symmetries of curved body & Symmetries of tangent space $T_m ({G/H})$ at point $m$ \\ \hline

Breaking of fore-aft symmetry of pressure distribution on a sphere when accelerated & Gauge symmetry $G$ spontaneously breaks to $H$ when scalars are charged under $G$ and $\langle \phi \rangle \ne 0$.  \\ \hline

`Benign' flow around body moving uniformly through inviscid fluid & Goldstone mode \\ \hline

Different boundary conditions on body surface & Different (non-minimal) couplings between scalars and gauge fields \\ \hline

Newton's law $F_i - ma_i \; = \; \mu_{ij} \; a_j$ & Proca equation $-j^{\nu} + \partial_{\mu}F^{\mu\nu} \; = \; g^2 \eta^2 A^{\nu}$ \\ \hline

Longest wavelength mode in compressible flow around an accelerated body & Higgs boson - longest wavelength mode of Higgs scalar field \\ \hline

Compressional waves in otherwise constant density flow & Quantum fluctuations around constant strength of Higgs condensate \\ \hline

Expansion in powers of Mach number describing effects of compressibility & Semi-classical loop expansion in powers of $\hbar$ \\ \hline

\end{tabular}
\normalsize

\caption{Analogies between the added mass effect and the Higgs mechanism.}
\label{t:HAM}

\end{center}

\end{table}
\normalsize
We now mention some striking analogies between the added-mass effect and Higgs mechanism. They are summarized in Table~\ref{t:HAM}. The rigid body plays the role of gauge bosons, both can gain mass. The fluid plays the role of the scalar field. When the body is accelerated, some energy goes into the flow. Figuratively, the body carries fluid, adding to its mass. Similarly, gauge bosons gain mass by carrying Goldstone modes. The analogy relates the space of fluid flow, to the Lie algebra $\underline{G}$ (the location of the body provides an origin for the flow domain and it is the space of directions in which the body can move that corresponds to the gauge Lie algebra). The dimension $d$ of the fluid container ($\mathbb{R}^d$ for simplicity) equals the number $\dim \underline{G}$ of gauge bosons. The added-mass tensor $\mu_{ij}$ and the vector boson mass-squared matrix $M_{ab}$ are both $d \times d$ matrices. A direction of acceleration relative to the body is equivalent to a direction in $\underline{G}$. Just as accelerating the body in different directions could result in different added masses, various directions in the Lie algebra could correspond to gauge bosons with possibly different masses. Zero modes of $\mu_{ij}$ are directions in which the acceleration reaction force vanishes. These are like directions of residual symmetry in the Lie algebra $\underline{H}$. A thin disk accelerated along its surface gains no added-mass when moving in a 3d fluid, just as we have a massless photon along an unbroken symmetry generator of $G$. In general, the number of flat directions of the body is equal to the number of massless vectors.

Given a pattern of gauge boson masses, one may associate with it a rigid body moving through an ideal fluid. We say that a particular spontaneous symmetry breaking pattern {\it corresponds} to a particular rigid body if the vector boson masses coincide with the added-mass eigenvalues. The latter do not, generally, determine the body. A sphere and cube of appropriate sizes have identical added-mass eigenvalues\footnote{This is similar to how a cube and a sphere of appropriate sizes can both be spherical tops with the same principal moments of inertia.}, just as appropriate SU$(2)$ and U$(1)^3$ gauge theories share vector boson mass spectra. So the correspondence, at this level, relates a class of classical gauge theories to a family of rigid bodies. Among these rigid bodies with a given added mass tensor, there is an `ideal' one with maximal symmetry group. Similarly, we have the notion of an ideal SSB pattern being the `simplest' among all SSB patterns corresponding to a given mass-squared matrix as discussed in \S\ref{s:ideal-SSB-pattern}. The identification of $\underline{G}$ with the space of fluid flow, allows us to relate symmetries of the `mass' metric at any point of $G/H$ to those of the curved factor of the corresponding ideal rigid body (see \S \ref{s-symm-GmodH-and-rigid-body}).

Consider a bounded rigid body that moves at constant velocity through an infinite, inviscid, incompressible, irrotational potential flow without the formation of vortex sheets, wakes or cavities. It feels no added-mass (this is part of d'Alembert's `paradox' \cite{Batchelor}). However, it is associated with a `benign' flow not requiring energy input. For example, the flow field around a uniformly moving sphere of radius $a$, instantaneously centered at $U t \hat z$ is
	\beq
	\bfv(\bfr,t) = \frac{a^3 U}{2r'(t)^3} \left[2\cos \tht'(t) \,\hat \bfr'(t) + \sin \tht'(t) \, \hat\tht'(t) \right]
	\label{e:vel-around-sphere}
	\eeq
where $\bfr' = \bfr - U t \hat z$ is the position vector of the observation point relative to the center of the sphere. So a body moving steadily is not coupled to the fluid through energy exchange. Similarly, if the scalar vacuum expectation value $\bra \phi \ket$ is non-zero but the gauge coupling $g$ is zero, then we have spontaneous symmetry breaking and Goldstone modes, but massless gauge bosons. The Goldstone modes are analogous to the above benign flow.

Is there a broken symmetry in the added-mass effect? When a sphere moves uniformly, from (\ref{e:vel-around-sphere}) and Bernoulli's equation (\ref{e-bernoulli-eqn}), the pressure distributions on the front and rear hemispheres are identical. This front-back symmetry is broken upon accelerating the sphere. It is a discrete analog of the broken gauge symmetry. Moreover, spontaneous symmetry breaking is caused by a non-zero vacuum expectation value $|\bra \phi \ket| = \eta$. The density $\rho$ is its counterpart. Both occur as pre-factors in mass matrices ($\mu^{\rm sphere}_{ij} \propto \rho a^3 \del_{ij}$, $M^{{\rm SU(2)}}_{ab} \propto \eta^2 g^2 \del_{ab}$) and are exclusively properties of the fluid and scalar field (i.e., not having to do with the rigid body or gauge fields).

Our analogy extends to the dynamical equations of the body ($F_i - ma_i = \mu_{ij} a_j $) and massive vector boson ($-j^{\nu} + \partial_{\mu}F^{\mu\nu} = g^2 \eta^2 A^{\nu}$). The added-mass $\mu_{ij}a_j$ is like the Proca mass. The external force $F_i$ and current $j^\nu$ are both sources in otherwise homogeneous equations. $\partial_{\mu}F^{\mu\nu} = 0$ is the analog of $m a_i = 0$: free propagation of electromagnetic waves is like uniform motion of a rigid body. The impenetrable body-fluid boundary condition is analogous to gauge-scalar minimal coupling. Other boundary conditions would correspond to non-minimally coupled scalars.

From the spontaneously broken U$(1)^d$ models of \S \ref{s:ssb-patters-rigid-bodies}, we obtain further analogies. There are $3$ ways to prevent spontaneous gauge symmetry breaking: (a) set the gauge couplings $g_i$ to zero, (b) make the scalars uncharged ($n_{ij} \to 0$) under U$(1)^d$ and (c) let the scalar vacuum expectation value $\eta \to 0$. Similarly, there are $3$ ways to make the added force/mass vanish: (a) set the acceleration components $a_i = 0$, (b) shrink the body to a point and (c) let fluid density $\rho \to 0$.

Just as an accelerating body `carries' a flow and acquires an added mass, it is as if the $W$ boson `carries' the Goldstone mode and becomes massive. Remarkably, the Higgs particle has a simple interpretation in this analogy. Indeed, the Higgs particle which is the longest wavelength mode of oscillation of the Higgs field, may be thought of as analogous to a long-wavelength wave\footnote{We would like to clarify that we do not relate the movement of the vector boson in physical space to movement of a rigid body in a fluid. Instead we relate directions of acceleration of the body in the fluid to directions in the gauge Lie algebra. We are not suggesting a direct correspondence between the fluid waves produced around a rigid body and the particles radiated by an accelerated gauge boson.

Among the fluids modes generated by a moving rigid body, we believe it is natural to consider the longest wavelength mode to correspond to the Higgs, which is the lightest scalar particle. However, the wavelengths of fluid modes are not related to the mass of the moving body, but depend on its shape and motion. This is different from the situation in particle physics: an accelerated vector boson or fermion cannot radiate particles more massive than itself. However, we speculate that just as there could be shorter wavelength modes in a fluid, there could also be scalar particles heavier than the Higgs. But we are not suggesting that these heavier modes are excited by the acceleration of vector bosons. They could be created by other means like collision of particles, etc.} in the fluid around an accelerated body. More generally, quantum fluctuations around the scalar vacuum expectation value are analogous to density fluctuations (e.g. sound waves) around constant density flow. In fact, the semi-classical `loop expansion' in powers of Planck's constant ($\hbar$) that accounts for quantum fluctuations is analogous to an expansion in powers of the Mach number that is used to describe effects of compressibility around the limit of incompressible (constant density) flow.

\chapter[Dispersive regularization of gas dynamics]{Nonlinear dispersive regularization of inviscid gas dynamics}

The motivation for studying a conservative regularization of gas dynamics was given in the Introduction (Chapter \ref{s:gas-dyn-intro}). This chapter is based on \cite{r-gas-dynamics}.

\section[Hamiltonian and Lagrangian formulations]{Hamiltonian and Lagrangian formulations of 3d R-gas dynamics}
\label{s:3d-hamil-form-R-gas-dyn}

It is well-known \cite{whitham} that adiabatic dynamics of an ideal gas with constant specific heat ratio $\g = c_p/c_v$ is governed by the continuity, momentum and internal energy equations 
	\beq
	\rho_t + \grad \cdot (\rho \bfv) = 0, \quad
	(\rho v_i)_t + \pdr_j \left( p \del_{ij} + \rho v_i v_j \right) = 0 \quad \text{and}\quad
	\left( \frac{p}{\g - 1} \right)_t + p \grad \cdot \bfv  + \grad \cdot\left( \frac{p \bfv}{\g - 1}\right) = 0,
	\label{e:ideal-mass-mom-int-egy-3d}
	\eeq
with the temperature in energy units given by $T = m p/\rho$ for a molecular mass $m$. In adiabatic flow, specific entropy (per unit mass) is advected ($D_t s \equiv \pdr_t s +  \bfv \cdot \grad s = 0$), while the entropy per unit volume is locally conserved, $\pdr_t (\rho s) + \grad \cdot (\rho s \bfv) = 0$. Though the terms `reversibly adiabatic' and `isentropic' are often used interchangeably, in this thesis we use adiabatic for $D_t s = 0$ and isentropic for the special case where $s$ is a global constant. For adiabatic flow, $\rho$ and $p$ may be taken as independent variables with $s$ being a function of them. For a polytropic gas, $s = c_v \log\left( (p/\bar p)/ (\rho/ \bar \rho)^\g \right)$ where $\bar p, \bar \rho$ are reference values. These equations follow from the Hamiltonian
	\beq
    H_{\rm ideal} = \int \left[ \half \rho \bfv^2 + \frac{p}{\g-1} \right] d\bfr
    \label{e:3d-hamiltonian-ideal}
	\eeq
and Hamilton's equations $\dot f = \{ f, H \}$ using the (non-zero) non-canonical Poisson brackets (PB) \cite{landau, morrison-greene}
	\beq
	\{ \bfv(\bfx), \rho(\bfy) \} = \grad_y \del(\bfx - \bfy), \;\;
	\{ \bfv (\bfx), s(\bfy) \} = \frac{\grad s}{\rho} \del ( \bfx - \bfy) \quad
	\text{and} \quad \{ v_i(\bfx), v_j(\bfy) \} = \frac{\eps_{ijk} w_k}{\rho} \del (\bfx - \bfy).\quad
	\label{e:PB-3d}
	\eeq
where $\bfw = \grad \times \bfv$ is the vorticity. Our conservative regularization involves adding a density gradient term to the Hamiltonian while retaining the same PBs:
	\beq
    H = \int {\cal E} \; d\bfr \equiv  \int \left[ \half \rho \bfv^2 + \frac{p}{\g-1} + \frac{\beta_*}{2} \frac{(\grad \rho)^2}{\rho} \right] d\bfr.
    \label{e:3d-hamiltonian}
	\eeq
The density gradient energy, which could arise from capillarity \cite{vdW,korteweg}, has been chosen $\propto (\grad \rho)^2$ to ensure positivity, parity conservation and to prevent discontinuities in density, so as to conservatively regularize shock-like discontinuities. It involves the capillarity coefficient $\beta(\rho) = \beta_*/\rho$, where $\beta_*$ is a constant with dimensions $L^4 T^{-2}$. $\beta_*$ can be taken as $\la^2 c^2$ where $\la$ is a short-distance cut-off and $c$ a typical speed. This is the simplest form for $\beta(\rho)$ that ensures the mass, momentum and energy equations are all in conservation form for the ideal mass and momentum densities. It also leads to other nice properties such as a transformation to the NLSE for isentropic potential flow. 

The continuity and entropy equations following from (\ref{e:3d-hamiltonian}) and (\ref{e:PB-3d}) are as in the ideal model. The momentum and consequently the velocity equation however, differ due to the presence of a capillary force term $\beta_* \bfF$:
	\beqs
	\bfv_t + \bfv \cdot \grad \bfv + \frac{\grad p}{\rho} 
	&=& \beta_* \bfF
	= \beta_* \grad \left[ \half \frac{(\grad \rho)^2}{\rho^2} + \grad \cdot \left( \frac{\grad \rho}{\rho} \right) \right] \cr
	&=& \beta_* \grad \left[ \frac{\grad^2 \rho}{\rho} - \half \frac{(\grad \rho)^2}{\rho^2}\right] 
	= 2 \beta_* \grad \left( \frac{\grad^2 \sqrt{\rho}}{\sqrt{\rho}} \right).
	\label{e:3d-vel-eqn}
	\eeqs
Remarkably, $\beta_* \bfF = \beta_* \grad \Phi$ is a gradient, so that for barotropic flow ($\grad p/\rho = \grad h$), it augments the specific enthalpy $h \to h + \beta_* \Phi$. Thus, the vorticity evolves exactly as in ideal gas dynamics (in other words, we only regularize the `potential' part of the velocity and don't deal with vortical singularities as in \cite{thyagaraja, govind-sonakshi-thyagaraja-pop}). Thus, Kelvin's theorem would apply in R-gas dynamics, unchanged.
The momentum and velocity equations may be expressed in terms of a regularized stress tensor: 
	\beqs
	&& \pdr_t (\rho v_i) + \pdr_j \left( \rho v_i v_j + \sig_{ij} \right) = 0\quad
	\text{and} \quad \pdr_t v_i + v_j \pdr_j v_i = - \ov{\rho} \pdr_j \sig_{ij} \cr
	&&\text{where}\quad \sig_{ij} = p \, \del_{ij} + \beta_* \left( \frac{(\pdr_i \rho) (\pdr_j \rho)}{\rho}  - \pdr_i \pdr_j \rho \right).
	 \label{e:stress-R-gas-dyn-3d}
	\eeqs
The scalar part of $\sig$ defines a regularized pressure $p_*$ which includes the Gross `quantum pressure' \cite{gross}:
	\beq 
	p_* = \ov{d} \tr \sig = p + \frac{\beta_*}{d} \left( \frac{(\grad \rho)^2}{\rho} - \grad^2 \rho \right) \;\; \text{where} \;\; d=3.
	\label{e:pstar-3d}
	\eeq
The energy equation for the energy density $\cal E$ defined in (\ref{e:3d-hamiltonian}) is given by:
	\beq
	{\cal E}_t + \grad \cdot \left( \frac{\rho \bfv^2}{2} \bfv + \frac{\g}{\g - 1} p\bfv\right) +
	\beta_* \grad \cdot\left[\frac{\grad \rho}{\rho} \grad \cdot (\rho \bfv) - \rho \bfv \grad \cdot \left( \frac{\grad \rho}{\rho} \right)  - \frac{\rho \bfv}{2}\frac{ (\grad \rho)^2}{\rho^2}  \right] = 0.
	\label{e:energy-eqn-3d}
	\eeq
The fact that (\ref{e:energy-eqn-3d}) is in local conservation form follows from the PB formulation. Indeed, $\{ H, H \} = 0$ implies that ${\cal E}_t = \{ {\cal E}, H \}$ must be a divergence. The internal energy per unit volume is therefore
	\beq
	\rho \varepsilon_* = \rho \varepsilon + \frac{\beta_*}{2} \frac{(\grad \rho)^2}{\rho}
	\;\; \text{where} \;\:\; \varepsilon = \frac{p}{\rho(\g - 1)} = \frac{T}{(\g - 1) m}.
	\label{e:spec-int-energy}
	\eeq
These regularization terms in the pressure, enthalpy and internal energy depend upon density gradients and are therefore not strictly thermodynamic properties of the gas, any more than the regularized stress tensor. They are conservative analogues of the viscous stress tensor which depends on velocity gradients in dissipative gas dynamics.

Interestingly, the potential $\Phi$ in (\ref{e:3d-vel-eqn}) is also the Bohm potential $U$ \cite{Bohm-1} that arises as a correction to the classical potential $V$ in the quantum-corrected Hamilton-Jacobi equation for the Schr\"odinger wavefunction $\psi = \sqrt{\rho} e^{i S/\hbar}$: \small
	\beqs
	\rho_t + \grad \cdot \left( \rho \frac{\grad S}{m} \right) &=& 0 \quad \text{and} \quad 
	S_t + \frac{(\grad S)^2}{2m}+ V + U = 0 \cr
	\text{where} \quad U &=& - \frac{\hbar^2}{2m} \frac{\grad^2 \sqrt{\rho}}{\sqrt{\rho}} = - \frac{\hbar^2}{4 m} \left( \frac{\grad^2 \rho}{\rho} - \half \frac{(\grad \rho)^2}{\rho^2}  \right).
	\eeqs \normalsize
Our regularized stress $\sig$ also resembles the Korteweg stress $\sig^{\rm Kor}$ of Ref. \cite{korteweg,gorban-karlin}. Indeed, if $\beta = \beta_*/\rho$,
	\beqs
	\sig^{\rm Kor}_{ij} 
	&=&  p \del_{ij} - \rho \left[\pdr_k \left( \beta(\rho) \pdr_k \rho \right) \right] \del_{ij} + \beta(\rho) \pdr_i \rho \pdr_j \rho \cr
	&=& p \del_{ij} -\beta_* \left[ \left( \grad^2 \rho - \ov{\rho} (\grad \rho)^2 \right) \del_{ij} -  \frac{\pdr_i \rho \; \pdr_j \rho}{\rho} \right]. \quad
	\label{e:korteweg-stress}
	\eeqs
However, though $\sig^{\rm Kor}_{ij}$ has the term $({\beta_*}/{\rho}) (\pdr_i \rho) (\pdr_j \rho)$ in common with $\sig_{ij}$ (\ref{e:stress-R-gas-dyn-3d}), they are not quite equal. Thus, though the qualitative physical features of our equations may be similar to those of the Korteweg equations, ours additionally possess some remarkable mathematical properties facilitating the analysis in this thesis.

Finally, if the flow domain is all of $\mathbb{R}^3$, then $\beta_*$ can be scaled out by defining ${\bf R} = {\bf r}/\sqrt{\beta_*}$ and $T = t/\sqrt{\beta_*}$, just as we may eliminate the dispersion coefficient in KdV on the whole real line. By contrast, in the presence of a characteristic length scale $l$, $\beta_*$ {\it cannot} be scaled out and $l/\la$ serves as a conservative analogue of the Reynolds number.

\subsection{Lagrangian formulation via Clebsch variables}

To obtain a Lagrangian for R-gas dynamics, we use the Clebsch representation \cite{clebsch,zakharov-kuznetsov}  $\bfv = \grad \phi + (\la \grad \mu + \al \grad s)/\rho$. The PBs in (\ref{e:PB-3d}) are recovered by postulating canonical PBs among Clebsch variables:
    \beq
    \{\rho(\bfr), \phi(\bfr') \} = \{\alpha(\bfr), s(\bfr') \} = \{\la(\bfr), \mu(\bfr') \} = \del(\bfr - \bfr').
    \eeq
The Hamiltonian density in terms of Clebsch variables:
    \beq
    \mathcal{H} = \frac{\rho}{2} \left( \grad \phi + \frac{(\la \grad \mu + \al \grad s)}{\rho} \right)^2 + \rho \,\varepsilon(\rho, s) + \frac{\beta_*}{2} \frac{(\grad \rho)^2}{\rho}
    \eeq
where $\varepsilon(\rho, s)$ is the ideal specific internal energy (\ref{e:spec-int-energy}). The EOM (\ref{e:3d-vel-eqn}) follow as the Euler-Lagrange equations (EL) for the Bateman-Thellung \cite{bateman,thellung} Lagrangian density linear in velocities \cite{sudarshan-mukunda} augmented by the density gradient energy: 
    \beq
    \mathcal{L}_1 \; = \; \rho_t \phi + \la_t \mu + \alpha_t s - {\cal H}.
    \label{e:non-barotropic-lag}
    \eeq
The EL equations for $\al$ and $\la$ imply the advection of $s$ and $\mu$, while that for $\phi$ is the continuity equation and that for $s$ and $\mu$ are the evolution equations $\al_t + \grad \cdot (\al \bfv) = \rho T$ and $\la_t + \grad \cdot (\la \bfv) = 0$. The regularization only affects the EL equation for $\rho$. Upon using $p = \rho^2 \partial \varepsilon /\partial \rho$, it becomes the time-dependent Bernoulli equation for adiabatic R-gas dynamics:
    \beq
    \phi_t - \frac{\bfv^2}{2} + \bfv \cdot \grad \phi + \varepsilon(\rho, s) + \frac{p}{\rho} -\beta_* \left( \frac{\grad^2 \rho}{\rho} - \half \frac{(\grad \rho)^2}{\rho^2}\right) = 0.
    \eeq
Using these, one obtains (\ref{e:3d-vel-eqn}) for $\bfv$. There are of course related Lagrangians for the same EOM, e.g.,
	\beqs
	\mathcal{L}_2 &=& -\rho \phi_t - \la \mu_t - \alpha s_t - \mathcal{H} \quad \text{and} \cr
	\mathcal{L}_3 &=& \rho \left(\frac{\bfv^2}{2} - \varepsilon \right)
	- \frac{\beta_*}{2} \frac{(\grad \rho)^2}{\rho} + \phi \left( \rho_t + \grad \cdot (\rho \bfv) \right)  - \la \frac{D \mu}{Dt} - \al \frac{D s}{Dt}.
	\label{e:alternate-lagrangians}
	\eeqs
Thus, we may interpret $\phi,\la$ and $\al$ as Lagrange multipliers enforcing the EOM for $\rho, \mu$ and $s$.

\section{Formulation of 1D regularized gas dynamics}
\label{s:formulation-r-gas-dynm}

\subsection{Hamiltonian and equations of motion}

In what follows, we will primarily be interested in 1d adiabatic R-gas dynamics where $\rho, s$ and $p$ are independent of two of the Cartesian coordinates and $\bfv = (u(x,t),0,0)$. The non-zero PBs (\ref{e:PB-3d}) simplify as $\bfw = 0$: $\{u , u \} = 0$ and 
	\beq
	\{ u(x), s(y) \} = \frac{s'}{\rho} \del(x-y) \;\; \text{and} \;\;  \{ \rho(x), u(y) \} = \pdr_y \del(x-y).
	\label{e:PB-u-rho-s}
	\eeq
The total mass $(\int \rho \: dx)$, entropy  $(\int \rho \, s \: dx)$ and more generally $\int \rho \Sigma(s) dx$ for any $\Sigma(s)$ are Casimirs of this algebra. As before, the dynamics is generated by a Hamiltonian that involves a capillary energy
	\beq
	H = \int \left[\half \rho u^2 + \frac{p}{\g - 1} + \half \beta(\rho) \: \rho_x^2 \right] dx,
	\label{e:beta-reg-hamiltonian}
	\eeq
where $\beta(\rho)$ will be chosen by requiring that the momentum equation be in conservation form. The continuity and entropy equations are as in the ideal model:
	\beq
	\rho_t + (\rho u )_x = 0, \;\;
	s_t + u s_x = 0 \;\; \text{with} \;\;
	s = c_v \log\left( \frac{p \bar \rho^\g}{\bar p \rho^\g} \right).
	\label{e:cont-entropy-eqn} 
	\eeq
Thus, even with our regularization we continue to have $D_t p = c_s^2 D_t \rho$ where $c_s^2 = (\pdr p/\pdr \rho)_s = \g p/\rho$.
The regularized momentum  and velocity equations are
	\beqs
	(\rho u)_t + (\rho u^2 + p)_x &=& \rho \left[ \left( \beta \rho_x  \right)_x - \half \beta' \rho_x^2 \right]_x
	\quad
	\text{and} \cr
	u_t + u u_x &=& - \frac{p_x}{\rho} + \left[ \left( \beta \rho_x  \right)_x - \half \beta' \rho_x^2 \right]_x.
	\label{e:vel-mom-r-gas-dynm}
	\eeqs
The simplest way for the former to be in conservation form is for the momentum density to equal $\rho u$ and for the regularization term to be a divergence. $\beta = \beta_*/\rho$ is the simplest capillarity coefficient that ensures this, giving 
	\beqs
	(\rho u )_t + \left[ \rho u^2 + p - \beta_* \left(\rho_{xx} - \frac{\rho_x^2}{\rho} \right)\right]_x &=& 0
	\quad \text{and} \cr
	\frac{Du}{Dt} + \frac{p_x}{\rho} 
	= \beta_* f 
	=  \beta_* \left[ \frac{\rho_{2x}}{\rho} - \frac{\rho_x^2}{2 \rho^2} \right]_x 
	&=& 2 \beta_* \left[ \frac{(\sqrt{\rho})_{xx}}{\sqrt{\rho}}\right]_x. \quad
	\label{e:reg-vel-eqn-gas}
	\eeqs 
We note that the apparently simpler choice of constant $\beta$ leads to a KdV-like $\rho_{xxx}$ term in the velocity equation but prevents the momentum equation from being in conservation form. Our regularization amounts to modifying the  pressure $p \to p_*$ in the momentum and velocity equations:
	\beqs
	(\rho u)_t + (\rho u^2 + p_*)_x &=& 0
	\quad \text{and} \quad 
	u_t + u u_x = - \frac{{p_*}_x}{\rho}
	\cr
	\text{where} \quad p_* &=& p - \beta_* (\rho_{xx} - \rho_x^2/\rho).
	\label{e:reg-vel-mom-eqn-p_*}
	\eeqs
It is instructive to compare our velocity equation with Korteweg's. For capillarity coefficient $\beta = \beta_*/\rho$, the 1d Korteweg velocity equation following from (\ref{e:korteweg-stress}) is
	\beq
	u_t + u u_x + \frac{p_x}{\rho} = \beta_* f^{\rm Kor} = \frac{\beta_*}{\rho} \left( \rho_{xx} -  \frac{2 \rho_x^2}{\rho}\right)_x.
	\eeq	
Unlike our force per unit mass $\beta_* f$ which is a gradient, $\beta_* f^{\rm Kor}$ is not. Thus, in the barotropic case of \S\ref{s:barotropic-r-gas-dyn} where $p_x/\rho = h_x$, our velocity equation (\ref{e:reg-vel-eqn-gas}) (but not Korteweg's) comes into conservation form as in ideal gas dynamics. Finally, our energy equation is also in local conservation form,
	\beq
	\left[ \half \rho u^2 + \frac{p}{\g -1} +  \frac{\beta_*}{2} \frac{\rho_x^2}{\rho} \right]_t + \left[ \half \rho u^2 u + \frac{\g}{\g - 1} pu\right]_x
	+ \: \beta_* \left[ \frac{\rho_x}{\rho} (\rho u)_x - \rho u \left[ \frac{\rho_x}{\rho} \right]_x  - \half \frac{u \rho_x^2}{\rho} \right]_x = 0.
	\label{e:reg-total-energy-PB}
	\eeq
It takes a compact form in terms of a regularized specific internal energy $\varepsilon_*$ and enthalpy $h_*$: 
	\beqs
	&& \left( \half \rho u^2 + \rho \varepsilon_* \right)_t + \left( \rho \left(\half u^2 +  h_* \right) u + \beta_* u_x \rho_x  \right)_x = 0
	\cr
	&& \text{where} \quad \rho \varepsilon_* = \frac{p}{\g -1} + \frac{\beta_*}{2} \frac{\rho_x^2}{\rho} 
	\;\; \text{and} \;\;
	\rho h_* = \rho \varepsilon_* + p_*.
	\label{e:reg-total-energy-eps_*-enthalpy-PB}
	\eeqs 
The internal energy equation may be interpreted as the $1^{\rm st}$ law of thermodynamics for adiabatic flow:
	\beq
	D_t \varepsilon_* + p_*  D_t \left( \frac{1}{\rho} \right) + \frac{\beta_*}{\rho} (u_x \rho_x)_x = T D_t s = 0.
	\label{e:reg-1st-law-adiab-specific}
	\eeq
Evidently, the gas does work against the pressure $p_*$ as well as a new type of reversible, non-dissipative work due to the regularization while ensuring that the specific entropy $s$ is constant along the flow.


The action corresponding to ({\ref{e:non-barotropic-lag}}) possesses three obvious symmetries: (a) constant shift in $\phi$, (b) space translation and (c) time translation, leading to the local conservation laws for mass $(\rho)$, momentum $(\rho u)$ and energy $(\half \rho u^2 + \rho \varepsilon_*)$ densities (\ref{e:cont-entropy-eqn}, \ref{e:reg-vel-eqn-gas}, \ref{e:reg-total-energy-PB}). In addition, under an infinitesimal Galilean boost ($t \to t$, $x \to x-ct$), the fields transform as
    \beqs
    \del\phi &=& c(t\phi_x - x),\quad \del u = c(t u_x -1) \quad \text{and} \cr 
    \del \Upsilon &=& ct\Upsilon_x \quad \text{for} \quad \Upsilon = \rho, \al, s, \la \quad \text{and} \quad \mu,
    \eeqs 
leading to a change in the Lagrangian (\ref{e:alternate-lagrangians}) by a spatial derivative $\del \mathcal{L}_2 = ct \left( p - \beta_* \rho_{xx} \right)_x$. The corresponding Noether charge and flux densities are
    \beq
    j^t = \sum_{\chi} \dd{\mathcal{L}_2}{\dot \chi} \del\chi \; \; \; \text{and} \;\; \; j^x = \sum_\chi \dd{\mathcal{L}_2}{\chi_x}\del\chi - ct (p-\beta_* \rho_{xx}).
    \eeq
Here, we sum over $\chi = \rho, \phi, \al, s, \la$ and $\mu$. The resulting conservation law $\pdr_t j^t + \pdr_x j^x = 0$ or $(\rho (x - t u) )_t + ( x \rho u - t F^{\rm p})_x = 0$ involves an explicitly time-dependent Galilean charge and flux, where $F^{\rm p} = \rho u^2 + p - \beta_* \rho \left(\log \rho \right)_{xx}$ is the regularized momentum flux (\ref{e:reg-vel-eqn-gas}). Thus, $G = \int (x - tu) \, \rho \, dx$ is conserved even though $\{ G, H \} = P$ where $P = \int \rho u \, dx$ is the total momentum. $P$, $G$ and $H$ satisfy a 1d Galilei algebra with the total mass $M$ furnishing a central extension: $\{ G, P \} = M$.

\subsection{Isentropic R-gas dynamics of a polytropic gas}
\label{s:barotropic-r-gas-dyn}

Sans entropy sources/sinks and boundaries, one is mainly interested in cases where $s = \bar s$ is initially constant and by (\ref{e:cont-entropy-eqn}), independent of time. Thus, we consider isentropic flow where $p$ and $\rho$ satisfy the barotropic relation
	\beq
	p = (\g - 1) K \rho^\g \quad \text{where} \quad
	K = \frac{e^{\bar s/c_v}}{\g - 1} \frac{\bar p}{\bar \rho^\g} \: > \:  0
	\label{e:entropy-barotropic}
	\eeq
is a constant that encodes the constant value of entropy and labels isentopes. A feature of isentropic flow is that in addition to the continuity, momentum and energy equations, the velocity equation is {\it also} in conservation form:
	\beq
	\rho_t + F^{\rm m}_x = 0, \quad
	(\rho u)_t + F^{\rm p}_x = 0,
	\left[ \half \rho u^2 + \rho \eps + \frac{\beta_*}{2} \frac{\rho_x^2}{\rho} \right]_t + F^{\rm e}_x = 0 \;\;\;\text{and} \;\;\; u_t + F^{\rm u}_x = 0.
	\label{e:unsteady-barotropic-eqns}
	\eeq
Here $\varepsilon = K \rho^{\g-1}$ and $h = \g K \rho^{\g-1}$ are specific internal energy and enthalpy. The corresponding fluxes are 
	\beqs
	F^{\rm m} &=& \rho u, \quad F^{\rm p} = \rho u^2 + p - \beta_* \left( \rho_{xx} - \frac{\rho_x^2}{\rho} \right), \cr
	F^{\rm e} &=&\left(\frac{u^2}{2} + h \right) \rho u + \; \beta_* \left( \frac{\rho_x}{\rho} (\rho u)_x - \rho u \left( \frac{\rho_x}{\rho} \right)_x  - \frac{u \rho_x^2}{2\rho} \right) \quad \text{and} \cr
	F^{\rm u} &=& \half u^2 + h - \beta_* \left( \frac{\rho_{xx}}{\rho} - \half \frac{\rho_x^2}{\rho^2} \right).
	\label{e:barotropic-curr}
	\eeqs
In ideal gas dynamics $F^{\rm u} = F^{\rm e}/F^{\rm m}$, but no such algebraic relation holds when $\beta_* \ne 0$. The emergence of a $4^{\rm th}$ conservation law in the isentropic case is tied to the global constancy of entropy. These equations follow from the degenerate Landau PBs $\{ \rho, \rho \} = \{ u, u \} = 0$ and $\{ \rho(x), u(y) \} = \pdr_y \del(x-y)$ whose Casimirs include $M = \int \rho \: dx$ and $\int u \: dx$. These PBs become canonical $[\{ \rho(x), \phi(y) \} =\del (x - y)]$ upon introducing a velocity potential $u(y) = \phi_y$. The corresponding EOM follow the Lagrangian ${\cal L}_1 = \rho_t \phi - {\cal H}$ where ${\cal H} = \half \rho \phi_x^2 + \rho \varepsilon(\rho) + \half \beta_* \rho_x^2/\rho$. As above, the local conservation laws for mass, momentum, energy and Galilei charge follow from Noether's theorem. However, the conservation law $u_t + F^{\rm u}_x = 0$ (\ref{e:unsteady-barotropic-eqns}) does not arise from a symmetry via Noether's theorem. This is because $C = \int u \: dx$ is a Casimir, it acts trivially on all observables: $\del \phi = \{ C, \phi \} = 0$, etc.


\section{Dispersive sound, steady and traveling waves}
\label{s:dispersive-sound-steady-trav}

\subsection{Dispersive sound waves}
\label{s:dispersive-sound-waves}

To discuss sound waves it is convenient to nondimensionalize the variables in (\ref{e:cont-entropy-eqn}) and (\ref{e:reg-vel-mom-eqn-p_*}):
	\beq
	x = l \hat x,  \quad
	t = \frac{l}{\bar c} \hat t, \quad
	\rho = \bar \rho \hat \rho, \quad p = \bar p \hat p, \quad {\bar c}^2 = \frac{\g \bar p}{\bar \rho},
	\quad u = \bar c \hat u, \quad
	\hat s = \frac{s}{c_v} = \log\left(\frac{\hat p}{\hat \rho^\g} \right).
	\label{e:non-dim-var}
	\eeq
Here, $l$ is a macroscopic length. The nondimensional (hatted) variables satisfy
	\beq
	\hat \rho_{\hat t} + (\hat \rho \hat u)_{\hat x} = 0, \quad
	\hat s_{\hat t} + \hat u \hat s_{\hat x} = 0 \quad \text{and}\quad
	\hat u_{\hat t} + \hat u \hat u_{\hat x} = - \ov{\g} \frac{\hat p_{\hat x}}{\hat \rho} + \frac{\eps^2}{\hat \rho} \left( \hat \rho_{\hat x \hat x} - \frac{\hat \rho_{\hat x}^2}{\hat \rho} \right)_{\hat x}.
	\label{e:non-dim-eoms}
	\eeq
Here $\beta_* = {\bar c}^2 \la^2$ where $\la$ is a regularization length and $\eps = \la/l$ its nondimensional version.

\begin{figure}
	\begin{center} 
 \includegraphics[height = 4cm]{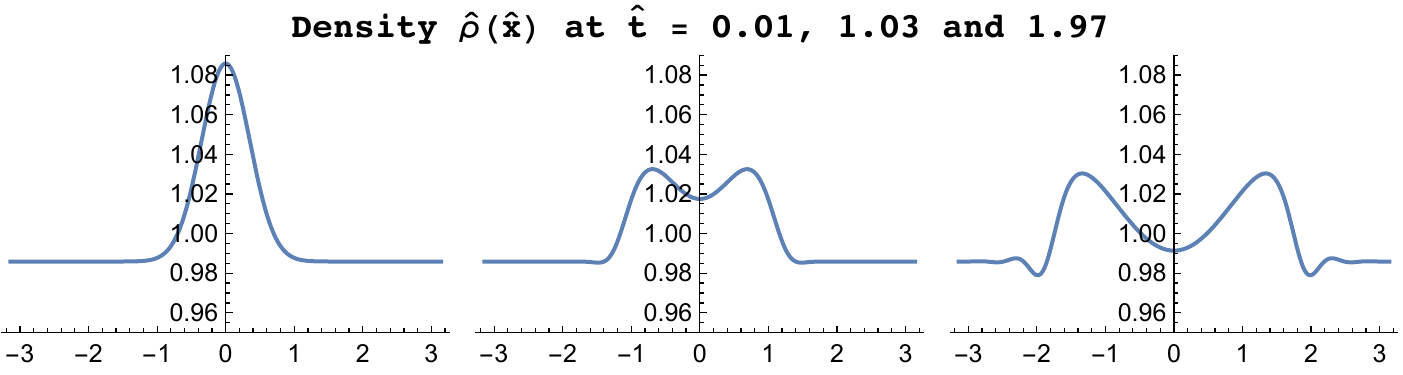} 
 \caption{R-gas dynamic evolution of a pulse ($\hat \rho(\hat x,0) = 1 + 0.1 e^{-4 \hat x^2}$ and $\hat u(\hat x, 0) = 0$) showing d'Alembert-like splitting of the pulse. Dispersion and nonlinearity modify the shape and produce `forerunners' and `backrunners'. The evolution is for $\g = 2$ and $\eps= 0.1$ with periodic BCs using the scheme of \S\ref{s:IVP-numerical} with $n_{\rm max} = 20$ Fourier modes, a time step $\D = 0.01$ and `nonlinearity strength' $\delta = 0.1$.}
 \label{f:gaussian-rho-compare-dalembert}
	\end{center}
	\end{figure}

A homogenous, stationary fluid [$\hat \rho = 1$, $\hat p = 1$, $\hat u = 0$ and $\hat s = 0$] is a solution of (\ref{e:non-dim-eoms}). To study sound, we consider linear perturbations $\hat p = 1 + \del \tl p$, $\hat \rho = 1 + \del \tl \rho$, $\hat s = \del \tl s$ and $\hat u = \del \tl u$ around this solution where $\del \ll 1$. The entropy equation upon linearization gives $\tl s_t = 0$. If we initially choose $\tl s(x,0) \equiv 0$ then $\tl s(x,t) \equiv 0$ and the entropy $\hat s(x,t) = \log\left({\hat p}/{\hat \rho^\g} \right) = 0$. Linearizing this we get $\tl p = \g \tl \rho$. The linearized continuity and velocity equations are 
	\beq
	\tl \rho_{\hat t} + \tl u_{\hat x} = 0 \quad \text{and} \quad
	\tl u_{\hat t} = - \tl \rho_{\hat x} + \eps^2 \tl \rho_{\hat x \hat x \hat x}.
	\label{e:sound-wave-eqn-dispersive}
	\eeq
Thus, we arrive at an equation for dispersive sound $\tl \rho_{\hat t \hat t} = \tl \rho_{\hat x \hat x} - \eps^2 \tl \rho_{4\hat x}$. The $4^{\rm th}$ derivative is reminiscent of elasticity, so our regularization force is like a tension. Fig.~\ref{f:gaussian-rho-compare-dalembert} shows the splitting of a pulse in density into two smaller pulses including effects of dispersion and weak nonlinearity.
Eqns. (\ref{e:sound-wave-eqn-dispersive}) have several conserved quantities including  
	\beqs
	 M &=& \int \tl\rho \, d\hat x,\;\; P = \int \tl u \: d\hat x, \;\; H_1 = \half \int [\tl u^2 + \tl \rho^2 + \eps^2 \tl \rho_{\hat x}^2] d \hat x\quad
	\text{and} \cr
	H_2 &=& \half \int \left[ \tl \rho_t^2 + \tl \rho_{\hat x}^2 + \eps^2 \rho_{\hat x \hat x}^2  \right] d \hat x.
	\label{e:sound-waves-cons-qtys}
	\eeqs
Putting $\tl \rho \propto e^{i(k \hat x - \om \hat t)}$ we get a dispersion relation akin to that of linearized KdV ($u_t + uu_x = \epsilon u_{3x}$), $\om_{\rm KdV} = k + \eps k^3$:
	\beq
	\om^2 = k^2( 1 + \eps^2 k^2) \quad \text{or} \quad \om = \pm \left(k + \half \eps^2 k^3 + \cdots \right).
	\eeq
The phase velocity is $v_p = \om/k = \pm (1 + \eps^2 k^2)^{1/2} \approx \pm( 1 + \half \eps^2 k^2)$, while the group velocity is
	\beq
	v_g = \dd{\om}{k} = \pm \frac{1 + 2 \eps^2 k^2}{\sqrt{(1 + \eps^2 k^2)}}
	\approx \pm (1 + 2 \eps^2 k^2)(1 - \half \eps^2 k^2 + ..)
	= \pm \left( 1 + \frac{3}{2} \eps^2 k^2 + \cdots \right).
	\eeq
Note that the regularization increases the phase speed while $|v_g|$ always exceeds $|v_p|$.

\subsection{Steady and traveling waves in one-dimension}
\label{s:steady-trav-quadrature}

Traveling waves are those where $\rho, u, p$ and $s$ are functions only of $(x-ct)$, where $c$ is the velocity of the wave. The entropy equation $s_t + u s_x = 0$ becomes $(u-c) \,s' = 0$. Thus, either $s = \bar s$ is a constant in space and time or $u = c$. In the former case, we have isentropic flow. In the latter, $s$ can be an arbitrary function of $(x - ct)$, but the fluid is at rest (`aerostatic') in a frame moving at velocity $c$. We will focus on the first possibility and look at steady solutions, subsequently `boosting' them to get traveling waves.


\subsubsection{Isentropic steady solutions}
\label{s:set-up-steady-barotropic}

For steady flow $(c=0)$ the mass, momentum and velocity fluxes (\ref{e:barotropic-curr}) are constant:
	\beqs
	F^{\rm m} &=& \rho u, \quad 
	F^{\rm p} = \rho u^2 + (\g -1 ) K \rho^\g - \beta_* \left( \rho_{xx} - \frac{\rho_x^2}{\rho} \right)\quad \text{and}\cr
	\quad F^{\rm u} &=& \half u^2 + \g K \rho^{\g-1} - \beta_* \left( \frac{\rho_{xx}}{\rho} - \half \frac{\rho_x^2}{\rho^2} \right).
	\label{e:fluxes-j-Pi-B}
	\eeqs
Moreover, the steady continuity equation $u \rho_x + u_x \rho = 0$ implies that the constant energy flux of Eqn. (\ref{e:barotropic-curr}) is not independent: $F^{\rm e} = F^{\rm m} F^{\rm u}$. Eliminating $u = F^{\rm m}/\rho$ we get two expressions for $\rho_{xx}$:
	\beqs
	\beta_* \rho_{xx} &=& - F^{\rm p} + \frac{{(F^{\rm m})}^2}{\rho} + (\g - 1) K \rho^\g + \beta_* \frac{\rho_x^2}{\rho} \quad \text{and}
	\cr
	\beta_* \rho_{xx} &=& - F^{\rm u} \rho + \frac{{(F^{\rm m})}^2}{2 \rho} + \g K \rho^{\g} + \frac{\beta_*}{2} \frac{\rho_x^2}{\rho}.
	\label{e:rhoxx-steady-eqn-two-versions}
	\eeqs
Taking a linear  combination allows us to eliminate the $\rho^\g$ term and arrive at the second order equation
	\beqs
	\beta_* \rho_{xx} &=& - V'(\rho) + \frac{(\g + 1) \beta_*}{2} \frac{\rho_x^2}{\rho}, \quad  \text{where} \cr 
	V'(\rho) &=& F^{\rm p} \g - F^{\rm u}(\g -1) \rho - \frac{(\g + 1) {(F^{\rm m})}^2}{2\rho}.
	\label{e:steady-reg-gas-eqn-rho}
	\eeqs
In Appendix \ref{a:parabolic-embed-LJ-id}, a different linear combination that eliminates the $\rho_x^2/\rho$ term is considered, leading to additional results. The current choice makes it easier to treat all values of $\g$ in a uniform manner. Interpreting $x$ and $\rho$ as time and position, this describes a Newtonian particle of mass $\beta_*$ moving in a (linear + harmonic + logarithmic) potential $V$ on the positive half-line subject also to a `velocity-dependent' force $\propto \rho_x^2/\rho$. This ensures that the motion is `time-reversal' $(x \to - x)$ invariant. The qualitative nature of trajectories is elucidated via a $\rho$-$\rho_x$ phase plane analysis in Appendix \ref{a:vect-fld-phase-portrait}. There are only two types of non-constant bounded solutions for $\rho(x)$: solitary waves of depression (cavitons) and periodic waves. The latter correspond to closed trajectories around an elliptic fixed point (O-point) in the phase portrait while cavitons correspond to the homoclinic separatrix orbit that encircles an O-point and begins and ends at a hyperbolic X-point to its right. The location of these fixed points are determined (for any $\gamma$) by the roots of the quadratic $V'(\rho) = 0$ whose discriminant $\D = \g^2 (F^{\rm p})^2 - 2(\g^2 -1) F^{\rm u}{(F^{\rm m})}^2$ must therefore be positive. In the generic non-aerostatic situation (i.e. $u \not \equiv 0$ or equivalently $F^{\rm m} \neq 0$), the only cases when we get non-constant bounded solutions for $\rho$ are (a) $F^{\rm p}, F^{\rm u} > 0$: both periodic solutions and cavitons and (b) $F^{\rm u} < 0$: only periodic solutions.

{\fl \bf Remark:} Eqn.~(\ref{e:steady-reg-gas-eqn-rho}) is a generalization of the Ermakov-Pinney equation \cite{ermakov, pinney} which corresponds to $V(\rho) = \rho^4/2$, $\g = 2$ and $\beta_* = 1$. This leads to an alternate approach to understanding (\ref{e:steady-reg-gas-eqn-rho}), since the transformation $z^2 = 1/\rho$ converts it into a Newton equation with sextic potential and no velocity dependent force for any $\g$:
	\beq
	z_{xx} =  \half (\g - 1) F^{\rm u} z - \g F^{\rm p} \, z^3 + \frac{(\g +1)}{4(F^{\rm m})^2} z^5.
	\eeq
{\fl \bf Reduction to quadrature:} Subtracting the two equations in (\ref{e:rhoxx-steady-eqn-two-versions}), we get a first order ordinary differential equation (ODE) for $\rho$:
	\beqs
	\frac{\beta_*}{2} \rho_x^2 &=& -\frac{(F^{\rm m})^2}{2} + F^{\rm p} \rho - F^{\rm u} \rho^2 + K  \rho^{\g + 1} \equiv  \rho^{\g + 1} (K - U) \quad \text{where} \cr
	K &=& \half \left( \frac{\beta_*}{\rho^{\g + 1}} \right) \rho_x^2 + U
	\quad \text{and} \quad U (\rho)= \ov{\rho^{\g + 1}}\left[ \frac{(F^{\rm m})^2}{2} - F^{\rm p} \rho + F^{\rm u} \rho^2 \right].
	\label{e:rhox-vs-rho-diff-eqn}
	\eeqs
The `potential energy' $U$ is related to the potential $V$ via $\rho^{\g + 1} U'(\rho) = V' (\rho)$. This allows us to reduce the determination of the steady density profile $\rho(x)$ to quadrature:
	\beq
	dx = \frac{d \rho}{\sqrt{(2/\beta_*)\rho^{\g + 1}(K - U(\rho))}}
	= \frac{d \rho}{\sqrt{(2/\beta_*) ( K  \rho^{\g + 1} - F^{\rm u} \rho^2 + F^{\rm p} \rho  -((F^{\rm m})^2/2))}}.
	\label{e:quadrature-steady-rho}
	\eeq
For integer $\g \geq 1$ (\ref{e:quadrature-steady-rho}) is a hyperelliptic integral though it reduces to an elliptic integral when $\g = 2$ (see \S \ref{s:exact-cavitons-cnoidal-g-2}). For other values of $\g$, steady solutions may be found via the parabolic embedding of Appendix \ref{a:parabolic-embed-LJ-id}.

\subsubsection{Nondimensionalizing the steady equation}
\label{s:non-dim-moduli}

\begin{figure}	
\begin{center}
		\includegraphics[height=4cm]{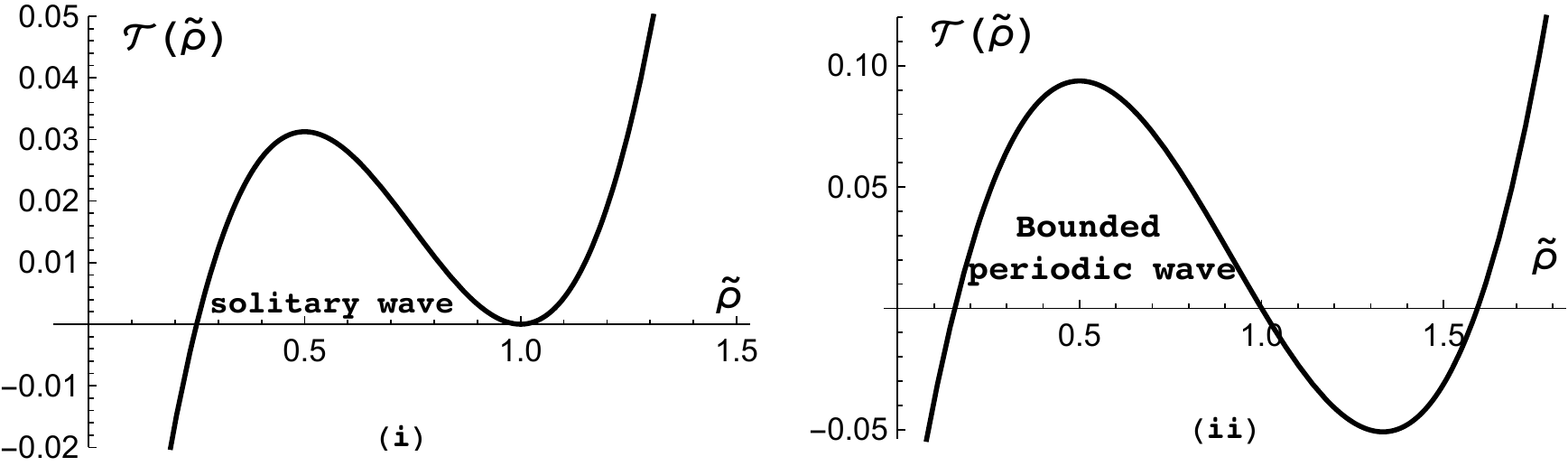}
		\end{center}
	\caption{(i) Density of caviton lies between simple and double root of ${\mathcal T}(\tl \rho)$
	 (\ref{e:non-linear-ode-for-rho-xi-M1-param}). (ii) Periodic wave density lies between simple roots where $\T > 0$. Here $\g = 2$ and in (i) $\ka_\circ = 0$, $M_\circ = 0.5$ while in (ii) $\ka_\circ = -0.25$ and $M_\circ = 0.5$.}
	\label{f:T-rho-vs-rho}
\end{figure}

To integrate (\ref{e:rhox-vs-rho-diff-eqn}), it is
convenient to replace the four constants $(F^{\rm m}, F^{\rm p}, F^{\rm u}, K)$ with two dimensionless shape 
parameters $(\ka_\pt, M_\pt)$ and two dimensional ones 
$(\rho_\pt, c_\pt)$ that set scales. These parameters are 
adapted to the solutions one seeks to find: $\rho_\pt$ is 
the density at a point $x_\pt$ where $\rho_x = 0$. For a 
caviton, $\rho_\pt$ can be the asymptotic or trough 
density while for a periodic wave, it can be the trough 
or crest density (or the trough density of an unbounded 
solution with the same $K$). This choice will simplify 
the expressions for the constant fluxes 
(\ref{e:fluxes-j-Pi-B}) and $K$ when evaluated at 
$x_\pt$. For example, $K = U(\rho_\pt)$ gives
	\beq
	K = \ov{\rho_\pt^{\g + 1}}\left[ \frac{(F^{\rm m}_\pt)^2}{2} - F^{\rm p}_\pt \rho_\pt + F^{\rm u}_\pt \rho^2_\pt \right] = \frac{p_\pt \rho_\pt^{-\g}}{\g - 1}	= \frac{c_\pt^2 \rho_\pt^{1 -\gamma}}{\gamma (\gamma - 1)}.
\label{e:K-at-rhostar}
	\eeq
where $p_\pt$ and $c_\pt$ are the pressure and sound speed at $x_\pt$. We may use $c_\pt^2$ to trade $\beta_*$ for a regularization length $\la_\pt = \sqrt{\beta_*}/c_\pt$ which is used to define the nondimensional position $\xi = x/\la_\pt$. Next, let $M_\pt^2 = {u_\pt^2}/{c_\pt^2}$ be the square of the Mach number at $x_\pt$. Positive $M_\pt$ corresponds to rightward flow at $x_\pt$ and vice-versa. We will take $M_\pt \geq 0$ with the remaining steady solutions obtained by taking $M_\pt \to - M_\pt$. Using these definitions, we rewrite (\ref{e:rhox-vs-rho-diff-eqn}) as a 1st order ODE for the nondimensional density $\tl \rho(\xi) = \rho(\la_\pt \xi)/\rho_\pt$:
	\beqs
	\half \left( \frac{d \tl \rho}{d\xi}\right)^2 = {\cal T}(\tl \rho) &\equiv& \frac{\tl \rho^{\g + 1}}{\g (\g - 1)} + \left(\ka_\pt - \frac{M_\pt^2}{2} - \ov{\g - 1} \right) \tl \rho^2
	+ \left(M_\pt^2 + \ov{\g} - \ka_\pt \right) \tl \rho - \half M_\pt^2 \cr
	&=& \frac{\tl \rho^{\g +1}}{\g(\g - 1)} - \half M_\pt^2 ( \tl \rho - 1)^2 - \frac{\tl \rho^2}{\g - 1} + \frac{\tl \rho}{\g} - \ka_\pt \, \tl \rho \left( 1 - \tl \rho \right).
	\label{e:non-linear-ode-for-rho-xi-M1-param}
	\eeqs
Here, $\ka_\pt = \la_\pt^2 \rho''(x_\pt)/\rho_\pt = \tl \rho''(\xi_\pt)$ measures the curvature of the density profile at $x_\pt = \la_\pt \xi_\pt$. A virtue of this nondimensionalization is that unlike the four parameters in (\ref{e:rhox-vs-rho-diff-eqn}), only two parameters $\ka_\pt$ and $M_\pt$ appear in (\ref{e:non-linear-ode-for-rho-xi-M1-param}). Eqn. (\ref{e:non-linear-ode-for-rho-xi-M1-param}) describes zero energy trajectories $\tl \rho(\xi)$ of a unit mass particle in the potential $- {\cal T}(\tl \rho)$. Evidently, the allowed values of $\tl \rho$ must lie between adjacent positive zeros of ${\cal T}$ with ${\cal T} > 0$ in between (Fig.~\ref{f:T-rho-vs-rho}). To obtain (\ref{e:non-linear-ode-for-rho-xi-M1-param}), we used the following expressions for the constant fluxes and entropy: 
	\beqs
	(F^{\rm m})^2 &=& \rho_\pt^2 c_\pt^2 M_\pt^2, \quad
	F^{\rm p} = \rho_\pt c_\pt^2 \left( M_\pt^2 + \ov{\gamma} - \ka_\pt \right), \cr
	F^{\rm u} &=& c_\pt^2 \left( \frac{M_\pt^2}{2} + \ov{\gamma - 1} - \ka_\pt \right) \quad \text{and} \quad 
	K = \frac{c_\pt^2 \rho_\pt^{1 -\gamma}}{\gamma (\gamma - 1)}.
	\label{e:steady-sol-currents}
	\eeqs 
Conversely, we may invert (\ref{e:steady-sol-currents}) by first determining $\rho_\pt$ by solving the algebraic equation
	\beq
	K = U(\rho_\pt) = \rho_\pt^{-(1 + \g)}\left[ \frac{(F^{\rm m})^2}{2} - F^{\rm p} \rho_\pt + F^{\rm u} \rho_\pt^2 \right]
	\label{e:determine-rho-star-from-K}
	\eeq
following from (\ref{e:rhox-vs-rho-diff-eqn}). The remaining new parameters follow from (\ref{e:steady-sol-currents}) and (\ref{e:fluxes-j-Pi-B}):
	\beq
	c_\pt^2 =  \frac{K \g(\g -1)}{\rho_\pt^{1 - \g}}, \;\; M_\pt^2 = \frac{(F^{\rm m})^2}{\rho_\pt^2 c_\pt^2} 
	\;\; \text{and} \;\; \ka_\pt = \frac{(F^{\rm m})^2 -  \rho_\pt F^{\rm p}}{\rho_\pt^2 c_\pt^2}  + \frac{1}{\g}.
	\eeq

\subsubsection{Exact cavitons and cnoidal waves for $\g = 2$}
\label{s:exact-cavitons-cnoidal-g-2}

\begin{figure*}	
	\centering
	\begin{subfigure}[t]{8cm}
	\centering
		\includegraphics[width=8cm]{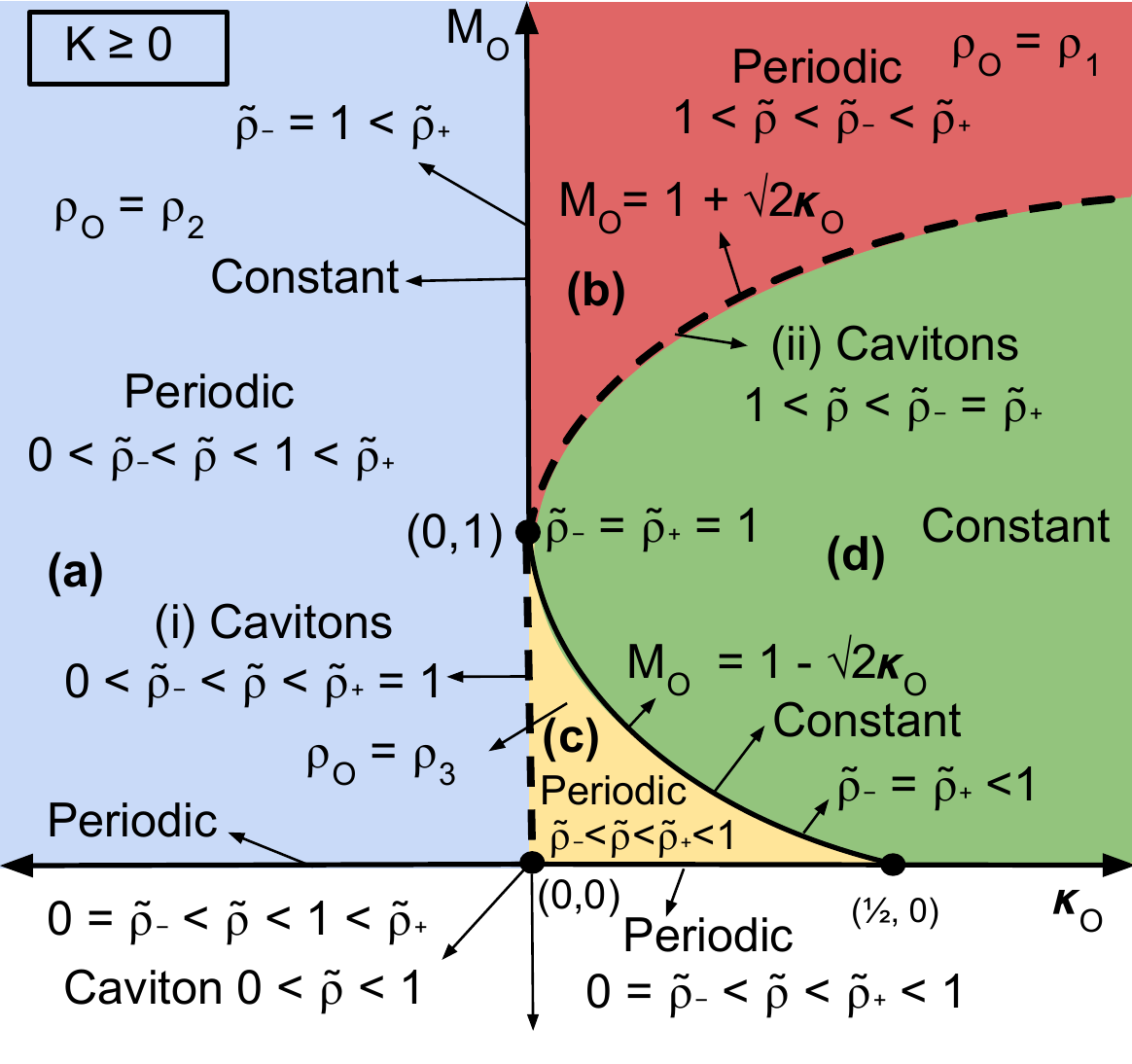}
	\caption{}
	\label{f:kappa-M-K>0}	
	\end{subfigure} \qquad 
	\begin{subfigure}[t]{7cm}
		\includegraphics[width=7cm]{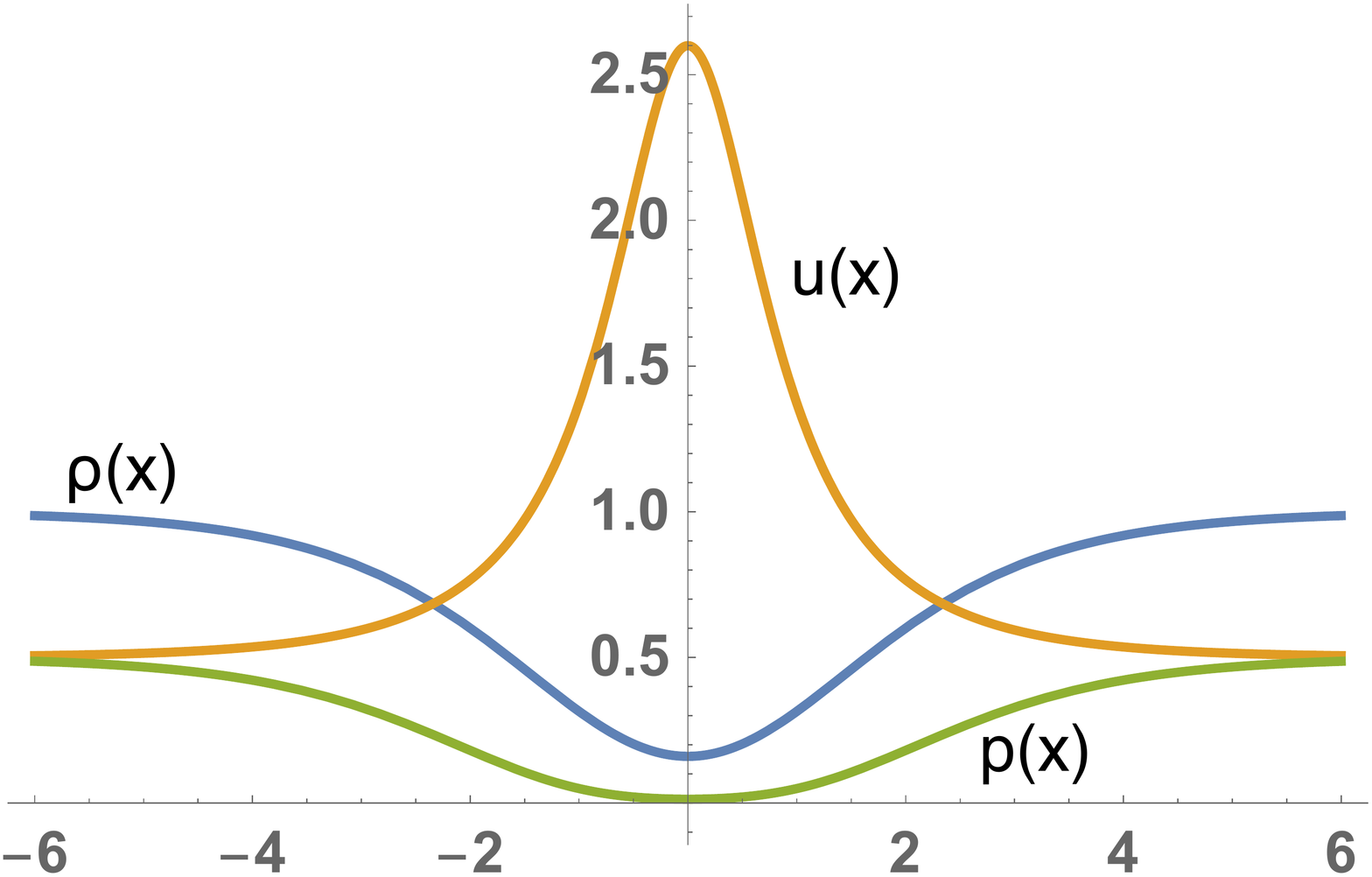}
		\caption{}	
	\end{subfigure}
	\qquad 
	\begin{subfigure}[t]{7cm}
	\centering
		\includegraphics[width=7cm]{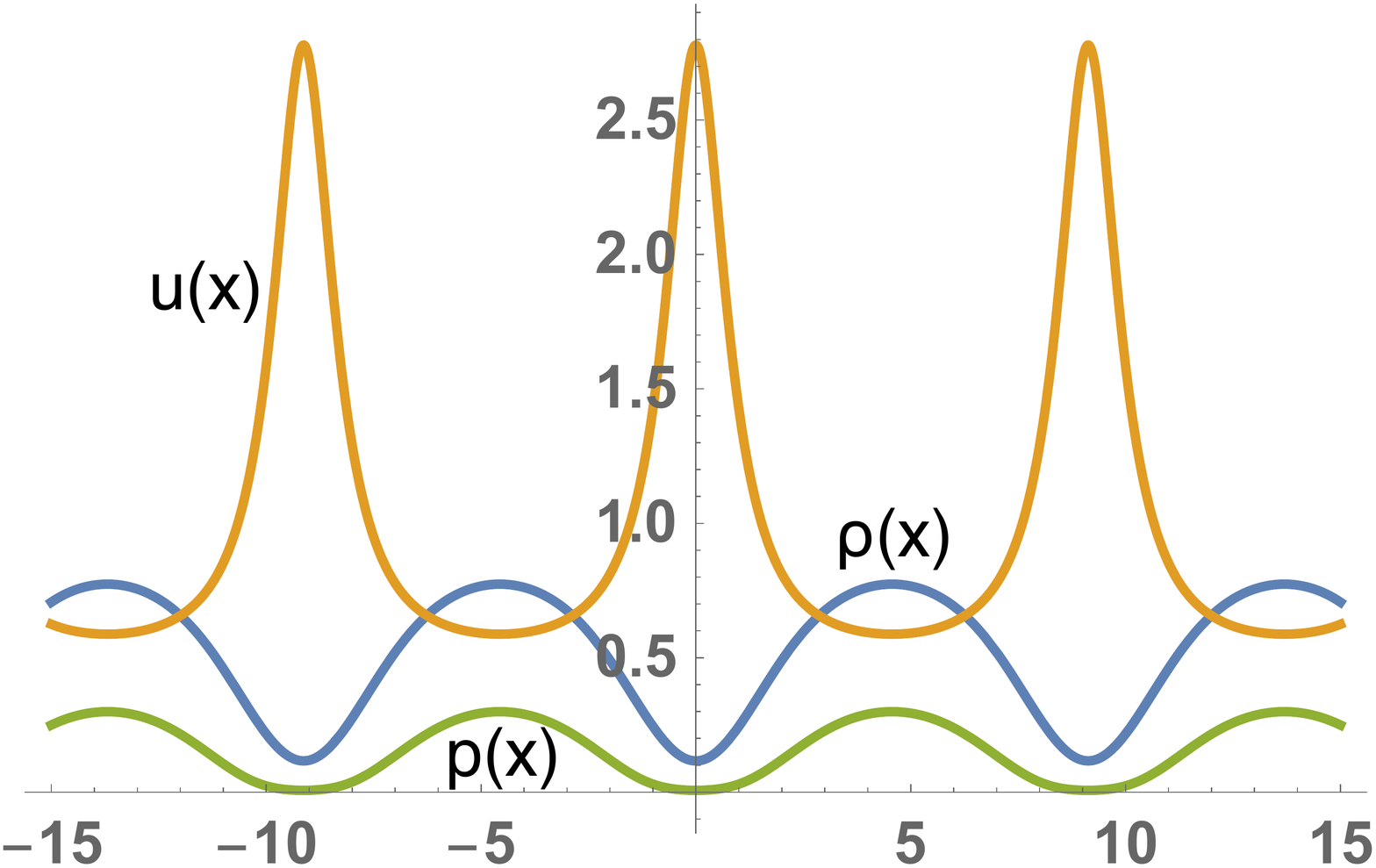}
		\caption{}
	\end{subfigure}
	\caption{\label{f:crho-u-p-caviton-traveling-cnoidal} (a) Nature of steady bounded solutions (\ref{e:non-linear-ode-for-rho-xi-M1-param}) for $\gamma = 2$ in $\kappa_\pt$-$M_\pt$ parameter space. (b)  and (c) Density, velocity and pressure for $K = 1/2$ for (b) traveling cavitons ($\rho_\pt =c_\pt =1, M_\pt = 0.5, \ka = 0$ and $\tl c = 0.1$) and (c) traveling periodic cnoidal waves $\rho_\pt = c_\pt = 1, M_\pt = 0.5, \ka = 0.1$ and $\tl c = 0.2$ (lower triangular region in $|M_\pt - \tl c|$-$\ka_\pt$ plane). Cavitons are solutions where the temperature drops isentropically in a small region of size $\la_\circ$.}
\end{figure*}

For $\g = 2$, ${\cal T}(\tl \rho)$ (\ref{e:non-linear-ode-for-rho-xi-M1-param}) becomes a cubic with roots $\tl \rho = 1 $ and 
	\beq
	\tl \rho_\pm  = \half \left(1 + M_\pt^2 - 2 \ka_\pt \pm \sqrt{ (1 + M_\pt^2 - 2 \ka_\pt)^2 -4 M_\pt^2} \right).
	\label{e:steady-rho-plus-minus}
	\eeq
The density of periodic and solitary waves must lie between adjacent positive roots with ${\cal T} > 0$ in between. Interestingly, it can be shown that if for $\g = 2$, all three roots of $\T$ are positive, then the same holds for any $1 < \g < 2$. So some qualitative features of solutions for $\g = 2$ are valid more generally. For $\g = 2$, the nature of solutions on the $\ka_\pt$-$M_\pt$ plane (Fig. \ref{f:kappa-M-K>0}) changes when the two relevant roots coalesce, i.e., when the discriminant of the cubic ${\cal T}$ vanishes:
	\beqs
	\Delta(\ka_\pt, M_\pt) &=& \left[(\tl\rho_+ - \tl\rho_-)(\tl\rho_- - 1)(1 - \tl\rho_+)\right]^2 \cr
	&=& 4\ka_\pt^2 \left[M_\pt^4 - 2 M_\pt^2 (1 + 2 \ka_\pt) + (1 - 2 \ka_\pt)^2\right] = 0.\qquad 
	\eeqs
$\D$ vanishes only along the vertical axis $\ka_\pt = 0$, the two parabolic curves $M_\pt = 1 \pm \sqrt{2\ka_\pt}$ and their reflections in the $\ka_\pt$-axis. In what follows, we restrict to rightward flow by taking $M_\pt \geq 0$. There are three regions in the upper half $\ka_\pt$-$M_\pt$ plane (pictured in Fig.~\ref{f:kappa-M-K>0}, colors online) admitting {\it periodic solutions}: (a) the blue second quadrant $\ka_\pt < 0$, (b) the red north-east region above $M_\pt = 1 + \sqrt{2 \ka_\pt}$ and (c) the yellow triangular region below the curve $M_\pt = 1 - \sqrt{2 \ka_\pt}$. In the green wedge (d) lying within the parabola but above the horizontal axis, non-constant solutions are unbounded since either $\tl \rho_\pm$ are negative (when $M_\pt < -1 + \sqrt{2 \ka_\pt}$) or not real (when $M_\pt > -1 + \sqrt{2 \ka_\pt}$).

{\it Solitary waves} (cavitons) occur only on the dashed boundaries (i) between (a) and (c) ($0 < M_\pt < 1$, $\ka_\pt = 0$) and (ii) between (b) and (d) ($M_\pt = 1 + \sqrt{2 \ka_\pt}$). When $\ka_\pt = M_\pt = 0$, ${\cal T} =  \tl \rho (\tl \rho - 1)^2/2$ so that we have an aerostatic ($u \equiv 0$) caviton with $0 \leq \tl \rho \leq 1$. {\it Constant solutions} occur when $\tl\rho \equiv$ any zero of ${\cal T}$. ${\cal T}$ has a double zero $(\tl \rho = 1)$ along the vertical axis, the double zero $\tl\rho_+ = \tl\rho_-$ along the curve $M_\pt = 1- \sqrt{2\ka_\pt}$ and the triple zero $\tl \rho = 1$ at $(\ka_\pt = 0, M_\pt = 1)$. At all other points, ${\cal T}$ has either one or three positive simple zeros.

Interestingly, when we reinstate dimensions, the periodic solutions from regions (a), (b) and (c) of the $\ka_\pt$-$M_\pt$ plane (Fig \ref{f:kappa-M-K>0}) are physically identical. They differ by the choice of nondimensionalizing density $\rho_\pt$ which could be any one of the roots of the cubic in (\ref{e:determine-rho-star-from-K}). Thus, the parameters $(\ka_\pt, M_\pt, c_\pt, \rho_\pt)$ generically furnish a 3-fold cover (redundancy) of the original space of constants $((F^{\rm m})^2, F^{\rm u}, F^{\rm p}, K)$. For solitary waves, it degenerates into a double cover: the two families of cavitons ((i) and (ii)) in Fig.~\ref{f:kappa-M-K>0} differ via the choice of $\rho_\pt$ as asymptotic or trough density. Moreover, $M_\pt$ is the asymptotic Mach number in (i) while it is the Mach number at the trough in (ii). In a caviton, the flow goes from asymptotically subsonic to supersonic at $x = 0$. A caviton is like a pair of normal shocks joined at the trough. Finally, the map between the two sets of parameters  becomes a 1-fold cover for the aerostatic cavitons at $\ka_\pt = M_\pt = 0$, since their trough densities vanish and $\rho_\pt$ can only be chosen as the asymptotic density.

In light of the above remarks, we now restrict attention to the yellow triangular region (c) where $\ka_\pt > 0$ and $0 < M_\pt < 1 - \sqrt{2 \ka_\pt}$. Here, the roots of ${\cal T}$ are $0 < \tl \rho_- < \tl \rho_+ < 1$ and (\ref{e:non-linear-ode-for-rho-xi-M1-param}) is reduced to quadrature:
	\beqs
	\xi(\tl\rho) - \xi(\tl\rho_-) &=& \int_{\tl\rho_-}^{\tl\rho} \frac{d\rho'}{\sqrt{(\rho' - \tl\rho_-)(\rho' - \tl\rho_+)(\rho' - 1)}} \cr
	&=& \frac{2}{\sqrt{1 - \tl\rho_-}} F\left( \arcsin \sqrt{\frac{\tl\rho - \tl\rho_-}{\tl\rho_+ - \tl\rho_-}} \;\vert\;  {\frac{\tl\rho_+ - \tl\rho_-}{1 - \tl\rho_-}} \right).
	\eeqs
Here, $F(\phi\vert m) = \int_0^\phi {(1 - m \sin^2\theta)^{-1/2}} d\theta$ is the incomplete elliptic integral of the first kind with $m$ the square of the elliptic modulus $k$ (see 17.4.62 of \cite{abramowitz-stegun}). Inverting, we write $\tl \rho (\xi)$ in terms of the Jacobi $\cn(u,m)$ function:
	\beq
	\tl \rho (\xi) = \tl \rho_+ - (\tl \rho_+ - \tl \rho_-) \: \cn^2 \left( \frac{\sqrt{1 - \tl \rho_-}}{2}(\xi - \xi_-), \frac{\tl \rho_+ - \tl \rho_-}{1 - \tl \rho_-}\right).
	\label{e:cnoidal-wave-rho-til-triangle}
	\eeq
Here $\xi_- = \xi(\tl \rho_-)$. The periodic wave extends from a trough at $\tl \rho_-$ to a crest at $\tl \rho_+$ with amplitude and wavelength 
	\beqs
	&& {\cal A}_{\ka_\pt > 0} = \sqrt{ (1 + M_\pt^2 - 2 \ka_\pt)^2 -4 M_\pt^2} \quad \quad\text{and} \cr 
	&& {\Lambda}_{\ka_\pt > 0} = \int_{\tl \rho_-}^{\tl \rho_+} \frac{2 \: d \tl \rho}{\sqrt{2 {\cal T}(\tl \rho)}} = \frac{4}{\sqrt{1 - \tl \rho_-}} K \left( \frac{\tl \rho_+ - \tl \rho_-}{1 - \tl \rho_-}\right). \quad
	\label{e:amplitude-waveleng-steady-cnoidal}
	\eeqs
Here $K(m) = F(\frac{\pi}{2}|m)$ is the complete elliptic integral of the $1^{\rm st}$ kind. When we approach the left boundary $\ka_\pt \to 0^+$ with $0 < M_\pt < 1$, the wavelength diverges ($K(1/2 \ka_\pt) \sim \sqrt{\ka_\pt} \log \ka_\pt$ for small $\ka_\pt$) and the periodic waves turn into cavitons which extend from a trough density $\tl \rho = \tl \rho_- = M_\pt^2 < 1$ to an asymptotic density $\tl \rho = \tl \rho_+ = 1$:
	\beq
	\tl \rho(\xi) = 1 - (1 - M_\pt^2) \sech^2 \left( \sqrt{\frac{1-M_\pt^2}{4}} \xi \right) \;\; \text{for} \;\; 0 \leq M_\pt \leq 1.
	\label{e:caviton-nondim-gamma=2}
	\eeq
On the other hand, when we approach the upper boundary $M_\pt = 1 - \sqrt{2 \ka_\pt}$, ${\cal A} \to 0$ and we get constant solutions. By contrast, on the lower boundary $(M_\pt = 0, 0< \ka_\pt < \half)$ we continue to have periodic solutions except that $\tl \rho$ vanishes at the trough ($\tl \rho_- = 0$) while the crest density $\tl \rho_+ = 1 - 2 \ka_\pt$:
	\beq
	 \tl \rho (\xi) = (1 - 2 \ka_\pt) \,\text{sn}^2 \left( \frac{\xi}{2}, 1 - 2\ka_\pt \right) \quad \text{and}\quad
	{\Lambda} =  2 i \left[ \sqrt{\frac{2}{\ka_\pt}} K \left( \ov{2\ka_\pt} \right) - 2 K(2\ka_\pt) \right].
	\label{e:cnoidal-wave-gamma=2}
	\eeq
When $\ka_\pt = M_\pt = 0$ in (\ref{e:caviton-nondim-gamma=2})/(\ref{e:cnoidal-wave-gamma=2}), we get an {\it aerostatic caviton} which reaches {\it vacuum conditions} at $\xi = 0$:
	\beq
	\tl \rho(\xi) =  \tanh^2(\xi/2) \quad \text{with} \quad u \equiv 0.
	\label{e:aerostatic-caviton-Mpt-0}
	\eeq
The dimensional $\rho, u$ and $p$, are got by reinstating the constants ($\la_\pt$, $\rho_\pt$, $c_\pt$). Writing $x = \la_\pt \xi$ we have
	\beq
	\rho(x) = \rho_\pt \tl \rho\left( \xi \right), \; 
	u(x) =  \frac{c_\pt M_\pt}{\tl \rho(\xi)} \quad \text{and} \quad
	p(x) = \frac{c_\pt^2 \rho_\pt}{\g} {\tl \rho(\xi)}^\g
	\eeq
for isentropic flow. Reversing the sign of $M_\pt$ reverses the flow direction leaving $p$ and $\rho$ unaltered. Moreover, since $\tl \rho \geq 0$ and $u = F^{\rm m}/\rho$, the flow is unidirectional with positive velocity solitary waves being waves of elevation in $u$ and vice-versa. A caviton is superficially a bifurcation of the constant solution $\rho(x) \equiv \rho({\pm\infty})$. However, though the caviton and constant solutions have the same constant specific entropy, they have different values of mass and energy (per unit length). For instance, the energy density (\ref{e:beta-reg-hamiltonian}) of an aerostatic caviton is less than that of the constant state:
    \beq
	{\cal E}_{\rm aerostatic \; caviton} = p_\infty \left[ 1 - \frac{2 \rho(x)}{\rho_\infty} \left( 1 - \frac{\rho(x)}{\rho_\infty}  \right) \right] < {\cal E}_{\rm const.} = p_\infty.
    \eeq
As a consequence, the uniform state cannot, by any isentropic time-dependent motion with fixed BCs at $\pm \infty$, in the absence of sources and sinks, evolve via R-gas dynamics to the caviton, or vice versa, since the two states have different invariants. However, one could imagine creating, say an aerostatic caviton, by starting with a uniform state and introducing a point sink at the origin, to suck fluid out. A symmetrical pair of expansion waves would travel to infinity on both sides, without affecting the conditions at infinity. When the density reaches $0$ at the origin, we stop the sink. The pressure gradient will then be balanced by the regularizing density gradient force, and the solution should tend to the aerostatic caviton as $t\to \infty$. Since temperature $T = K m (\g-1) \rho^{\g-1}$, the caviton corresponds to a region of width $\lambda$ where the temperature drops. Loosely, the regularizing force is like Pauli's exchange repulsion, capable of maintaining a depression in density with variable but isentropic temperature and pressure distributions.

\subsubsection{Traveling waves of isentropic R-gas dynamics}
\label{s:traveling-waves}

Here we generalize the above steady solutions to waves traveling at speed $c$: $(\rho,u,p)(x,t) = (\rho,u,p)(x-ct)$. The continuity, velocity and momentum equations (\ref{e:unsteady-barotropic-eqns}) are readily integrated, giving the constant fluxes: 
	\beqs
	F^{\rm m} &=& \rho({u - c}),\;\; 
	F^{\rm u} = \frac{u^2}{2} - c u - \beta_* \left[ \frac{\rho''}{\rho} -  \frac{\rho'^2}{2 \rho^2} \right] + \g K \rho^{\g-1}
	\quad \text{and} \cr
	F^{\rm p} &=& \rho u( u - c) + (\g - 1) K \rho^\g - \beta_* \left(\rho'' - \frac{\rho'^2}{\rho} \right).
	\eeqs 
Eliminating $u = c + F^{\rm m}/\rho$ and taking a linear combination leads us to a $2^{\rm nd}$ order nonlinear ODE for $\rho$: 
	\beqs
	\beta_* \rho'' &=& - V'(\rho) + \frac{(\g + 1) \beta_*}{2} \frac{\rho'^2}{\rho} \quad \text{where} \cr
	 V'(\rho) &=&   \g (c F^{\rm m}  - F^{\rm p}) - (\g -1) \left(\frac{c F^{\rm p}}{F^{\rm m}} - F^{\rm u}  - \frac{c^2}{2} \right) \rho
	+ \frac{(\g +1) (F^{\rm m})^2}{2 \rho}.
	\label{e:traveling-reg-gas-eqn-rho}
	\eeqs 
Proceeding as in \S \ref{s:set-up-steady-barotropic} and \S \ref{s:non-dim-moduli}, this may be reduced to the nonlinear first order equation
	\beqs
	&& \half \left( \DD{\tl \rho}{\xi} \right)^2 = {\cal T}( \tl \rho) \equiv\frac{\tl \rho^{\g + 1}}{\g(\g-1)} -  \half (\tl \rho - 1)^2 (\tl c - M_\pt)^2
	- \frac{\tl \rho^2}{\g - 1} + \frac{\tl \rho}{\g} - \ka_\pt \tl \rho ( 1 - \tl \rho)\quad \text{where} \cr
	&& \quad \beta_* = \la_\pt^2 c_\pt^2, \quad  \tl \rho = \frac{\rho}{\rho_\pt}, \quad \tl c = \frac{c}{c_\pt}, \quad M_\pt = \frac{u_\pt}{c_\pt}, \quad \ka_\pt = \tl \rho''(\xi_\pt),\quad \xi = \frac{x - ct}{\la_\pt}\quad
	\eeqs
with $\tl \rho'(\xi_0) = 0$. Comparing with (\ref{e:non-linear-ode-for-rho-xi-M1-param}), we see that the passage from steady to traveling waves involves only a shift in the square of the Mach number $M_\pt^2 \to (M_\pt - \tl c)^2$. Here, $\tl c$ is the speed of the traveling wave in units of the sound speed $c_\pt$ at the point where $\rho'(x - ct) = 0$. Thus, for each value of $\beta_*$, we have a 5-parameter ($\ka_\pt, M_\pt, \rho_\pt,c_\pt, \tl c$) space of traveling cavitons and periodic waves. The dimensionful $\rho, u$ and $p$ for any value of $(M_\pt - \tl c)$ are given by
	\beqs
	&& \rho(x,t) = \rho_\pt \tl \rho\left( \xi \right), \quad 
	u(x,t) = c_\pt \left( \frac{M_\pt - \tl c}{\tl \rho(\xi)} + \tl c \right) \quad \text{and} \cr
	&& p(x,t) = \frac{c_\pt^2 \rho_\pt}{\g} {\tl \rho(\xi)}^\g
\quad \text{where} \quad \xi = \frac{x- \tl c \,  c_\pt t}{\la_\pt}
	\eeqs


\vspace{.1cm}

{\fl \bf Traveling cavitons for $\g = 2$:} In these nondimensional variables, traveling cavitons have the profile
	\beq
	\tl \rho(\xi) = 	
	 1 - (1 - (\tl c - M_\pt)^2) \sech^2 \left( \sqrt{\frac{1-(\tl c - M_\pt)^2}{4}} \xi \right),
	\eeq
for $(\tl c - M_\pt )^2 < 1$. These cavitons are reminiscent of the solitary waves of depression/elevation of the KdV equation $u_t \mp 6 u u_x + u_{xxx}=0$ that move rightward with speed $c > 0$: $u(x,t)= \mp (c/2) \sech^{2}\,\left((\sqrt{c}/2)(x - ct) \right)$. Just as in KdV, narrower cavitons (width $\propto 1/\sqrt{|\tl c - M_\pt|-1}$) are taller (height $\propto |\tl c - M_\pt|-1$) and have a higher speed relative to the speed at the reference location where the density has an extremum ($\tl c - M_\pt \propto c - u_\pt$ is the speed of the traveling wave in the rest frame of the fluid at the reference location).


\vspace{.1cm}

{\fl \bf Traveling cnoidal waves for $\g = 2$:} The nondimensional density profile of traveling cnoidal waves is
	\beqs
	&& \tl \rho(\xi) = 
\tl \rho_+ - (\tl \rho_+ - \tl \rho_-) \;\text{cn}^2 \left( \frac{\sqrt{1 - \tl \rho_-}}{2}(\xi - \xi_-), \frac{\tl \rho_+ - \tl \rho_-}{1 - \tl \rho_-}\right) \cr
	&& \quad \text{for} \quad 0 < \ka_\pt < \half 
	\quad \text{and} \quad |M_\pt -\tl c| < 1 - \sqrt{2 \ka_\pt}.
	\eeqs
Here, $\tl \rho_\pm$ are got by replacing $M_\pt^2 \to (M_\pt - \tl c)^2$ in (\ref{e:steady-rho-plus-minus}).
Their wavelength is given by the same formula (\ref{e:amplitude-waveleng-steady-cnoidal}). Our cnoidal waves are very similar to those of KdV $(u_t - 6 u u_x + u_{xxx} = 0)$:
	\beq
	u= f(\xi) = f_2 \left[1 - \:\text{cn}^2 \,\left[ \sqrt{\frac{f_1 - f_2}{2}} (\xi - \xi_3), \: \frac{f_2 - f_3}{f_1 - f_3} \right]\right] + f_3.
	\eeq
Here $ \xi = x - ct$ and $f(\xi_3) = f_3$ and $f_{1,2,3}$ are the roots of $f^3 + \half c f^2 + A f + B$ with $A$ and $B$ constants of integration.

\section{Weak form and shock-like profiles}
\label{s:weak-and-shock-like-sol}

\subsection{Weak form of R-gas dynamic equations}
\label{s:weak-form}

The R-gas dynamic equations ((\ref{e:cont-entropy-eqn}) and (\ref{e:reg-vel-eqn-gas})) involve $u_x, p_x, \rho_x, \rho_{xx}$ and $\rho_{xxx}$. Thus, classical solutions need to be $C^1$ in $u$ and $p$ and $C^3$ in $\rho$. However, by multiplying the conservation equations by $C^\infty$ test functions $\phi, \psi$ and $\zeta$ and integrating by parts, we arrive at a weak form of the equations:
	\beqs
	&&\int [\rho_t \phi - (\rho u) \phi_x] dx = 0,\quad 
	\int \left[ (\rho u)_t \psi - \left(p + \rho u^2 + \beta_* \frac{\rho_x^2}{\rho} \right) \psi_x + \beta_* \rho \psi_{xxx} \right] dx = 0 \quad \text{and}\quad\quad \cr
	&&\int \left[\frac{\rho u^2}{2} + \frac{p}{\g-1} +
	\frac{\beta_*  \rho_x^2}{2 \rho} \right]_t \zeta dx
	= \int \left[ \left[\frac{\rho u^3}{2} + \frac{\g p u}{\g - 1} +  \frac{3\: \beta_* \:u }{2} \frac{\rho_x^2}{\rho} \right] \zeta_x - \beta_* u \rho_x \zeta_{xx} \right] dx.
	\label{e:weak-form-of-Rgas-dyn}
	\eeqs
Thus, for weak solutions, it suffices that $\rho$ be $C^1$ and $u$ and $p$ be merely continuous.

\subsection{Steady shock-like profile from half a caviton}
\label{s:patched-shock-weak-sol}

\begin{figure}	
\begin{center}
	\begin{subfigure}[t]{7cm}
		\includegraphics[height=4cm]{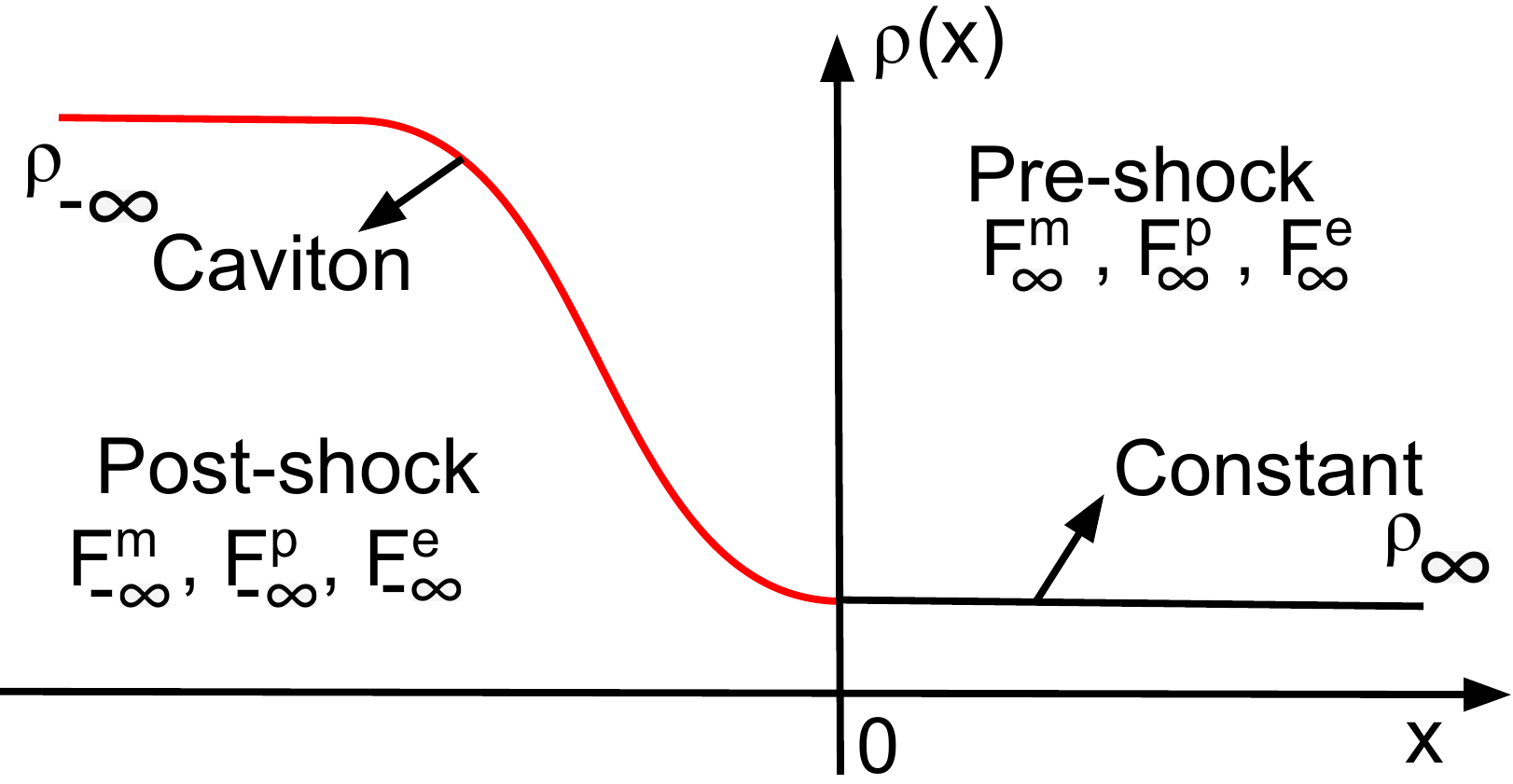}
	\end{subfigure}
		\end{center}
	\caption{Steady shock-like profiles from patching half a caviton with a constant state violate one or more Rankine-Hugoniot conditions.}
	\label{f:patched-shock}
\end{figure}

Here, we try to use the steady solutions of \S \ref{s:steady-trav-quadrature} to model the structure of a normal shock propagating to the right in the lab frame. As in Fig.~\ref{f:patched-shock}, in the shock frame, the shock is assumed to be located around $x = 0$. The undisturbed pre-shock medium is to the right $(x > 0)$ while the `disturbed' post-shock medium is to the left ($x<0$) \cite{whitham}. As $x \to \pm \infty$ the variables approach the asymptotic values $\rho_{\pm \infty}, u_{\pm \infty}$ and $p_{\pm \infty}$ with $\rho_{+ \infty} < \rho_{-\infty}$. The Rankine-Hugoniot (RH) conditions are obtained by equating the conserved fluxes $F^{\rm m}$ (\ref{e:cont-entropy-eqn}), $F^{\rm p}$ (\ref{e:vel-mom-r-gas-dynm}) and $F^{\rm e}$ (\ref{e:reg-total-energy-PB}) at $x = \pm \infty$:
	\beqs
	&& (\rho u )_{-\infty} = (\rho u )_{\infty}, \quad (\rho u^2 + p)_{-\infty} = (\rho u^2 + p)_{\infty} \quad \text{and} \cr
	&& \left( \half \rho u^2 u + \frac{\g}{\g - 1}pu \right)_{-\infty} = \left( \half \rho u^2 u + \frac{\g}{\g - 1}pu \right)_{\infty}.
	\eeqs 
In our $\g = 2$ cavitons (\ref{e:caviton-nondim-gamma=2}), the flow is subsonic at $x = \pm \infty$ and supersonic at $x = 0$. We exploit this in trying to find a shock-like steady solution by patching half a caviton with a constant solution. Thus, we seek a steady solution where $\rho(x) \equiv \rho_{\infty}$ in the pre-shock medium and is the left half of a caviton for $x < 0$. The half-caviton profile is got from (\ref{e:caviton-nondim-gamma=2}) by taking the reference location $x_\pt = -\infty$ so that $\rho_\pt = \rho_{-\infty}$ and $\ka_\pt = 0$:
	\beq
    \rho(x) = \rho_{-\infty} \left[1 - (1 - M_{-\infty}^2) \sech^2 \left[ \sqrt{\frac{1-M_{-\infty}^2}{4}} \frac{x}{\la_{-\infty}} \right] \right]
    \quad \text{for} \quad x < 0 \quad \text{with} \quad 0 \leq M_{-\infty}^2 \leq 1.
	\eeq
Here $\la_{-\infty}^2 = \beta_*/c_{-\infty}^2$. On the other hand, $\rho_{+\infty}$ must correspond to a constant solution with fluxes $F^{\rm m, p, e}_{\infty}$. It can take one of two density values corresponding to the X/O point (\ref{e:rho-pm-fixedpts-steady}):
	\beq
	 \rho_{\infty}^{\rm X,O} = \frac{F^{\rm m}_{\infty} (2 F^{\rm p}_{\infty} \pm \sqrt{\D_{\infty}})}{2 F^{\rm e}_{\infty}} \;\; \text{with} \;\; \D = 4 (F^{\rm p}_{\infty})^2 - 6 F^{\rm e}_{\infty} F^{\rm m}_{\infty}.
	\eeq
We now attempt to patch these two solutions requiring $\rho$ and $\rho_x$ to be continuous at $x = 0$. Since $\rho$ has a local minimum at the trough of a caviton, $\rho_x(0^-) = 0 = \rho_x(0^+)$ of the undisturbed medium. However, a difficulty arises in trying to ensure that $\rho$ is continuous at $x=0$. Indeed, suppose we impose the RH conditions, $F^{\rm m}_\infty = F^{\rm m}_{- \infty}$, $F^{\rm p}_{-\infty} = F^{\rm p}_{\infty}$ and $F^{\rm e}_{-\infty} = F^{\rm e}_{\infty}$, then both the pre- and post-shock regions correspond to a common phase portrait. We observe (see also Fig. \ref{f:phase-por-caviton}) that the caviton trough density $\rho(0^-) = \rho_{- \infty} M_{-\infty}^2$ is strictly less than both $\rho_{\infty}^{\rm X,O}$. Thus, the post-shock semi-caviton solution cannot be continuously extended into the pre-shock region.

Stated differently, in the above patched shock construction, if $\rho_{+\infty}$ is chosen to be equal to $\rho(0^-)$ in order to make $\rho$ continuous, then the RH conditions are violated. Let us illustrate this with a $\g = 2$ aerostatic example. We take the pre-shock region to be vacuum ($\rho = u = p \equiv 0$ for $x > 0$) and try to patch this at $x=0$ with the left half of the aerostatic caviton [$\rho(x) = \rho_{-\infty} \tanh^2(x/2 \la_{-\infty})$ and $u \equiv 0$] of (\ref{e:aerostatic-caviton-Mpt-0}). This caviton corresponds to the values $F^{\rm m}_{-\infty} = F^{\rm e}_{-\infty} = \ka_{-\infty} = M_{-\infty} = 0$ and has trough density $\rho(0^-) = 0$ with trough density gradient $\rho_x(0^-) = 0$ as well. Since the caviton is isentropic, its trough pressure $p(0^-) = K \rho(0^-)^2 = 0$. Thus, $\rho$ and $p$ are both $C^1$ at $x=0$, while $u \equiv 0$ so that this is a weak solution in the sense of \S\ref{s:weak-form}. However, it violates the RH conditions: while $F^{\rm m} \equiv 0$ is continuous, $F^{\rm p}$ and $F^{\rm u}$ are not. In fact, in the pre-shock vacuum region $F^{\rm p} = F^{\rm u} = 0$ while in the post-shock region they are non-zero, as evaluating them at  $x = - \infty$ shows:
	\beq
	F^{\rm p}_{-\infty} = p(-\infty) = K_- \rho_{-\infty}^2 \neq 0 
	\quad \text{and} \quad
	F^{\rm u}_{- \infty} = 2 K_- \rho_{-\infty} \neq 0.
	\eeq
In the post-shock region $K_- = F^{\rm p}_{- \infty}/\rho_{-\infty}^2 \neq 0$ whereas in the pre-shock vacuum $K$ is arbitrary since $p = \rho = 0$ for $x > 0$. In conclusion, there are no continuous steady shock-like solutions in the shock frame that satisfy the RH conditions. To see how initial conditions (ICs) that  would lead to shocks in the ideal model are regularized, we turn to a numerical approach.

\section{Numerical solutions to the initial value problem}
\label{s:IVP-numerical}

\subsection{Spectral method with nonlinear terms isolated}
\label{s:numerical-scheme}

In this section, we discuss the numerical solution of the isentropic R-gas dynamic initial value problem (IVP). It is convenient to work with the nondimensional variables $\hat \rho$, $\hat u$, $\hat s$, etc. of \S \ref{s:dispersive-sound-waves}. The continuity equation is $\hat \rho_{\hat t} + (\hat \rho \hat u)_{\hat x} = 0$ while for isentropic flow, $\hat s$ is a global constant which may be taken to vanish by adding a constant to entropy. Thus, we can eliminate $\hat p = \hat \rho^\g$ in the velocity and momentum equations, both of which are in conservation form (\ref{e:unsteady-barotropic-eqns}):
	\beqs
	\hat u_{\hat t} + \left[\half \hat u^2 + \ov{\g-1} \hat \rho^{\g - 1} - \eps^2 \left( \frac{\hat \rho_{\hat x \hat x}}{\hat \rho} - \half \frac{\hat \rho_{\hat x}^2}{\hat \rho^2} \right) \right]_{\hat x} &=& 0 \quad \text{or} \cr
	(\hat \rho \hat u)_{\hat t} + \left[ \hat \rho \hat u^2 + \ov{\g} \hat \rho^\g - \eps^2 \left( \hat \rho_{\hat x \hat x} - \frac{\hat \rho_{\hat x}^2}{\hat \rho} \right) \right]_{\hat x} &=& 0.
	\label{e:non-dim-barotropic-vel-momentum}
	\eeqs
The energy equation is $\pdr_{\hat t} \hat{\cal E} + \hat F^{\rm e}_{\hat x} = 0$ where the energy density and flux are
	\beqs
	\hat {\cal E} &=& \half \hat \rho \hat u^2 + \frac{\hat \rho^\g}{\g(\g - 1)} + \frac{\eps^2}{2} \frac{\hat \rho_{\hat x}^2}{\hat \rho}
	\quad \text{and} \cr 
	\hat F^{\rm e} &=& \left[\frac{\hat \rho \hat u^2}{2} + \frac{\hat \rho^\g}{\g-1} \right] \hat u - \eps^2 \left[ \hat u \hat \rho_{\hat x \hat x} - \frac{3}{2}  \frac{\hat u\hat \rho_{\hat x}^2}{\hat \rho} - \hat u_{\hat x} \hat \rho_{\hat x} \right].
	\eeqs
These equations follow from the PB $ \{ \hat \rho(\hat x), \hat u(\hat y) \} = \pdr_{\hat y} \del(\hat x - \hat y)$ and the Hamiltonian
	\beq
	\hat H = \int \left[\half \hat \rho \hat u^2 + \frac{\hat \rho^\g}{\g(\g - 1)} + \frac{\eps^2}{2} \frac{\hat \rho_{\hat x}^2}{\hat \rho} \right] d\hat x.
	\eeq
We will consider ICs that are fluctuations around a uniform state. For stability of the numerics, we separate the linear and nonlinear terms in the equations and treat the former implicitly and the latter explicitly. Introducing the book-keeping parameter $\del$ (which  will also enter through the ICs and may eventually be set to 1), we write
	\beq
	\hat \rho(\hat x, \hat t) = 1 + \del\, \tl \rho(\hat x, \hat t) \quad \text{and} \quad
	\hat u(\hat x, \hat t) = \thickbar u(\hat x, \hat t) + \del\, \tl u(\hat x, \hat t).
	\eeq
We will consider flow in the domain $-\pi \leq \hat x \leq \pi$ with periodic BCs and thus expand $\tl \rho$ and $\tl u$ as	\beq
 	\tl \rho = \sum_{ -\infty}^{\infty} \rho_n(\hat t) e^{in \hat x}, \quad
	\tl u = \sum_{ -\infty}^{\infty}  u_n(\hat t) e^{in \hat x} \;\; \text{with} \;\; {(\rho,u)}_{-n} = (\rho,u)_n^*.
	\eeq
Since $\int \hat \rho \: d \hat x$ is conserved, $\rho_0$ can be taken time-independent. Furthermore, we choose the constant $\bar \rho$ used to nondimensionalize $\rho$ as the (conserved) average density, so that $\rho_0 = 0$. We also suppose that the `background' flow velocity $\thickbar u$ is independent of position $\hat x$. Since $\int \hat u \: d\hat x$ is conserved and $\int \hat u \: d\hat x = 2\pi (\thickbar u(\hat t) + \del \: u_0(\hat t))$, we may absorb $\del \: u_0(\hat t)$ into $\thickbar u(\hat t)$ and thereby take $u_0 = 0$. 
Next, we write the continuity equation with nonlinear terms isolated:
	\beq
	\tl \rho_{\hat t} = - (\thickbar u \tl \rho + \tl u)_{\hat x} - \del\, {\calF^{\rm m}} \quad \text{where} \quad {\calF^{\rm m}} = (\tl \rho \tl u)_{\hat x}.
	\label{e:cont-linear-nonlinear-split}
	\eeq
The velocity equation (\ref{e:non-dim-barotropic-vel-momentum}) in conservation form is
	\beq
	\del \tl u_{\hat t} + \left( \frac{(\thickbar u + \del \tl u)^2}{2} + \ov{\g-1} (1 + \del \tl \rho)^{\g-1} \right)_{\hat x}
	- \eps^2 \del \left( \frac{\tl \rho_{\hat x \hat x}}{1 + \del \tl \rho} - \frac{\del}{2} \frac{\tl \rho_{\hat x}^2}{(1 + \del \tl \rho)^2} \right)_{\hat x} = 0.
	\eeq
Separating out the linear part we get
	\beqs
	\tl u_{\hat t} &=& - \left(\thickbar u \tl u + \tl \rho - \eps^2 \tl \rho_{\hat x \hat x} \right)_{\hat x} - \del \calF^{\rm u}
	\qquad \text{\rm where} \cr
	\calF^{\rm u} 
	&=& \left[ \frac{\tl u^2}{2} 
	+ \ov{\del^2}\ov{\g-1} \left\{ (1 + \del \tl \rho)^{\g-1} - \del (\g-1) \tl \rho \right\}\right]_{\hat x}\cr
	&&+ \left[\frac{\eps^2}{(1 + \del \tl \rho)^2} \left[  \tl \rho \tl \rho_{\hat x \hat x} ( 1 +  \del \tl \rho)  + \half \tl \rho_{\hat x}^2  \right]\right]_{\hat x}.
	\label{e:vel-linear-nonlinear-split} 
	\eeqs
	In (\ref{e:cont-linear-nonlinear-split}) and (\ref{e:vel-linear-nonlinear-split}) the linear terms are at ${\cal O}(\del^0)$ while the nonlinearities involve $\del$ to higher powers, depending on the value of $\g$. Interestingly, for $\g = 2$, the pressure gradient doesn't contribute to the nonlinear part of the acceleration:
	\beq
	\calF^{\rm u}_{\g=2} = \left[ \frac{\tl u^2}{2} 
	+ \frac{\eps^2 }{(1 + \del \tl \rho)^2}\left( \tl \rho \tl \rho_{\hat x \hat x}( 1+ \del \tl \rho)  + \half \tl \rho_{\hat x}^2  \right) \right]_{\hat x}.
	 \eeq
Expanding $\calF^{\rm m} = \sum_n \calF^{\rm m}_n e^{i n \hat x}$ and $\calF^{\rm u} = \sum_n \calF^{\rm u}_n e^{i n \hat x}$, the EOM in Fourier space become
	\beq
	\dot \rho_n = - i n (\thickbar u \rho_n + u_n) - \del \calF^{\rm m}_n \quad
	\text{and} \quad
	\dot u_n = - i n (\thickbar u u_n + (1 + \eps^2 n^2) \rho_n) - \del \calF^{\rm u}_n.
	\label{e:cont-vel-baro-fourier}
	\eeq
When nonlinearities ($\calF^{\rm m}$, $\calF^{\rm u}$) are ignored and we assume ($\rho_n(t), u_n(t) \propto e^{- i \om_n t}$), one finds the dispersion relation $(\om_n - n \thickbar u)^2 = n^2 (1 + \eps^2 n^2)$ familiar from \S \ref{s:dispersive-sound-waves}. To deal with the fully nonlinear evolution given by (\ref{e:cont-vel-baro-fourier}), we use a semi-implicit numerical scheme outlined in Appendix \ref{a:numerical-scheme}.

\subsection{Numerical results: Avoidance of gradient catastrophe, solitons and recurrence}
\label{s:numerical-results}

The above numerical scheme for $\g = 2$ is implemented by truncating the Fourier series after $n_{\rm max} = 16$ modes. The evolution is done for 750 time steps $(0 \leq \hat t \leq 15)$, each of size $\D = 0.02$, starting with the nondimensional ICs
	\beqs
	\hat \rho(\hat x,0) &=& 1 + \del \tl\rho(\hat x,0) \quad \text{and} \quad \hat u(\hat x,0) = 0 \quad \text{where} \cr
	\tl \rho(\hat x,0) &=& \sin \hat x \quad \text{or} \quad \cos 2 \hat x \quad\text{and} \quad \del = 0.1.
	\eeqs
We find that at early times, where $\hat u$ has a negative slope, its gradient increases and decreases where the slope is positive. Without the regularization $(\eps = 0)$, the higher Fourier modes can then get activated and the velocity and density profiles become highly oscillatory with steep gradients. Moreover, amplitudes begin to grow and the code eventually ceases to conserve energy and momentum.

\begin{figure}
\begin{center} 
 \includegraphics[width = 15cm]{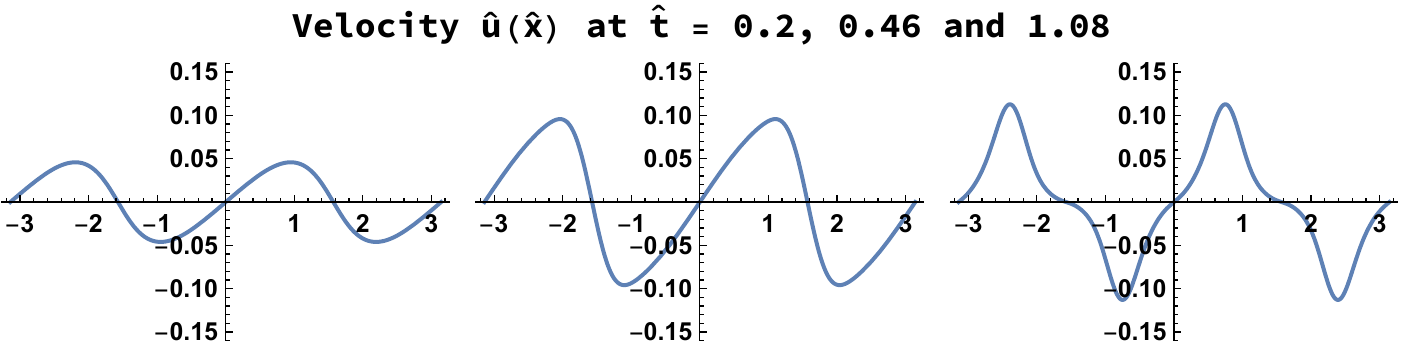} \caption{Evolution of velocity for IC $\hat \rho = 1 + 0.1 \cos{2 \hat x}$ and $\hat u = 0$ showing how the gradient catastrophe is averted through the formation of a pair of solitary waves in the velocity profile.}
 \label{f:u-reg-soliton-pair-forms}
 \end{center}
\end{figure}

\begin{figure}
\begin{center}
  \begin{subfigure}[t]{8cm}
    \includegraphics[width=7cm]{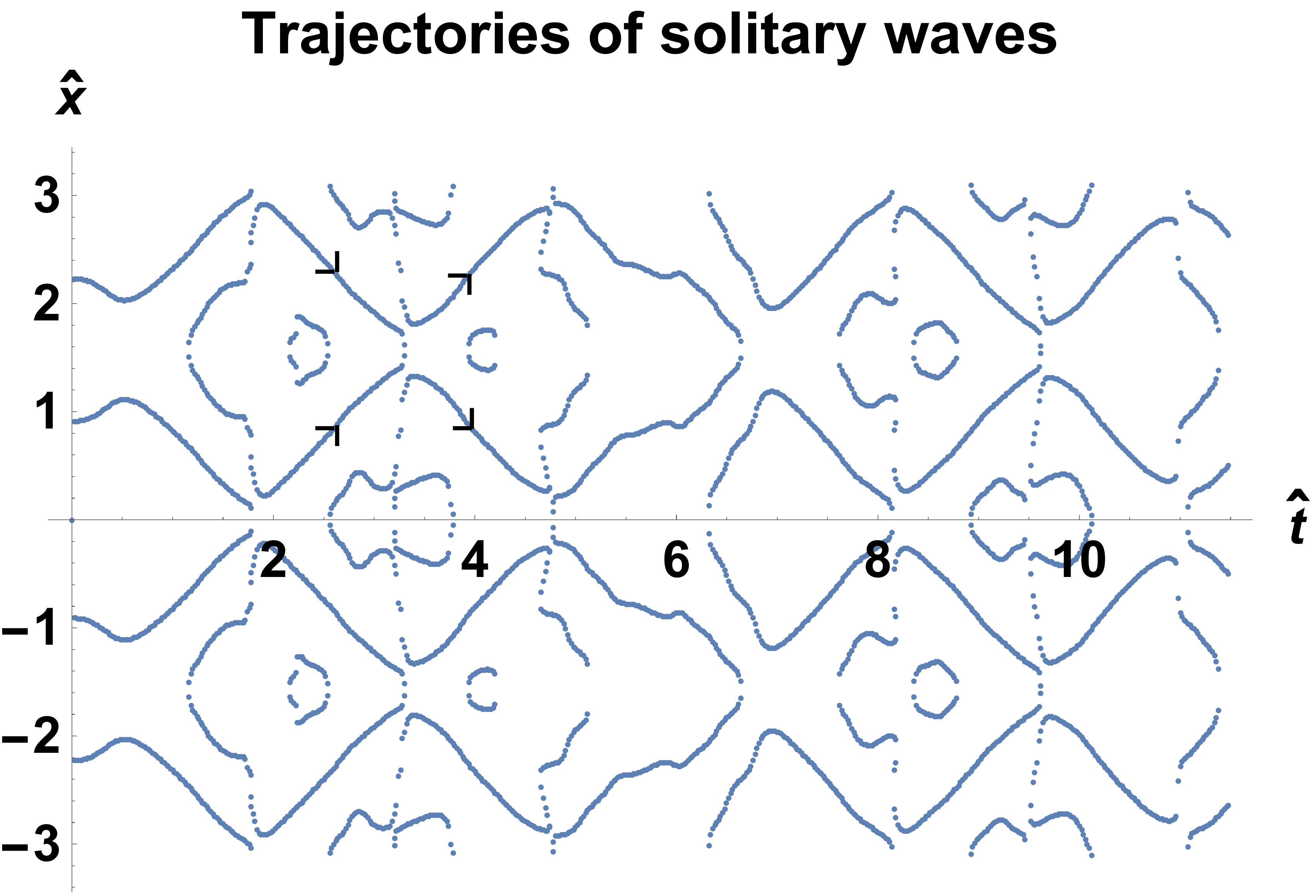}
    \caption{}
  \end{subfigure}
  \begin{subfigure}[t]{8cm}
    \includegraphics[width=8cm]{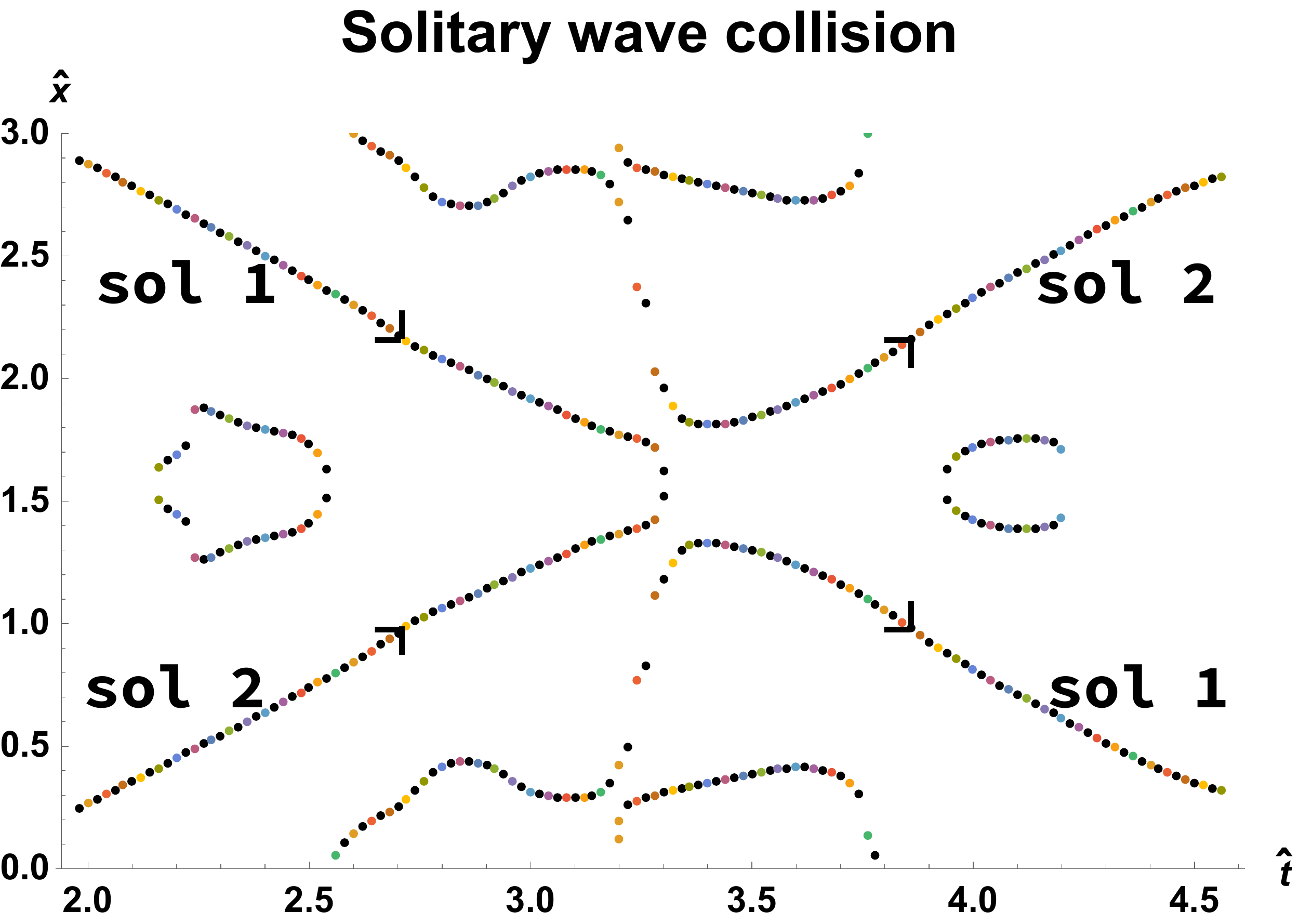}
    \caption{}
  \end{subfigure}
\caption{\label{f:solitary-wave-coll-cos2x} (a) Space-time trajectories of the `centers' of `solitary waves' in the velocity profile (for initial condition $\hat \rho = 1 + 0.1 \cos{2 \hat x}$ and $\hat u = 0$) showing several collisions. The locations of the `centers' are determined by finding the maxima/minima of $\hat u$ at each instant of time. (b) Close-up of one collision of solitary waves (sol 1 and sol 2), showing approximate asymptotic straight-line trajectories and phase shifts. The figures include the trajectories of the crests/troughs of small ripples that typically arise and go away in pairs but do not qualify as solitary waves.}
\end{center}
\end{figure}

By contrast, in the presence of the regularization (say $\eps = 0.2$), we find that the above gradient catastrophe is averted and energy is conserved (to about 3 parts in 1000) while momentum is conserved to machine precision. In fact, we find that the real and imaginary parts  $Q_r$ and $Q_i$ of the next conserved quantity (\ref{e:Qr-Qi}) are also conserved. Interestingly, when $u$ develops a steep negative gradient, a pair of solitary waves emerge at the top (wave of elevation) and bottom (wave of depression) of the $u$ profile and the gradient catastrophe is avoided (see Fig.~\ref{f:u-reg-soliton-pair-forms}). This mechanism by which the incipient shock-like discontinuity is regularized is to be contrasted with KdV, where an entire train of solitary waves can form \cite{kruskal-zabusky}. However, as with KdV, our solitary waves can suffer a head-on collision and pass through each other. After the collision, they re-emerge with roughly similar shapes and a phase shift. Fig. \ref{f:solitary-wave-coll-cos2x} shows the space-time trajectories of the centers of several of these solitary waves, showing their collisions. 

\begin{figure*}
\begin{center}
 \includegraphics[width = 7cm]{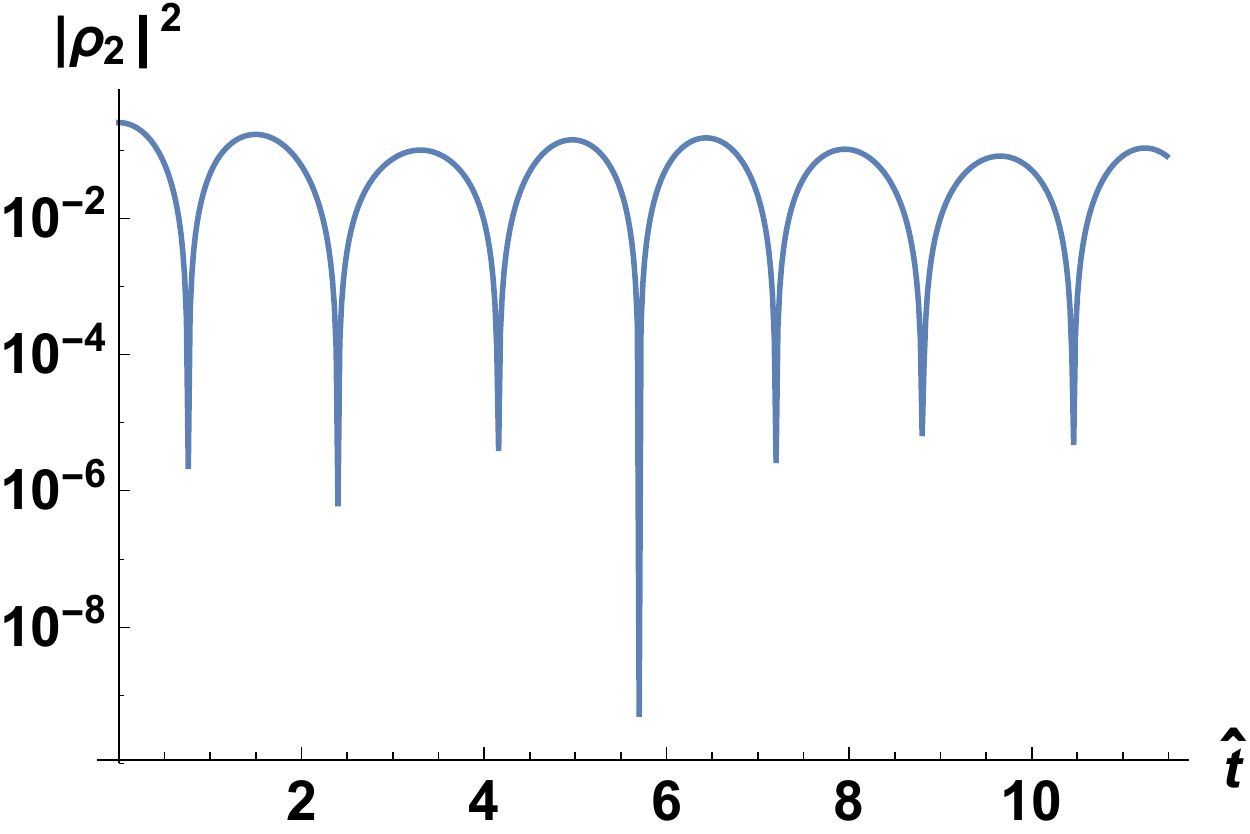}\qquad 
\includegraphics[width = 7cm]{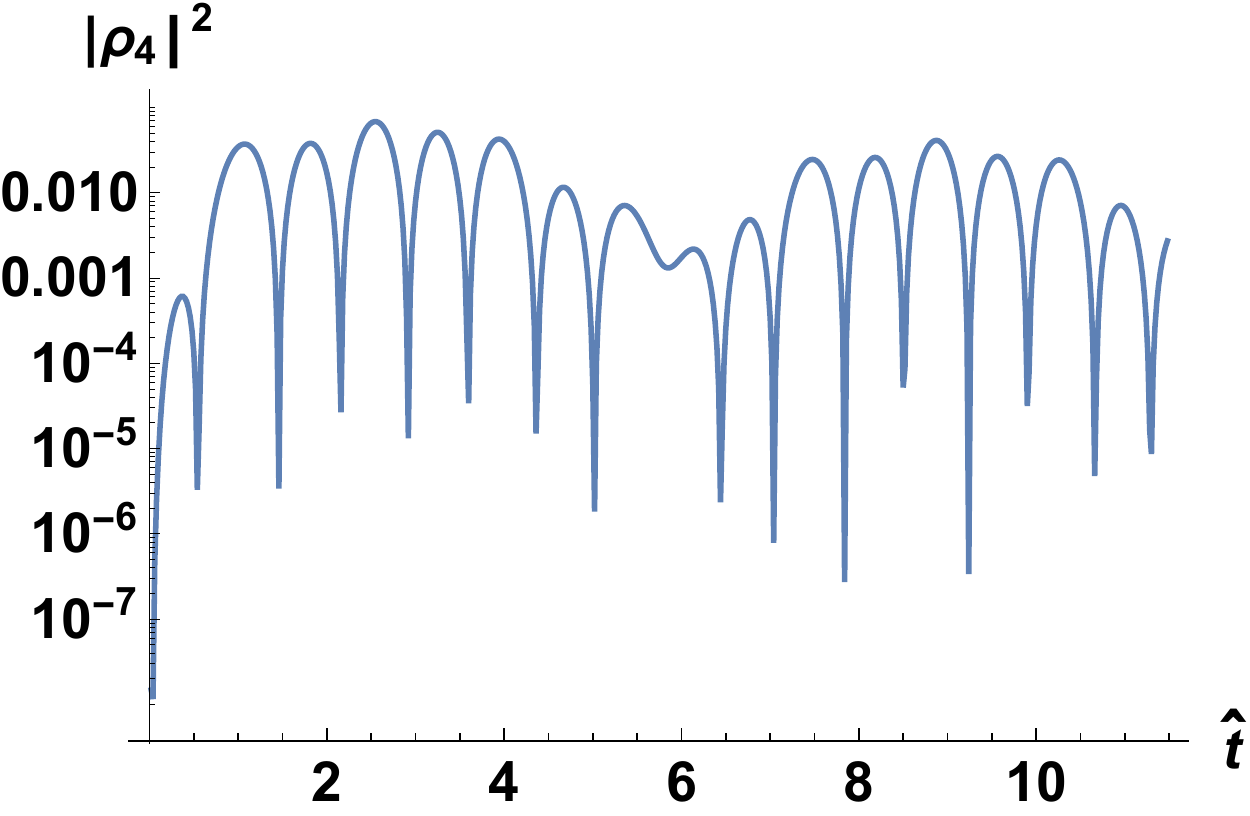}\qquad
 \includegraphics[width = 7cm]{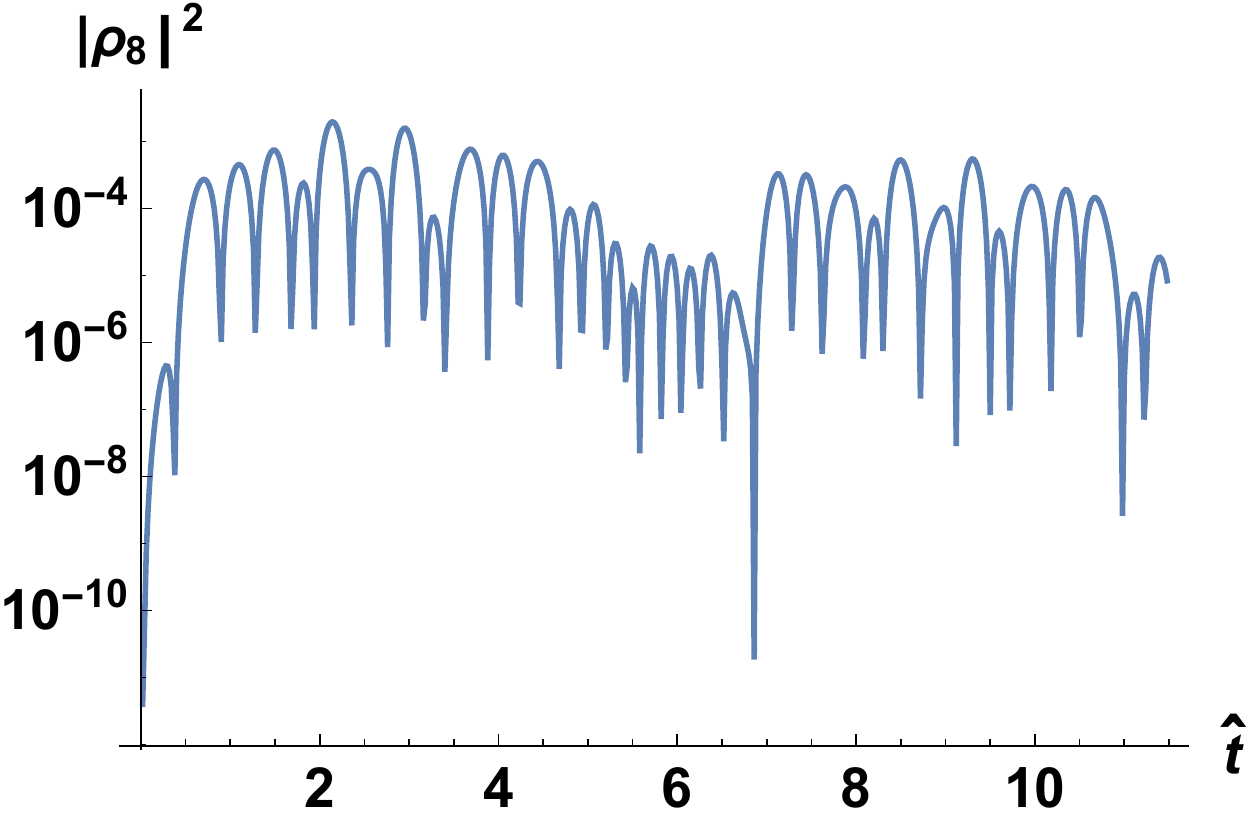}\qquad
 \includegraphics[width = 7cm]{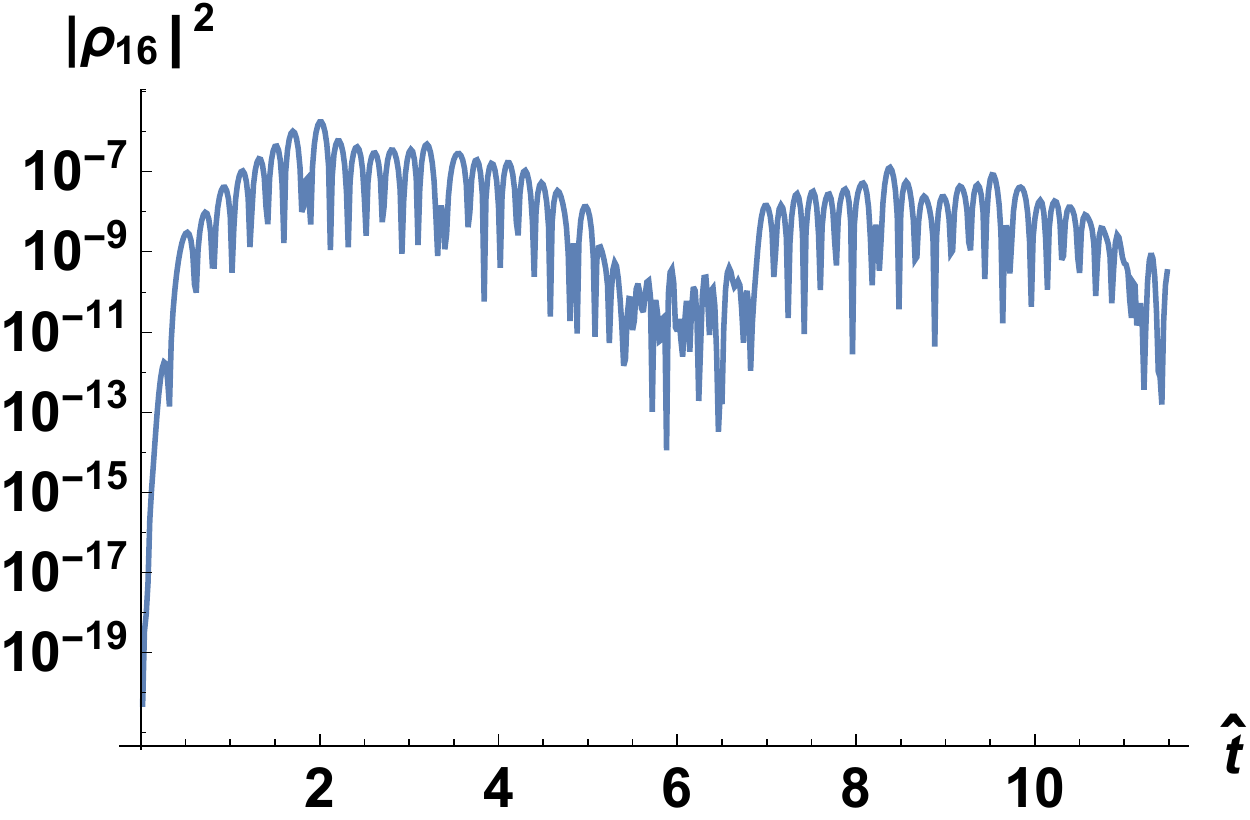}
\end{center}
 \caption{Time evolution of Fourier modes $|\rho_n|^2$ for $n = 2,4,8,16$ for IC $\tl \rho = \cos{2 \hat x}$ and $\tl u = 0$ showing that the higher modes remain uniformly small compared to the first few, justifying truncation of Fourier series.}
 \label{f:u-cos2x-four-modes-growth}
\end{figure*}

\begin{figure*}	
\begin{center}
	\begin{subfigure}[t]{7cm}
		\includegraphics[width=7cm]{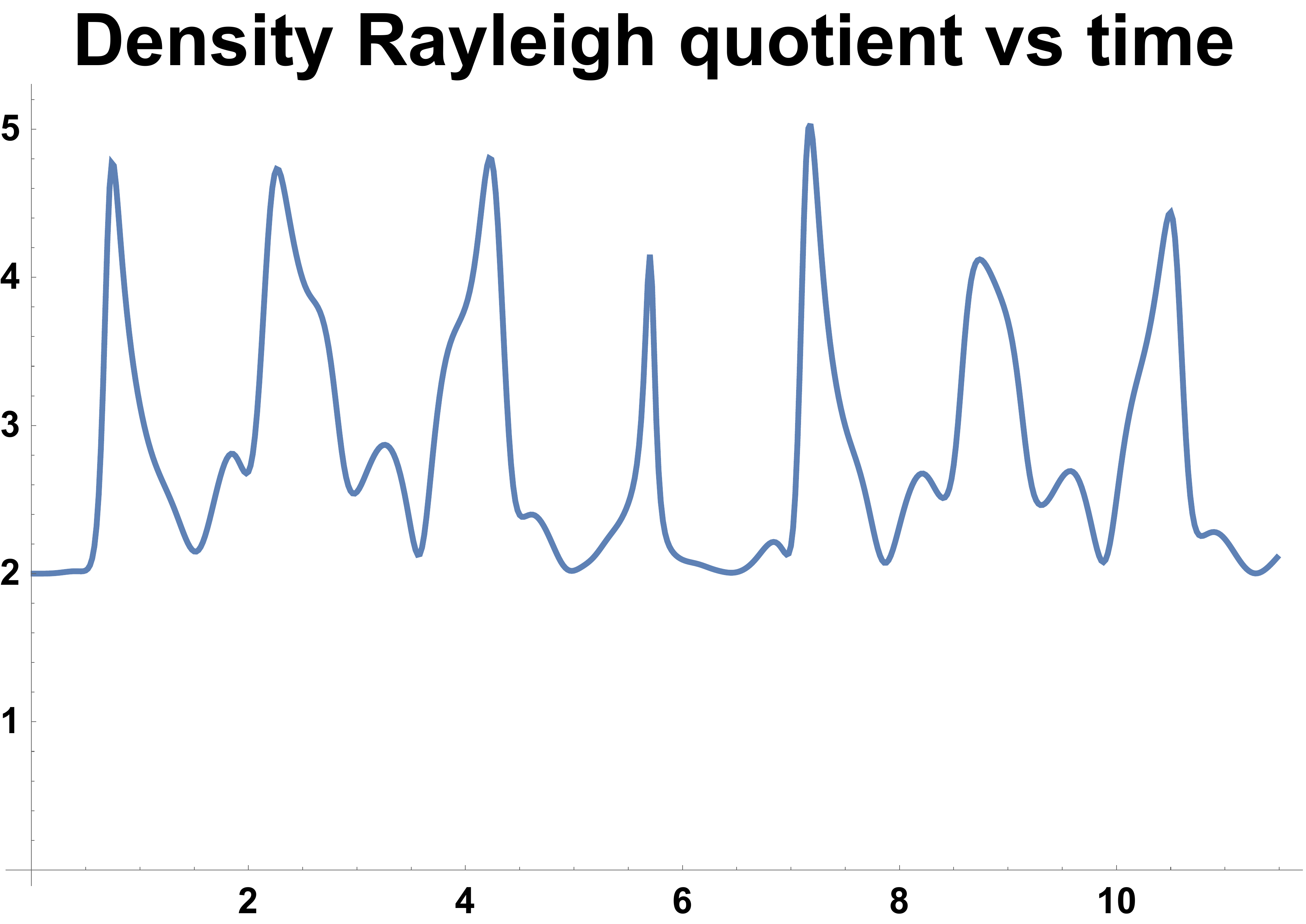}
		\caption{}
	\label{f:Rayleigh-quotient}
	\end{subfigure}
\qquad
	\begin{subfigure}[t]{8cm}
		\includegraphics[width=8cm]{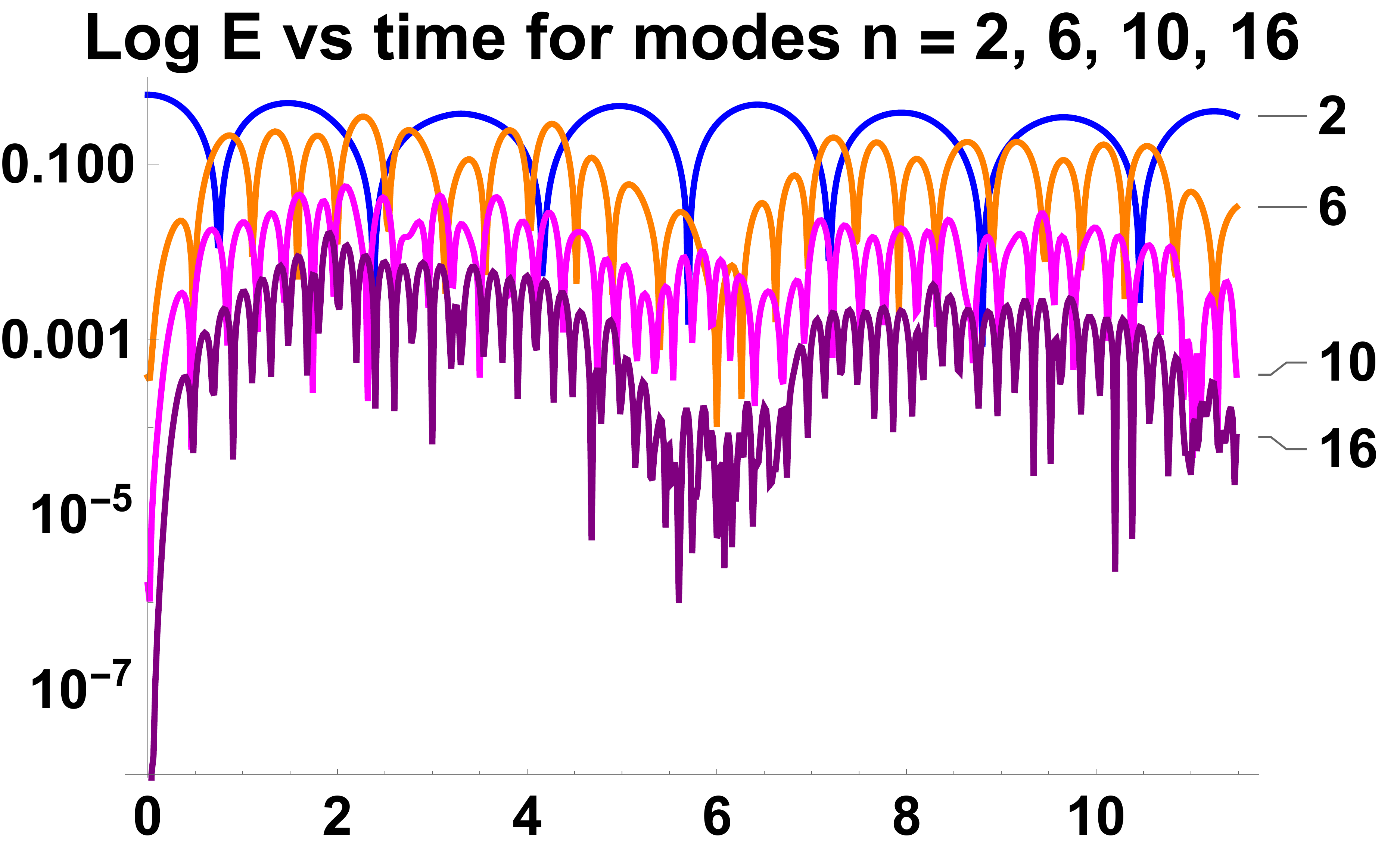}
		\caption{}
		\label{f:Four-time-evol-period}	
	\end{subfigure}
	\qquad 
	\begin{subfigure}[t]{8cm}
		\includegraphics[width=8cm]{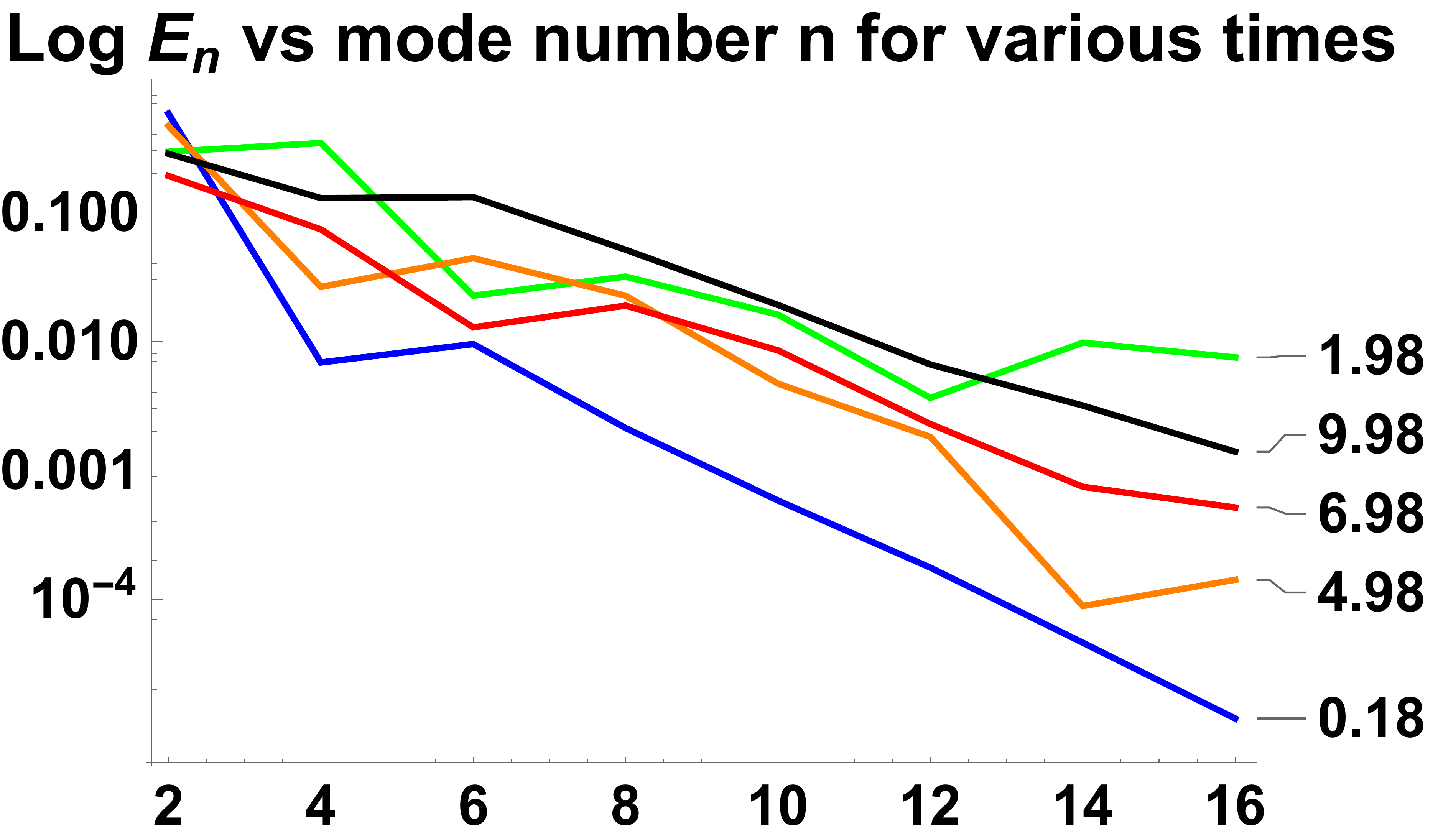}
		\caption{}
		\label{f:En-vs-n}	
	\end{subfigure}
\end{center}
	\caption{(a) Rayleigh quotient displays bounded oscillations indicating only a few modes are active. (b) Time evolution of $\log{E_n}$ for modes $n = 2,6,10,16$. The higher modes remain small with each one showing approximately periodic oscillations. (c) $\log{E_n}$ vs $n$ for a few values of $\hat t$ showing exponential drop with $n$. In all cases, the IC was $\tl \rho =\cos{2 \hat x}$ and $\tl u =0$.}
	\label{f:energy-spectral-distrib-plot}
\end{figure*}

We also find that higher modes $u_n$ and $\rho_n$ ($n \gtrsim 9$) grow from zero but soon saturate and remain a few orders of magnitude below the first few modes (see Fig.~\ref{f:u-cos2x-four-modes-growth}). This justifies truncating the Fourier series at $n_{\rm max} = 16$. The modes also display an approximate periodicity in time. This suggests recurrent motion \cite{thyagaraja-1979,thyagaraja-1983,SIdV}. This behavior is also captured in Fig.~\ref{f:Rayleigh-quotient} where we plot the Rayleigh quotient or mean square mode number
	\beq
	R =  \frac{\int |\hat \rho_{\hat x}|^2 d\hat x}{\int |\hat \rho|^2 d\hat x} = \frac{\sum_n n^2 |\rho_n|^2}{\sum_n |\rho_n|^2},
	\eeq
as a function of time. It is found to fluctuate between bounded limits indicating that effectively only a finite number of modes participate in the dynamics. This suggests that the system possesses additional conserved quantities (see \S~\ref{s:rayleigh-quotient-NLSE-conserved-quantities}). Another interesting statistic is the spectral distribution of energy $(E_n)$ and its dependence on time. Fig.~\ref{f:Four-time-evol-period}, shows the time evolution of $\log{E_n}$ for $n = 2,6,10,16$ for the IC $\tl \rho = \cos{2\hat x}$ and $\tl u = 0$ and demonstrates that the energy in the higher modes remains small. Moreover, each mode $\rho_n$ and $u_n$ oscillates between an upper and lower bound and shows approximate periodicity with differing periods. In Fig.~\ref{f:En-vs-n}, $\log{E_n}$ vs $n$ is plotted for a few values of $\hat t$. Unlike the power law decay $n^{-5/3}$ in the inertial range of fully developed turbulence, here we see that $E_n$ drops exponentially with $n$. In particular, there is no equipartition of energy among the modes.

\section[Connection to nonlinear Schr\"odinger]{Connection to nonlinear Schr\"odinger and generalizations}
\label{s:r-gas-to-nlse}

Our numerical results indicate recurrence and soliton-like scattering in 1d isentropic R-gas dynamics for $\g = 2$, suggesting integrability. Remarkably, in this case, R-gas dynamics 
is transformable into the defocussing (repulsive) cubic nonlinear Schr\"odinger equation (NLSE). More generally, adiabatic R-gas dynamics may be viewed as a novel generalization of the NLSE. Indeed, let us write the velocity field as $\bfv = \grad \phi + \frac{\la \grad \mu}{\rho} + \frac{\al \grad s}{\rho}$ where $s$ is the specific entropy and $\al,\la$  and $\mu$ are Clebsch potentials \cite{zakharov-kuznetsov}. As in treatments of superfluidity \cite{gross,pitaevskii} and `quantum hydrodynamics' \cite{madelung}, if we define the Madelung transform, $\psi(\bfr,t) = \sqrt{\rho} \exp{(i  \phi(\bfr)/2 \sqrt{\beta_*})}$, then the R-gas dynamic energy density (\ref{e:3d-hamiltonian}) becomes:
	\beqs \small
	{\cal E} &=& 2 \beta_* |\grad \psi|^2 + 2 \frac{\sqrt{\beta_*}}{|\psi|^2}\left( \al \grad s + \la \grad \mu\right) \cdot \left( \frac{\psi^* \grad \psi - \psi \grad \psi^* }{2i} \right) \cr
	&& + \frac{\la \al}{|\psi|^2} \grad \mu \cdot \grad s + \frac{\al^2 (\grad s)^2 + \la^2 (\grad \mu)^2 }{2|\psi|^2} + \frac{\bar p e^{s/c_V}}{(\g-1)} \frac{|\psi|^{2\g}}{\bar \rho^\g}.
	\label{e:hamil-3d-psi-s-alpha}
	\eeqs
Here $\bar p$ and $\bar \rho$ are constant reference pressure and density. The transformation from $(\rho, \phi)$ to $(\psi, \psi^*)$ is canonical. The corresponding equations of motion for $\psi, \al, \la, \mu$ and $s$ may be obtained using the canonical bosonic PBs $\{ \psi(\bfr), \psi^*(\bfr') \} = - (i/2 \sqrt{\beta_*}) \del(\bfr - \bfr')$, $\{ \la(\bfr), \mu(\bfr') \} = \del(\bfr - \bfr')$ and $\{ \al(\bfr), s(\bfr') \} = \del(\bfr - \bfr')$.

Specializing to isentropic potential flow where $s = \bar s$ is constant, $p = K (\g - 1) \rho^\g$ (\ref{e:entropy-barotropic}) and $\bfv = \grad \phi$, (\ref{e:hamil-3d-psi-s-alpha}) simplifies to
	\beq
	H = \int \left[ 2 \beta_* |\grad \psi|^2 + K |\psi|^{2\g} \right] d\bfr.
	\eeq
Using the above PBs for $\psi$, one finds  that $\psi$ satisfies the {\it defocusing} nonlinear Schr\"odinger equation
	\beq
	i \sqrt{\beta_*} \psi_t = - \beta_* \grad^2 \psi + \half \g K |\psi|^{2(\g-1)} \psi.
	\label{e:nlse-3d-gen-gamma}
	\eeq
Interestingly, in 1d, the R-gas dynamic form of the {\it focusing} cubic $(\g = 2)$ NLSE had been obtained in the context of the Heisenberg magnetic chain \cite{lakshmanan-ruijgrok,turski}. However, as noted in \cite{turski}, the Heisenberg chain leads to negative pressure! Returning to (\ref{e:nlse-3d-gen-gamma}), we see that the real and imaginary parts of the NLSE correspond to the Bernoulli and continuity equations. The $\grad^2\psi$ term leads to the divergence of the mass flux, $\bfv^2$ and regularization terms in the isentropic R-gas dynamic equations
	\beq
	\rho_t + \grad \cdot (\rho \bfv) = 0 \quad \text{and} \quad
	\phi_t = - \g K \rho^{\g-1} - \frac{\bfv^2}{2} + 2 \beta_* \frac{\grad^2 \sqrt{\rho}}{\sqrt{\rho}}.
	\eeq
Evidently, $\beta_*$ plays the role of $\hbar^2$. The nonlinear term $(\g K/2) |\psi|^{2 (\g-1)} \psi$ corresponds to the isentropic pressure $p = (\g-1) K \rho^\g$ whose positivity implies we get the defocusing/repulsive NLSE. Thus, our regularization term $2 \beta_*(\grad^2 \sqrt{\rho})/\sqrt{\rho}$ is like a quantum correction to the classical isentropic pressure. For $\g = 2$ we get the cubic NLSE or Gross-Pitaevskii equation (without an external trapping potential). Note that 1d isentropic flow on the line with $\bfv = (u(x),0,0)$ is always potential flow: $u = \phi_x$. So the above transformation takes 1d isentropic R-gas dynamics (\ref{e:unsteady-barotropic-eqns}) to the defocusing 1d NLSE, which for $\g = 2$ and periodic BCs admits infinitely many conserved quantities in involution \cite{faddeev-takhtajan}. This explains our numerical observations of approximate phase shift scattering of solitary waves and recurrence.

\subsection{NLSE interpretation of steady R-gas dynamic cavitons and cnoidal waves}
\label{s:caviton-to-nlse}

It turns out that steady solutions of 1d isentropic R-gas dynamics (\S \ref{s:steady-trav-quadrature}) correspond to NLSE wavefunctions $\psi$ with harmonic time dependence. For $\g = 2$, our aerostatic caviton corresponds to the dark soliton of NLSE. More generally, aerostatic steady solutions correspond to $\psi$ of the form $\sqrt{\rho(x)} \exp(-i F^{\rm u} t/2\sqrt{\beta_*})$ where $F^{\rm u}$ is the constant velocity flux (\ref{e:barotropic-curr}). Finally, non-aerostatic cnoidal waves correspond to interesting asymptotically plane wave NLSE wavefunctions modulated by a periodic cnoidal amplitude. Here, we consider the cavitons and $\text{cn}$ waves in increasing order of complexity.

\noindent{\bf Aerostatic caviton:} The simplest caviton solution of \S \ref{s:exact-cavitons-cnoidal-g-2} is aerostatic: 
    \beq
	\rho(x) = \rho_\pt \tanh^2\left( \frac{x}{2\la_\pt} \right), \quad u(x)=0 \quad \text{and}\quad p(x) = K \rho^2 \quad \text{where} \quad K = \frac{c_\pt^2}{2\rho_\pt},
    \eeq
Here, $\rho_\pt, \la_\pt$ and $c_\pt^2$ are positive constants. The corresponding specific enthalpy and velocity flux are 
    \beqs
    h &=& 2 K \rho = c_\pt^2 \tanh^2\left( \frac{x}{2\la_\pt}\right) \quad \text{and} \cr
    F^{\rm u} &=& \frac{u^2}{2} + h - 2\beta_*\frac{(\sqrt{\rho})_{xx}}{\sqrt{\rho}} = c_\pt^2 \quad \text{where} \quad \beta_* = \la_\pt^2 c_\pt^2.
    \eeqs
Thus, the Bernoulli equation $\phi_t + F^{\rm u} = 0$ (\ref{e:barotropic-curr}) is satisfied provided we take the velocity potential $\phi = -c_\pt^2\, t$ to be time-dependent. The resulting $\psi$ is the dark soliton solution of the defocusing NLSE (see \S 6.6 of \cite{ablowitz})
    \beq
    \psi = \sqrt{\rho} \;\exp\left( {\frac{i\phi}{2\sqrt{\beta_*}}} \right) =  \sqrt{\rho_\pt} \tanh\left(\frac{x}{2\la_\pt}\right) \exp\left({{-\frac{ic_\pt^2t}{2\sqrt{\beta_*}}} }\right).
    \eeq
\noindent{\bf Non-aerostatic caviton:} More generally, for a non-aerostatic caviton (for $0 < M_\pt < 1$ and $\rho_\pt, \la_\pt$ and $c_\pt^2$ positive constants)
    \beq
    \rho(x) = \rho_\pt \left( 1 - (1 - M_\pt^2) \sech^2\left( \sqrt{\frac{1 - M_\pt^2}{4}} \frac{x}{\la_\pt} \right) \right),\quad
    u = \frac{c_\pt M_\pt \rho_\pt}{\rho(x)} \quad \text{and} \quad p = \frac{c_\pt^2}{2\rho_\pt} \rho(x)^2.
    \eeq
In this case, the constant velocity flux is $F^{\rm u} = c_\pt^2 (2 + M_\pt^2)/2$. The resulting velocity potential is
    \beq
    \phi = - \half c_\pt^2 (2 + M_\pt^2)t + c_\pt  M_\pt x+ 2 c_\pt \la_\pt \arctan\left( \frac{\sqrt{1 - M_\pt^2}}{M_\pt} \tanh \left( \frac{\sqrt{1 - M_\pt^2}}{2\la_\pt}x \right) \right).\quad 
    \eeq
Thus, $\psi$ is asymptotically a plane wave with phase speed ${c_\pt (2 + M_\pt^2)/2M_\pt}$. This $\psi$ may be regarded as a high-frequency carrier wave modulated by a localized signal.

{\fl \bf Aerostatic snoidal waves:} The simplest steady periodic solutions is the aerostatic snoidal wave (\ref{e:cnoidal-wave-gamma=2}):
    \beq
    u \equiv 0, \quad \rho = \rho_\pt (1-2\ka_\pt)\; \text{sn}^2\left(\frac{x}{2\la_\pt}, 1-2\ka_\pt \right) \quad\text{for} \quad 0< \ka_\pt < \frac{1}{2}.
    \eeq
Here, $F^{\rm u} = c_\pt^2 (1 - \ka_\pt)$ and $\phi = - F^{\rm u} \,t$. The resulting $\psi$ is a snoidal wave with harmonic time dependence:
    \beq
    \psi = \sqrt{\rho_\pt(1-2\ka_\pt)} \; \text{sn}\left(\frac{x}{2\la_\pt}, 1-2\ka_\pt \right) \exp\left({-\frac{ i c_\pt^2 (1 - \ka_\pt)\,t}{2\sqrt{\beta_*}}}\right).
    \eeq
\noindent{\bf Non-aerostatic cnoidal waves:} Finally, in the general case, we have from (\ref{e:cnoidal-wave-rho-til-triangle}), 
    \beqs
    \rho(x) &=& \rho_\circ\left[ \tl\rho_+ - (\tl\rho_+ - \tl\rho_-) \,{\rm cn}^2\left( \frac{\sqrt{1 - \tl \rho_-}}{2} \frac{x}{\la_\circ}, \frac{\tl\rho_+ - \tl\rho_-}{1 - \tl\rho_-} \right) \right], \cr 
    u(x) &=& \frac{c_\circ M_\circ \rho_\circ}{\rho(x)} \quad \text{and} \quad
    p(x) = \frac{c_\circ^2}{2\rho_\circ}\rho(x)^2
    \eeqs
where $0 < \ka_\circ < 1/2$ and $0 < M_\circ < 1 - \sqrt{2\ka_\circ}$ (lower triangular region of Fig.~\ref{f:kappa-M-K>0}). Here $\tl\rho_{\pm}(\ka_\circ, M_\circ)$ are as in (\ref{e:steady-rho-plus-minus}). Furthermore, $\rho_\circ, c_\circ^2$ and $\la_\circ$ are positive constants that set scales. As before, $\phi = - F^{\rm u} t + \int_0^x u(x') \,dx'$, where the constant velocity flux $F^{\rm u}$ depends on $c_\circ, \ka_\circ$ and $M_\circ$, but is independent of $\rho_\circ$ and $\la_\circ$. Though an explicit formula for $\phi(x,t)$ is not easily obtainable, it is evident that for large $x$, $\phi$ grows linearly in $x$ with a subleading oscillatory contribution. Thus, $\psi$ has a purely harmonic time-dependence $\exp \left( -i F^{\rm u} t/2\sqrt{\beta_*} \right)$, a periodic cnoidal modulus $|\psi| = \sqrt{\rho(x)}$ and an argument that asymptotically grows linearly in $x$. Thus, asymptotically, $\psi$ is a plane wave modulated by the periodic amplitude $\sqrt{\rho(x)}$.

\subsection{Conserved quantities and Rayleigh quotient of NLSE and R-gas dynamics}
\label{s:rayleigh-quotient-NLSE-conserved-quantities}

The cubic 1d NLSE admits an infinite tower of conserved quantities. The first three are
	\beq
	N = \int |\psi|^2 \: dx, \quad P_{\rm NLSE} = \int \Im (\psi^* \psi_x) \: dx, \quad \text{and}\quad
	E_{\rm NLSE} = \int \left(|\psi_x|^2 + \frac{K}{2 \beta_*} |\psi|^4 \right) {\beta_*^{1/4}} dx.
	\eeq
These correspond to the mass $M = N$, momentum $P = 2 \sqrt{\beta_*} P_{\rm NLSE}$ and energy $H = 2 \beta_*^{3/4} E_{\rm NLSE}$ of R-gas dynamics. The next conserved quantity of NLSE (with periodic BCs) is \cite{faddeev-takhtajan}
	\beq
	Q = \int_{-L}^L \left[ \sqrt{\beta_*} \psi^* \psi_{xxx} - \frac{K |\psi|^2}{2 \sqrt{\beta_*}} (\psi \psi^*_x + 4 \psi^* \psi_x) \right]dx.
	\eeq
$\Re Q$ and $\Im Q$  correspond to the following in R-gas dynamics:
	\beq
	Q_r = \frac{3\sqrt{\beta_*}}{4} \int_{-L}^L \left( \frac{\rho_x^3}{2 \rho^2} - \frac{\rho_x \rho_{xx} }{\rho} \right) \: dx \quad \text{and}\quad
	Q_i = - \int_{-L}^L \left[ \frac{u_x \rho_x}{2} + \frac{3 u \rho_x^2}{8 \rho} + \frac{u^3 \rho}{8 \beta_*} + \frac{ 3 K u \rho^2}{4\beta_*} \right] dx.
	\label{e:Qr-Qi}
	\eeq
In \S \ref{s:numerical-scheme} we use the conservation of $Q_r$ and $Q_i$ to test our numerical scheme. In Ref.~\cite{thyagaraja-1979}, it was shown that for the cubic 1d NLSE, the Rayleigh quotient or mean square mode number for periodic BCs,
    \beq
    R_{\rm NLSE} = \frac{\int_{-L}^L |\psi_x|^2\,dx}{\int_{-l}^l |\psi|^2\, dx},
    \eeq
is related to the number of active degrees of freedom and is bounded in both focussing and defocussing cases. This has a simple interpretation in R-gas dynamics, since
    \beq
    R_{\rm NLSE} = \frac{1}{2\beta_* M} \int_{-L}^L \left( \half \rho u^2 + \frac{\beta_*}{2} \frac{\rho_x^2}{\rho} \right)\, dx \leq \frac{H}{2\beta_* M}.
    \eeq
Consequently, 1d isentropic R-gas dynamic potential flows are recurrent in the sense discussed in \cite{thyagaraja-1979}.

\subsection{Connection to vortex filament and Heisenberg chain equations}
\label{s:neg-pressure-vortex-filament}

Intriguingly, the {\it negative pressure} $\g = 2$ isentropic R-gas dynamic equations in 1d for $\rho$ and $u$ (\ref{e:unsteady-barotropic-eqns}) are equivalent to the vortex filament equation $\dot \xi = G\, \xi' \times \xi''$ where $G$ is a constant. Here the curve $\xi(x,t)$ represents a vortex filament with tangent vector $\xi'$. The equivalence is most transparent when the vortex filament equation is expressed in the Frenet-Serret frame, with curvature $\ka = \sqrt{\rho}$ and torsion $\tau = u$ \cite{lakshmanan-ruijgrok}. The FS frame ($\xi', n, b$) is an orthonormal basis along the filament, where $n = \xi''/\ka$ is the unit normal and $b = \xi'\times n$ the binormal. The FS equations describe how the frame changes along the filament.
	\beq
	\xi'' = \ka n, \quad n' = -\ka \xi' + \tau b \quad \text{and} \quad b' = -\tau n
	\eeq
We may use the vortex-filament and FS equations to find evolution equations for the FS frame: 
	\beqs
	&& \frac{\dot \xi'}{G} = - \ka \tau n + \ka' b, \quad 
	\frac{\dot n}{G} = \ka \tau \xi' + \left( \frac{\ka''}{\ka} - \tau^2 \right) b
	- \ov{\ka}\left( 2\ka' \tau + \ka \tau' + \frac{\dot \ka}{G} \right) n \cr
	&& \text{and} \quad 
	\frac{\dot b}{G} = - \left(\ka' \xi' +  \left( \frac{\ka''}{\ka} - \tau^2 \right) n\right).
	\label{e:evol-of-fs-frame}
	\eeqs 
Eqn.~(\ref{e:evol-of-fs-frame}) leads to evolution equations for $\ka$ (using $\dot n \cdot n = 0$ as $n\cdot n = 1$) and $\tau = n'\cdot b$:
	\beq
	\dot\ka = -G \left( 2\ka'\tau + \ka\tau' \right) \quad \text{and}\quad \dot \tau = G \left( \frac{\ka^2}{2} - \tau^2 + \frac{\ka''}{\ka}  \right)'.
	\label{e:ka-tau-evol}
	\eeq
Taking $G = 1/2$, $\ka = \sqrt{\rho}$ and $\tau = u$, (\ref{e:ka-tau-evol}) reduces to the continuity and velocity equations for $\g = 2$ isentropic R-gas dynamics with $\beta_* = 1/2$, but with a negative $p = -\rho^2/8$ \cite{turski}. Furthermore, it is well-known \cite{sgrajeev} that the vortex filament equation is related to the Heisenberg magnetic chain equation $\dot S = G \,S \times S''$ with $S = \xi'$. An open question is to give a geometric or magnetic chain interpretation of the {\it positive pressure} R-gas dynamic equations as well as those for general $\g$. It is noteworthy that negative pressures (relative to atmospheric pressure) can arise in real flows, for instance in the presence of strong currents \cite{ali-kalisch-2}.

\chapter{Conclusions and discussion}

\section{The Higgs added mass correspondence}
\label{s:discussion-ham}

In the first part of this thesis, we have proposed a new physical correspondence between the Higgs mechanism in particle physics and the added-mass effect in fluid mechanics. While plasmas and superconductors illustrate the abelian Higgs model, the Higgs added-mass correspondence provides a non-dissipative hydrodynamic analogy for the fully non-abelian Higgs mechanism. It encodes a pattern of gauge symmetry breaking in the shape of a rigid body accelerated through fluid. A dictionary relates symmetries and various quantities on either side. By identifying the gauge Lie algebra with the space of fluid flow, and relating added-mass eigenvalues to vector boson masses, we are able to specify when a particular pattern of spontaneous symmetry breaking {\it corresponds} to a particular rigid body accelerated through a fluid.  Besides possible refinements and generalizations (to compressible [see Appendix~\ref{a:extension-added-mass-effect-compressible}] and rotational flows or inclusion of fermion masses [see Appendix~\ref{a:ham-extension-fermions}]), the new viewpoint raises several interesting questions and directions for further research in both fluid mechanics and particle physics (1) The Higgs is the lightest scalar particle. We conjecture that the fluid analog is a characteristic fluid mode around an accelerating body, with wavelength comparable to the size of the body (rather than the container). There may be several such modes, which could suggest heavier scalar particles. We also wonder whether there is any relation between the dispersion relation of such a wave around the body and the Higgs mode in the field theory. (2) Understanding such modes requires extension of the added mass formalism to flows other than those usually studied in marine hydrodynamics (incompressible potential flow) \cite{newman}. This would allow for waves around the body, that could play the role of the Higgs particle. Perhaps the simplest such flows are compressible potential flow, incompressible flows with vorticity (even in two dimensions) and surface gravity waves in incompressible flow around an accelerated body. Moreover, density fluctuations in compressible flow around a rigid body should be analogous to quantum fluctuations around the scalar vacuum expectation value. Thus the HAM correspondence gives a new viewpoint and impetus to develop techniques to study the added mass effect in flows other than those studied so far. (3) We identified a discrete broken symmetry in the added-mass effect. Is there a continuous one, perhaps having to do with Galilean invariance? (4) The fluid flow affects the rotational inertia of a rigid body, giving it an added inertia tensor. Is there a particle physics analog consistent with the quantization of angular momentum? For instance, could motion through the scalar medium modify the magnetic moments of particles? (5) The HAM correspondence relates rigid body motion through $d$-dimensional flows (see \S~\ref{s:d-geq-3-added-mass-effect}) to SSB of gauge theories with $d$-dimensional gauge group. Given the importance and simplifications in the 't Hooft limit of multi-color gauge models, one wonders whether there are aspects of these fluid flows that simplify as $d \to \infty$. Could a suitable $d \to \infty$ limit provide a starting point for an approximation method for studying 3d flows? (6) How is the added-mass of a composite body related to the added masses of its constituents? Correspondingly, can one compute a small correction to the mass of a hadron, from Higgs interactions among a system of quarks [beyond the Higgs contribution to individual current quark masses]? This would be a small `Higgs force' correction to the mass of the proton in addition to the main contributions from strong and electromagnetic forces. (7) There is one difference between the added mass effect in fluids and mass generation via the Higgs mechanism. While gauge fields start out being massless (for renormalizability), the original mass of a rigid body is arbitrary. This difference does not affect our analogy, which only concerns the mass {\it gained} through interaction with the scalar field or fluid. Besides, the added mass is independent of the original mass of the rigid body. In this context, turning things around, we wonder if there is there an example of a massive particle which gains additional mass through the Higgs mechanism. (8) We would also like to find a fluid analog of a spontaneously broken gauge theory with multiple coupling constants corresponding to different simple factors of the gauge group (e.g. SU$(2)$ and U$(1)$ in the Weinberg-Salam model). In particular, is there a fluid analog of the Weinberg mixing angle?

\section{Dispersive regularization of gas dynamics}
\label{s:discussion-r-gas-dyn}

It is a significant feature of our attempt to conservatively regularize singularities in gas dynamics that it has led us (in the case of isentropic potential flow) to the {\it defocusing} NLSE. Heuristically, the defocusing interaction tends to amplify linear dispersive effects and thereby prevent blowups. For $\gamma = 2$, this connection to the cubic NLSE should provide powerful tools [including the inverse scattering transform in 1d (see \S 9.10 of \cite{ablowitz}, \cite{faddeev-takhtajan} and \cite{novikov})] to deal with the initial-boundary value problems in various dimensions as well as alternative numerical schemes. Moreover, the bound on the NLSE Rayleigh quotient obtained in \cite{thyagaraja-1979} (see also \S\ref{s:rayleigh-quotient-NLSE-conserved-quantities}) generalizes to 2d and 3d as well as to nonlinearities other than cubic ($\gamma \ne 2$). This would have implications for recurrence in more general R-gas dynamic isentropic potential flows, even in the absence of integrability. The techniques of dispersive shock wave theory \cite{whitham,El-Hoefer,miller} could provide additional tools to address R-gas dynamic flows.

In \cite{tao-nonlinear-dispersive,tao-global-behav}, a classification of semilinear PDEs [perturbations of linear equations by lower order nonlinear terms] into subcritical, critical and supercritical, based on conserved quantities (mass, energy etc.), scaling symmetries and regularity of initial data is described. Though the equations of R-gas dynamics given in \S \ref{s:3d-hamil-form-R-gas-dyn} are not semilinear, in the special case of isentropic potential flow, the transformation to NLSE makes them semilinear. It is thus interesting to examine the implications of this classification for R-gas dynamics. For example, the critical scaling regularity \cite{tao-nonlinear-dispersive} of NLSE in $d$ spatial dimensions is $s_c = d/2 - 1/(\g-1)$. Thus, if the initial data is such that the number of particles and energy are finite (so that $\psi, \grad \psi \in L^2$ and $\psi \in H^1$), then according to the scaling heuristic, the NLSE is subcritical for any $\g > 1$ in 1d and 2d and also for $1 < \g < 3$ in 3d.

In \S\ref{s:patched-shock-weak-sol} we argued that 1d R-gas dynamics does not admit any smooth or continuous shock-like steady solutions. In fact, we found that if we try to patch half a caviton at its trough density with a constant state, then the mass, momentum and energy fluxes in the pre-shock region cannot all match their values in the post-shock region, so that the Rankine-Hugoniot conditions are violated. We conjecture that this absence of steady shock-like solutions is a general feature of a wide class of conservatively regularized gas dynamic models. Loosely, this is like d'Alembert's `theorem' that continuous solutions of Euler's equations cannot ever lead to drag, although possibly to lift. On the other hand, inclusion of viscosity {\it does} permit drag as well as steady shock-like solutions as in the Burgers equation \cite{whitham, ablowitz}. Allowing for non-steady solutions, we find that in R-gas dynamics, the gradient catastrophe is averted through the formation of a pair of solitary waves (see \S\ref{s:numerical-results}). It would be interesting to see if this mechanism is observed in any physical system, say, one where dissipative effects are small as in nonlinear optics, weak shocks, cold atomic gases or superfluids. For further discussion on steepening gradients and criteria for detecting wave breaking in dispersive hydrodynamics, see \cite{hatland-kalisch,wilkinson-banner}.

Though we have formulated R-gas dynamics in 3d, our analytic and numerical solutions have been restricted to 1d. It would be interesting to extend this work to higher dimensional problems such as oblique shocks and the Sedov-Taylor spherical blast wave problem. The mechanical and thermodynamic stability of our traveling caviton and periodic wave solutions is also of interest. One also wishes to examine whether the capillarity energy considered here arises from kinetic theory in a suitable scaling limit of small Knudsen number as for the Korteweg equation \cite{gorban-karlin,huang-wang-wang-yang}. Finally, our Hamiltonian and Lagrangian formulations of R-gas dynamics can be used as starting points in formulating the quantum theory. The transformation to NLSE provides another approach to quantization for isentropic potential flow especially when $\g = 2$.
Though we have focused on the conservatively regularized model, a more complete and realistic treatment would have to include viscous dissipation just as in the KdV-Burgers equation.

Our attempt to generalize KdV to include the adiabatic dynamics of density, velocity and pressure has led to a interesting link between KdV and NLSE that is quite different from the well known ones (see E.g. \cite{ablowitz}). In fact, we may view R-gas dynamics as a natural generalization of both. While it extends the KdV idea of a minimal conservative dispersive regularization to adiabatic gas dynamics in any dimension and shares with it the cnoidal and $\sech^2$ solutions, it also reduces to the defocusing cubic NLSE for isentropic potential flow of a gas with adiabatic index two. Thus, the cubic 1d NLSE is a very special member of a larger class of R-gas dynamic equations that make sense in any dimension and for nonlinearities other than cubic while also allowing for adiabatically evolving entropy and vorticity distributions.



\appendix

\chapter{Symmetry of the added mass tensor in potential flow}
\label{a:symmetry-of-mu}

To show the symmetry of $\mu_{ij}$(\ref{e:added-mass-tensor}) for incompressible potential flow around a rigid body, we begin by noting that on the surface of the body, the impenetrable BC leads to
\begin{equation}
 n_i U_i  = n_j \partial_j \phi = n_j \partial_j (U_i \Phi_i) =  (n_j \partial_j \Phi_i) U_i \quad \textmd{for all}\; \; U_i.
\end{equation}
Hence, the potential vector field $\bf \Phi$ satisfies the identity
	\beq
	n_j \,(\partial_j \Phi_i)= n_i
	\label{e:mu-symmetry-identity}
	\eeq
which states that the unit normal at every point on the body surface is a left-eigenvector of the matrix $\pdr_j \Phi_i$ with eigenvalue $1$. Thus, the antisymmetric part of $\mu_{ij}$ becomes
	\beqs
    \frac{\mu_{ji} - \mu_{ij}}{\rho} &=&
    \int_A \Phi_j n_i\,dA - \int_A \Phi_i n_j \, dA = \int_A \left( \Phi_j \partial_k \Phi_i - \Phi_i \partial_k \Phi_j \right) n_k dA\cr
	&=& \int_S \left( \Phi_j \partial_k \Phi_i - \Phi_i \partial_k \Phi_j \right) n_k dS - \int_V \left( \Phi_j \nabla^2 \Phi_i - \Phi_i \nabla^2 \Phi_j \right)\, dV.
	\label{e:mu-antisymmetric-part}
	\eeqs
Here we have used the divergence theorem to express the integral over body surface $A$ as the difference between (i) an integral over the outer surface $S$ enclosing $A$ and (ii) a volume integral over the region $V$ between $S$ and $A$. Since $\phi$ (and consequently $\bf \Phi$) satisfies the Laplace equation in the region $V$, the volume integral vanishes. Assuming the fluid extends to infinity, we choose the surface $S$ to be a sphere of large radius $R$. Thus, the surface integral also vanishes\footnote{For incompressible flow, by the multipole expnasion (\ref{e:multipole-expansion}), $\phi, \bfPhi \sim 1/R^2$, so the integral decays as $1/R^{3}$.}. Hence the RHS of (\ref{e:mu-antisymmetric-part}) vanishes and $\mu_{ij}$ is a symmetric tensor for a fluid extending to infinity in all directions.

\section[BC on container to ensure symmetry of $\mu_{ij}$]{Boundary condition on container to ensure symmetry of $\mu_{ij}$}

While we have imposed impenetrable boundary conditions on the surface of the body, one may wonder what the appropriate BCs on the container enclosing the fluid may be. In the absence of any container, we have used the decaying BC $\phi \to 0$ as $r \to \infty$. Thus, in the presence of an outer boundary, one wonders whether to impose the BC $\phi = 0$ or the impenetrable BC. It turns out that the requirement that $\mu_{ij}$ be symmetric is satisfied with the impenetrable boundary condition on the outer surface and not with the $\phi = 0$ (`penetrable no-slip') condition. To show this, we make use of the identity $(\bfn \cdot \grad)\,\bfPhi = \bfn$ valid on the surface of the rigid body $A$, which is a consequence of the impenetrability of $A$: 
	\beq
	n_j \partial_j \phi = n_j \partial_j (\Phi_i \,U_i) = n_i\,U_i \imply n_j \partial_j \Phi_i = n_i
	\eeq
as in (\ref{e:mu-symmetry-identity}). Thus the antisymmetric part of $\mu_{ij}$ is
\begin{eqnarray}
 \mu_{ij} - \mu_{ji} &=& \int_A \Phi_j\,n_i\,dA - \int_A \Phi_i\,n_j\,dA = \int_A \left[ \Phi_j\,\partial_k\Phi_i - \Phi_i\,\partial_k\Phi_j \right] \, n_k\,dA\cr
 &=& \int_{A'}\left[ \Phi_j\,\partial_k\Phi_i - \Phi_i\,\partial_k\Phi_j \right] \, n_k\,dA - \int_{V_*}\left( \Phi_j\grad^2\Phi_i - \Phi_i\grad^2\Phi_j \right)\,dV.
\end{eqnarray}
Using the divergence theorem, we have written an integral over the body surface $A$ as the difference between an integral over outer surface $A'$ and an integral over $V_*$, the annular region between $A'$ and $A$. The integral over $A'$ vanishes due to impenetrable boundary conditions on that surface: $n_k\partial_k \phi = U_j n_k\partial_k \Phi_j= 0$ for any $U_j$, which implies $n_k\partial_k\Phi_j = 0$. The integral over $V_*$ vanishes because $\phi$ (and consequently $\bfPhi$) satisfies the Laplace equation in $V_*$: $\grad^2\phi = U_j\grad^2\Phi_j = 0$. Hence, $\mu_{ij}$ is symmetric for impenetrable boundary conditions on the outer boundary $A'$. 

If $\phi = 0$ on $A'$ (Dirichlet condition), then all we know is that $\bfU \cdot \bfPhi = 0$ on $A'$ and the integral over $A'$ need not vanish. Consequently, $\mu_{ij}$ is not necessarily symmetric.

\chapter{The non-acceleration reaction force}
\label{a:non-acceleration-reaction-force}

In this appendix, we study the part of the fluid force on the body that is not proportional to the body's acceleration. Apart from buoyancy, it may be written as a sum of contributions from the body surface $A$ and the inner surface of the container $S$. While the former does no work, the latter can. However, when the fluid occupies all of $\mathbb{R}^3$ outside the body, both contributions vanish.

Recall from (\ref{e:G-G-prime}) of \S\ref{s:pressure-force-added-mass-effect} that the pressure force on a rigid body moving at velocity $\bfU$ through an incompressible, inviscid fluid of density $\rho$ may be expressed as a sum
	\beq
	{\bf F} = \rho \int_A \dot {\bf U} \cdot {\bf \Phi} \: {\bf n} \: dA
	+ \int_A \left[ \half \rho v^2 + \rho g z - \rho {\bf U} \cdot {\bf v} \right] \: {\bf n} \: dA = {\bf G} + {\bf G}'.
	\eeq
The first term $\bf G$ is the acceleration reaction force and we called ${\bf G}'$ the `non-acceleration reaction force'. In this section, we show that the contribution of the presence of the body surface to $\bfG'$ does no work, and that that in an infinite fluid domain, $\bfG'$ does no work and vanishes.

$\bfG'$ includes buoyant force ${\bf F}_{\rm buoy} = \int_A \rho g z\, {\bf n}\, dA$, which, using the divergence throrem, can be written as an integral over the volume of the body
	\beq
 	\left(F_{\rm buoy}\right)_i = \rho g \int_A z\, n_i \,dA = \rho g \int_{V_{\rm body}} (\partial_i z)\, dV = \rho\, g\, V_{\rm body}\, \del_{i3}.
	\eeq
Hence, the buoyant force points upwards and its magnitude is equal to the weight of displaced fluid: ${\bf F}_{\rm buoy} = \rho \, g \, V_{\rm body} \, {\bf \hat z}$.

The fluid kinetic part of ${\bf G}'$ can be rewritten using the divergence theorem for a volume V of fluid bounded between the body surface $A$ and outer (container) surface $S$
	\beq
	G'_i - (\bfF_{\rm buoy})_i + \rho \int_A {\bf U} \cdot {\bf v} n_i dA =  \rho \int_A \half v^2  n_i dA = - \half \rho \int_V \pdr_i (v_j v_j) dV + 
	\rho \int_S \half v^2 n_i dA.
	\eeq
Now by incompressibility $\pdr_i v_i = 0$ and irrotationality $\pdr_i v_j = \pdr_j v_i$, $\pdr_i (v_j v_j) = \pdr_j (v_i v_j)$. The volume integral is now rewritten as a surface integral
	\beq
	G'_i - (\bfF_{\rm buoy})_i + \rho \int_A {\bf U} \cdot {\bf v} n_i dA =
	\rho \int_S \left( \half v^2 n_i - v_i v_j n_j \right) dS + \rho \int_A v_i v_j \: n_j \: dA.
	\eeq
Thus, apart from the buoyancy force, the non-acceleration-reaction force is
	\beq
	G'_i - (\bfF_{\rm buoy})_i = \rho \int_S \left( \half v^2 n_i - v_i v_j n_j \right) dS + \rho U_j  \int_A \left(- v_j n_i + v_i n_j \right) \: dA
	\label{G-prime-S-A}
	\eeq
where we have used the boundary condition $v_j n_j = U_j n_j$ on the body surface. The part of $\bfG'$ coming from the body surface does no work since 
	\beq
	\rho U_i U_j \int_A (v_i n_j - v_j n_i) dA = 0.
	\eeq
What is more, we may show that $\bfG' - \bfF_{\rm buoy}$ vanishes if the fluid extends to infinity. Supposing the outer surface/container is a large sphere of radius $R$, the integrand for the outer surface integral ($S$) goes like $1/R^4$ since\footnote{For large $r = |\bfr|$, $\phi$ admits the multipole expansion: $\phi = c_0 + c_1/r + c_2/r^2 + \cdots$. $c_0$ can be chosen to vanish as it does not affect the velocity. The monopole coefficient $c_1$ must vanish since there are no sources or sinks in the fluid, just as it does for the electrostatic potential around a neutral body. Therefore, $\phi$ can be at most of order $1/r^2$ asymptotically.} $\phi \sim 1/R^2, |\bfv| = |\grad\phi| \sim 1/R^3, dA \sim R^2$. So the integral over $S$ in (\ref{G-prime-S-A}) can do work\footnote{The contribution to $\bfG'$ in (\ref{G-prime-S-A}) coming from the outer surface $S$ is a boundary effect - figuratively, it is as if some fluid is hitting the container and returning to push the body.} but vanishes as $1/R^2$ when $R \to \infty$. As for the body surface integral, applying the divergence theorem,
	\beq
	\rho U_j \int_A (v_i n_j - v_j n_i) dA 
	= \rho U_j \int_S (v_i n_j - v_j n_i) dS - \rho U_j \int_V (\pdr_j v_i - \pdr_i v_j) \: dV.
	\eeq
The volume integral vanishes since the flow is irrotational. As before, the $S$-surface integral goes like $1/R$ and vanishes as $R \to \infty$. Thus, ${\bf G}' - \bfF_{\rm buoy} = 0$ for a fluid that fills all of $\mathbb{R}^3$ outside the body.

\chapter[$\mu_{ij}$ in $d$ dimensions from flow energy]{Added mass tensor in $d$ dimensions from energy of induced flow}
\label{a:derivation-energy-added-mass}

In this appendix we generalize Kambe's method \cite{Kambe} of using the energy of induced flow to obtain an expression for the added mass tensor of a rigid body moving in fluid in a $d$ dimensional container at velocity ${\bf U}(t)$. Let $\phi$ be the velocity potential of this induced flow, which must satisfy the Laplace equation $\grad^2 \phi = 0$, with boundary conditions which prevent the fluid from penetrating the surface of the body, i.e., on the surface $S_b$ of the body, ${\bf \hat{n}}\cdot \grad \phi = {\bf \hat{n}}\cdot {\bf U}(t)$, where ${\bf \hat{n}}$ is the outward unit normal on the body's surface. Additionally, we assume that the fluid is at rest at infinity, i.e, as $r \to \infty$, $\phi({\bf r}) \to 0$.
	
Now, $1/r^{d-2}$ and its spatial derivatives of all orders are solutions to the Laplace equation away from $r = 0$ which vanish as $r \rightarrow \infty$. The general solution is a linear combination of these, leading to a multipole expansion
	\beq
	\phi(\bfr) = \frac{c}{r^{d-2}} + c_i \partial_i \left( \frac{1}{r^{d-2}} \right) + c_{ij} \partial_i \partial_j \left( \frac{1}{r^{d-2}} \right) + \ldots.
	\eeq
However, for incompressible flow, $c = 0$, as it corresponds to a point source at the origin. We refer to the second term above as the dipole term. The components of $c_i, c_{ij}$ etc. are constants in space. 

Consider a body moving through the incompressible, irrotational fluid at a fixed instant in time. Let the origin of the coordinate system lie inside the body at this instant. This is the instant we will concentrate on in the rest of this appendix. Taking the density $\rho$ to be constant, the total energy of the fluid can be calculated to be (see \S \ref{TotalEnergyInFluidFlow})
	   \beq
	 E = \frac{1}{2}\,\rho \int_{V_*} |{\bf v}|^2 \, dV = \frac{1}{2}\,\rho\,\left(\frac{4\pi^{d/2}}{\Gamma\left( \frac{d}{2} - 1 \right)} {\bf c}\cdot{\bf U} - V_b U^2 \right)
	 \label{e:flow-energy}
	   \eeq
   where $V_b$ is the volume occupied by the body and $V_*$ the volume of fluid. The expression for energy depends only on the dipole term
	\beq
	 \phi_{\rm dipole} = {\bf c}\cdot{\nabla}\frac{1}{r^{d-2}} = -(d-2)\frac{{\bf c}\cdot{\bf \hat{r}}}{r^{d-1}}.
	 \label{e:phi-dipole-d-dims}
	\eeq
The boundary condition now becomes
   \begin{equation}
	-(d-2){\bf \hat{n}}\cdot {\nabla}\frac{{\bf c}\cdot{\bf \hat{r}}}{r^{d-1}} = {\bf \hat{n} }\cdot{\bf U}.
   \end{equation}
   From this BC, we see that ${\bf c}$ can only depend linearly on ${\bf U}$ (${\bf U}$ is a constant in space since it is the velocity of a rigid body). Therefore, we write $c_i = d_{ij}U_j$ where $d_{ij}$ is independent of position and time (even if ${\bf U}$ changes with time in magnitude and direction). We can now express the total energy of the fluid as
   \beq
	E = \frac{1}{2} \mu_{ij} U_i U_j\quad \text{where}\quad
	\mu_{ij} = \rho \left( \frac{2(d-2)\pi^{d/2}}{\Gamma\left( \frac{d}{2}\right)} d_{ij} - V_{\rm b} \del_{ij} \right)
   \eeq
is the added mass tensor. Since $E \geq 0$ for all $\bf U$, the tensor $\mu_{ij}$ is positive semidefinite. Also, only the symmetric part of $\mu_{ij}$ contributes in the above equation and hence the tensor can be chosen to be symmetric.
   
   We will now show that if the body is accelerating through the fluid, it experiences an added force. Let ${\bf f}$ be the force exerted by the moving body on the fluid. The resulting change in momentum of the fluid over an infinitesimal time interval $dt$ is $ d{\bf P} = {\bf f}\,dt$. Dotting both sides with the velocity of the body $\bf U$, we obtain ${\bf U}\cdot d{\bf P} = {\bf f}\cdot {\bf U}\,dt$. But, ${\bf f}\cdot {\bf U}\,dt$ is the work done on the fluid by the body, and is hence equal to the increase in the energy of the fluid $dE = \mu_{ij}U_i \,dU_j = U_i dP_i$. Since this equation holds for all $U_i$, we have $dP_i = \mu_{ij}\,dU_j$ and hence $\dot P_i =  \mu_{ij}\dot U_j$. Therefore, the total external force ${\bf f}$ applied on the body is a sum of two parts - one which accelerates the body, and a second which accelerates the fluid
   \beq
	f_i = (m \delta_{ij} + \mu_{ij}) \dot U_j.
   \eeq
   Therefore, for a given desired acceleration of the body, the force to be applied on it when immersed in a fluid is larger than that in vacuum by an amount $\mu_{ij} \dot U_j$.
   
   \section{Expression for energy of induced flow}\label{TotalEnergyInFluidFlow}
   
   We now obtain formula (\ref{e:flow-energy}) for the energy of the flow induced by accelerating the body. If ${\bf v}$ is the flow velocity and $V_*$ the region occupied by the fluid, then
	   \beq
		E = \frac{1}{2}\rho\,\int_{V_*} |{\bf v}|^2 \, dV.
	   \eeq
   Let the velocity of the body be ${\bf U}$. Using the identity $|{\bf v}|^2 = |{\bf U}|^2 + ({\bf v} - {\bf U})({\bf v} + {\bf U})$, we get
   \beq
	 \int_{V_*} |\bfv|^2 \, dV = U^2 (V_R - V_b) + \int_{V_*} (\bfv - \bfU) \cdot (\bfv + \bfU)\,dV
	\eeq
	where $V_*$ is the volume enclosed between a large sphere $S_R$ of radius $R$ (with volume $V_R$) and the surface of the body. To evaluate the second integral above, we use irrotationality to write ${\bf v} = {\nabla}\phi$ and the identity ${\bf U} = {\nabla}({\bf U}\cdot{\bf r})$:
   \begin{equation}
	({\bf v} - {\bf U}) ({\bf v} + {\bf U}) \;=\; ({\bf v} - {\bf U})\cdot{\nabla} (\phi + {\bf U}\cdot{\bf r}) \;=\; {\nabla}\cdot[(\phi + {\bf U}\cdot{\bf r})({\bf v} - {\bf U})].
   \end{equation}
   In the second equality, we used incompressibility. Thus, the second integral can be converted to an integral over the surface of $V_*$, namely, an integral over the body surface $S_b$ and another over the surface of the container $S_R$ (taken to be a sphere of large radius $R$):
   \beq
	\int_{V_*} |{\bf v}|^2\, dV = U^2 (V - V_b) + \oint_{S_R - S_b} (\phi + {\bf U}\cdot{\bf r})({\bf v} - {\bf U})\cdot{\bf \hat{n}}dS
   \eeq
   On $S_b$, we have ${\bf v}\cdot{\bf \hat{n}} = {\bf U}\cdot{\bf \hat{n}} \Rightarrow \oint_{S_b} = 0$. On $S_R$, (\ref{e:phi-dipole-d-dims}) implies
   \beq
   \phi_{\rm dipole} = -(d-2)\frac{{\bf c}\cdot{\bf \hat{r}}}{R^{d-1}}\quad \text{and}\quad
    {\bf v} = (d-2)\left[ \frac{ d(\bfc \cdot \hat r )\hat r - \bfc}{R^d}\right].
    \eeq
Therefore,
   	\footnotesize
	\beqs
	 \int_{V_*} |\bfv|^2 \, dV &=& U^2 (V_R - V_{\rm b}) + \int_{S_R} \left( (2-d) \frac{{\bf c}\cdot {\hat r}}{R^{d-1}} + R \bfU\cdot {\hat r} \right) \left( (2-d)(1-d)\frac{{\bf c}\cdot {\hat r}}{R^d} - \bfU \cdot {\hat r} \right)\,R^{d-1}\,d\Omega \cr
	 &=& U^2 (V_R - V_{\rm b}) + \int_{S_R} \left[ d(d-2)({\bf c}\cdot {\hat r}) (\bfU \cdot {\hat r}) - R^d (\bfU \cdot {\hat r})^2 \right] \, d\Omega
	\eeqs
	\normalsize
	where we have ignored all terms which vanish as $R\to \infty$. The higher multipole terms do not contribute to the energy of induced flow. Using the identity
	\begin{equation}
	 \int ({\bf A} \cdot {\hat r}) ({\bf B} \cdot {\hat r})\, d\Omega = \frac{\pi^{d/2}}{\Gamma\left( \frac{d}{2} + 1 \right)} \, ({\bf A} \cdot {\bf B})\quad \textmd{for } \quad d \geq 3
	\end{equation}
	we get
	\begin{equation}
	 \int_{V_*} |\bfv|^2\, dV = U^2 (V_R - V_{\rm b}) + \frac{\pi^{d/2}}{\Gamma\left( \frac{d}{2} + 1 \right)} \, \left( d(d-2) {\bf c} \cdot \bfU - R^d U^2 \right).
	\end{equation}
	Noting that $V_R = {2 \pi^{d/2}R^d}/{d\,\Gamma\left({d}/{2}\right)}$ we finally obtain
	\beq
	 E = \ov{2}\rho \int_{V_*}|\bfv|^2\,dV = \frac{1}{2}\,\rho\,\left(\frac{4\pi^{d/2}}{\Gamma\left( \frac{d}{2} - 1 \right)} {\bf c}\cdot{\bf U} - V_b U^2 \right).
	\eeq

\chapter{Added mass from the momentum of induced flow}
\label{a:derivation-mom-added-mass}

Just as the added mass tensor appears in the formula for the kinetic energy of induced flow, it also appears in the formula for the flow momentum. Indeed, this gives an alternate way of formulating the added mass effect. Unlike computations based on the pressure force (\S \ref{s:pressure-force-added-mass-effect}) on the body or on the energy of the induced flow \cite{Kambe}, which involve absolutely convergent integrals, $\bf P$ is not absolutely convergent in $\mathbb{R}^3$. Here we define a regularized $\dot {\bf P}$ as the difference between surface integrals on a `container' $A'$ and on the body $A$. The virtue of this formulation is that the integral over $A$ gives the standard added mass force $\mu_{ij} \dot U_j$. The contribution of $A'$ is generally non-zero, even as the surface is withdrawn, leading to an apparent addition to the added mass tensor computed in \S~\ref{s:pressure-force-added-mass-effect}. This additional contribution depends on the shape of $A'$ and on how it is sent to infinity. Physically, this apparent paradox may be resolved by accounting for the force exerted by a second external agent $E'$ to hold $A'$ fixed while fluid pushes against it.

\section{Cylindrical regularization to compute flow momentum}

Consider a rigid body executing purely translational motion in potential flow. In general, in the region V$_*$ outside the body, the velocity potential and velocity field admit multipole expansions (\ref{e:multipole-expansion}) beginning with the dipole term (involving the source doublet vector $\bf c$):
	\beq
	\phi = - \frac{{\bf c} \cdot \hat r}{r^2} + \cdots \quad
	\text{and} \quad {\bf v} = \grad \phi =  3 \frac{{\bf c} \cdot \hat r}{r^3} \hat r - \frac{\bf c}{r^3} + \cdots.
	\eeq
No particular boundary condition at large distances has been imposed here. Since the dipole velocity field decays like $1/r^3$, the momentum of induced flow ${\bf P} = \rho \int_{V_*} \bfv \: d^3r$ is logarithmically divergent if one performs the radial integral first (corresponding to a regularizing spherical outer surface that is eventually removed). However, suppose one integrates over angles first, holding the radius $R$ of the regularizing outer surface fixed with the body instantaneously located at its origin. Then the integral vanishes since the two dipolar terms in $\bf v$ contribute equal and opposite amounts $\pm 4\pi \frac{\bf c}{r^3} r^2 dr$. Thus the volume integral for momentum is {\it not} absolutely convergent and we are faced with an ambiguity to define the flow momentum with a spherical outer boundary. We consider a physically motivated way of regularizing this divergence. Suppose, for simplicity, the rigid body executes purely translational motion along the axis (pointing in the $\hat z$ direction) of a cylindrical tube of length $L$ and radius $R'$ containing fluid. We will eventually let $R',L \to \infty$. We choose a cylindrical container since (unlike the spherical container) it can be kept stationary as the rigid body accelerates along the axis. The flow momentum is evaluated by converting the volume integral into surface integrals 
	\begin{equation}
	 \bfP \;=\; \rho \int_{V_*} \bfv \, dV \;=\; \rho \int_{V_*} \grad\phi \, dV \;=\; \rho \left[ \int_{A'} - \int_{A}\right] \phi \,\bfn\, dA = \bfP^{\rm out} + \bfP^{\rm in}.
	\end{equation}
The fluid region $V_*$ is bounded on the inside by the surface of the body $A$ and on the outside by the cylindrical surface $A'$ enclosing the fluid. $\bfn$ is the unit normal to the surface under consideration directed into the fluid. $\dot \bfP^{\rm in}$ may be viewed as the force on the body exerted by agent $E$, while $\dot \bfP^{\rm out}$ is the external force on $A'$ exerted by agent $E'$. An advantage of this formulation is that $\bfP^{\rm in}$ leads directly to the same time-independent added mass tensor (\ref{e:added-mass-tensor}) obtained by computing the pressure force on the body
	\beq
	P^{\rm in}_i = - \rho \int_A \phi \: n_i \: dA = - \rho \int_A \Phi_j \cdot U_j n_i = \mu_{ij} U_j
	\label{e:inner-surface-momentum}
	\eeq
where $\bf \Phi$ is the potential vector field (\ref{e:potential-vector-field}). The outer surface integral is in general non-zero, as the following example illustrates:

{\bf Spherical body in a spherical container:} For a spherical outer surface of radius $R$, ${\bf P}^{\rm out}$ is independent of $R$
	\beq
	P^{\rm out}_i = - \rho \int_{A'} \frac{{\bf c} \cdot \hat n}{R^2} n_i \, dA = - \rho c_j \int_{S^2} n_i n_j \, d\Omega = - \frac{4}{3} \pi \rho c_i.
	\eeq
The higher multipoles do not contribute. This leads to the force $\dot P_i = \left( \mu_{ij} - \frac{4}{3} \pi \rho d_{ij} \right) \dot U_j$ where $c_i = d_{ij} U_j$ and $d_{ij}$ is the constant dipole tensor of (\ref{e:source-doublet-vector}). $\mu_{ij} \dot U_j$ is the added mass force applied on the body by $E$ while $- \frac{4}{3} \pi \rho d_{ij} \dot U_j$ is the force applied by $E'$ on the fluid via the container. E.g., for a spherical body of radius $a$, $d_{ij} = (a^3/2) \del_{ij}$ (\ref{e-sphere-dipole-moment}), so that $P^{\rm in} = P^{\rm out}$, provided the body and container are concentric. So the added force supplied by $E$ in this case is completely conveyed to the outer surface by the flow.

What is more, the value\footnote{Note that even if there was a monopole contribution to the integral over $A'$, it would vanish due to the angular integral being zero.} of $\bfP^{\rm out}$ depends on the order in which we take the dimensions $R',L$ of the cylinder to infinity. The outer surface is the union of two `caps' at $z = \pm L$ and the cylindrical surface at $s = R'$, so ${\bf P}^{\rm out} = {\bf P}^{\rm caps} + {\bf P}^{\rm cyl}$. Let us work in cylindrical coordinates $(s = \sqrt{x^2 + y^2},\, \tht = \arctan(y/x),\, z)$

{\bf (a)} If $R' \to \infty$ before $L \to \infty$, then the cylindrical outer surface integrand 
	\small
	\beq
	\int_{A'} \phi \,\bfn \,dA = \int_{A'} \left[ -\frac{ {\bf c}\cdot{\hat r} }{r^2} + \mathcal{O}\left( \frac{1}{r^3} \right) \right] \,\bfn \,dA =  
	\int \left[ -\frac{{\bf c}\cdot (R' \,{\hat s} + z \, {\hat z})}{(R'^2 + z^2)^{\frac{3}{2}} } + \mathcal{O}\left(\frac{1}{(R'^2 + z^2)^{\frac{3}{2}}}\right) \right] \bfn\, R'\,d\tht\,dz
	\label{e:outer-surface-integral-momentum}
	\eeq
	\normalsize
vanishes like $1/R'$, with the second term giving a $1/R'^2$ subleading contribution. Under these circumstances, we may ignore the cylindrical outer surface integral. The two caps contribute equally: 
	\beq
	\bfP^{\rm caps} = -2\rho \hat z \int \frac{\bfc\cdot \hat r}{r^2} s\,ds\,d\tht = -\hat z \int_0^{2\pi} d\tht \int_0^{R'}s\,ds\, \frac{(c_1 s \cos\tht + c_2 s \sin\tht + c_3 L)}{(s^2 + L^2)^{3/2}}.
	\eeq
The integral over $\tht$ eliminates the first two terms, leading to
	\beq
	\bfP^{\rm caps} = -4\pi c_3\rho\left( 1 - \frac{L}{\sqrt{L^2 + R'^2}} \right) \hat z.
	\label{e:p-caps}
	\eeq
Thus, if we take $R' \to \infty$ holding $L$ fixed, we get a non-zero contribution to the fluid momentum from $\bfP^{\rm caps} = -4\pi c_3 \rho \hat z$. Introducing the source doublet tensor (\ref{e:source-doublet-vector}), the fluid momentum is
	\beq
	P_i = \mu_{ij} U_j - 4\pi \rho c_3 \del_{i3} = (\mu_{ij} - 4\pi \rho d_{3j} \del_{i3}) U_j.
	\eeq
The added force that the external agents $E$ and $E'$ must supply to impart an acceleration $\dot \bfU$ to the body while holding the caps in place is thus,
	\beq
	 F_i^{\rm add} = \dot{P_i} = (\mu_{ij} - 4\pi \rho d_{3j} \del_{i3}) \, \dot{U_j}.
	\eeq
Though the usual added mass tensor enters the formula for the added force, there is an extra contribution from the outer caps. This extra contribution can also do work. Interestingly, there is another way of taking the limits which results in a contribution from the container surface to flow momentum, that does no work.

{\bf (b)} On the other hand, if we let $L \to \infty$ before $R' \to \infty$, then we see from (\ref{e:p-caps}) that $\bfP^{\rm caps} = 0$, though $\bfP^{\rm cyl}$ is non-zero, resulting in an additional contribution to the flow momentum, beyond $\mu_{ij} U_j$. This additional fluid momentum is present only for anisotropic (non-spherical) bodies, though its rate of change does no work on a body translated along the axis of the cylinder. To see this, we proceed from (\ref{e:outer-surface-integral-momentum}); the integral over the outer cylindrical surface $A'$ gives
	\footnotesize
	\beqs
	\int_{A'} \phi \,\bfn \,dA
	&=& \int \left[ -\frac{(R'\,c_1 \,\cos\tht \, + R'\,c_2\,\sin\tht\, + z\,c_3)}{(R'^2 + z^2)^{\frac{3}{2}} } + \mathcal{O}\left(\frac{1}{(R'^2 + z^2)^{\frac{3}{2}}}\right) \right] (\cos\tht \,{\hat x} + \sin\tht\,{\hat y})\, R'\,d\tht\,dz\cr
	&=& - {\hat x} \int_{-\infty}^{\infty} \frac{R'^2\,c_1\,\pi}{(R'^2 + z^2)^{\frac{3}{2}}}\,dz \; - \; {\hat y} \int_{-\infty}^{\infty} \frac{R'^2\,c_2\,\pi}{(R'^2 + z^2)^{\frac{3}{2}}}\,dz\cr
	&=& - 2\pi\, (c_1\,{\hat x} \;+\; c_2\,{\hat y})
	= - 2\pi\,[{\bf c} - ({\bf c}\cdot {\hat z})\,{\hat z}]
	\eeqs
	\normalsize
where ${\bf c} = c_1 \,{\hat x} + c_2 \,{\hat y} + c_3 \,{\hat z}$. The quadrupole and higher multipole terms do not contribute in the limit when $R' \to \infty$, even after the $z$ integral is performed. For a spherical rigid body, the above integral is zero since the source doublet vector ${\bf c} = \frac{a^3}{2} {\bf U}$ for a sphere points along $\bfU$, which has been chosen along $\hat z$. Including the contribution of the inner surface (\ref{e:inner-surface-momentum}), the total momentum of the flow is
\begin{equation}
 P_i = 2\pi\rho\, (c_i - c_3\,\del_{i3}) \; + \mu_{ij} U_j \;=\; [2\pi\rho\, (d_{ij} - d_{3j}\,\del_{i3}) \;+\; \mu_{ij}]\, U_j.
\end{equation}
The rate of change of flow momentum is
	\begin{equation}
	 \dot{P_i} = [2\pi\rho\, (d_{ij} - d_{3j}\,\del_{i3}) \;+\; 	\mu_{ij}]\,\dot{U_j}.
	\end{equation}
However, only the $\mu_{ij}\dot{U}_j$ part of $\dot P_i$ does work on the fluid. The extra added force from the outer boundary, $2\pi\rho\, (d_{ij} - d_{3j}\,\del_{i3})\,\dot U_j$ points in the $x$-$y$ plane perpendicular to the body's velocity and does no work for longitudinal translation. We may interpret this extra added force $2\pi \rho (\dot c_1 \hat x + \dot c_2 \hat y)$ as the force the external agent $E'$ must apply on the curved surface to hold the cylinder in place.

In conclusion, though the flow momentum for flow in the whole of $\mathbb{R}^3$ (outside the body) is strictly divergent, by considering a cylindrical regularizing outer surface and taking the length of the cylinder to infinity before its radius, we are able to compute the momentum of the fluid flow and arrive at the standard added mass tensor, through the rate of work done by the added force.

\chapter{Added mass for an ellipsoid}
\label{a:ellipsoid-added-mass}

Consider an ellipsoidal body whose semiaxes have lengths $a, b$ and $c$. Let the body be oriented such that its axes are along the coordinate axes. When such a body is moved along the $z$ axis, through an incompressible, irrotational and inviscid 3d fluid, it is known (see Art. 114 of \cite{Lamb}) that the kinetic energy of the induced fluid flow is
	\beq
    T_z = \half\left(\frac{\gamma}{2 - \gamma}\right)\, \frac{4}{3}\pi \, abc \, \rho\,U^2
	\eeq
where $({4}/{3})\,\pi\,abc$ is the volume of the ellipsoid and the dimensionless parameter
	\beq
    \gamma = abc\, \int_{0}^{\infty} \frac{d\lambda}{(c^2 + \lambda)\, \Delta}
    \quad \text{where}\quad
    \Delta = \sqrt{(a^2 + \lambda)\,(b^2 + \lambda)\,(c^2 + \lambda)}.
	\eeq
Comparing the above with the expression for the flow kinetic energy $T_z = \frac{1}{2}\mu_{zz}U^2 $, we read-off the added mass of the ellipsoid when accelerated along the $z$ axis:
	\beq
	\mu_{zz} = \frac{\gamma}{2 - \gamma} \frac{4}{3} \pi \, abc\, \rho.
	\eeq
The expressions for $\mu_{xx}$ and $\mu_{yy}$ are given by replacing $\gamma$ in $T_z$ above, by $\alpha$ and $\beta$ respectively, given by
	\beq
    \alpha = abc\, \int_{0}^{\infty} \frac{d\lambda}{(a^2 + \lambda)\, \Delta}\;\;\;\;\textmd{and}\;\;\;\; \beta = abc\, \int_{0}^{\infty} \frac{d\lambda}{(b^2 + \lambda)\, \Delta}.
    \eeq
We will now argue that the off-diagonal entries of the added mass tensor are zero in the principal axis basis. The ellipsoid is symmetric with respect to reflections about the $xy, yz$ and $zx$ planes when its principal axes lie along the coordinate axes. Therefore, when accelerated along the $z$ direction, if the ellipsoid experiences an added force in the positive $x$ direction, there \emph{must} simultaneously be an equal force acting on it along the negative $x$ direction. The same argument holds for forces along the $y$ direction. Therefore, the net added force that an ellipsoid experiences when accelerated along the $z$ direction is along the $z$ direction alone. Taken together with similar arguments about acceleration along the $x$ and $y$ directions, we deduce that in its principal axis basis, the added mass tensor of an ellipsoid is diagonal\footnote{The same argument holds for any rigid body which possesses reflection symmetry with respect to three mutually perpendicular planes, e.g. a cuboid. Hence, the added mass tensor of a cuboid is diagonal when the basis vectors point from its center to the centers of its three faces. In particular, the added mass tensor of a cube is a multiple of the identity, just as for a sphere. \label{CubeSphereSameMu}}. Thus, the added mass tensor of an ellipsoid in this basis is
	\beq
    \mathbf{\mu}= \frac{4}{3}\pi\, abc\, \rho
      \left[ {\begin{array}{ccc}
       \frac{\alpha}{2-\alpha} & 0 & 0\\
      0 & \frac{\beta}{2-\beta} & 0\\
     0 & 0 & \frac{\gamma}{2-\gamma}
	  \end{array} } \right].
	\label{e:ellipsoid-mu}
	\eeq
Though the pre-factor is proportional to the volume of the ellipsoid, the added mass eigenvalues are not functions of the volume alone. This is because $\al/(2- \al)$ etc. depend on $a,b,c$ not just through their product $abc$. But in the case of a sphere, the added mass is proportional to the volume of the sphere (see \S\ref{s:examples-added-mass-tensors}).

\section{Limit of an elliptical disk}

Taking the limit $c\rightarrow 0$ of the ellipsoid would result in an infinitesimally thin elliptical disk with semiaxes $a$ and $b$. We find that $\mu_{xx}$ and $\mu_{yy}$ vanish, and $\mu_{zz} \neq 0$. Thus, the disk has no added mass when accelerated along its surface. Recall that $\mu_{xx} = \frac{4}{3}\,\pi\,abc\,\rho\,\frac{\alpha}{2 - \alpha}$. Now,
	\beq
    \lim_{c \to 0} \alpha = \lim_{c \to 0}abc\, \int_{0}^{\infty} \frac{d\lambda}{(a^2 + \lambda)\, \Delta}
      = ab \, (\lim_{c \to 0}c) \, \int_{0}^{\infty} \frac{d\lambda}{(a^2 + \lambda)^{3/2}\sqrt{(b^2 + \lambda)\lambda}}.
	\eeq
The above integrand decays like $\lambda^{-5/2}$ as $\lambda \to \infty$ and diverges as $\lambda^{-1/2}$ as $\lambda \to 0$. Therefore, the integral is finite and
	\beq
	\lim_{c \to 0} \alpha = 0 \quad \text{and similarly}\quad
	\lim_{c \to 0} \beta = 0.
	\eeq
Therefore, $\mu_{xx} = \mu_{yy} = 0$ when $c \to 0$. To facilitate taking the $c \to 0$ limit, we expand $\gamma$ in a Taylor series around $c = 0$:
	\beq
    \gamma = abc\, \int_{0}^{\infty} \frac{d\lambda}{(c^2 + \lambda)^{3/2}\, \sqrt{(a^2 + \lambda)(b^2 + \lambda)}}
     = 2 - \frac{2c}{b} E\left( 1 - \frac{b^2}{a^2} \right) + {\cal O}(c^2).
	\eeq
Here $E(m) = \int_0^{\pi/2} \sqrt{1 - m \sin^2 \tht} \: d\tht$ is the complete elliptic integral of the second kind. Therefore we may compute the non-zero entry in (\ref{e:ellipsoid-mu}) to be
	\beq
 \lim_{c \to 0} \frac{4}{3}\pi \rho\, abc\,\frac{\gamma}{2 - \gamma} = \frac{4}{3}\pi \rho\,(\lim_{c \to 0} \gamma) \cdot \left(\lim_{c \to 0} \frac{abc}{\frac{2c}{b} E\left( 1 - \frac{b^2}{a^2}\right)}\right) =  \frac{4}{3}\pi \rho\,\frac{ab^2}{E\left( 1 - \frac{b^2}{a^2} \right)}.
	\eeq
Consequently, the added mass tensor for an elliptical disk is
	\beq
    \mathbf{\mu}(a,b)= \frac{4}{3}\pi \rho  \frac{ab^2}{E\left( 1 - \frac{b^2}{a^2} \right)}
    \left[ {\begin{array}{ccc}
    0 & 0 & 0\\
    0 & 0 & 0\\
    0 & 0 & 1
    \end{array} } \right].
    \label{e:elliptical-disk-mu}
	\eeq
Therefore, an elliptical disk has no added mass when accelerated in any direction along its plane. Despite appearances, $\mu(a,b)$ is a symmetric function of the lengths of the semi-axes $a$ and $b$ of the elliptical disk. To see this, we write
	\beq
	\left(\frac{4}{3} \pi \rho \right)^{-1} \mu_{33} 
	= \frac{ab^2}{\int_0^{\pi/2} \sqrt{1 - (1-\frac{b^2}{a^2}) \sin^2 \tht} \: d\tht}
	= \frac{a^2 b^2}{\int_0^{\pi/2} \sqrt{a^2 \cos^2 \tht + b^2 \sin^2 \tht} \: d\tht}.
	\eeq
The integral in the denominator is symmetric under $a \leftrightarrow b$. This is seen by using the $\tht \to \pi/2 - \tht$ change of variables. 

{\fl\bf Remarks:}
\begin{enumerate}
\item We notice that the added mass $\mu_{33}$ is not a function of $ab$ alone, so it is not a function of the area alone.
\item When the elliptical disk shrinks to a flat, thin strip (by taking $b \ll a$, for instance) we see that $\mu_{33}$ vanishes as the numerator vanishes and the denominator is non-zero. For small $b$, holding $a$ fixed, we find
	\beq
	\mu_{33} = \frac{4}{3} \pi \rho a b^2 + {\cal O}(b^3).
	\eeq
Here we used the series expansion
	\beq
	E(1-x) = 1 + \left(\log 2 - \ov{4} - \ov{4}\log(x) \right)x + \cdots
	\eeq
for small x. Even though the strip has zero volume, it has an added mass. It presents an area of $\pi ab$ to the fluid. It is as if it also has a depth proportional to the shorter dimension $b$ so as to displace fluid in a volume $\propto \pi a b^2$.
\end{enumerate}

\section{Limit of a circular disk}

Finally, we obtain the added mass tensor for a circular disk by putting $a = b$ in (\ref{e:elliptical-disk-mu}):
	\beq
    \mathbf{\mu}= \frac{8}{3}\,\rho a^3
      \left[ {\begin{array}{ccc}
       0 & 0 & 0\\
      0 & 0 & 0\\
     0 & 0 & 1
      \end{array} } \right]
	\eeq
where we have used $E(0) = \pi/2$. In this case, for acceleration orthogonal to the disk, the added mass scales as the $3/2$ power of the area of the circular disk.

\chapter{Added mass effect in 2d}
\label{a:2d-added-mass-effect}

In \S~\ref{s:added-mass-cylinder-2d}, the added mass for 2d flow in a plane perpendicular to the axis of a right circular cylinder was obtained. Here we summarize the general theory \cite{Batchelor} paying attention to some features special to 2d, such as the possibility of having circulation around a body even for a curl-free velocity field, and the possibility to use conformal transformations to calculate the added force.

\section{Multipole expansion for velocity potential}

\begin{figure}
\centering  
 \includegraphics[scale=0.35]{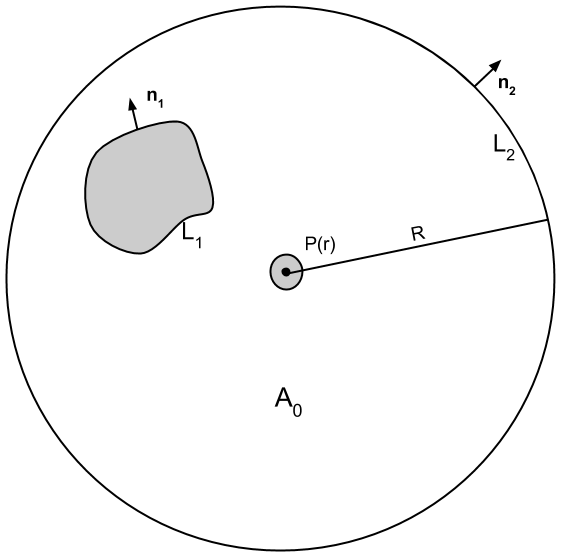}
 \caption{Rigid body moving through 2D fluid. The point of observation $P({\bf r})$ and the contours and area over which integration is performed; the grey portions are regions excluded from area $A_0$. The excluded circular region of radius $\epsilon$ around the point of observation $P({\bf r})$ will subsequently be taken to zero.} \label{fig-2d-flow}
\end{figure}

By Green's theorem, two functions $f(\bfr)$ and $g(\bfr)$, single valued and continuous in a region $A_0$ (see Fig.~\ref{fig-2d-flow}) bounded by the curves $L_1$ and $L_2$, satisfy
\begin{equation}
 \left[\oint_{L_2} + \oint_{L_1}\right] (f\,{\bf \nabla}g - g\,{\bf \nabla}f) \cdot {\bf n}\,dl = \int_{A_0} (f\,\nabla^2 g - g\,\nabla^2 f)\,dA
\end{equation}
where the contour integrals are all counter-clockwise, and the normals point as shown in the figure. Suppose $\phi(\bfr')$ is the velocity potential for incompressible and irrotational flow around a rigid body in $\mathbb{R}^2$. We choose $f(\bfr') = \phi(\bfr') - \frac{\kappa \theta}{2\pi}$ where $\ka$ is the circulation around the body, and $g(\bfr') = \ov{2\pi}\log(|\bfr - \bfr'|)$, which is Green's function for the Laplacian in $\mathbb{R}^2$ satisfying ${\grad^\prime}^2 g(\bfr') = 2\pi\delta(\bfr - \bfr')$. Here, $\bfr$, $\bfr'$ and $\tht$ are all measured with respect to an arbitrary common origin. Since both $\phi$ and $\tht$ are harmonic in $A_0$, we have
{\small
	\beq
	\ov{2\pi}\left[\oint_{L_2}+\oint_{L_1}\right] \left\{ \left(\phi(\bfr')-\frac{\kappa \theta}{2\pi}\right) \grad' \log|\bfr - \bfr'| - \log|\bfr - \bfr'|\,\grad'\left(\phi(\bfr') - \frac{\kappa\theta}{2\pi}\right)\right\}\cdot {\bf n}\,dl = \phi(\bfr) - \frac{\kappa\tht}{2\pi}.
	\eeq
}
As shown in figure \ref{fig-2d-flow}, $L_2$ is a circle of large radius $R'$, centered at $P$ whose position vector is $\bfr$. The $L_2$ integral vanishes for $R'\to\infty$, assuming $\phi \sim 1/R'$ for large $R'$:
\begin{equation}
 \ov{2\pi}\oint_{L_2}\left[\ldots\right] = \ov{2\pi}\int_{0}^{2\pi} \left[ \left(\phi(\bfr')-\frac{\kappa\tht}{2\pi}\right)\ov{R'} - \log R'\, \grad'\left( \phi(\bfr')-\frac{\kappa\tht}{2\pi} \right)\right]\,R'\,d\tht \xrightarrow{R'\to\infty} 0.
\end{equation}
Hence, we are left with
\begin{equation}
 \phi(\bfr) - \frac{\kappa\tht}{2\pi} = \ov{2\pi}\oint_{L_1} \left[ \left(\phi(\bfr')-\frac{\kappa \theta}{2\pi}\right) \grad' \log s - \log s\,\grad'\left(\phi(\bfr') - \frac{\kappa\theta}{2\pi}\right)\right]\cdot {\bf n}\,dl
 \label{e:greens-theorem-output}
\end{equation}
where $s = |\bfr - \bfr'|$. We wish to give a multipole expansion for $\phi$ in terms of derivatives of $\log s$ with respect to $\bfr$. Hence, we expand $\log s$ about $\bfr' = 0$, noting that $\grad'\log s = \grad \log s$:
	\beq
	\log s = \log r + r'_i \frac{\partial}{\partial r_i}\log r + r'_i r'_j \frac{\partial^2}{\partial r_i\,\partial r_j}\log r + \ldots.
	\eeq
Plugging this into (\ref{e:greens-theorem-output}) we obtain
	\beqs
	\phi(\bfr) - \frac{\kappa\tht}{2\pi} &=& \ov{2\pi} \oint_{L_1}\partial'_k\left( \log r + r'_i \pdr_i\log r + r'_i r'_j \pdr_i\pdr_j\log r + \ldots \right)\left(\phi(\bfr') - \frac{\kappa\tht}{2\pi}\right){n_k}\,dl\cr
 && - \ov{2\pi}\oint_{L_1} \left( \log r + r'_i \pdr_i\log r + \ldots \right) \partial'_k\left(\phi(\bfr') - \frac{\kappa\tht}{2\pi}\right) {n_k}\,dl.
	\eeqs
Collecting terms proportional to $\log r$, $\partial_i \log r$, $\partial_i\partial_j \log r$, etc, we get the multipole expansion
\begin{equation}
 \phi(\bfr) - \frac{\kappa \tht}{2\pi} = c\log r + c_i \partial_i \log r + c_{ij} \partial_i \partial_j \log r + \ldots
\end{equation}
where
\begin{eqnarray}
 && c = -\ov{2\pi} \oint_{L_1} \grad \left(\phi-\frac{\kappa\tht}{2\pi}\right)\cdot{\bf n}\,dl,\quad c_i = -\ov{2\pi} \oint_{L_1}\left[ r_i ({\bf n}\cdot \grad)\left( \phi - \frac{\kappa\tht}{2\pi} \right) - n_i \left( \phi - \frac{\kappa\tht}{2\pi} \right) \right] \, dl \cr 
 && c_{ij} = -\ov{2\pi}\oint_{L_1}\left[ r_i r_j ({\bf n}\cdot \grad)\left( \phi - \frac{\kappa\tht}{2\pi} \right) - n_i r_j - n_j r_i \right]\,dl,\ldots.
\label{e:2d-multipole-coeffs-phi}
\end{eqnarray}
For an incompressible fluid, $c = 0$.

{\bf Multipole expansion for a circular disk:} The multipole coefficients $c_i, c_{ij}$, etc. depend only on the shape of the rigid body and its velocity $U_i$. For a circular disk of radius $a$ there is no preferred direction supplied by the body shape. Hence, all the multipole coefficients must be constructed out of the single vector $U_i$. In addition, the impenetrable boundary condition on the surface of the body ${\bf n} \cdot \grad\phi = {\bf n}\cdot {\bf U}$ implies that all the multipole coefficients must be linear in $U_i$. Thus in this case  the dipole coefficient $c_i$ is the only non-vanishing one. Hence, for flow around a moving circular disk,
\begin{equation}
 \phi(\bfr) - \frac{\kappa\tht}{2\pi} = \frac{c_i r_i}{r^2}.
\label{e:phi-circular-disk}
\end{equation}
We plug this into the expression for $c_i$ in (\ref{e:2d-multipole-coeffs-phi}), choose the origin to lie at the centre of the disk so that $\bfn\cdot \grad \tht = 0$ and use the impenetrable BC to get
	\beqs
    c_i &=& -\ov{2\pi} \oint_{L_1}\left[ r_i n_j U_j - n_i \left(\frac{c_j r_j}{a^2}\right) \right] \, dl = -\ov{2\pi}\int_A \left(\partial_j(r_i U_j) - \ov{a^2} \pdr_i \left(c_j r_j\right) \right)\,dA\cr
    &=& -\ov{2\pi} \left( U_i - \frac{c_i}{a^2} \right) \text{Area}(\text{disk}).
    \label{e:dipole-vector-circular-disk}
	\eeqs
Thus, $c_i = -a^2 U_i$.

\section{Added mass tensor in $\mathbb{R}^2$}

As in \S~\ref{s:pressure-force-added-mass-effect}, the total pressure-force on a rigid body moving through 2d potential flow is
\begin{equation}
 {\bf F} = - \oint_C p\, {\bfn} \,dl = \sigma \oint_C \left( \frac{\partial \phi}{\partial t} + \half v^2 \right) \, \bfn \, dl
\end{equation}
where $C$ is the perimeter of the rigid body, $\bfn$ the unit normal to the body on its perimeter, $\sigma$ the constant density per unit area of the fluid, $\phi$ the velocity potential of the flow and $v = |\grad\phi|$ the magnitude of the flow velocity. Proceeding as in \S~\ref{s:pressure-force-added-mass-effect} the total pressure-force on the body is
\begin{equation}
 {\bf F} = \sigma \oint_C ({\bf \Phi} \cdot \dot{\bfU}) \, \bfn \, dl \;+\; \sigma \oint_C \left(\half v^2 - \bfv \cdot \bfU \right)\, \bfn \, dl = {\bf G} + {\bf G}'.
\end{equation}
Here the acceleration reaction force is
	\beq
 	G_i = -\mu_{ij} \dot{U_j} \quad
	\text{where} \quad
	\mu_{ij} = -\sigma \oint_C \Phi_j\, n_i\,dl.
	\eeq
The added mass tensor $\mu_{ij}$ can be expressed in terms of the source doublet tensor $d_{ij}$ (\ref{e:source-doublet-vector}). Indeed, using the impenetrable boundary condition on the surface of the body $\bfn \cdot \grad \phi = \bfn \cdot \bfU$ and the factorization $\phi = \bfU \cdot \bfPhi$, the expression for the source doublet vector (\ref{e:2d-multipole-coeffs-phi}) becomes
	\beqs
    c_i &=& -\ov{2\pi}\oint_C\left[ r_i(\bfn \cdot \bfU) - n_i (\bfU \cdot {\bf \Phi})\right]\,dl = -\ov{2\pi} U_j \oint_C \left( r_i n_j - n_i \Phi_j \right)\, dl\cr
     &=&-\ov{2\pi}U_j\int_{\rm body} \partial_j r_i \, dA - \ov{2\pi \sigma}\,\mu_{ij}\,U_j = -\left( A_{\rm body}\del_{ij} -\frac{\mu_{ij}}{\sig} \right)\frac{U_j}{2\pi}
	\eeqs
where $A_{\rm body}$ is the area of the body. Putting $c_i = d_{ij}U_j$, and noting that the equation holds for any $\bfU$, we deduce that
	\beq
	\mu_{ij} = -\sigma (2\pi d_{ij} + A_{\rm body}\del_{ij}).
	\eeq

{\bf Added mass tensor for a circular disk:} In this case, from (\ref{e:dipole-vector-circular-disk}) $c_i = -a^2 U_i = d_{ij} U_j$ so that $\phi = -a^2 \bfU\cdot \hat r/r$ with the origin at the centre of the disk. Since $d_{ij}$ is independent of the arbitrary body velocity $\bfU$, it follows that $d_{ij} = -a^2\del_{ij}$ and $\mu_{ij} = \sig \pi a^2 \delta_{ij}$.

\subsection{Added mass tensor for an elliptical disk}
\label{a:mu-elliptical-disk}

To find the velocity potential for flow around an elliptical disk, we transform the known result for a circular disk (\ref{e:phi-circular-disk}) using a conformal map $w(z)$ that takes a circle of radius $a$ to an ellipse of semi-axes $(a+s)$ and $(a-s)$ along the $\hat X$ and $\hat Y$ directions respectively, where $s$ is a measure of the eccentricity of the ellipse. Denoting $z = x + i y$ and $w = X + i Y$, we take
\begin{equation}
 w(z) = z + \frac{as}{z} \quad \text{or conversely} \quad z(w) = \frac{w + \sqrt{w^2 - 4 a s}}{2}.
\end{equation}
The velocity potential (\ref{e:phi-circular-disk}) for flow (with $\ka = 0$; see \cite{Batchelor} \S6.6 for inclusion of circulation) around a circular disk moving with velocity $\bfU = U \hat x$ is
\begin{equation}
 \chi(x, y) = -\frac{a^2 U \cos \tht'}{\sqrt{x^2 + y^2}} = \Re \left[ f(z) \right] \quad \text{where} \quad f(z) = - \frac{a^2 U}{z}
\end{equation}
and $\tht' = \arctan{(y/x)}$. Since conformal maps take harmonic functions to harmonic functions, the velocity potential for flow around the ellipse is
	\beqs
    \phi(X,Y) &=& \Re[f(z(w))] = \Re \left[ f\left( \frac{w + \sqrt{w^2 - 4 a s}}{2} \right) \right] = -\frac{a U}{2s} \Re [w - \sqrt{w^2 - 4 a s}]\cr
    &=&\begin{cases}
                      \phi_+(X,Y) & \textmd{if } X \geq 0\\
                      - \phi_+(-X,Y) & \textmd{if } X < 0
                     \end{cases}
   \quad\quad \text{where}\cr
	\phi_+(X,Y) &=& -\frac{a U}{2s}\left( X - \sqrt{\frac{(X^2 - Y^2 -4as) + \sqrt{(X^2 - Y^2 -4as)^2 + 4 X^2 Y^2}}{2}} \right).\quad \quad  
	\eeqs
It is verified that $\phi$ satisfies the Laplace equation. An easy way to obtain the velocity of the elliptical disk $\bfU_{\rm el}$, is to impose the impenetrable BC $\bfn \cdot \grad \phi = \bfn \cdot \bfU_{\rm el}$ at the `perigee' and `apogee', leading to
	\beq
	\bfU_{\rm el} = \left( {\hat X} \cdot \grad \phi (a+s, 0) \;,\; {\hat Y} \cdot \grad \phi (0, a-s) \right) = \left(\frac{a \,U}{a-s}\;,\; 0\right).
	\eeq
Using the expression for the normal to the ellipse at the point $(X = (a+s) \cos\tht$, $Y = (a-s) \sin \tht)$
	\beq
	\bfn = \left(\frac{\cos^2 \tht}{(a+s)^2} + \frac{\sin^2 \tht}{(a - s)^2}\right)^{-1/2} \left(\frac{\cos \tht}{a+s}, \frac{\sin \tht}{a - s}\right)
	\eeq
it is verified that the impenetrable BC is satisfied all along the perimeter of the ellipse.

To obtain the added mass tensor, we need the potential vector field $\bfPhi$ defined via $\phi = \bfPhi \cdot \bfU_{\rm el}$. Its $X$-component is given by ${\bf \Phi} \cdot {\hat X} = \phi/U_{\rm el}$ where $U_{\rm el} = |\bfU_{\rm el}|$. To obtain its $Y$-component, we consider the problem of a circular disk moving with velocity $\bfU' = U {\hat y}$, and repeat the above steps to obtain the velocity potential $\psi(X, Y)$ for flow around the same elliptical disk of semi-axes $(a+s)$  and $(a-s)$, this time moving with a velocity $\bfU'_{\rm el} = \left(0 \;,\; {a\,U}/{(a + s)}\right)$:
	\beqs
    \psi(X, Y) &=& \begin{cases}
                          \psi_+(X,Y) & \textmd{if }\quad Y \geq 0 \quad \text{and}\\
                          - \psi_+(X,-Y) & \textmd{if }\quad Y < 0 \quad \text{where}
                         \end{cases}\cr
     \psi_+(X, Y) &=& \frac{a\, U}{2s}\left( Y - \ov{\sqrt{2}} \sqrt{-(X^2 -Y^2 -4a s) + \sqrt{(X^2 -Y^2 -4a s)^2 + 4X^2 Y^2}} \right).\quad\quad\quad
	\eeqs
Thus, ${\bf \Phi}(X,Y) = \left({\phi(X,Y)}/{U_{\rm el}} \;,\; {\psi(X,Y)}/{U'_{\rm el}}\right)$
and we can calculate the added mass tensor:
\begin{equation}
 \mu_{ij} = -\sigma \int \Phi_i n_j\,dl = -\sigma \int_{0}^{2\pi} \Phi_i n_j \sqrt{(a+s)^2 \sin^2\tht + (a-s)^2 \cos^2\tht} \; d\tht.
\end{equation}
Here $dl = \sqrt{dX^2 + dY^2}$. Thus, we obtain $\mu = \pi\,{\rm diag}((a-s)^2, (a+s)^2)$ for an elliptical disk moving through 2d fluid.

\chapter[Added mass effect in compressible potential flow]{Extension of the added mass effect to compressible potential flow}
\label{a:extension-added-mass-effect-compressible}

Treatments of the added mass effect assume for simplicity that the flow is inviscid, incompressible and irrotational. However, it is physically plausible that the effect is present even in compressible or rotational flow. Indeed, according to our correspondence, density fluctuations around incompressible flow should correspond to quantum fluctuations around a constant vev for the scalar field. Moreover, to look for a fluid analogue of the Higgs particle, i.e., a `Higgs wave' around an accelerated rigid body, we need a generalization of the added mass effect to compressible flow. As is well known, the resulting flows can be very complicated. Here we take a small step by formulating the added mass effect for compressible potential flow ($\rho$ not necessarily constant) around a rigid body executing purely translational motion at velocity $\bfU(t)$. We assume the flow is isentropic so that $\grad p/\rho = \grad h$ where $h$ is specific enthalpy. Euler's equation $\dd{\bfv}{t} + \bfv \cdot \grad \bfv = - \grad h$ then implies an unsteady Bernoulli equation for the velocity potential $\phi$,
	\beq
	\pdr_t \phi + (1/2) \bfv^2 + h = {\rm constant}(t).
	\eeq
For concreteness, we consider an ideal gas with barotropic pressure $(p/p_0) = (\rho/\rho_0)^\gamma$ where $\gamma = c_p/c_v$ is the adiabatic index and $p_0,\rho_0$ are a reference pressure and density. Then $h = [\gamma/(\gamma -1)] p/\rho$. Of course, $\phi$ and $\rho$ are to be determined by solving the Euler and continuity equations subject to initial and boundary conditions. To identify the added force on the body, it helps to regard the continuity equation and impenetrable boundary condition on the body, namely
	\beq
	(\grad \rho \cdot \grad + \rho \grad^2) \phi = - \pdr_t \rho \quad \text{and} \quad
	\hat n \cdot \grad \phi = \hat n \cdot \bfU,
	\label{e:continuity-and-impenetrable-bc}
	\eeq
as a system of inhomogeneous linear equations for $\phi$ given $\rho$ and $\bf U$. The RHS of this system is linear in $\bfU$ (and $\rho$), so formally, the solution of this equation can be expressed as $\phi = \bfU \cdot {\bf \Phi}(\bfr,t) + \psi(\bfr,t)$ where the potential vector field $\bf \Phi$ and the supplementary potential $\psi$ are $\bfU$-independent but depend on $\rho$\footnote{To see why this is true, discretize the system as a matrix equation $A(\rho) \phi = b$. The upper rows of the matrix $A$ encode the operator $\grad \rho \cdot \grad + \rho \grad^2$ while the lower rows encode $\hat n \cdot \grad$. The upper rows of the column vector $b$ represent $- \pdr_t \rho$ and the lower rows contain $\hat n \cdot \bfU$, so that we may write $b  = b_1(\rho) + b_2(\bfU)$ where $b_2$ is linear in $\bf U$. Inverting $A$ gives the desired decomposition.}.

With the aid of Bernoulli's equation, the force on the body $-\int_A p \, \hat n \, dA$ becomes
	\beq
	F_i = \left(\frac{\gamma -1}{\gamma} \right) \int_A \rho \left[ \pdr_t \phi + \half \bfv^2 - {\rm const}(t) \right] n_i dA.
	\eeq
Using our factorization $\phi = \bfU \cdot {\bf \Phi} + \psi$, the force on the body is the sum of an acceleration reaction force $G_i = - \mu_{ij} \dot U_j$ and a non-acceleration force ${\bf G}'$:
	\beq
	G_i = \frac{(\gamma-1)}{\gamma} \dot U_j \int_A \rho \Phi_j n_i \, dA \quad \text{and} \quad G'_i = \frac{\gamma-1}{\gamma} \int_A \rho \left[U_j \dot \Phi_j + \dot \psi + \frac{\bfv^2}{2} - {\rm const}(t) \right] n_i \, dA.
	\eeq
The added mass tensor $\mu_{ij} = -\frac{(\gamma-1)}{\gamma} \int_A \rho \Phi_j n_i \, dA$. To find $\mu_{ij}$ for a given body, we need to solve for $\rho$ and $\bfv$ using the equations of motion. Unlike for constant density, where $\mu_{ij}$ is constant, here it could change with time and location of the body, due to density variations arising from the acceleration of the body. Corrections to the added mass due to density fluctuations are analogous to corrections to the $W$ and $Z$ boson masses due to quantum fluctuations around a constant scalar vev.

\section{The added mass tensor may not be symmetric}
To examine the symmetry of the added mass tensor, we compute its anti-symmetric part
	\beqs
	\Delta_{ij} \equiv -\frac{\gamma}{\gamma - 1}\left( \mu_{ij} - \mu_{ji} \right) &=& \int_A \rho (\Phi_j n_i - \Phi_i n_j)\, dA.
	\eeqs
In the incompressible case the impenetrable boundary condition led to an identity (\ref{e:mu-symmetry-identity}) ($n_i = n_j \partial_j \Phi_i$) which helped simplify $\Delta_{ij}$ (see Appendix \ref{a:symmetry-of-mu}). In compressible potential flow, the impenetrable boundary condition is different:
	\beq
	n_k U_k = n_j \partial_j \left( \Phi_k U_k + \psi \right) = n_j \left( \partial_j \Phi_k \right) U_k + n_j \partial_j \psi \quad \text{or} \quad
	\left( n_k - n_j \partial_j \Phi_k \right) U_k = n_j \partial_j \psi
	\eeq
for all $\bfU$. Since the RHS is independent of $\bfU$ and the LHS is dependent on $\bfU$, we conclude that both sides should be identically zero. Hence,
	\beq
	n_i = n_k \partial_k \Phi_i.
	\eeq
Using this in the above expression for $\Delta_{ij}$, we find by the divergence theorem,
\footnotesize
	\beq
	\Delta_{ij} = \int_{S^2_\infty} \rho \left( \Phi_j \partial_k\Phi_i - \Phi_i \partial_k\Phi_j \right) n_k\, dS - \int_V \left(\partial_k \rho\right) \left( \Phi_j \partial_k\Phi_i - \Phi_i \partial_k\Phi_j \right) \, dV - \int_V \rho \left( \Phi_j \nabla^2 \Phi_i - \Phi_i \nabla^2\Phi_j \right).
	\eeq
\normalsize
The first term is an integral over a spherical surface at infinity, which vanishes. The second term would be zero for constant density and the third term would also vanish for incompressible flow since $\phi$ and $\bfPhi$ are harmonic in that case. But for compressible potential flow the continuity equation (\ref{e:continuity-and-impenetrable-bc}) does not reduce to the Laplace equation for $\bfPhi$ and we have not been able to show that $\D_{ij}$ vanishes and do not have any reason to believe it is generally zero. It would be nice to find an example of a flow where $\D$ is non-zero.

\subsection{Existence of principle directions for $\mu_{ij}$ in the non-symmetric case}

It is interesting to see what possible canonical forms, a not necessarily symmetric $\mu_{ij}$ can take, and their physical consequences. Since $\mu_{ij}$ is a 3d real matrix (when the body moves through $\mathbb{R}^3$), it has at least one real eigenvalue and a corresponding real eigenvector. Hence, there is at least one principle direction and $\mu_{ij}$ can fall into one of the following categories:
\begin{enumerate}
\item All three eigenvalues real and a full set of eigenvectors: In this case, $\mu_{ij}$ is diagonalisable and has three principal directions; when the body is accelerated along one of these, it gains an added mass equal to the corresponding eigenvalue.
\item One real and two non-real complex conjugate eigenvalues: In this case $\mu_{ij}$ can be brought to its Jordan form which can be taken as
	\beq
	\mu = 	
	\begin{bmatrix}
    	a 	& 	b 	& 	0 \\
   	-b 	& 	a 	& 	0 \\
	0 	& 	0 	& 	\lambda_3
	\end{bmatrix}
	\eeq
where the eigenvalues are $(\lambda_1 = a + ib, \lambda_2 = \la_1^*, \lambda_3)$ with $a,b$ and $\lambda_3$ real. Thus, when the body is accelerated along the third axis, the added force points along the same direction and the body experiences a definite added mass. On the other hand, if its acceleration has a component in the 1-2 plane, the added force would not point along its direction of acceleration.
\item All three eigenvalues real but only two or one independent eigenvectors: $\mu_{ij}$ is not diagonalizable, and hence the added force points in a different direction when accelerated along a direction other than an eigenvector. This can happen if either two or all eigenvalues coincide.
\end{enumerate}

It is interesting to note that in the gauge theory, cases 2 and 3 would correspond to a non-diagonalizable mass-squared matrix.

\chapter{Extending the HAM correspondence to fermions}
\label{a:ham-extension-fermions}

In this Appendix we consider fermion mass generation in various models of fermions coupled to scalars. We then attempt to develop a correspondence between fermion mass generation in each model and the added mass effect for a body accelerated through a fluid. In what follows, we ignore the gauge bosons. If included in a naive way, we would be treating the gauge bosons and fermions as two different bodies moving in two different fluids. Evidently, it would be preferable to treat them in a more unified manner.

\section{Single fermion coupled to a real scalar}

Consider a single Dirac fermion with mass $M$ coupled to a real scalar $\phi$. Let the scalar be subject to a Mexican-hat potential $V(\phi) = -m^2\phi^2 + \lambda \phi^4$ so that when $m>0$, $\phi$ has a non-zero vacuum expectation value $\langle\phi\rangle = \eta = \sqrt{\frac{m^2}{2\lambda}} \neq 0$. The Lagrangian for the system is
	\beq
	\mathcal{L} = \bar\psi (i \gamma^\mu \partial_\mu - M)\psi + \left( \partial_\mu \phi \right)^2 - V(\phi) -  g \phi \bar \psi \psi
	\eeq
where the coupling $g$ between the scalar field and the fermion is a real constant. The fermion gets a mass in addition to its Dirac mass $M$, due to its coupling with $\phi$. The new fermion mass is $M' = M + g\eta$.

This corresponds to a body of mass $M$ and extent $l$ accelerated through a 1D fluid along the circumference of a circle of radius $r$, which gains an added mass $(2\pi r - l) \sigma$ where $\sigma$ is the fluid density along the circumference of the circle and $l < 2\pi r$. The scalar vev $\eta$ corresponds to the fluid density $\sig$.

It is worth noting that the additional mass of the fermion could be negative, resulting in a decreased fermion mass due to coupling with a real scalar. In the added mass effect however, the added mass is always positive.

\section{Two fermions in two different generations}

\subsection{Massless fermions coupled to a real scalar}

Now consider a system of two massless fermions, where each fermion belongs to a different generation. Let the fermions be coupled to a real scalar subject to the same Mexican-hat potential as before. Let us also allow for the mixing of generations in the fermion-scalar coupling term, via a Yukawa-coupling matrix $g_{ab}$:
	\beqs
	\mathcal{L} &=& i \bar \psi_a\gamma^{\mu}\partial_\mu  \psi_a + \left( \partial_\mu \phi \right)^2 - V\left( \phi \right) - g_{ab} \phi \bar \psi_a \psi_b - \left(g^\dagger\right)_{ab}\phi \bar\psi_a \psi_b  \cr
	&=& i \bar \psi_a\gamma^{\mu}\partial_\mu  \psi_a + \left( \partial_\mu \phi \right)^2 - V\left( \phi \right) - \left(g + g^\dagger \right)_{ab}\phi \bar\psi_a \psi_b
	\eeqs
where $a = 1,2$ labels the generation of the fermion. The Yukawa coupling term is a mass term for the fermions. To find the fermion mass eigenstates and their corresponding masses, we must diagonalise the hermitian Yukawa coupling matrix $\tilde g = g + g^\dagger$. The fermions therefore become massive. The eigenvalues of $\tilde g$ are the masses of the fermions in units of $\langle \phi \rangle = \eta$: $M_a = \tilde g_a\eta$, where $\tilde g_a$, $a = 1,2$ are the eigenvalues of $\tilde g$.

This system corresponds to an elliptical disk accelerated through a fluid of density $\rho$ (which corresponds to the scalar vev $\eta$) distributed on a plane, whose added mass in the principal directions $a=1,2$ are $\eta \tilde g_a$. The eigenvalues $\tilde g_a$ of the Yukawa coupling matrix $\tilde g$ are inversely proportional to the principal axes of the elliptical disk.

\subsection{Massive fermions coupled to a real scalar}

Let us consider the two fermions in the previous section to be massive, i.e, having non-zero Dirac masses, the rest of the system being identical:
	\beqs
	\mathcal{L} &=& \sum_{a=1}^{2} \bar \psi_a\left( i\gamma^{\mu}\partial_\mu  - M_a \right) \psi_a + \left( \partial_\mu \phi \right)^2 - V\left( \phi \right) - g_{ab} \phi \bar \psi_a \psi_b - \left(g^\dagger\right)_{ab}\phi \bar\psi_a \psi_b  \cr
	&=& \sum_{a=1}^{2} \bar \psi_a\left( i\gamma^{\mu}\partial_\mu  - M_a \right) \psi_a + \left( \partial_\mu \phi \right)^2 - V\left( \phi \right) - \left(g + g^\dagger \right)_{ab}\phi \bar\psi_a \psi_b
	\eeqs
where $a = 1,2$ labels the generation of the fermion. If the vev of the scalar $\langle \phi \rangle = \eta \neq 0$, we obtain an additional mass term for the fermions. Adding the Dirac mass term $\sum_{a=1}^2 M_a \bar \psi_a \psi_a$ to the Yukawa mass term $\eta \tilde g_{ab}\bar \psi_a \psi_b$, where $\tilde g = \left(g + g^\dagger \right)$, we have reduced the present system to the previous system with two massless fermions, having the following augmented Yukawa mass term (we assume that $g$ and $g^\dagger$ are given such that $\tilde g$ turns out to be a positive matrix):
	\beq
	G_{ab}\bar\psi_a \psi_b = \left( M_a \del_{ab} + \eta \tilde g_{ab} \right) \bar \psi_a \psi_b.
	\eeq
The fermions therefore gain mass. The eigenvalues of $G_{ab}$ are the new masses of the fermions.

This corresponds to an elliptical disk accelerated through a fluid of density $\rho$ distributed on a plane, whose added mass in the principal directions $a=1,2$ are $G_a$, the corresponding eigenvalues of $G$. The eigenvalues $G_a$ are inversely proportional to the principal axes of the elliptical disk, and $\eta$ corresponds $\rho$.

An interesting feature of this fermion model that we have not captured on the fluid side of the correspondence is that the fermions have different bare masses - this would mean that the elliptical disk would have different inertial mass when accelerated in different directions.

\subsection[Massless fermions coupled to a complex scalar]{Massless\footnote{If the fermions were massive, their Dirac mass term $M \left( \bar\psi_R \psi_L + \bar\psi_L\psi_R \right)$ would force $\psi_L$ and $\psi_R$ to have the same charge and the scalar to be uncharged under the global U$(1)$.} fermions coupled to a  complex scalar with global U$(1)$ symmetry}

Now consider two massless fermions belonging to different generations coupled to a complex scalar with global U$(1)$ symmetry. Under a global U$(1)$ transformation,
	\beq
	\phi \to e^{iY\theta}\phi, \quad \psi_L \to e^{iY_L\theta}\psi_L \quad \text{and} \quad \psi_R \to e^{iY_R\theta}\psi_R
	\eeq
where $Y$ is the charge of the scalar field and $Y_L, Y_R$ are those of the left and right handed projections of the fermion states of either generation. A term of the form $\phi \bar \psi \psi = \phi (\bar \psi_R \psi_L + \bar \psi_L \psi_R )$ in the Lagrangian would not preserve the global U$(1)$ symmetry, unless the charges of $\phi, \psi_L$ and $\psi_R$ satisfy the relation $Y + Y_L - Y_R = 0$. Imposing this relation, we see that all terms in the following Lagrangian are invariant under global U$(1)$ transformations:
	\beq
	\mathcal{L} = \bar \psi_a i\gamma^{\mu}\partial_\mu \psi_a + \left( \partial_\mu \phi \right)^2 - V(|\phi|) - g_{ab}\phi \bar \psi_R^a \psi_L^b - \left(g^\dagger\right)_{ab}\phi^* \bar\psi_L^a\psi_R^b.
	\eeq
If the scalar has non-zero vacuum expectation value $\langle \phi \rangle = \eta \neq 0$, then the Yukawa coupling terms become mass terms for the fermions. 

In order to find the fermion mass eigenstates and their masses, we diagonalise $g$ and $g^\dagger$ through singular value decomposition as follows. $g$ can be expressed as the product of a positive matrix $P$ (hermitian with non-negative eigenvalues) and a unitary matrix $\Theta$: $g = P\Theta$. Let $U$ be the unitary matrix which diagonalises $P$: $U^\dagger P U = D$. Taking $V = \Theta^\dagger U$, we see that $U^\dagger g V = D$. Similarly, $V^\dagger g^\dagger U = D$. The diagonal values of $D$ are the singular values of both $g$ and $g^\dagger$: $g_a$, $a = 1, 2$ where $g_a$ are real and non-negative.

With $\langle \phi \rangle = \eta > 0$, the mass term for the fermions is:
	\beqs
	-\eta\left( \bar\psi_R g \psi_L + \bar \psi_L g^\dagger \psi_R\right) &=& -\eta \left[ \bar\psi_R U \left( U^\dagger g V \right) V^\dagger \psi_L + \bar\psi_L V \left( V^\dagger g^\dagger U \right) U^\dagger \psi_R \right]  \cr
	&=& -\eta\left[ \bar\psi_R' D \psi_L' + \bar\psi_L' D \psi_R' \right] = -\eta \sum_{a=1}^2 g_a\left( \bar\psi_R^{\prime a} \psi_L^{\prime a} + \bar\psi_L^{\prime a} \psi_R^{\prime a} \right) \cr
	&=& -\eta \sum_{a=1}^2 g_a \bar\psi^{\prime a} \psi^{\prime a}
	\eeqs
where $\psi_L' = V^\dagger \psi_L$ and $\psi_R' = U^\dagger \psi_R$. Thus the masses of the two fermions are $\eta g_a$, $a = 1,2$.

As before, this system corresponds to an elliptical disk accelerated through a fluid of density $\rho$ distributed on a plane, whose added mass in the principal directions $a=1,2$ are $\eta g_a$. The eigenvalues $g_a$ of the Yukawa coupling matrix $g_{ab}$ determine the dimensions of the axes of the elliptical disk, and $\eta$ corresponds to $\rho$.

A similar calculation could be done for one massless fermion in each of $N_g$ different generations. In this case, the Yukawa coupling matrices $g$ and $g^\dagger$ would be $N_g \times N_g$ complex matrices whose singular values $g_a$, $a = 1,\ldots,N_g$ would be the masses of the fermions in units of $\langle \phi \rangle = \eta$. This system would correspond to a rigid ellipsoid of codimension $1$ accelerated through fluid in $\mathbb{R}^{N_g}$. The singular values $g_a$ would be inversely proportional to the semi-axes of the ellipsoid.

\chapter{The Casimir added mass effect}
\label{a:casimir-added-mass}

There is an intriguing parallel between the added mass effect and the Casimir effect. Here, we summarize results from \cite{machado-neto}, which show that there exists an extra inertial force between two parallel moving plates, which turns out to be very similar to the added mass effect in fluids.

Consider a massless scalar field in $D$-dimensional space. The field obeys Dirichlet boundary conditions on the surface of two parallel hypersurfaces (of codimension $1$) separated by distance $a$ (when $D = 3$ we have a system of parallel plates separated by distance $a$). In other words, the scalar field vanishes on the two surfaces. If the plates are allowed to oscillate in a direction perpendicular to their surfaces, their equations of motion are
	\beq
	m_\alpha \ddot{\delta q_\alpha} (t) = -\sum_{\beta=1}^2 \mu_{\alpha \beta} \ddot{\delta q_\beta}(t) + F_\alpha
	\eeq
where $F_\alpha$ represents the sum of all forces acting on plate $\alpha$, excluding the inertial force itself. The coefficients $\mu_{\alpha \beta}$ go to zero as $a^{-D}$ when the plates are widely separated, hence $m_{\alpha}$ is the free-space mass of plate $\alpha$.

When the two plates oscillate together as a rigid `cavity', i.e, when $\delta q_1 = \delta q_2 \equiv \delta q$, we can add the equations of motion of the two plates to obtain
	\beq
	M \ddot{\delta q} = F_{\rm ext}
	\eeq
where $F_{\rm ext}$ is the sum of all external forces (the static Casimir forces and the rigidity-enforcing stresses cancel), and 
	\beq
	M = m_1 + m_2 + 2(\mu_{11} + \mu_{12}). \label{e:global-mass-correction}
	\eeq
Here, $(\mu_{11} + \mu_{12})$ cannot be taken as a mass correction to each individual plate, since $\mu_{12}$ is in fact a coupling between the two plates. $\Delta M = 2(\mu_{11} + \mu_{12})$ is thus a global mass correction introduced by the Casimir effect. We may write the mass correction as
	\beq
	\Delta M = (D + 1) \frac{E_0}{c^2}
	\eeq
where $E_0$ is the static Casimir energy of the system for a cell of area $L^{D-1}$ on either plate:
	\beq
	E_0 = \frac{\Gamma\left(\frac{D+1}{2}\right) \zeta(D+1)}{(4\pi)^{(D+1)/2}} \frac{\hbar L^{D-1}}{a^D}.
	\eeq
In view of recent measurements of the Casimir force, we can estimate the mass correction for parallel plates in $3$ dimensional space, assuming that (\ref{e:global-mass-correction}) holds in the electromagnetic case. The mass per unit area is
	\beq
	\frac{\Delta M}{L^2} = -\frac{\pi^2 \hbar}{180 \,c a^3}.
	\eeq
where $c$ is the speed of light. For a plate separation $a = 100 {\rm nm}$, $\Delta M/ L^2 = -2 \times 10^{-24}$g/cm$^2$, which is a very small correction.

\chapter[Vector field and phase portrait for steady flow]{Vector field and phase portrait for steady R-gas dynamics}
\label{a:vect-fld-phase-portrait}

With $x$ and $\rho$ viewed as time and position, steady, isentropic density profiles must satisfy the Newtonian ODE (\ref{e:steady-reg-gas-eqn-rho}):
    \beqs
    \beta_* \rho_{xx} &=& - V'(\rho) + \frac{(\g + 1) \beta_*}{2} \frac{\rho_x^2}{\rho} \quad\quad \text{for} \quad\quad \rho > 0 \quad\quad \text{where} \cr
    V(\rho) &=&  \g F^{\rm p} \rho - \frac{ (\g -1)F^{\rm u} \rho^2}{2}  - \frac{(\g + 1) ( F^{\rm m})^2}{2}\log \rho.
    \label{e:steady-newtons-law}
    \eeqs 
Ignoring the velocity-dependent force, two cases arise: (a) for $ F^{\rm u} < 0$, $V$ has only bound states, corresponding to periodic $\rho$ and (b) for $ F^{\rm u} > 0$, there are periodic waves and a caviton, provided $V$ has a local minimum [this happens when $F^{\rm p}$ and $\D =  (F^{\rm p})^2\g^2 - 2(\g^2 - 1) F^{\rm u}  (F^{\rm m})^2$ are both positive]. The velocity-dependent force $\propto \rho_x^2/\rho$ is reminiscent of the drag force $\propto -|\rho_x| \rho_x$ on a body at high Reynolds number. However, while air drag tends to decrease the speed, this force is positive and tends to increase the velocity $\rho_x$. Moreover, though the naive `energy' $\beta_*\rho_x^2/2 + V$ is {\it not} conserved, Eqn.~(\ref{e:steady-newtons-law}) is non-dissipative and in fact Hamiltonian (see Appendix \ref{a:hamil-lagr-for-steady-sol}). Since the velocity-dependent force tends to push the particle outwards, it cannot convert a scattering state into a bound state. We also find that the qualitative nature of the phase portraits is not significantly altered by this force.

\begin{figure*}	
	\begin{center}
			\begin{subfigure}[t]{6cm}
		\includegraphics[width=6cm]{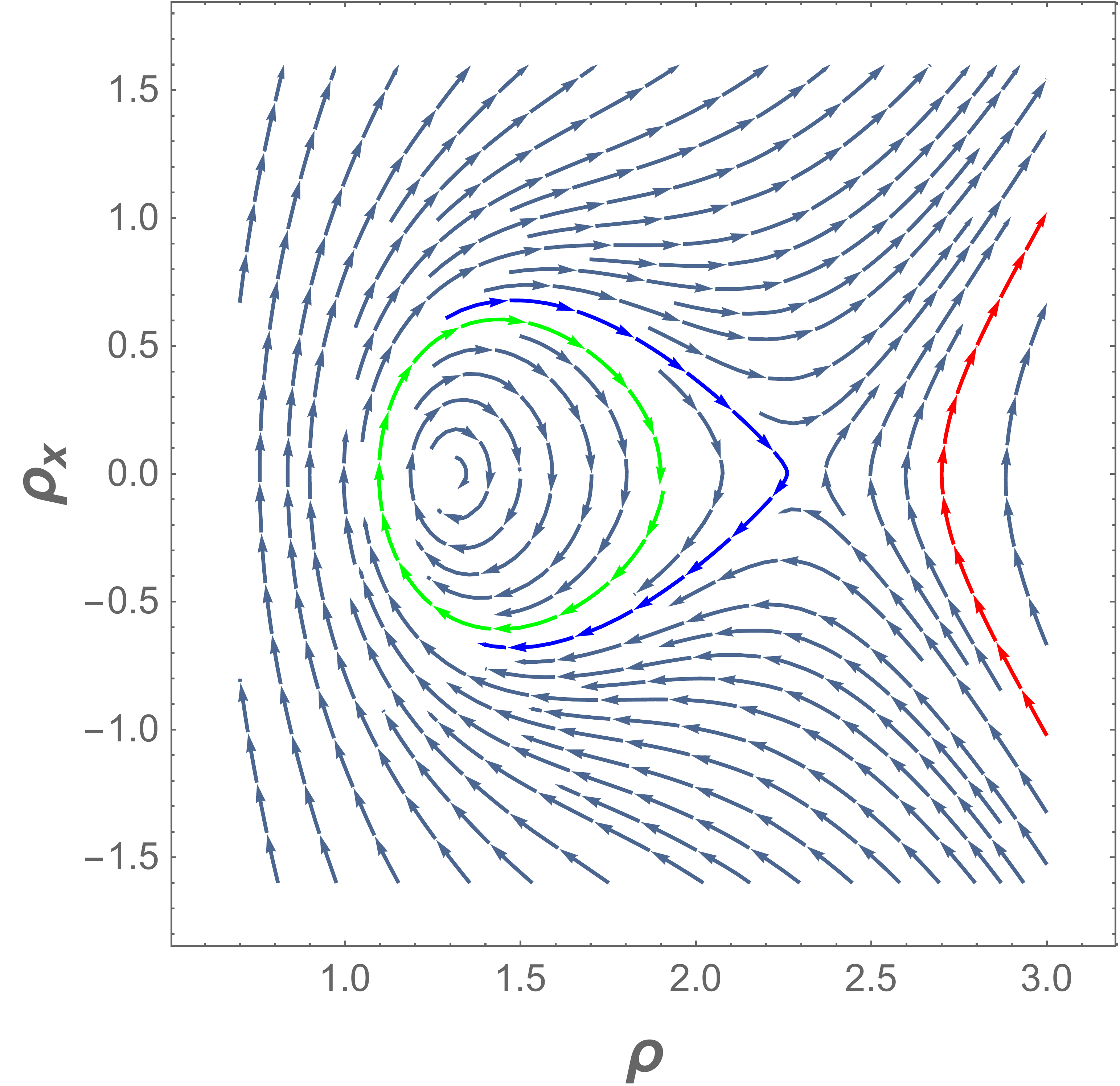}
		\caption{}
		\label{f:phase-por-caviton}	
	\end{subfigure}
	\quad
	\begin{subfigure}[t]{6cm}
		\includegraphics[width=6cm]{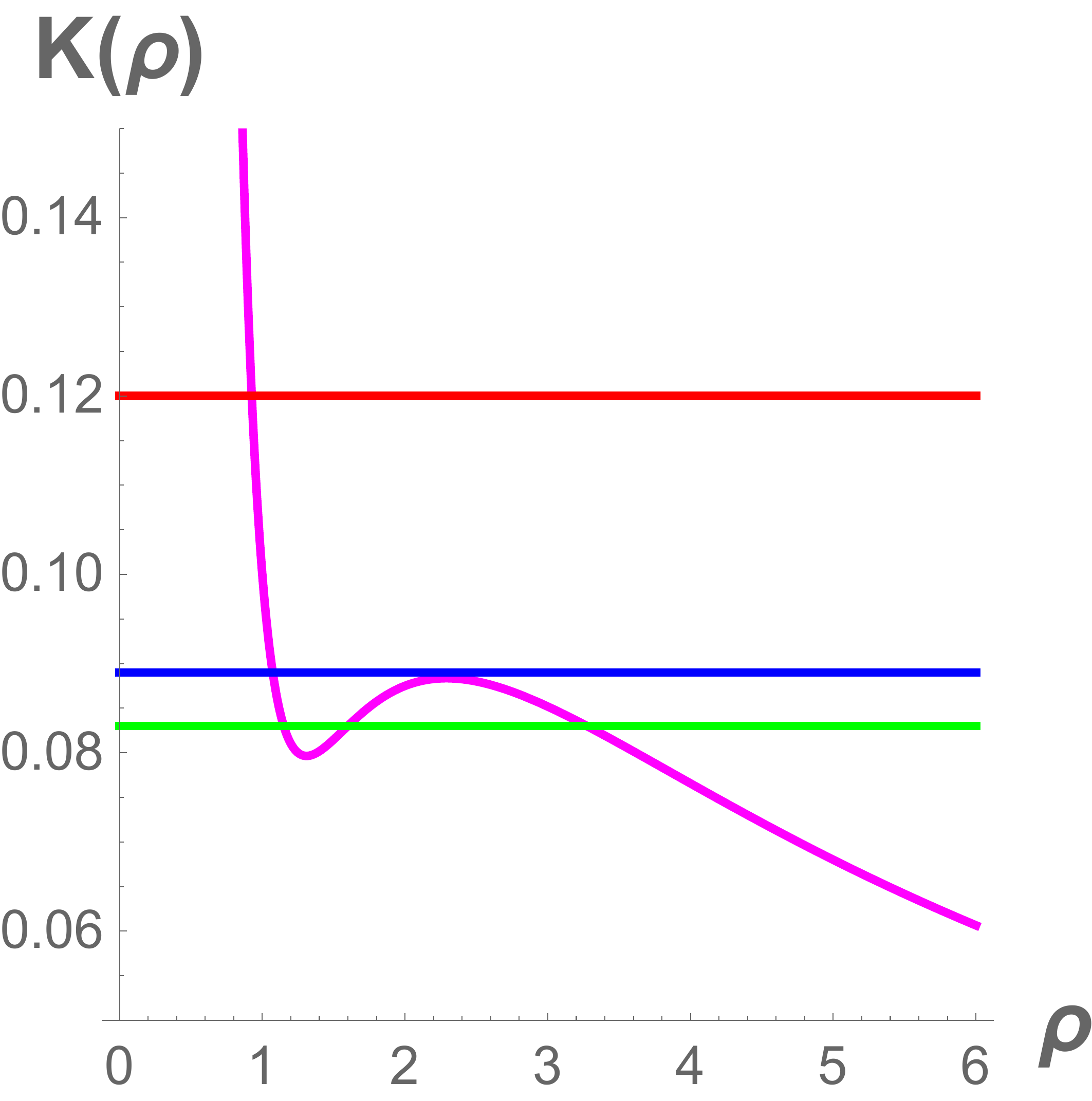}
		\caption{}
	\end{subfigure} 
	\quad
	\begin{subfigure}[t]{6cm}
		\includegraphics[width=6cm]{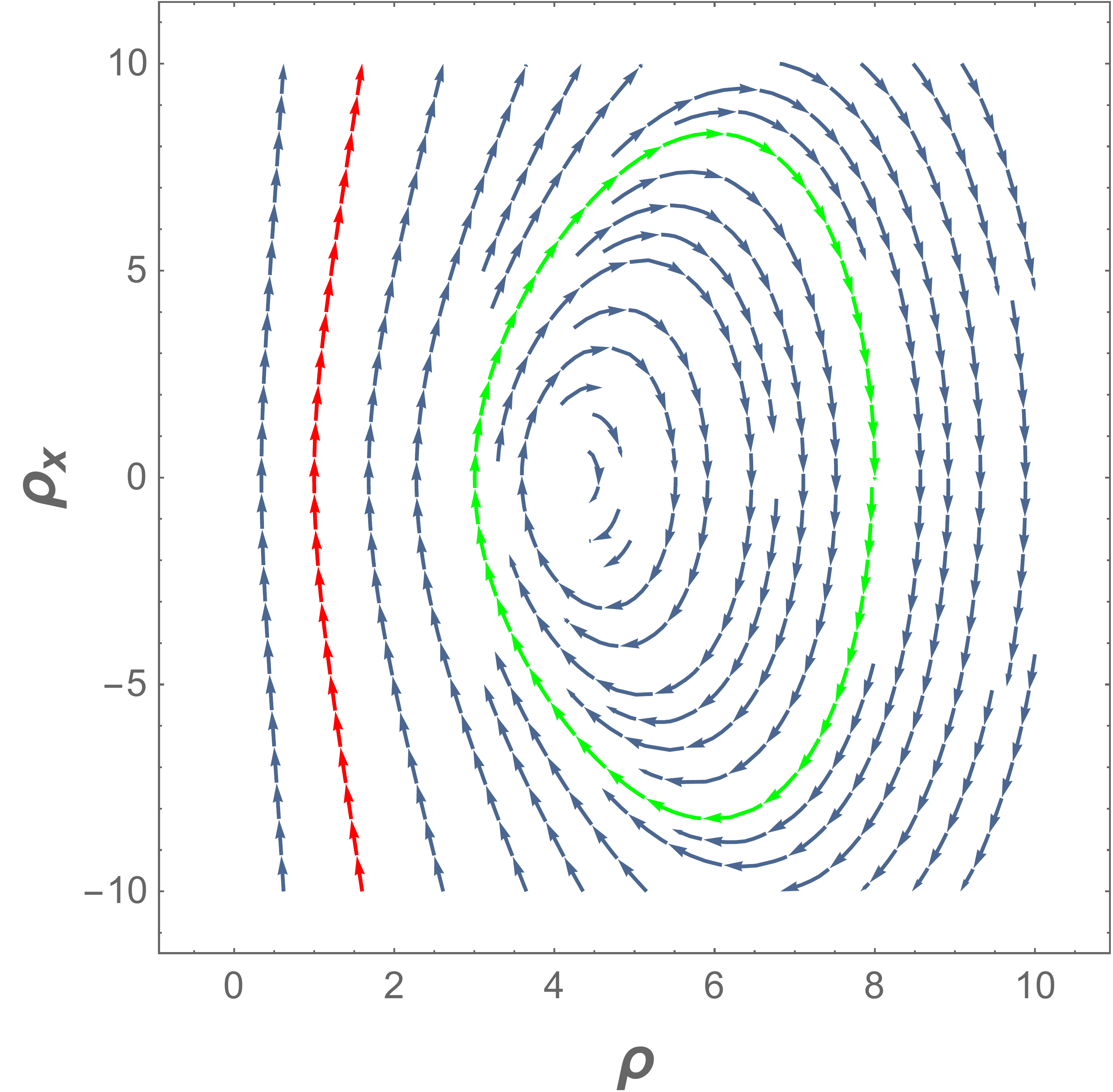}
		\caption{}
	\end{subfigure} 
	\quad
	\begin{subfigure}[t]{6cm}
		\includegraphics[width=6cm]{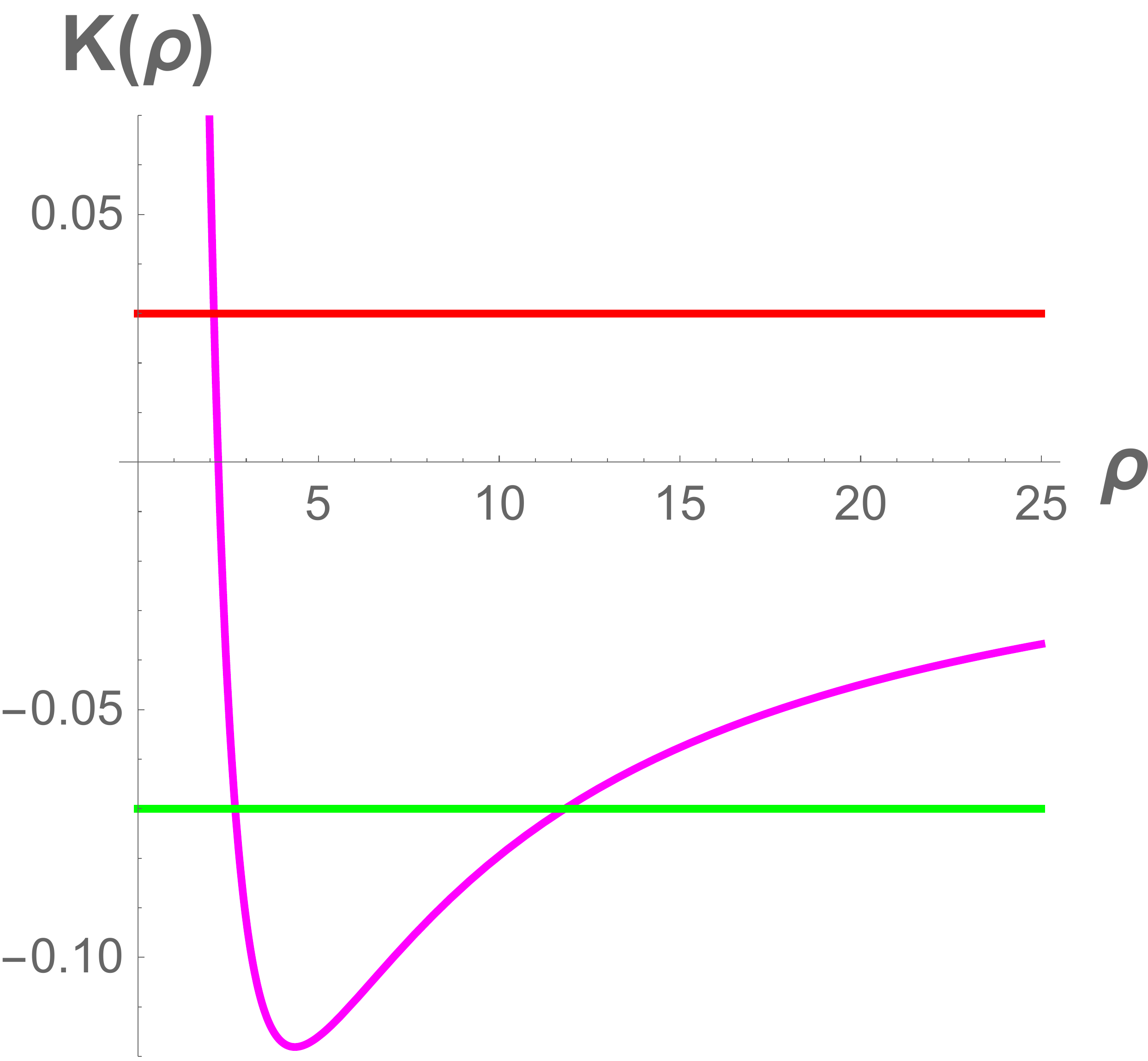}
		\caption{}
	\end{subfigure}
	\end{center}
	\caption{(a) and (b): The vector field $W$ on the $\rho$-$\rho_x$ phase plane for $\g =2$ and $\beta_* = 0.1$ and the entropy constant $K(\rho_\pt)$ (\ref{e:K-at-rhostar}) (which labels trajectories) for $F^{\rm m} = 1$, $F^{\rm p} = 0.9$ and $F^{\rm u} = 0.5$. As $K$ is decreased from infinity, we encounter unbounded solutions followed by a bounded caviton separatrix emanating from the X-point. The caviton encircles periodic orbits around the O-point. The X and O points are the only constant solutions. (c) and (d): The vector field $W$ and corresponding $K$ for $F^{\rm m} = 1$, $F^{\rm p} = - 2$ and $F^{\rm u} = -1$. Scattering states for $K > 0$ are followed by periodic orbits around the O-point with $K < 0$ and an infinite caviton separatrix at $K = 0$. Solutions with $K <0$ have negative pressure.}
	\label{f:vec-fld}
\end{figure*}


Introducing $\eta = \rho_x$, (\ref{e:steady-newtons-law}) defines a vector field $W$ on the right half $\rho$-$\rho_x$ phase plane:
    \beq
    \colvec{2}{\rho}{\eta}_x = W \equiv \colvec{2}{\eta}{- \frac{V'(\rho)}{\beta_*}+ \frac{(\g + 1)}{2} \frac{\eta^2}{\rho}}.
    \label{e:vect-field-W}
    \eeq
Though $\eta^2/\rho$ is singular along the $\rho = 0$ axis, it is `shielded' by the repulsive logarithmic potential in $V$. Bounded integral curves of $W$ correspond to bounded steady densities. Fixed points (FPs) of $W$ correspond to constant density solutions. They are located at $(\rho_{p,m}, 0)$ where $\rho_{p,m}$ are the extrema of $V$:
    \beq
    \rho_{p,m} = \frac{ \g F^{\rm p} \pm \sqrt{\D}}{2 (\g-1)  F^{\rm u}}.
    \label{e:rho-pm-fixedpts-steady}
    \eeq 
    
\begin{table*}
	\small
    \parbox{.5\linewidth}{
    \centering
    \caption*{\bf (a) Non-aerostatic steady solutions \vspace{-.2cm}}
    \begin{tabular}{|c|c|c|c|}
    \hline
        $ F^{\rm p}$ &   $F^{\rm u}$ & Fixed point & Bounded solutions  \\
        \hline
        $+$ & $+$ & O and X point & periodic, caviton \\
        \hline
        $+$ & $-$ & O point & periodic  $(K < 0)$\\ 
        \hline
        $-$ & $-$  & O point & periodic  $(K < 0)$ \\
        \hline
        $-$ & $+$  & none & none \\
        \hline
    \end{tabular}
    }
    \hfill
    \parbox{.5\linewidth}{
    \centering
    \caption*{\bf (b) Aerostatic steady solutions \vspace{-.3cm}}
    \begin{tabular}{|c|c|c|c|}
        \hline
        $ F^{\rm p}$ &  $F^{\rm u}$ & Fixed point & Bounded solutions  \\
        \hline
        $+$ & $+$ & X point & periodic, caviton \\
        \hline
        $-$ & $-$  & O point & periodic $(K < 0)$\\
        \hline
        $+$ & $-$ & none & periodic $(K < 0)$\\ 
        \hline
        $-$ & $+$  & none & none \\
        \hline
        0 & + & none & none \\
        \hline
        0 & $-$ & none & elevatons $(K < 0)$ \\
     \hline
     0 & 0 & none & none \\
     \hline
    \end{tabular}
    }
    \caption{Nature of fixed points and bounded solutions on $\rho$-$\rho_x$ half-plane. (a) General non-aerostatic case, when $ F^{\rm m}, F^{\rm e} \ne 0$ and $\D > 0$. (b) Aerostatic limit $(u \equiv 0,  F^{\rm m} =  F^{\rm e} = 0)$. $K < 0$ corresponds to solutions with negative pressure.}
	\label{t:nonaero-aero-steady}
	\normalsize
\end{table*}

There may be two, one or no FPs in the physical region $\rho > 0$. We are interested in the cases where there is at least one FP in the physical region, as otherwise $\rho$ is unbounded. This requires $\D > 0$. Assuming this is the case and also assuming that the flow is not aerostatic $( F^{\rm m} \neq 0$ or $u \not \equiv 0)$, we find that there are two physical FPs if $ F^{\rm p}$ and $F^{\rm u}$ are both positive, one fixed point if $F^{\rm u} < 0$ and none otherwise. The character of these FPs may be found by linearizing $W$ around them. Writing $\rho = \rho_{p,m} + \del\rho$ and $\eta = 0 + \del\eta$, we get
    \beq
	\DD{}{x} \colvec{2}{\del \rho}{\del \eta} = A \colvec{2}{\del \rho}{\del \eta} \quad \text{where} \quad A = \colvec{2}{0 & 1}{- V''(\rho_{p,m})/\beta_* & 0}. 
    \eeq
The eigenvalues of the coefficient matrix $A$ are
    \beq
	\la = \pm \sqrt{\frac{- V''(\rho_{p,m})}{\beta_*}} \quad \text{where} \quad
	V''(\rho_{p,m}) = \mp \frac{\sqrt{\D}}{\rho_{p,m}}.
	\label{e:Vprpr-formula}
    \eeq
Thus, the physical FPs of $W$ must either be X or O points (saddles or centers in the linear approximation) according as $V'' < 0$ (real eigenvalues) or $V'' > 0$ (imaginary eigenvalues). The Hartman-Grobman Theorem guarantees that the linear saddles remain saddles even upon including nonlinearities. Moreover, the linear O-point at $(\rho_m,0)$ is always a true O-point since we may verify that $(\rho_m,0)$ is a minimum of the conserved quantity $K$ (\ref{e:rhox-vs-rho-diff-eqn}). Thus, as summarized in Table \ref{t:nonaero-aero-steady}(a), there are two types of phase portraits leading to bounded solutions $\rho(x)$: (i) if $ F^{\rm p}$ and $ F^{\rm u}$ are both positive, then $W$ has an O-point at $(\rho_m, 0)$ and an X-point at $(\rho_p, 0)$ to its right and (ii) if $ F^{\rm u}<0$, $W$ has only one physical fixed point, an O-point at $(\rho_m, 0)$. As shown in Fig.~\ref{f:vec-fld} (a,b), in case (i) we have two types of bounded solutions: periodic waves corresponding to closed curves around the O-point $(\rho_m,0)$ and a solitary wave corresponding to the separatrix orbit that begins and ends at $(\rho_p,0)$ and encircles the O-point. Since $\rho_p > \rho_m$, a solitary wave must be a caviton. In case (ii) the only bounded solutions $\rho(x)$ are periodic waves corresponding to closed curves encircling the O-point $(\rho_m,0)$ as shown in Fig.~\ref{f:vec-fld} (c,d).  


{\fl \bf Isentropic aerostatic steady solutions:} In the aerostatic limit $(u \equiv 0)$, both the fluxes $F^{\rm e}$ and $F^{\rm m}$ vanish, though their ratio $F^{\rm e}/F^{\rm m} = F^{\rm u}$ is finite. Eqn. (\ref{e:steady-reg-gas-eqn-rho}) for steady solutions becomes
    \beq
	\beta_* \rho_{xx} = - V'(\rho) + \frac{(\g + 1) \beta_*}{2} \frac{\rho_x^2}{\rho} \quad \text{where} 
	\quad V(\rho) = \g  F^{\rm p}\rho - (\g - 1) \frac{ F^{\rm u}}{2} \rho^2.
    \eeq
In this limit, the small-$\rho$ logarithmic barrier in $V$ (\ref{e:steady-newtons-law}) is absent, and the singularity along $\rho = 0$ becomes `naked'. One of the FPs in Eqn.~(\ref{e:rho-pm-fixedpts-steady}) tends to $(0,0)$, while the other one tends to $(\g  F^{\rm p}/(\g - 1) F^{\rm u}, 0)$. Table \ref{t:nonaero-aero-steady}(b) summarizes the nature of physical fixed points and bounded solutions for various possible signs of $ F^{\rm p}$ and $ F^{\rm u}$. Interestingly, for $ F^{\rm p} = 0$ and $ F^{\rm u} < 0$, there is a family of solitary waves of {\it elevation}, though with negative pressure.

\chapter{Canonical formalism for steady R-gas dynamics}
\label{a:hamil-lagr-for-steady-sol}

The equation for steady solutions (\ref{e:steady-reg-gas-eqn-rho}) describes a mechanical system with 1 degree of freedom and conserved quantity $K(\rho,\rho_x)$ (\ref{e:rhox-vs-rho-diff-eqn}). Here we give a canonical formulation for (\ref{e:steady-reg-gas-eqn-rho}) by taking $K$ to be the Hamiltonian. We seek a suitable PB $\{\rho, \rho_x \}$ so that Hamilton's equations $\rho_x = \{\rho, K \}$ and $\rho_{xx} = \{\rho_x, K \} $ reproduce (\ref{e:steady-reg-gas-eqn-rho}). The former gives
	\beq
	\{\rho, K\} = \left\{ \rho, \half  \frac{\beta_* \rho_x^2}{\rho^{\g +1}} \right\} = \rho_x  \quad \text{or} \quad  \frac{\beta_* \rho_x}{\rho^{\g + 1}} \{\rho, \rho_x \} = \rho_x
	\quad\imply \quad  \{\rho, \rho_x \} = \frac{\rho^{\g + 1}}{\beta_*}.
	\eeq
Using this PB, $\rho_{xx} = \{\rho_x , K \}$ reproduces  (\ref{e:steady-reg-gas-eqn-rho}). This PB is not canonical, but if we define $\varpi = \beta_* \rho_x/\rho^{\g + 1}$ then $\{ \rho , \varpi \} = 1$ so that $\varpi$ is the momentum conjugate to $\rho$. The corresponding Hamiltonian is
	\beq
	K(\rho,\varpi) = \half  \frac{\rho^{\g + 1}}{\beta_*} \varpi^2 + U(\rho),
	\eeq
with $U$ as in (\ref{e:rhox-vs-rho-diff-eqn}). The `mass' factor in the kinetic term is `position' $(\rho)$ dependent. In terms of the contravariant `mass metric' $m^{-1}(\rho) = {\rho^{\g + 1}}/{\beta_*}$, $K = (1/2) m^{-1} \varpi^2 + U$. The corresponding Lagrangian is
	\beq
	L = \text{ext}_\varpi ( \varpi \rho_x  - K) =  \half \frac{\beta_*}{\rho^{\g + 1}} \rho_x^2 - U  = \half m(\rho) \rho_x^2 - U.
	\eeq
The Euler-Lagrange equation reduces to (\ref{e:steady-reg-gas-eqn-rho}). Thus, we have Hamiltonian and Lagrangian formulations for both the full R-gas dynamic field equations and their reduction to the space of steady solutions.

\chapter[Parabolic embedding and Lagrange-Jacobi identity]{Parabolic embedding and Lagrange-Jacobi identity for steady flow}
\label{a:parabolic-embed-LJ-id}

For $\g \ne 2$ the quadrature in (\ref{e:quadrature-steady-rho}) cannot be done using elliptic functions, but could be done numerically. An alternative approach is to take a linear combination of the equations in (\ref{e:rhoxx-steady-eqn-two-versions}) to obtain a form of the steady equation for $\rho$ without the velocity dependent term but with a generally non-integral power of $\rho$:
	\beq
	\beta_* \rho_{xx} = (\g + 1) K \rho^\g - 2 F^{\rm u} \rho +  F^{\rm p}.
	\label{e:steady-eqn-rhoxx-no-vel-dep-term}
	\eeq
If we introduce a pseudo-time $\tau$, then steady solutions can be obtained via a parabolic embedding in a nonlinear heat equation with a source:
	\beq
	\rho_{\tau}-\beta_{*}\rho_{xx}
 	= - (\g + 1) K \rho^\g + 2 F^{\rm u} \rho - F^{\rm p}.
	\eeq
By prescribing suitable BCs and starting with an arbitrary initial condition, the solution of this PDE should relax to the stable solutions of (\ref{e:steady-eqn-rhoxx-no-vel-dep-term}).

(\ref{e:steady-eqn-rhoxx-no-vel-dep-term}) may also be used to derive additional virial/Lagrange-Jacobi-type identities by multiplying by $\rho$ using (\ref{e:rhox-vs-rho-diff-eqn}) and repeating the process. The first two such identities are 
	\beqs
	\frac{\beta_*}{2} (\rho^2)_{xx} &=& (\g+3) K \rho^{\g+1} - 4 F^{\rm u} \rho^2 + 3 F^{\rm p} \rho - (F^{\rm m})^2 \quad\text{and}\cr
	\frac{\beta_*}{3} (\rho^3)_{xx} &=& (\g+5) K \rho^{\g+2} - 6 F^{\rm u} \rho^3 + 5 F^{\rm p} \rho^2 - 2 (F^{\rm m})^2 \rho.
	\label{e:lagrange-jacobi-idnetities}
	\eeqs 
Integrating (\ref{e:steady-eqn-rhoxx-no-vel-dep-term}) and (\ref{e:lagrange-jacobi-idnetities}) with periodic BCs we get a hierarchy of integral invariants for steady soultions 
	\beqs
	\int_{-L}^L \left[ (\g + 1) K \rho^\g - 2 F^{\rm u} \rho +  F^{\rm p} \right] dx
	&=& \int_{-L}^L \left[ (\g+3) K \rho^{\g+1} - 4 F^{\rm u} \rho^2 + 3 F^{\rm p} \rho - (F^{\rm m})^2 \right] dx \cr
	&=& \int_{-L}^L \left[ (\g+5) K \rho^{\g+2} - 6 F^{\rm u} \rho^3 + 5 F^{\rm p} \rho^2 - 2 (F^{\rm m})^2 \rho \right] dx\cr
	&=& 0 \quad \text{etc.}  
	\eeqs
These integral identities can provide valuable checks on any numerics used to obtain steady solutions.

\chapter{Semi-implicit spectral scheme for time evolution}
\label{a:numerical-scheme}

Here, we describe the scheme used to solve the IVP for the 1d $\g = 2$ isentropic R-gas dynamic equations (\ref{e:cont-vel-baro-fourier}). To include the effects of the nonlinear terms in (\ref{e:cont-vel-baro-fourier}), we discretize time $\hat t = j \D$, $j = 0,1,2, \ldots$, denote the Fourier modes $\rho_n (j \D) = \rho_n^j$ etc. and use a centered difference scheme for the linear part:
	\beqs
	\rho_n^{j+1} &=& \rho_n^j - \frac{i n \D}{2} \left( \thickbar u (\rho_n^{j} + \rho_n^{j+1}) + u_n^j + u_n^{j+1} \right) - \D \del (\calF^{\rm m})_n^{j} \cr
	u_n^{j+1} &=& u_n^j - \frac{i n \D}{2} \left( \thickbar u (u_n^j + u_n^{j+1}) + (1 + \eps^2 n^2) (\rho_n^j + \rho_n^{j+1}) \right)
	- \D \del (\calF^{\rm u})_n^{j}.
	\eeqs
For simplicity, the nonlinear terms are treated explicitly; treating the linear terms explicitly leads to numerical instabilities. In matrix form, the equations read
	\beqs
	A \colvec{2}{\rho_n^{j+1}}{u_n^{j+1}}
	&=& B \colvec{2}{\rho_n^{j}}{u_n^{j}} 
	- \D \del \colvec{2}{(\calF^{\rm m})_n^{j}}{(\calF^{\rm u})_n^{j}} \cr 
	\text{where}\quad
	A &=& I + \frac{i n \D}{2} \colvec{2}{ \thickbar u & 1}{p_n^2 & \thickbar u} \quad \text{and} \quad
	B =  I - \frac{i n \D}{2} \colvec{2}{ \thickbar u & 1}{p_n^2 & \thickbar u}.
	\eeqs
with $p_n^2 = (1 + \eps^2 n^2)$. $B$ is related to $A$ via conjugation or by $\D \to - \D$. Thus, the variables at the $j+1^{\rm st}$ time step are
	\beqs
	\colvec{2}{\rho_n^{j+1}}{u_n^{j+1}}
	&=& U \colvec{2}{\rho_n^{j}}{u_n^{j}} 
	- \D \del A^{-1} \colvec{2}{(\calF^{\rm u})_n^{j}}{(\calF^{\rm u})_n^{j}} \quad \text{where}  \cr
	U &=& A^{-1} B = \ov{\det A} \left[ I + n \D \colvec{2}{ \frac{\D}{4} n (\thickbar u^2 - p_n^2 ) & -i}{-i p_n^2 &  \frac{\D}{4} n(\thickbar u^2 - p_n^2 )} \right], 
	\cr
	(\det A) \, A^{-1} &=& I +  \frac{i n \D}{2} \colvec{2}{\thickbar u & - 1}{- p_n^2 &\thickbar u } \quad \text{and} \quad
	\det A = \left(1 + \frac{i n \thickbar u \D}{2} \right)^2 + \frac{n^2 \D^2 p_n^2}{4}.\quad
	\eeqs
$U$ is unitary with respect to the inner product $\left\langle (\rho, u), (\tl \rho, \tl u)  \right\rangle = p_n^2 \rho^* \tl \rho + u^* \tl u$, which ensures that the linear evolution conserves $H_1$ of (\ref{e:sound-waves-cons-qtys}). An advantage of this scheme is that conservation of $\int \rho \, dx $ and $\int u \, dx$ are automatically satisfied to round-off accuracy. We also find that $Q_r$ and $Q_i$ (\ref{e:Qr-Qi}) are quite accurately conserved in our numerical evolution for the ICs in \S \ref{s:numerical-results}. Moreover, since $\calF^{\rm m}$ (\ref{e:cont-linear-nonlinear-split}) and $\calF^{\rm u}$ (\ref{e:vel-linear-nonlinear-split}) are divergences, their Fourier coefficients can be calculated by integration by parts without any differentiation.

\end{document}